\documentclass[10pt,draftcls,onecolumn]{IEEEtran}
\usepackage{cite}
\usepackage[dvipsnames]{xcolor}
\usepackage{graphicx}
\usepackage{amsmath,amssymb,amsthm,bm}
\usepackage{dsfont}
\usepackage{mathrsfs}
\usepackage{mathtools}
\usepackage{longtable}
\usepackage{hyperref}
\usepackage{stmaryrd}
\usepackage{enumerate}
\usepackage[caption=false,font=footnotesize]{subfig}
\usepackage[margin=2.2cm]{geometry}
\usepackage{tikz,pgfplots}
\pgfplotsset{compat=newest}
\usetikzlibrary{graphs,matrix,positioning,calc,fit,backgrounds,chains}
\usepackage[breakable]{tcolorbox}
\tcbuselibrary{breakable}

\newcommand{\proofbox}[2]{\begin{tcolorbox}[
	breakable,
	sharpish corners,
	opacitybacktitle=0.5,
	size=small,
	left=0mm,
	lefttitle=0cm,
	title=#1]{#2}\end{tcolorbox}}
\newcommand{\lemmabox}[1]{\begin{tcolorbox}[
	colback=blue!10!white,
	breakable,
	sharpish corners,
	opacitybacktitle=0.5,
	size=small,
	left=0mm,
	lefttitle=0cm,
	]{#1}\end{tcolorbox}}
\newcommand{\theorembox}[1]{\begin{tcolorbox}[
	colback=red!10!white,
	breakable,
	sharpish corners,
	opacitybacktitle=0.5,
	size=small,
	left=0mm,
	lefttitle=0cm,
	]{#1}\end{tcolorbox}}
\newcommand{\definitionbox}[1]{\begin{tcolorbox}[
	colback=green!10!white,
	breakable,
	sharpish corners,
	opacitybacktitle=0.5,
	size=small,
	left=0mm,
	lefttitle=0cm,
	]{#1}\end{tcolorbox}}

\DeclareMathOperator{\diag}{diag}%
\DeclareMathOperator{\Span}{span}%
\DeclareMathOperator{\rank}{rank}%
\DeclareMathOperator{\vol}{vol}%
\DeclareMathOperator{\vectorize}{vec}%
\DeclareMathOperator{\mymod}{mod}%

\DeclareMathOperator{\mynull}{\mathbf{0}}%

\newcommand{\LMAC}{\mathrm{LMAC}}
\newcommand{\MAC}{\mathrm{MAC}}
\newcommand{\Frob}{\mathcal{F}}

\newcommand{\q}{\mathsf{q}}

\newcommand{\Ac}{\mathcal{A}}
\newcommand{\Bc}{\mathcal{B}}

\newcommand{\Dc}{\mathcal{D}}

\newcommand{\Gc}{\mathcal{G}}
\newcommand{\Hc}{\mathcal{H}}
\newcommand{\Ic}{\mathcal{I}}

\newcommand{\Lc}{\mathcal{L}}

\newcommand{\Nc}{\mathcal{N}}

\newcommand{\Pc}{\mathcal{P}}
\newcommand{\Qc}{\mathcal{Q}}
\newcommand{\Rc}{\mathcal{R}}
\newcommand{\Sc}{\mathcal{S}}
\newcommand{\Tc}{\mathcal{T}}

\newcommand{\Vc}{\mathcal{V}}
\newcommand{\Wc}{\mathcal{W}}
\newcommand{\Xc}{\mathcal{X}}
\newcommand{\Yc}{\mathcal{Y}}

\newcommand{\hb}{\mathbf{h}}

\newcommand{\av}{\bm{a}}

\newcommand{\dv}{\bm{d}}

\newcommand{\iv}{\bm{i}}

\newcommand{\kv}{\bm{k}}

\newcommand{\uv}{\bm{u}}
\newcommand{\vv}{\bm{v}}
\newcommand{\wv}{\bm{w}}
\newcommand{\xv}{\bm{x}}
\newcommand{\yv}{\bm{y}}
\newcommand{\zv}{\bm{z}}

\newcommand{\Uv}{\bm{U}}

\newcommand{\Xv}{\bm{X}}

\newcommand{\Deltav}{\bm{\Delta}}

\newcommand{\xiv}{\bm{\xi}}

\newcommand{\Bf}{\mathsf{B}}
\newcommand{\Cf}{\mathsf{C}}

\newcommand{\Qf}{\mathsf{Q}}

\DeclareMathOperator\E{\sf E}
\let\P\relax
\DeclareMathOperator\P{\sf P}

\DeclarePairedDelimiter{\floor}{\lfloor}{\rfloor}
\DeclarePairedDelimiter{\bigfloor}{\big\lfloor}{\big\rfloor}

\DeclareMathOperator*{\argmin}{\arg\min}

\newtheorem{theorem}{Theorem}
\newtheorem{lemma}{Lemma}

\newtheorem{corollary}{Corollary}

\newtheorem{example}{Example}
\newtheorem{remark}{Remark}
\newtheorem{definition}{Definition}

\newcommand{\As}{{\mathbf{A}}}
\newcommand{\Bs}{{\mathbf{B}}}
\newcommand{\Cs}{{\mathbf{C}}}

\newcommand{\Fs}{{\mathbf{F}}}
\newcommand{\Gs}{{\mathbf{G}}}
\newcommand{\Hs}{{\mathbf{H}}}
\newcommand{\Is}{{\mathbf{I}}}

\newcommand{\Ks}{{\mathbf{K}}}

\newcommand{\Ps}{{\mathbf{P}}}
\newcommand{\Qs}{{\mathbf{Q}}}
\newcommand{\Rs}{{\mathbf{R}}}
\newcommand{\Ss}{{\mathbf{S}}}
\newcommand{\Ts}{{\mathbf{T}}}

\newcommand{\Xs}{{\mathbf{X}}}

\newcommand{\Zs}{{\mathbf{Z}}}
\newcommand{\Sigmas}{{\mathbf{\Sigma}}}
\newcommand{\as}{\mathbf{a}}
\newcommand{\bs}{\mathbf{b}}
\newcommand{\cs}{\mathbf{c}}
\newcommand{\ds}{\mathbf{d}}

\newcommand{\fs}{\mathbf{f}}
\newcommand{\gs}{\mathbf{g}}
\newcommand{\hs}{\mathbf{h}}
\newcommand{\is}{\mathbf{i}}
\newcommand{\js}{\mathbf{j}}

\newcommand{\qs}{\mathbf{q}}

\newcommand{\ts}{\mathbf{t}}
\newcommand{\us}{\mathbf{u}}
\newcommand{\vs}{\mathbf{v}}
\newcommand{\ws}{\mathbf{w}}
\newcommand{\xs}{\mathbf{x}}

\newcommand{\Omegas}{\bm{\Omega}}

\newcommand{\omegas}{\bm{\omega}}
\newcommand{\betas}{\bm{\beta}}

\newcommand{\T}{\mathsf{T}}

\newcommand{\intd}{{\,\operatorname{d}}}

\begin{document}

\title{A Unified Discretization Approach to Compute--\\Forward: From Discrete to Continuous Inputs}

\author{Adriano Pastore, Sung Hoon Lim, Chen Feng, Bobak Nazer, Michael Gastpar
\thanks{This paper was presented in part at the 2021 IEEE International Symposium on Information Theory.}
}

\maketitle

\begin{abstract}
Compute--forward is a coding technique that enables receiver(s) in a network to directly decode one or more linear combinations of the transmitted codewords. Initial efforts focused on Gaussian channels and derived achievable rate regions via nested lattice codes and single-user (lattice) decoding as well as sequential (lattice) decoding. Recently, these results have been generalized to discrete memoryless channels via nested linear codes and joint typicality coding, culminating in a simultaneous-decoding rate region for recovering one or more linear combinations from $K$ users. Using a discretization approach, this paper translates this result into a simultaneous-decoding rate region for a wide class of continuous memoryless channels, including the important special case of Gaussian channels. Additionally, this paper derives a single, unified expression for both discrete and continuous rate regions via an algebraic generalization of R\'enyi's information dimension.


\end{abstract}

\begin{IEEEkeywords}
Linear codes, compute--forward, physical-layer network coding, multiple-access channel, information dimension, entropy.
\end{IEEEkeywords}

{
\section{Introduction}
Consider a network information theory problem where one or more transmitters wish to communicate with one or more receivers, and our goal is to establish an achievable rate region for a wide class of memoryless sources and channels, both discrete and continuous. It is by now well-understood that there are many interesting network configurations where random codes with algebraic structure can be used to establish achievable rates that lie outside the rate regions available to i.i.d.~random codes. This phenomenon was discovered by K\"orner and Marton in the context of a two-help source coding problem~\cite{KoMa79} where linear codes can be employed to compress the \textit{sum} of dependent binary sources in a distributed fashion. Recent efforts have generalized this technique and applied it to a wide range of scenarios including multiple-access~\cite{Philosof--Zamir2009,Philosof--Zamir--Erez--Khisti2011,Wang2012,PaPr17,KhKoEr17,HePr20}, broadcast~\cite{PaPr18}, and interference channels~\cite{Bresler--Parekh--Tse2010,Motahari--Gharan--Maddah-Ali--Khandani2014,Niesen--Maddah-Ali2013,Ordentlich--Erez--Nazer2014,Shomorony--Avestimehr2014,Ntranos--Cadambe--Nazer--Caire2013b,PaSaPr16}, physical-layer security~\cite{He--Yener2014,Vatedka--Kashyap--Thangaraj2015,Xie--Ulukus2014}, distributed source coding~\cite{Krithivasan--Pradhan2009,Krithivasan--Pradhan2011,Wagner2011,Maddah-Ali--Tse2010,Lalitha--Prakash--Vinodh--Kumar--Pradhan2013, Yang--Xiong2014}, and relay networks~\cite{Wilson--Narayanan--Pfister--Sprintson2010,Nam--Chung--Lee2010,NaGa11,Niesen--Whiting2012,Song--Devroye2013,Hong--Caire2013,Ren--Goseling--Weber--Gastpar2014,Cheng--Yuan--Tan2018}. Broadly speaking, these efforts have handled discrete and continuous settings separately, employing (nested) linear codes~\cite{ZaShEr02} for the former and (nested) lattice codes~\cite{Zamir2014} for the latter.

In this paper, we present a unified approach for handling discrete and continuous channels in  the compute--forward framework.  In this context, the goal is to reliably communicate one or more \textit{linear combinations of codewords} across a multiple-access channel. This framework was initially proposed for Gaussian relay networks~\cite{NaGa11}, motivated by the potential gains of reliably implementing network coding directly on the physical layer. Subsequent efforts demonstrated that the compute--forward framework can serve as a building block in many other scenarios, such as multiple-antenna transceiver architectures~\cite{Zhan--Nazer--Erez--Gastpar2014,Ordentlich--Erez--Nazer2013,Ordentlich--Erez2015,He--Nazer--Shamai2018}, interference alignment~\cite{Ordentlich--Erez--Nazer2014}, multiple-access strategies~\cite{Ordentlich--Erez--Nazer2014,Ordentlich--Erez--Nazer2013,NaCaNtCa16, Zhu--Gastpar2015}, and distributed source coding~\cite{Ordentlich--Erez2016}. The initial Gaussian compute--forward framework~\cite{NaGa11} utilized ``single-user'' decoding (i.e., decoding to the closest nested lattice codeword), and follow-up work generalized this framework to allow for sequential decoding~\cite{NaCaNtCa16} (i.e., utilizing recovered linear combinations as side information to reduce the effective noise). Naturally, one might expect further gains by generalizing the Gaussian compute--forward framework to allow for simultaneous decoding of multiple combinations. However, determining the simultaneous-decoding rate region for the Gaussian setting has remained an open problem, partly due to the challenge of characterizing this region via effective noise volume arguments for nested lattice codes. (The single-user and sequential arguments only require arguing about the effective noise variances.) 

The foundations of network information theory can be derived using i.i.d.~random encoding coupled with joint typicality decoding, as elegantly demonstrated in the textbook of El Gamal and Kim~\cite{ElKi11}. However, for schemes that exploit the algebraic structure of the code, it was until recently unclear how to establish achievability results via joint typicality. The work of Padakandla and Pradhan opened the door to such results by deriving joint typicality encoding and decoding methods for nested linear codes, and demonstrating gains for discrete memoryless multiple-access~\cite{PaPr17}, broadcast~\cite{PaPr18}, and interference channels~\cite{PaSaPr16}. Our own recent efforts have expanded this toolkit by developing a compute--forward framework for discrete memoryless channels that allows for simultaneous decoding, starting with the two-user setting~\cite{LiFePaNaGa18}, and moving to the general case of $K$ users and a receiver that wants $L$ linear combinations~\cite{LiFePaNaGa20}. One of the main ideas underlying these approaches is that, although the marginal distribution of a random linear codebook is uniform, multicoding (i.e., joint typicality encoding) can be used to select codewords with the desired type from a random nested linear codebook while preserving the overall algebraic structure. (This idea has also previously appeared in other contexts, such as trellis shaping~\cite{Forney1992}, nested lattice codes~\cite{Gariby--Erez08}, and sparse linear codes~\cite{Miyake2010}.)

This paper presents a discretization approach that can translate the simultaneous-decoding, compute--forward rate region for the discrete memoryless case into a rate region for a wide class of continuous channels. The basic idea is to quantize the continuous channel inputs and outputs, evaluate the rate region for the resulting discrete channel, and finally take suitable limits of the quantization resolution in order to obtain a rate region for the continuous case. The primary obstacle that we need to surmount is that the simultaneous-decoding rate region for the discrete case is not expressed in terms of mutual information terms, but rather in terms of differences of discrete entropies. For mutual informations, the limits follow directly from standard arguments (see, e.g., \cite[Sec.~3.4.1]{ElKi11}), whereas in our setting establishing the limits is considerably more challenging. To illustrate this point a bit more finely, consider a two-user discrete memoryless multiple-access channel with output $Y$ where the $k^{\text{th}}$ user transmits a (nested linear) codeword with marginal distribution $p(u_k)$ over the finite field $\mathbb{F}_\q$ and the receiver attempts to decode the linear combination $a_1 U_1 + a_2 U_2$ (over the same finite field). From~\cite{LiFePaNaGa18}, it follows that the following rate region is achievable via \textit{single-user decoding}\footnote{By single-user decoding, we mean that the receiver searches for a unique (nested linear) codeword that is jointly typical with the channel output and selects this codeword as its estimate of the linear combination. Simultaneous decoding yields a larger rate region, as described in detail in Section~\ref{subsec:simultaneous_joint_decoding}, but the additional terms would obfuscate the simple contrast between discrete and differential entropies.}
\begin{align}
    R_k < H(U_k) - H\bigl( a_1 U_1 + a_2 U_2 | Y\bigr),~~~k=1,2.
\end{align}
Notice that these rates are unchanged if the receiver demands any other linear combination that lies in the span of the vector $[a_1~a_2]$. This is a desirable property, e.g., the receiver can simply obtain $2U_1 + 2U_2$ by scaling $U_1 + U_2$, and this should be reflected in the rate expression. However, as we move to the continuous case, we must keep in mind that differential entropies are not invariant under linear transformations. For instance, say that the channel inputs $U_k$ are Gaussian and that the receiver wishes to decode the integer-linear combination $a_1 U_1 + a_2 U_2$ over $\mathbb{R}$. As a na\"ive first attempt, we might expect the (single-user decoding) rate region to take the form $R_k < h(U_k) - h(a_1 U_1 + a_2 U_2|Y)$. Of course, this fails to account for the fact that differential entropy is not scale invariant, e.g., $h(2U_1 + 2U_2) = h(U_1 + U_2) + \log(2)$. Overall, for the special case of decoding a single linear combination, this scaling issue can be handled by adding a correction term of the form $\log(\gcd(|a_1|,|a_2|))$ where $\gcd$ represents the greatest common divisor. It then follows from our prior work~\cite{LiFePaNaGa18} that the following rate region is achievable via single-user decoding for the \textit{two-user} Gaussian channel:
\begin{align}
    R_k < h(U_k) - h(a_1 U_1 + a_2 U_2 | Y) + \log(\gcd(|a_1|,|a_2|))~~~k=1,2.
\end{align} For decoding multiple linear combinations, the correction terms as well as the limit arguments become more complex. 

To handle the general case of $K$ users and $L$ linear combinations, we introduce an \textit{algebraic generalization} of R\'enyi's information dimension and $d$-dimensional entropy~\cite{Renyi1959} that allows us to express the simultaneous-decoding rate region for both the discrete and continuous setting via a single, compact expression. This leads to our main result in Theorem~\ref{thm:continuous_CF}: a simultaneous-decoding, compute--forward rate region for any continuous multiple-access channel of the form $Y = \sum_k h_k u_k + Z$ as well as for any  channel law with a finite input constellation. Notably, this theorem includes the Gaussian setting (with multiple antennas at the receiver) as a special case, and the resulting rate region is worked out in Corollary~\ref{cor:gaussian_CF}, thus resolving the open problem described above. For completeness, we  demonstrate how this framework recovers our earlier results for finite-field compute--forward in Theorem~\ref{thm:discrete_CF} and integer compute--forward in Theorem~\ref{thm:integer_CF}. For comparison, we also derive sequential-decoding versions of these results in Theorems~\ref{thm:discrete_CF_sequential},~\ref{thm:integer_CF_sequential}, and~\ref{thm:continuous_CF_sequential}. The algebraic information dimension also enables us to organize and simplify the proofs of the required limit arguments, and may be useful in other settings, especially when attempting to map algebraic network information theory results from the discrete to the continuous case.

The paper is structured as follows: Section~\ref{sec:notation} defines notation, Section~\ref{sec:problem_statement} describes the problem statement, Section~\ref{sec:preliminaries} introduces several notions (a generalized definition of entropy, matroids, lattices) that are needed to state the main results, Section~\ref{sec:main_results} states our three main results, Section~\ref{sec:special_cases} deals with particularizing our main results to some instructive special cases (sequential decoding, the two-user case, decoding a single linear combination, the Gaussian case), Section~\ref{sec:conclusion} concludes our paper. The proofs of our main results are relegated to Appendices.

}

{
\section{Notation}   \label{sec:notation}
Several notational conventions differ from those in our previous publications~\cite{LiFePaNaGa18,LiFePaNaGa20}.

We denote the probability of an event $\Ac$ by $\P\{\Ac\}$ and use $P_X$, $p_X(\cdot)$, $f_X(\cdot)$, and $F_X(\cdot)$ to denote the probability measure, the probability mass function (pmf), the probability density function (pdf), and the cumulative distribution function (cdf) of the random variable $X$, respectively.

The (positive) natural numbers, the integers, the real numbers, and prime numbers are denoted as $\mathbb{N}$, $\mathbb{Z}$, $\mathbb{R}$, and $\mathbb{P}$, respectively. A field is generally denoted as $\mathbb{F}$ and a finite field of order $\q$ is denoted as $\mathbb{F}_\q$. Only finite fields of prime order are relevant in this publication, hence the symbol $\q$ stands for a prime number throughout.

We denote column vectors with lowercase bold font whereas matrices are denoted with uppercase bold font. By convention, random vectors or matrices will be put in italic shape. For example, we would write $\P\{\xv = \xs\}$ or $\P\{\Xv = \Xs\}$.

For $\As$ a matrix (or column vector) and $\Sc$ a set of row indices, $[\As]_\Sc$ denotes the submatrix of $\As$ comprising only those rows indexed $\Sc$.

For any positive integer $p$, we define the bracket $[p]$ as the set of integers $\{1,\dotsc,p\}$.\footnote{Note that we deviate from the convention in~\cite{LiFePaNaGa18,LiFePaNaGa20}, in which $[p]$ stood for $\{0,\dotsc,p-1\}$.} For any odd positive integer $p$, we define the double bracket
\begin{IEEEeqnarray*}{rCl}
	\left\llbracket p \right\rrbracket
	&\triangleq&
	\left\{-\frac{p-1}{2},\dotsc,-1,0,1,\dotsc,\frac{p-1}{2}\right\}.   \IEEEeqnarraynumspace\IEEEyesnumber\label{def:double_bracket_notation}
\end{IEEEeqnarray*}
In other words, for $p$ an odd number, $\llbracket p \rrbracket$ is obtained by shifting $[p]$ to be centered around zero.

For any odd prime number $\q \geq 3$, we shall define the centered modulo-$\q$ reduction
\begin{equation}   \label{def:mod_q}
	\mymod_\q(x)
	\triangleq x - \q \left\lfloor \frac{x}{\q} + \frac{1}{2} \right\rfloor.
\end{equation}
In other words, for $x$ an integer (or real scalar), $\mymod_\q(x)$ is the unique element of $\llbracket \q \rrbracket$ such that the difference $x-\mymod_\q(x)$ is a multiple of $\q$. When endowed with the modulo-$\q$ arithmetic, where addition `$\oplus_\q$' and multiplication `$\otimes_\q$' operations on two integers $a,b \in \mathbb{Z}$ are defined as
\begin{subequations}
\begin{IEEEeqnarray}{rCl}
	a \oplus_\q b
	&=& \mymod_\q(a + b) = \mymod_\q(\mymod_\q(a) + \mymod_\q(b))   \label{finite_field_summation} \\
	a \otimes_\q b
	&=& \mymod_\q(ab) = \mymod_\q(\mymod_\q(a)\mymod_\q(b)),   \label{finite_field_multiplication}
\end{IEEEeqnarray}
\end{subequations}
the finite ring $(\llbracket \q \rrbracket,\oplus_\q,\otimes_\q)$ is isomorphic to the finite field $\mathbb{Z}/\q\mathbb{Z}$. Therefore, we shall from now on identify the finite field $\mathbb{F}_\q$ with
\begin{equation}
	\mathbb{F}_\q
	= (\llbracket \q \rrbracket, \oplus_\q, \otimes_\q)
\end{equation}
without loss of generality.
Additionally, we will denote as
\begin{equation}   \label{def:natural_mapping}
	\phi_\q \colon \mathbb{Z} \to \mathbb{F}_\q
\end{equation}
the natural mapping that modulo-reduces an integer $x \in \mathbb{Z}$ via the function $\mymod_\q$, and then maps $\mymod_\q(x) \in \llbracket \q \rrbracket$ to its corresponding element in $\mathbb{F}_\q$.
The function $\phi_\q$ (and occasionally, inverse images $\phi_\q^{-1}(\cdot)$) are invoked whenever it is important to explicitly distinguish between integer numbers and their finite-field counterparts. When applied on vectors or matrices, these functions act entrywise.

Finally, if $\mathbb{U}$ denotes a ring and $\mathbb{A} \subseteq \mathbb{U}$ denotes a discrete additive subgroup of $\mathbb{U}$. We shall denote the \emph{lattice} over $\mathbb{A}$ generated by the columns of the matrix $\Qs \in \mathbb{U}^{n \times d}$ as
\begin{equation}
	\Lambda_\mathbb{A}(\Qs)
	= \bigl\{ \Qs\vs \colon \vs \in \mathbb{A}^d \bigr\}.
\end{equation}
For the common special case $\mathbb{U} = \mathbb{R}$ and $\mathbb{A} = \mathbb{Z}$ (for which lattices are often defined) we will generally omit the subscript `$\mathbb{Z}$' and simply write $\Lambda(\Qs)$ for notational brevity. Such lattices are called Euclidean lattices.

By contrast, if the coefficients $\vs$ are not restricted to a discrete additive subgroup $\mathbb{A}$, but instead run over the entire alphabet $\mathbb{U}$, then we refer to the resulting set as the \emph{span} of $\Qs$, and denote it as
\begin{equation}
	\Span(\Qs)
	= \bigl\{ \Qs\vs \colon \vs \in \mathbb{U}^d \bigr\}.
\end{equation}
In principle, whenever $\mathbb{A}=\mathbb{U}$ is discrete, the concepts of \emph{span} and \emph{lattice} could be used interchangeably, but we will generally invoke the notion of \emph{lattice} to comply with standard terminology.

Norms are denoted using the widespread double bar notation. That is, for a real-valued matrix (or vector) $\As = [\As_{i,j}]$, the $p$-norm is denoted as $\lVert \As \rVert_p = (\sum_{i,j}\As_{i,j}^p)^{(1/p)}$ and the infinity norm is denoted as $\lVert \As \rVert_\infty = \max_{i,j} |A_{i,j}|$. The $n$-dimensional unit ball of a $p$-norm is denoted as $\mathscr{S}_p^n = \{ \xv \in \mathbb{R}^n \colon \lVert \xv \rVert_p \leq 1 \}$ and the $n$-dimensional unit ball of the infinity norm is denoted as $\mathscr{S}_\infty^n = [-1,1]^n$.

Given an indexed collection of sets $\{\mathscr{A}_n\}_{n \in \mathbb{N}}$, the \emph{inner set limit} (or limit inferior) and \emph{outer set limit} (or limit superior) of this collection are defined respectively as
\begin{subequations}
\begin{IEEEeqnarray}{rCl}
	\varliminf_{n \to \infty} \mathscr{A}_n
	&=& \bigcup_{\{n_i\}_{i \in \mathbb{N}}} \bigcap_{i \in \mathbb{N}} \mathscr{A}_{n_i}   \label{def:liminf} \\
	\varlimsup_{n \to \infty} \mathscr{A}_n
	&=& \bigcap_{\{n_i\}_{i \in \mathbb{N}}} \bigcup_{i \in \mathbb{N}} \mathscr{A}_{n_i}   \label{def:limsup} 
\end{IEEEeqnarray}
\end{subequations}
where the union in~\eqref{def:liminf} and the intersection in~\eqref{def:limsup} are over all growing sequences $\{n_i\}_{i \in \mathbb{N}}$ of natural numbers. If inner and outer set limits coincide, we say that $\{\mathscr{A}_n\}_{n \in \mathbb{N}}$ has a set limit
\begin{equation}
	\lim_{n \to \infty} \mathscr{A}_n
	 = \varliminf_{n \to \infty} \mathscr{A}_n
	 = \varlimsup_{n \to \infty} \mathscr{A}_n.
\end{equation}
In addition, note that for a double indexed collection $\{\mathscr{A}_{n,m}\}_{(n,m) \in \mathbb{N}^2}$,
\begin{subequations}   \label{lims_of_union_and_intersections}
\begin{IEEEeqnarray}{rClCrCl}
	\varliminf_{n \to \infty} \bigcup_{m \in \mathbb{N}} \mathscr{A}_{n,m}
	&\supseteq& \bigcup_{m \in \mathbb{N}} \varliminf_{n \to \infty} \mathscr{A}_{n,m} &\qquad&
	\varliminf_{n \to \infty} \bigcap_{m \in \mathbb{N}} \mathscr{A}_{n,m}
	&\subseteq& \bigcap_{m \in \mathbb{N}} \varliminf_{n \to \infty} \mathscr{A}_{n,m} \\
	\varlimsup_{n \to \infty} \bigcup_{m \in \mathbb{N}} \mathscr{A}_{n,m}
	&\subseteq& \bigcup_{m \in \mathbb{N}} \varlimsup_{n \to \infty} \mathscr{A}_{n,m} &\qquad&
	\varlimsup_{n \to \infty} \bigcap_{m \in \mathbb{N}} \mathscr{A}_{n,m}
	&\supseteq& \bigcap_{m \in \mathbb{N}} \varlimsup_{n \to \infty} \mathscr{A}_{n,m}.
\end{IEEEeqnarray}
\end{subequations}

}

{
\section{General problem statement of compute--forward}   \label{sec:problem_statement}
Consider the $K$-user memoryless multiple-access channel (MAC)
\begin{equation}
(\Xc_1\times\cdots\times \Xc_K, P_{Y|X_1,\ldots,X_K}, \Yc)
\end{equation}
which consists of $K$ sender alphabets $\Xc_k$, $k\in[K]$, one receiver alphabet $\Yc$, and a conditional probability distribution $P_{Y|X_1,\ldots,X_K}$.

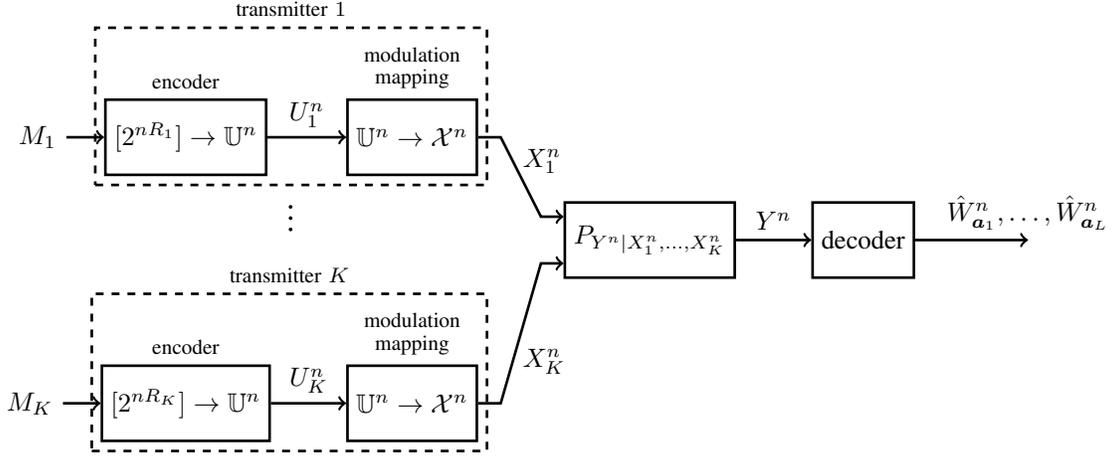
\begin{figure}[ht!]
\begin{center}
\begin{tikzpicture}[start chain, line width=1pt]
	\matrix[matrix of nodes, nodes={draw, anchor=center}, on chain] (chain-1) {
		\node[align=center,minimum height=10mm,label={[name=encoder1_label,font=\footnotesize]above:encoder}] (encoder1) {$[2^{n R_1}] \to \mathbb{U}^n$}; &[1cm] |[minimum height=10mm,label={[name=mapping1_label,font=\footnotesize,align=center]above:modulation\\[-\baselineskip+3mm]mapping}]| {$\mathbb{U}^n \to \Xc^n$} \\[1.7cm]
		\node[align=center,minimum height=10mm,label={[name=encoderK_label,font=\footnotesize]above:encoder}] (encoderK) {$[2^{n R_K}] \to \mathbb{U}^n$}; &[1cm] |[minimum height=10mm,label={[name=mappingK_label,font=\footnotesize,align=center]above:modulation\\[-\baselineskip+3mm]mapping}]| {$\mathbb{U}^n \to \Xc^n$} \\
	};
	\node[on chain, draw, minimum height=10mm] (channel) {$P_{Y^n|X_1^n,\dotsc,X_K^n}$};
	\node[on chain, draw, minimum height=10mm] (decoder) {decoder};
	\draw[<-] (encoder1.west) --+ (-5mm,0) node[left] {$M_1$};
	\draw[<-] (encoderK.west) --+ (-5mm,0) node[left] {$M_K$};
	\draw[->] (encoder1.east) -- node[above] {$U_1^n$} (chain-1-1-2.west);
	\draw[->] (encoderK.east) -- node[above] {$U_K^n$} (chain-1-2-2.west);
	\draw[->] (chain-1-1-2.east) --+ (3mm,0) -- node[pos=0.3, right] {$X_1^n$} ($(channel.north west)!.2!(channel.south west)+(-3mm,0)$) -- ($(channel.north west)!.2!(channel.south west)$);
	\draw[->] (chain-1-2-2.east) --+ (3mm,0) -- node[pos=0.3, right] {$X_K^n$} ($(channel.north west)!.8!(channel.south west)+(-3mm,0)$) -- ($(channel.north west)!.8!(channel.south west)$);
	\draw[->] (channel.east) -- node[above] {$Y^n$} (decoder.west);
	\draw[->] (decoder.east) --+ (15mm,0) node[above] {$\hat{W}^n_{\av_1},\ldots,\hat{W}^n_{\av_L}$};
	\node[fit=(encoder1)(chain-1-1-2)(encoder1_label)(mapping1_label), label={[font=\footnotesize]above:transmitter $1$}, draw, dashed] (enc1) {};
	\node[fit=(encoderK)(chain-1-2-2)(encoderK_label)(mappingK_label), label={[font=\footnotesize]above:transmitter $K$}, draw, dashed] (encK) {};
	\coordinate (dots) at ($(enc1.south)!.3!(encK.north)$);
	\node[circle, fill, inner sep=.5pt] at (dots) {};
	\node[circle, fill, inner sep=.5pt] at ($(dots)+(0,+1.5mm)$) {};
	\node[circle, fill, inner sep=.5pt] at ($(dots)+(0,-1.5mm)$) {};
\end{tikzpicture}
\end{center}
\caption{System model}
\end{figure}

Consider a ring $\mathbb{U}$ and a subset $\mathbb{A} \subseteq \mathbb{U}$ thereof. Let $\av_1,\ldots,\av_L \in \mathbb{A}^K$ denote $L$ coefficient vectors, assume $L \leq K$, and define the coefficient matrix
\begin{equation}
	\As
	= \begin{bmatrix} \av_1 \\ \vdots \\ \av_L \end{bmatrix} \in \mathbb{A}^{L \times K}.
\end{equation}
A $(2^{n R_1}, \dotsc, 2^{n R_K},n)$ code for compute--forward over $(\mathbb{U},\mathbb{A})$ with coefficient matrix $\As$ consists of
\begin{itemize}
	\item $K$ message sets $[2^{n R_k}]$, $k \in [K]$;
	\item $K$ encoders, where encoder $k$ maps a message $m_k \in [2^{n R_k}]$ to a codeword $u^n_k(m_k) \in \mathbb{U}^n$ such that $u^n_k(m_k)$ is {\em one-to-one};
	\item $K$ modulation mappings $\mathbb{U}^n \to \Xc^n$ that map\footnote{Modulation mappings may not necessarily be one-to-one mappings in general.} the entries of a codeword $u^n_k(m_k) \in \mathbb{U}^n$ to a \emph{physical codeword} $x^n_k(m_k) \in \mathbb{U}^n$;
	\item $L$ linear combinations (where $L \leq K$) for each message tuple $(m_1,\ldots, m_K)$
	\begin{equation*}
		\begin{bmatrix}
			w^n_{\av_1}(m_1,\ldots,m_K)\\
			\vdots\\
			w^n_{\av_L}(m_1,\ldots,m_K)
		\end{bmatrix}
		= \As \begin{bmatrix}
			u^n_{1}(m_1)\\
			\vdots \\
			u^n_{K}(m_K)
		\end{bmatrix},
	\end{equation*}
	where additions and multiplications are defined over the vector space $\mathbb{U}^n$, and
	\item a decoder that assigns estimates $(\hat{w}^n_{\av_1},\ldots,\hat{w}^n_{\av_L}) \in \mathbb{U}^n \times \cdots \times \mathbb{U}^n$ to each received sequence $y^n \in \Yc^n$.
\end{itemize}

Each message $M_k$ is independently and uniformly drawn from $[2^{n R_k}]$. The average probability of error is defined as $P_e^{(n)} = \P\big\{(\hat{W}^n_{\av_1},\ldots,\hat{W}^n_{\av_L}) \neq (W^n_{\av_1},\ldots,W^n_{\av_L})\big\}$. We say that a rate tuple $(R_1,\ldots,R_K)$ is achievable for computing the linear combinations with coefficient matrix $\As$ if there exists a sequence of $(2^{n R_1},\ldots,2^{n R_K},n)$ compute--forward codes such that $\lim_{n\rightarrow \infty} P_e^{(n)} = 0$.

}

{
\section{Preliminaries}   \label{sec:preliminaries}
In this Section, we lay out some important concepts, notations and auxiliary results that will be needed in later Sections. These include
\begin{itemize}
	\item	a new definition of entropy, dubbed \emph{algebraic entropy}, that generalizes beyond the classic Shannon (discrete) entropy and differential entropy, and incorporates a matrix-valued parameter (Section~\ref{ssec:algebraic_entropy});
	\item	a collection of definitions relative to Euclidean lattices and some classic fundamental results from the geometry of numbers (Section~\ref{ssec:euclidean_lattices});
	\item	a definition of matroids and some key results from matroid theory (Section~\ref{ssec:matroids}).
\end{itemize}
These preliminaries will be necessary for a complete understanding of the main Theorems presented in Section~\ref{sec:main_results}, as well as their proofs. The present Section~\ref{sec:preliminaries} may be skipped at first reading.
\subsection{Algebraic entropy and information dimension}   \label{ssec:algebraic_entropy}

For a probability space $(\Omega,\Sigma,\mu)$ defined on the reals $\Omega \subseteq \mathbb{R}^n$ and a reference measure $\rho \gg \mu$, we define the functional
\begin{equation}   \label{def:entropy}
	D(\mu \Vert \rho)
	= \int_\Omega \frac{\mathrm{d}\mu}{\mathrm{d}\rho} \log \left( \frac{\mathrm{d}\mu}{\mathrm{d}\rho} \right) \, \mathrm{d}\rho, 
\end{equation} which is often referred to as the Kullback-Leibler (KL) divergence, when $\rho$ and $\mu$ are probability measures.

If $\uv \sim \mu$ is discrete with support set $\mathcal{U}$ and $\rho$ is the counting measure, then $-D(\mu \Vert \rho)$ becomes equal to the discrete entropy (also known as the Shannon entropy)
\begin{equation}   \label{def:Shannon_entropy}
	H(\uv) = - \sum_{\mathbf{u} \in \mathcal{U}} \P\{\uv=\mathbf{u}\} \log \P\{\uv=\mathbf{u}\}.
\end{equation}
If instead $\uv \sim \mu$ is absolutely continuous and $\rho$ is the Lebesgue measure, then $-D(\mu \Vert \rho)$ becomes equal to the differential entropy
\begin{equation}
	h(\uv)
	= - \int f_{\uv}(\mathbf{u}) \log f_{\uv}(\mathbf{u}) \, \mathrm{d}\mathbf{u}
\end{equation}
where $f_{\uv} = \mathrm{d}\mu/\mathrm{d}\lambda$ ($\lambda$ denoting the Lebesgue measure) is referred to as the probability density function of $\uv \sim \mu$.

In his 1959 paper~\cite{Renyi1959}, R\'enyi elucidates some interesting connections between the discrete entropy of quantized variables and differential entropies. In particular, he introduces the concepts of \emph{information dimension} and \emph{$d$-dimensional entropy}, which we recall in the following definitions.

\definitionbox{
\begin{definition}[Information dimension and $d$-dimensional entropy]   \label{def:Renyi}
For a random variable $\uv \in \mathbb{R}^n$, the information dimension $d(\uv)$ and the $d$-dimensional entropy $\Hc(\uv)$ are defined as~\cite{Renyi1959}
\begin{subequations}
\begin{IEEEeqnarray}{rCl}
	d(\uv)
	&=& \lim_{\nu \to \infty} \frac{H(\lfloor \nu \uv \rfloor)}{\log(\nu)}   \label{def:information_dimension} \\
	\Hc(\uv)
	&=& \lim_{\nu \to \infty} \Bigl\{ H(\lfloor \nu \uv \rfloor) - d(\uv) \log(\nu) \Bigr\}    \label{Renyi_identity}
\end{IEEEeqnarray}
\end{subequations}
provided the limits $d(\uv)$ and $\Hc(\uv)$ exist.
\end{definition}
}

We now generalize these definitions by R\'enyi to the effect of incorporating a coefficient matrix as an additional parameter.

\definitionbox{
\begin{definition}[Algebraic information dimension and algebraic entropy]   \label{def:algebraic_entropy}
For a random vector $\uv \in \mathbb{R}^n$, the algebraic information dimension $d_\Qs(\uv)$ and the algebraic entropy $\Hc_\Qs(\uv)$ with coefficient matrix $\Qs \in \mathbb{R}^{m \times n}$ are defined respectively as
\begin{subequations}
\begin{IEEEeqnarray}{rCl}
	d_\Qs(\uv)
	&=& \lim_{\nu \to \infty} \frac{H(\Qs\lfloor \nu \uv \rfloor)}{\log(\nu)} \\
	\Hc_\Qs(\uv)
	&=& \liminf_{\nu \to \infty} \Bigl\{ H(\Qs \lfloor \nu \uv \rfloor) - d_\Qs(\uv) \log(\nu) \Bigr\}
\end{IEEEeqnarray}
\end{subequations}
if the limit $d_\Qs(\uv)$ exists. As a general convention, whenever $\Qs$ is the identity matrix, we may omit the subscript and write $d(\uv)$ and $\Hc(\uv)$ instead, so as to recover the quantities from Definition~\ref{def:Renyi}.
\end{definition}
}

Some remarks are in order:
\begin{enumerate}
	\item	The quantities $d_\Qs(\uv)$ and $\Hc_\Qs(\uv)$ depend on the matrix $\Qs$ only via its row span, since for any square full-rank $\Rs \in \mathbb{R}^{m \times m}$, we have $d_{\Rs\Qs}(\uv) = d_\Qs(\uv)$ and $\Hc_{\Rs\Qs}(\uv) = \Hc_\Qs(\uv)$.
	\item	The quantities $d(\uv)$ and $\Hc(\uv)$ from Definition~\ref{def:Renyi} correspond to special cases of $d_\Qs(\uv)$ and $\Hc_\Qs(\uv)$ when $m=n$ and $\Qs$ is some square full-rank matrix (e.g., the identity matrix).
	\item	Conversely, $d_\Qs(\uv)$ and $\Hc_\Qs(\uv)$ can only ever be different from $d(\uv)$ and $\Hc(\uv)$, respectively, if the row span of $\Qs$ is a strict subspace of $\mathbb{R}^m$, i.e., if $\Qs$ has rank less than its number of columns.
Whenever an information dimension or entropy in this paper is denoted with a subscript matrix like $\Qs$, the latter typically stands for a strictly broad matrix with full row rank.
\end{enumerate}

To complement Definition~\ref{def:algebraic_entropy}, we further define conditional versions of the quantities $d_\Qs(\uv)$ and $\Hc_\Qs(\uv)$ as follows.

\definitionbox{
\begin{definition}[Conditional algebraic information dimension and entropy]   \label{def:conditional_algebraic_entropy}
For a pair of random variables $(\uv,Y) \in \mathbb{R}^n \times \Yc$, the conditional algebraic information dimension $d_\Qs(\uv|Y)$ and the conditional algebraic entropy $\Hc_\Qs(\uv|Y)$ with coefficient matrix $\Qs \in \mathbb{R}^{m \times n}$ are defined respectively as
\begin{subequations}
\begin{IEEEeqnarray}{rCl}
	d_\Qs(\uv|Y)
	&=& \lim_{\nu \to \infty} \frac{H(\Qs\lfloor \nu \uv \rfloor | Y)}{\log(\nu)} \\
	\Hc_\Qs(\uv|Y)
	&=& \liminf_{\nu \to \infty} \Bigl\{ H(\Qs \lfloor \nu \uv \rfloor | Y) - d_\Qs(\uv|Y) \log(\nu) \Bigr\}
\end{IEEEeqnarray}
\end{subequations}
if the limit $d_\Qs(\uv|Y)$ exists. If $\rank(\Qs) = n$, we omit the subscript $\Qs$ and denote them as $d(\uv|Y)$ and $\Hc(\uv|Y)$, respectively. 
\end{definition}
}

\definitionbox{
\begin{definition}[Joint algebraic information dimension and entropy]   \label{def:joint_algebraic_entropy}
For a pair of random variables $(\uv_1,\uv_2) \in \mathbb{R}^{n_1} \times \mathbb{R}^{n_2}$, the joint algebraic information dimension $d_{\Qs_1,\Qs_2}(\uv_1,\uv_2)$ and the joint algebraic entropy $\Hc_{\Qs_1,\Qs_2}(\uv_1,\uv_2)$ with coefficient matrices $\Qs_1 \in \mathbb{R}^{m_1 \times n_1}$ and $\Qs_2 \in \mathbb{R}^{m_2 \times n_2}$ are defined respectively as
\begin{subequations}
\begin{IEEEeqnarray}{rCl}
	d_{\Qs_1,\Qs_2}(\uv_1,\uv_2)
	&=& \lim_{\nu \to \infty} \frac{H(\Qs_1\lfloor \nu \uv_1 \rfloor, \Qs_2\lfloor \nu \uv_2 \rfloor)}{\log(\nu)} \\
	\Hc_{\Qs_1,\Qs_2}(\uv_1,\uv_2)
	&=& \liminf_{\nu \to \infty} \Bigl\{ H(\Qs_1\lfloor \nu \uv_1 \rfloor, \Qs_2\lfloor \nu \uv_2 \rfloor) - d_{\Qs_1,\Qs_2}(\uv_1,\uv_2) \log(\nu) \Bigr\}
\end{IEEEeqnarray}
\end{subequations}
if the limit $d_{\Qs_1,\Qs_2}(\uv_1,\uv_2)$ exists. If $\rank(\Qs_1) = n_1$ and $\rank(\Qs_2) = n_2$, we omit the subscript $\Qs_1, \Qs_2$ and write $d(\uv_1,\uv_2|Y)$ and $\Hc(\uv_1,\uv_2|Y)$ instead, respectively.
\end{definition}
}

In addition, the notion of joint and conditional algebraic information dimension, as well as joint and conditional algebraic entropy, are defined by straightforward combination of Definitions~\ref{def:conditional_algebraic_entropy} and \ref{def:joint_algebraic_entropy}. In the following two lemmata, we provide two elementary relationships which will be used in later proofs, namely, a chain rule of information dimension and entropy, as well as a connection to mutual information.

\lemmabox{
\begin{lemma}[Chain rule for conditional algebraic information dimension and entropy]   \label{lem:chain_rule}
For a triple of random variables $(\uv_1,\uv_2,Y) \in \mathbb{R}^{n_1} \times \mathbb{R}^{n_2} \times \Yc$ such that the conditional information dimensions $d(\uv_1|Y)$ and $d(\uv_2|Y,\uv_1)$ exist and the mutual information $I(\uv_1,\uv_2;Y)$ is finite,\footnotemark\ the conditional algebraic information dimension $d(\uv|Y)$ and the conditional algebraic entropy $\Hc(\uv|Y)$ obey the chain rule
\begin{subequations}
\begin{IEEEeqnarray}{rCl}
	d(\uv_1,\uv_2|Y)
	&=& d(\uv_1|Y) + d(\uv_2|Y,\uv_1) \\
	\Hc(\uv_1,\uv_2|Y)
	&=& \Hc(\uv_1|Y) + \Hc(\uv_2|Y,\uv_1).
\end{IEEEeqnarray}
\end{subequations}
\end{lemma}
}

\proofbox{}{
\begin{IEEEproof}
The proof is deferred to Appendix~\ref{app:proof:chain_rule}.
\end{IEEEproof}
}
\footnotetext{Note that, in addition to the existence of $d(\uv_1|Y)$ and $d(\uv_2|Y,\uv_1)$, the finiteness assumption on $I(\uv_1,\uv_2;Y)$ is essential, as evidenced by Example~1 in \cite{GeKo19}.}

\lemmabox{
\begin{lemma}[Algebraic entropy and mutual information]   \label{lem:mutual_information}
For a pair of random variables $(\uv,Y) \in \mathbb{R}^n \times \Yc$ with finite mutual information $I(\uv;Y)$, we have
\begin{equation}
	I(\uv;Y)
	= \Hc(\uv) - \Hc(\uv|Y).
\end{equation}
\end{lemma}
}
\proofbox{}{
\begin{IEEEproof}
The proof is deferred to Appendix~\ref{app:proof:mutual_information}.
\end{IEEEproof}
}

As we have already hinted, algebraic entropy is closely connected to discrete entropy (for discrete variables) and to differential entropy (for continuous variables). These connections are elicited in the following two subsections.

\subsubsection{Discrete distributions}

The next lemma characterizes the algebraic information dimension and entropy of discretely supported distributions.
\lemmabox{
\begin{lemma}   \label{lem:discrete_algebraic_entropy}
For an integer matrix $\Qs \in \mathbb{Z}^{m \times n}$ and an integer random variable $\uv \in \mathbb{Z}^n$ with finite discrete entropy $H(\uv)$, the algebraic information dimension and algebraic entropy are given by
\begin{subequations}
\begin{IEEEeqnarray}{rCl}
	d_\Qs(\uv) &=& 0    \label{algebraic_information_dimension_discrete} \\
	\Hc_\Qs(\uv) &=& H(\Qs\uv).   \label{algebraic_entropy_discrete}
\end{IEEEeqnarray}
\end{subequations}
By convention, we extend~\eqref{algebraic_information_dimension_discrete}--\eqref{algebraic_entropy_discrete} to also hold for vectors over a finite field $\uv \in \mathbb{F}_\q^n$ with a coefficient matrix $\Qs \in \mathbb{F}_\q^{m \times n}$.\footnotemark
\end{lemma}
}
\footnotetext{This can be justified by the fact that $\mathbb{F}_\q$ is isomorphic to $(\llbracket \q \rrbracket, \oplus_\q, \otimes_\q)$ or $\mathbb{Z}/\q\mathbb{Z}$, which live on finite (and thus discrete) subsets of the real field $\mathbb{R}$.}

\proofbox{}{
\begin{IEEEproof}
The proof is deferred to Appendix~\ref{app:proof:discrete_algebraic_entropy}.
\end{IEEEproof}
}

\subsubsection{Continuous distributions}

Prior to stating an important identity, which is the counterpart to Lemma~\ref{lem:discrete_algebraic_entropy} for continuous distributions, in that it relates algebraic entropy to differential entropy, we need some preliminary definitions.

\definitionbox{
\begin{definition}[Unimodular matrix]   \label{def:unimodular}
A square integer matrix $\Qs \in \mathbb{Z}^{n \times n}$ is said to be \emph{invertible} or \emph{unimodular} if there exists an integer matrix $\Qs^{-1} \in \mathbb{Z}^{n \times n}$ (called the \emph{inverse}) such that $\Qs\Qs^{-1} = \Qs^{-1}\Qs = \Is$.
The following statements are equivalent:
\begin{enumerate}
	\item	$\Qs$ is unimodular
	\item	$\left|\det(\Qs)\right| = 1$.
\end{enumerate}
\end{definition}
}

\proofbox{}{
\begin{IEEEproof}
If $\Qs$ is unimodular, then $\Qs\Qs^{-1}  = \Is$ and so $\det(\Qs) \det(\Qs^{-1}) = 1$. Note that both $\det(\Qs)$ and $\det(\Qs^{-1})$ are integers, since $\Qs$ and $\Qs^{-1}$ are integer matrices. This implies that $\left|\det(\Qs)\right| = \left|\det(\Qs^{-1})\right| = 1$. On the other hand, by Cramer's rule, for any matrix $\Qs \in \mathbb{R}^{n \times n}$, $\Qs^{-1} = (\det(\Qs))^{-1}\mbox{adj}(\Qs)$, where $\mbox{adj}(\Qs)$ is the adjugate matrix of $\Qs$. In particular, if $\Qs$ is an integer matrix, so is $\mbox{adj}(\Qs)$ by the definition of the adjugate matrix. Therefore, if $\left|\det(\Qs)\right| = 1$, $\Qs^{-1}$ is an integer matrix as well.
\end{IEEEproof}
}

\definitionbox{
\begin{definition}[Right-invertible and left-invertible matrices]   \label{def:right-invertible}
A strictly broad integer matrix $\Qs \in \mathbb{Z}^{n \times m}$ with $n < m$ is said to be \emph{right-invertible} if there exists a tall integer matrix $\Qs^\sharp \in \mathbb{Z}^{m \times n}$ (called the \emph{right-inverse}) such that $\Qs\Qs^\sharp = \Is_n$.
\end{definition}
}

\definitionbox{
\begin{definition}[Smith normal form and elementary divisors]   \label{def:SNF}
For any integer matrix $\Qs \in \mathbb{Z}^{m \times n}$, there exists a non-negative integer diagonal matrix $\Sigmas(\Qs) = \diag(\sigma_1(\Qs), \sigma_2(\Qs), \ldots, \sigma_r(\Qs))$ with $r = \rank(\Qs) \leq \min\{n,m\}$ such that\footnotemark\ $\sigma_1(\Qs) \mid \sigma_2(\Qs) \mid \dotso \mid \sigma_r(\Qs)$ and
\begin{equation}   \label{SNF_decomposition}
	\Qs = \tilde{\Ss}(\Qs) \begin{bmatrix} \Sigmas(\Qs) & \mynull \\ \mynull & \mynull \end{bmatrix} \tilde{\Ts}(\Qs)
\end{equation}
for some unimodular matrices $\tilde{\Ss}(\Qs) \in \mathbb{Z}^{m \times m}$ and $\tilde{\Ts}(\Qs) \in \mathbb{Z}^{n \times n}$.
Equivalently, there exists a left-invertible $\Ss(\Qs) \in \mathbb{Z}^{m \times r}$ and a right-invertible $\Ts(\Qs) \in \mathbb{Z}^{r \times n}$ such that $\Qs = \Ss(\Qs) \Sigmas(\Qs) \Ts(\Qs)$. Here, $\Sigmas(\Qs)$ is called a \emph{reduced Smith normal form} of $\Qs$ and its diagonal entries $\sigma_i(\Qs)$ are called \emph{elementary divisors}. They can be computed as
\begin{equation}   \label{elementary_divisors}
	\sigma_i(\Qs)
	= \frac{d_i(\Qs)}{d_{i-1}(\Qs)}
\end{equation}
where $d_i(\Qs)$ is the greatest common divisor of the $i \times i$ minors\footnotemark\ of the matrix $\Qs$ and $d_0(\Qs) = 1$.
\end{definition}
Note that for a square full-rank $\Qs$, we have $\det(\Sigmas(\Qs)) = \prod_{i=1}^r \sigma_i(\Qs) = \left|\det(\Qs)\right|$.
The reader is referred to~\cite{Brown1993} for additional details on Smith normal forms.
}
\footnotetext{$a \mid b$ means $a$ divides $b$.}
\footnotetext{A minor of a matrix $\Qs$ is the determinant of a square submatrix of $\Qs$.}

Using the Smith normal decomposition, it becomes straightforward to prove the following lemma.
\lemmabox{
\begin{lemma}   \label{lem:right_invertible_matrices}
The following statements about a strictly broad integer matrix $\Qs \in \mathbb{Z}^{n \times m}$ (with $n < m$) are equivalent:
\begin{enumerate}
	\item	$\Qs$ is right-invertible
	\item	$\Qs^\T$ is left-invertible
	\item	$\Qs$ has elementary divisors equal to $\sigma_1(\Qs) = \dotso = \sigma_n(\Qs) = 1$
	\item	$\Qs$ can be completed to a unimodular matrix $\bigl[ \Qs^\T \ \ \Rs^\T \bigr]$ with some $\Rs \in \mathbb{Z}^{(m-n) \times m}$.
	\item	The $n \times n$ minors of $\Qs$ are coprime.
\end{enumerate}
\end{lemma}
}

\proofbox{}{
\begin{IEEEproof}
The proof is deferred to Appendix~\ref{app:proof:right_invertible_matrices}.
\end{IEEEproof}
}

\lemmabox{
\begin{lemma}   \label{lem:continuous_algebraic_entropy}
For an integer matrix\footnotemark\ $\Qs \in \mathbb{Z}^{m \times n}$ and an absolutely continuous random vector $\uv \in \mathbb{R}^n$ with finite differential entropy $h(\Qs\uv)$ and finite $H(\lfloor \uv \rfloor)$, the algebraic information dimension and algebraic entropy with parameter $\Qs$ are given respectively by
\begin{subequations}
\begin{IEEEeqnarray}{rCl}
	d_\Qs(\uv) &=& \rank(\Qs)   \label{algebraic_dimension_for_continuous} \\
	\Hc_{\Qs}(\uv) &=& h(\Ts(\Qs)\uv).   \label{algebraic_entropy_for_continuous}
\end{IEEEeqnarray}
\end{subequations}
\end{lemma}
}
\footnotetext{More generally, a matrix with rationally dependent entries. Since generalizing this lemma so as to cover \emph{algebraic (in)dependence} is not the focus of our work, we limit the statement to integer matrices.}

Given its central importance, and its reliance on several auxiliary lemmata, we provide the proof of Lemma~\ref{lem:continuous_algebraic_entropy} in a separate subsection further below (Section~\ref{subsec:proof_continuous_algebraic_entropy}).
Note that if $\Qs$ has full row rank, i.e., $\Qs$ is broad with $\rank(\Qs) = m$, then we have that $\Ss(\Qs)$ is unimodular and $\Sigmas(\Qs)$ is full-rank diagonal, in which case~\eqref{algebraic_entropy_for_continuous} can be alternatively expressed in terms of its Smith normal form $\Sigmas(\Qs)$ rather than $\Ts(\Qs)$, namely,
\begin{equation}   \label{algebraic_entropy_for_full_rank_Q}
	\Hc_{\Qs}(\uv)
	= h(\Qs\uv) - \log\left|\det(\Sigmas(\Qs))\right|.
\end{equation}
If in addition to having full row rank, $\Qs$ is also right-invertible, then $\Sigmas(\Qs)$ is unimodular. In this case, \eqref{algebraic_entropy_for_full_rank_Q} further simplifies to $\Hc_{\Qs}(\uv) = h(\Qs\uv)$.\footnote{In fact, in this case one can set $\Ss(\Qs)$ and $\Sigmas(\Qs)$ to be identity matrices, and $\Ts(\Qs) = \Qs$.}

\begin{remark}
Note that the right-hand side of~\eqref{algebraic_entropy_discrete} can be replaced without loss of generality by $H(\Ts(\Qs)\uv)$. Doing so will make Lemma~\ref{lem:discrete_algebraic_entropy} (for discrete distributions) closely resemble the formulation of its continuous counterpart (Lemma~\ref{lem:continuous_algebraic_entropy}).
\end{remark}

\subsection{Proof of Lemma~\ref{lem:continuous_algebraic_entropy}}   \label{subsec:proof_continuous_algebraic_entropy}

We will need the auxiliary Lemmata~\ref{lem:quantized_entropy_lower_bound}, \ref{lem:Renyi}, \ref{lem:entropy_finiteness}, \ref{lem:quantized_entropy_difference} and \ref{lem:Makkuva_Wu}, all stated below. Lemma~\ref{lem:Renyi} is mostly a classic result due to R\'enyi~\cite[Thm.~1]{Renyi1959}. The proof of Lemma~\ref{lem:continuous_algebraic_entropy} hinges mainly on Lemma~\ref{lem:Makkuva_Wu}, which is a generalization of a recent result due to Makkuva and Wu~\cite[Lem.~1]{MaWu18}.

\lemmabox{
\begin{lemma}   \label{lem:quantized_entropy_lower_bound}
Let $\uv \in \mathbb{R}^K$ be an absolutely continuous random vector with finite entropies $h(\uv)$ and $H(\floor{\uv})$. Consider a countable $\sigma$-algebra $\Qc$ on $\mathbb{R}^K$ and let $\lceil \uv \rfloor_\Qc$ denote a quantization of $\uv$ with respect to $\Qc$ (wherein the atoms of $\Qc$ represent the quantization cells). Assume that the atoms of $\Qc$ have Lebesgue measure uniformly upper-bounded by $\lambda_\mathrm{max} > 0$. Then
\begin{IEEEeqnarray}{rCl}
	h(\uv)
	&\leq& H(\lceil \uv \rfloor_\Qc) + \log(\lambda_\mathrm{max}).
\end{IEEEeqnarray}
\end{lemma}
}

\proofbox{}{
\begin{IEEEproof}
The proof is deferred to Appendix~\ref{app:proof:quantized_entropy_lower_bound}.
\end{IEEEproof}
}

\lemmabox{
\begin{lemma}   \label{lem:Renyi}
Consider an absolutely continuous random vector $\uv \in \mathbb{R}^n$ with finite entropies $h(\uv)$ and $H(\floor{\uv})$. Let $\zv$ be uniformly distributed on the cube $[0,1)^n$ and independent of $\uv$. The sequence
\begin{IEEEeqnarray*}{rCl}
	S_\nu
	&=& h(\floor{\nu\uv} + \zv) - n\log(\nu) \\
	&=& H(\floor{\nu\uv}) - n\log(\nu)
\end{IEEEeqnarray*}
indexed by $\nu \in \mathbb{N}$ is bounded as
\begin{equation}   \label{sandwich_bounds}
	h(\uv)
	\leq S_\nu
	\leq H(\floor{\uv}),
\end{equation}
satisfies $S_{m\nu} \leq S_\nu$ for any $m,\nu \in \mathbb{N}$ and has limit
\begin{equation}   \label{diff_entropy_as_a_limit}
	\lim_{\nu \to \infty} S_\nu
	= h(\uv).
\end{equation}
\end{lemma}
}

\proofbox{}{
\begin{IEEEproof}
The proof is deferred to Appendix~\ref{app:proof:Renyi}.
\end{IEEEproof}
}

\lemmabox{
\begin{lemma}   \label{lem:quantized_entropy_difference}
For any full row-rank integer matrix $\Ts \in \mathbb{Z}^{m \times n}$ and any random vector $\vv \in \mathbb{R}^n$, it holds that
\begin{equation}
	\bigl| H(\Ts \floor{\vv}) - H(\floor{\Ts\vv}) \bigr|
	\leq m \log\left( \frac{\lVert \Ts \rVert_1}{m} \right).
\end{equation}
\end{lemma}
}

\proofbox{}{
\begin{IEEEproof}
The proof is deferred to Appendix~\ref{app:proof:quantized_entropy_difference}.
\end{IEEEproof}
}

\lemmabox{
\begin{lemma}   \label{lem:entropy_finiteness}
For any absolutely continuous random vector $\vv = [V_1 \dotso V_n]^\T \in \mathbb{R}^n$, assume that $H(\floor{\vv})$ and $h([\vv]_\Ic)$ for all $\Ic \subset [n]$ are finite. Then, for any full row-rank matrix $\Qs \in \mathbb{R}^{m \times n}$, the entropies $H(\floor{\Qs\vv})$ and $h(\Qs\vv)$ are finite.\footnotemark
\end{lemma}
}
\footnotetext{For all entropy finiteness assumptions in the statement of Lemma~\ref{lem:entropy_finiteness} to be satisfied, it would suffice to assume that $H(\floor{\vv}) < +\infty$ and $h([\vv]_\Ic) > -\infty$ for all $\Ic \subset [n]$. In fact, $H(\floor{\vv}) > -\infty$ is obvious from the non-negativity of discrete entropy, whereas $h([\vv]_\Ic) < +\infty$ would ensue from $h([\vv]_\Ic) \leq H(\floor{[\vv]_\Ic}) \leq H(\floor{\vv}) < +\infty$, wherein the first inequality holds due to Lemma~\ref{lem:Renyi}.}

\proofbox{}{
\begin{IEEEproof}
The proof is deferred to Appendix~\ref{app:proof:entropy_finiteness}.
\end{IEEEproof}
}

\lemmabox{
\begin{lemma}   \label{lem:Makkuva_Wu}
For a right-invertible $\Ts \in \mathbb{Z}^{m \times n}$ and an absolutely continuous variable $\uv \in \mathbb{R}^n$ with finite entropies $H(\floor{\uv})$ and $h(\uv)$, we have
\begin{equation}
	\lim_{\nu \to \infty} \Bigl\{ H(\Ts\floor{\nu\uv}) - H(\floor{\nu\Ts\uv}) \Bigr\}
	= 0.
\end{equation}
\end{lemma}
}

\proofbox{}{
\begin{IEEEproof}
The proof is deferred to Appendix~\ref{app:proof:Makkuva_Wu}.
\end{IEEEproof}
}

With the help of the above Lemmata~\ref{lem:Renyi}, \ref{lem:entropy_finiteness}, \ref{lem:quantized_entropy_difference} and \ref{lem:Makkuva_Wu}, we can now proceed to proving Lemma~\ref{lem:continuous_algebraic_entropy} straightforwardly.
Let $r$ be shorthand for $\rank(\Qs)$ in what follows.
Recall that, according to Definition~\ref{def:SNF}, one can factorize $\Qs$ as $\Ss(\Qs) \Sigmas(\Qs) \Ts(\Qs)$ with left-invertible $\Ss(\Qs) \in \mathbb{Z}^{m \times r}$, full-rank diagonal $\Sigmas(\Qs) \in \mathbb{Z}^{r \times r}$ and right-invertible $\Ts(\Qs) \in \mathbb{Z}^{r \times n}$.
Using Lemma~\ref{lem:Makkuva_Wu}, we can exchange $H(\Ts(\Qs)\lfloor \nu \uv \rfloor)$ and $H(\lfloor \nu \Ts(\Qs)\uv \rfloor)$ in the limit as $\nu \to \infty$. Hence,
\begin{IEEEeqnarray*}{rCl}
	d_\Qs(\uv)
	&=& \lim_{\nu \to \infty} \frac{H(\Qs \lfloor \nu \uv \rfloor)}{\log(\nu)} \\
	&=& \lim_{\nu \to \infty} \frac{H(\Ts(\Qs)\lfloor \nu \uv \rfloor)}{\log(\nu)} \\
	&=& \lim_{\nu \to \infty} \frac{H(\lfloor \nu \Ts(\Qs)\uv \rfloor)}{\log(\nu)} \\
	&=& d(\Ts(\Qs)\uv).   \IEEEeqnarraynumspace\IEEEyesnumber
\end{IEEEeqnarray*}
Here, the first equality follows from the left-invertibility of $\Ss(\Qs)\Sigmas(\Qs)$.
Since $\uv \in \mathbb{R}^n$ is absolutely continuous and $\Ts(\Qs)$ is right-invertible, the vector $\Ts(\Qs)\uv \in \mathbb{R}^r$ is absolutely continuous and consequently, has information dimension $r$. Hence $d_\Qs(\uv) = d(\Ts(\Qs)\uv) = r$, which proves~\eqref{algebraic_dimension_for_continuous}.

Regarding~\eqref{algebraic_entropy_for_continuous}, recall that by its definition, the algebraic entropy is
\begin{equation}   \label{algebraic_entropy_from_definition}
	\Hc_\Qs(\uv)
	= \liminf_{\nu \to \infty} \Bigl\{ H(\Ts(\Qs) \lfloor \nu \uv \rfloor) - r \log(\nu) \Bigr\}.
\end{equation}
Again, Lemma~\ref{lem:Makkuva_Wu} allows us to exchange $H(\Ts(\Qs)\lfloor \nu \uv \rfloor)$ with $H(\lfloor \nu \Ts(\Qs)\uv \rfloor)$ inside the limit, leading to
\begin{equation}   \label{algebraic_entropy_from_definition_bis}
	\Hc_\Qs(\uv)
	= \liminf_{\nu \to \infty} \Bigl\{ H(\lfloor \nu \Ts(\Qs) \uv \rfloor) - r \log(\nu) \Bigr\}.
\end{equation}
Since $H(\floor{\uv})$ is finite by assumption, we have that $H(\floor{\Ts(\Qs)\uv})$ is also finite, because by Lemma~\ref{lem:quantized_entropy_difference}, $H(\floor{\Ts(\Qs)\uv}) \leq H(\Ts(\Qs)\floor{\uv}) + r \log\frac{\lVert \Ts(\Qs) \rVert_1}{r} \leq H(\floor{\uv}) + r \log\frac{\lVert \Ts(\Qs) \rVert_1}{r}$.
Hence, since $H(\floor{\Ts(\Qs)\uv})$ and $h(\Ts(\Qs)\uv)$ are both finite, we conclude from Lemma~\ref{lem:Renyi}
that the right-hand side of~\eqref{algebraic_entropy_from_definition_bis} equals $h(\Ts(\Qs)\uv)$, which concludes the proof of~\eqref{algebraic_entropy_for_continuous}.

\begin{remark}
Without much additional effort, one could prove a more general result, similar to~\cite[Thm.~3]{Renyi1959}, that has Lemmata~\ref{lem:discrete_algebraic_entropy} and~\ref{lem:continuous_algebraic_entropy} as corollaries. Namely, that for any non-singular $\uv \in \mathbb{R}^n$, which by Lebesgue's decomposition theorem can be uniquely decomposed as $\uv \sim \mu = \alpha \mu_\mathrm{d} + (1-\alpha) \mu_\mathrm{c}$ (where $\mu_\mathrm{d}$ is a measure supported on a discrete subset of $\mathbb{R}^n$, and $\mu_\mathrm{c}$ is an absolutely continuous measure on $\mathbb{R}^n$), the algebraic information dimension and entropy can be expressed as
\begin{subequations}
\begin{IEEEeqnarray}{rCl}
	d_\Qs(\uv)
	&=& (1-\alpha) \rank(\Qs)   \label{algebraic_information_dimension_mixed} \\
	\Hc_\Qs(\uv)
	&=& \alpha H(\Qs \uv_\mathrm{d}) + (1-\alpha) h(\Ts(\Qs) \uv_\mathrm{c}) - H_{\mathsf{b}}(\alpha)   \label{algebraic_entropy_mixed}
\end{IEEEeqnarray}
\end{subequations}
provided that the entropies $H(\uv_\mathrm{d})$, $H(\lfloor \uv_\mathrm{c} \rfloor)$ and $h([\uv_\mathrm{c}]_\Ic)$ are finite for every index set $\Ic \subset [n]$, where $\uv_\mathrm{d} \sim \mu_\mathrm{d}$ and $\uv_\mathrm{c} \sim \mu_\mathrm{c}$, and where $H_{\mathsf{b}}(\alpha) = -\alpha \log(\alpha) -(1-\alpha)\log(1-\alpha)$ denotes the binary entropy function. Setting $\alpha=1$ in~\eqref{algebraic_information_dimension_mixed}--\eqref{algebraic_entropy_mixed} recovers Lemma~\ref{lem:discrete_algebraic_entropy}, whereas setting $\alpha=0$ recovers Lemma~\ref{lem:continuous_algebraic_entropy}.
\end{remark}

\subsection{Euclidean lattices}   \label{ssec:euclidean_lattices}

In the following we will state some frequently used results from lattice theory.
Let $\mathscr{K}$ denote any closed convex body of positive volume in $\mathbb{R}^n$ that is symmetric around the origin, in the sense that $\xs \in \mathscr{K}$ implies $-\xs \in \mathscr{K}$.

\definitionbox{
\begin{definition}[Euclidean lattice and lattice dimension]
In Euclidean space, the lattice $\Lambda(\Qs) \subset \mathbb{R}^n$ generated by the linearly independent columns of a matrix $\Qs \in \mathbb{R}^{n \times d}$ is defined as
\begin{equation}
	\Lambda(\Qs)
	= \bigl\{ \Qs\vs \colon \vs \in \mathbb{Z}^d \bigr\}.
\end{equation}
The matrix $\Qs$ is called a \emph{generator matrix} or \emph{basis} of the lattice, and by the above linear independence assumption, has full column rank.
A more general definition of lattices (beyond Euclidean space) was given in Section~\ref{sec:notation}.

The column rank $d$ of the generator matrix $\Qs$ is the \emph{rank} of the lattice $\Lambda(\Qs)$. It represents the largest number of linearly independent vectors that can be constructed with lattice elements.
\end{definition}
}

\definitionbox{
\begin{definition}[Lattice determinant]
For an $n$-dimensional lattice $\Lambda(\Qs) \subset \mathbb{R}^n$ of rank $d$ with a basis $\Qs = \begin{bmatrix} \qs_1 & \dotso & \qs_d \end{bmatrix}$, the parallelepiped
\begin{equation}
	\left\{ \sum_{i=1}^d \alpha_i \qs_i \colon 0 \leq \alpha_i < 1 \right\}
\end{equation}
is called the \emph{fundamental domain} of $\Lambda(\Qs)$ associated to the basis $\Qs$.
The volume of this parallelepiped is called the \emph{determinant} $\det(\Lambda(\Qs))$ of the lattice. It is invariant with respect to the choice of basis and is equal to
\begin{equation}
	\det(\Lambda(\Qs))
	= \sqrt{\det(\Qs^\T\Qs)}.
\end{equation}
For a full-rank lattice, this expression simplifies to $\det(\Lambda(\Qs)) = \left| \det(\Qs) \right|$.
\end{definition}
}

\definitionbox{
\begin{definition}[Gauge, successive minima and minimal basis]
The gauge $g \colon \mathbb{R}^n \to \mathbb{R}$ of $\mathscr{K}$ is defined as the function
\begin{equation}
	g(\xs)
	= \min\bigl\{ \lambda \in \mathbb{R}_+ \colon \xs \in \lambda\mathscr{K} \bigr\}.
\end{equation}
A $d$-dimensional lattice $\Lambda(\Qs) \in \mathbb{R}^n$ has $d$ \emph{successive minima}. The $k$-th successive minimum $\lambda_k(\Qs)$ of $\Lambda(\Qs)$ with respect to $\mathscr{K}$ (or with respect to the gauge $g$) is defined as the infimum of positive numbers $\lambda > 0$ such that $\lambda\mathscr{K}$ contains at least $k$ linearly independent vectors of $\Lambda(\Qs)$. We have
\begin{equation}
	\lambda_1(\Qs) \leq \dotso \leq \lambda_d(\Qs).
\end{equation}
The symmetric convex body $\mathscr{K}$ with respect to which successive minima are defined, will always be clear from context. To make notation unambiguous, we may sometimes write $\lambda_{p,i}(\Qs)$ (instead of $\lambda_i(\Qs)$) when the successive minima are with respect to the $p$-norm ball $\mathscr{S}_p^n = \{ \xv \colon \lVert \xv \rVert_p \leq 1 \}$. 
\end{definition}
}

\definitionbox{
\begin{definition}[Korkin--Zolotarev reduced basis]   \label{def:KZ_bases}
Let $\Qs = \begin{bmatrix} \qs_1 & \dotsc & \qs_d \end{bmatrix} \in \mathbb{Z}^{n \times d}$ be an integer lattice basis and let us define $\mu_{i,j} = {\qs_i^\T \qs_j}/{\lVert \qs_j \rVert_2^2}$.
The basis $\Qs$ is reduced in the sense of Korkin and Zolotarev if it satisfies the following set of recursive conditions:
\begin{itemize}
	\item	$\qs_1$ is a shortest non-zero vector of $\Lambda(\Qs)$ in the Euclidean norm;
	\item	$\left| \mu_{i,1} \right| \leq \frac{1}{2}$ for $2 \leq i \leq d$;
	\item	if $\Lambda^{(n-1)}(\Qs)$ denotes the orthogonal projection\footnote{Note that $\Lambda^{(n-1)}(\Qs)$ is not necessarily an integer lattice.} of $\Lambda(\Qs)$ onto the orthogonal complement of $\Span(\qs_1)$, then the projected basis $\begin{bmatrix} \qs_2 - \mu_{2,1}\qs_1 & \dotso & \qs_d - \mu_{d,1}\qs_1 \end{bmatrix}$ is reduced in the sense of Korkin and Zolotarev.
\end{itemize}
Any subset of vectors $\begin{bmatrix} \qs_1 & \dotso & \qs_\ell \end{bmatrix}$ (for any $\ell \in [d]$) form a sublattice basis that is reduced in the sense of Korkin and Zolotarev.

If $\Qs = \begin{bmatrix} \qs_1 & \dotsc & \qs_d \end{bmatrix}$ is a Korkin--Zolotarev reduced basis and $\lambda_i(\Qs), i=1,\dotsc,d$ denote the successive minima with respect to the Euclidean norm (i.e., with respect to the unit sphere $\mathscr{K} = \{ \vs \in \mathbb{R}^n \colon \lVert \vs \rVert_2 \leq 1 \}$), then we have~\cite[Thm.~2.1]{Lag90}
\begin{equation}   \label{KZ_slack}
	\frac{4}{i+3} \lambda_i(\Qs)^2
	\leq \lVert \qs_i \rVert_2^2
	\leq \frac{i+3}{4} \lambda_i(\Qs)^2
	\quad \textnormal{(for $1 \leq i \leq d$).}
\end{equation}
\end{definition}
}

\definitionbox{
\begin{definition}[Integer lattice]
A lattice $\Lambda(\Qs) \subset \mathbb{R}^n$ is called an \emph{integer} lattice if it only contains points with integer coordinates, i.e., $\Lambda(\Qs) \subset \mathbb{Z}^n$. A lattice $\Lambda(\Qs)$ is an integer lattice if and only if $\Qs$ is an integer matrix.
\end{definition}
}

\definitionbox{
\begin{definition}[Orthogonal lattice and orthogonal basis]
Let $\ker(\Qs) \in \mathbb{R}^{n \times (n-d)}$ denote the nullspace of the column space $\Span(\Qs)$ of $\Qs \in \mathbb{R}^{n \times d}$.

The \emph{orthogonal lattice} $\Lambda^\perp(\Qs) \subset \mathbb{Z}^n$ of a rank-$d$ integer lattice $\Lambda(\Qs) \subset \mathbb{Z}^n$ is an integer lattice of rank $n-d$ defined as
\begin{equation}
	\Lambda^\perp(\Qs)
	= \ker(\Qs) \cap \mathbb{Z}^n.
\end{equation}

The transpose of the Korkin--Zolotarev reduced basis of $\Lambda^\perp(\Qs)$ shall be denoted as $\Qs_\perp \in \mathbb{Z}^{(n-d) \times n}$, so we have $\Lambda^\perp(\Qs) = \Lambda(\Qs_\perp^\T)$. By extension, we shall agree that $(\Qs^\T)_\perp = \Qs_\perp^\T$.

Note that if $\Qs_\perp$ is broad (resp.~tall) then it is right-invertible (resp.~left-invertible). As a consequence, when it is applied on a right-invertible or left-invertible $\Qs$, the $(\cdot)_\perp$ operation is involutive, i.e, $(\Qs_\perp)_\perp = \Qs$.

Furthermore, if $\Qs$ is left-invertible, then by~\cite[Thm.~1]{Ngu97} we have the determinant identity
\begin{equation}
	\det(\Qs^\T\Qs)
	= \det(\Qs_\perp\Qs_\perp^\T)
\end{equation}
or equivalently, $\det(\Lambda(\Qs)) = \det(\Lambda^\perp(\Qs))$.

\end{definition}
}

\lemmabox{
\begin{lemma}[{Van der Corput's Convex Body Theorem~\cite[Thm.~3.3]{LagariasGGL96}}]   \label{thm:van_der_Corput}
Let $\Lambda(\Qs) \subset \mathbb{R}^n$ denote a full-rank $n$-dimensional lattice. If
\begin{equation}
	\vol(\mathscr{K})
	\geq m 2^n \det(\Lambda(\Qs))
\end{equation}
then $\mathscr{K}$ contains at least $m$ symmetric pairs of non-zero lattice points in $\Lambda(\Qs)$.
This Theorem is due to van der Corput~\cite{vanderCorput1935} and generalizes the special case $m=1$, which is well known as Minkowski's First Theorem.

In its contrapositive formulation, Lemma~\ref{thm:van_der_Corput} states that, if $\mathscr{K}$ contains at most $m$ pairs of non-zero lattice points, then
\begin{equation}
	\vol(\mathscr{K})
	< (m+1) 2^n \det(\Lambda(\Qs)).
\end{equation}
Note that, $\mathscr{K}$ being a symmetric body that contains the origin, it contains $m$ pairs of non-zero lattice points if and only if it contains $2m+1$ lattice points (including the origin), that is, if and only if $\left| \Lambda(\Qs) \cap \mathscr{K} \right| = 2m+1$. Hence, a corollary of Lemma~\ref{thm:van_der_Corput} is the following inequality, relating the volume of any convex closed symmetric body $\mathscr{K}$ to the number of lattice points it encloses:
\begin{equation}   \label{van_der_Corput_inequality}
	\vol(\mathscr{K})
	< \left( \left| \Lambda(\Qs) \cap \mathscr{K} \right| + 1 \right) 2^{n-1} \det(\Lambda(\Qs))
\end{equation}
\end{lemma}
}

\lemmabox{
\begin{lemma}[{Minkowski's Second Theorem~\cite[Thm.~3.4]{LagariasGGL96}}]   \label{thm:Minkowski_Second_Theorem}
Let $\Lambda(\Qs) \subset \mathbb{R}^n$ be an $n$-dimensional Euclidean lattice of rank $d$. Let $\mathscr{K}$ be a symmetric convex body and let $\lambda_i(\Qs)$ denote the successive minima of $\Lambda(\Qs)$ with respect to $\mathscr{K}$. Then\footnotemark
\begin{equation}   \label{Minkowski_second_theorem}
	\frac{2^d}{d!} \det(\Lambda(\Qs))
	\leq \prod_{i=1}^d \lambda_i(\Qs)
	\leq 2^d \det(\Lambda(\Qs)).
\end{equation}
\end{lemma}
}
\footnotetext{Most references like~\cite[Thm.~3.4]{LagariasGGL96} state Minkowski's Second Theorem for the case $n=d$. Our statement~\eqref{Minkowski_second_theorem} is for $d \leq n$, which can be easily derived from the case $n=d$ by restricting $\mathbb{R}^n$ to its $d$-dimensional linear subspace that contains the rank-$d$ lattice.}

\subsection{Matroids}   \label{ssec:matroids}

We define matroids and state some fundamental results from matroid theory that will be leveraged in the proofs of Theorems~\ref{thm:integer_CF} and~\ref{thm:continuous_CF}.

\definitionbox{
\begin{definition}[Matroids]
A matroid $M$ is a pair $(\Gc,\Ic)$ consisting of a finite set $\Gc$ and a collection of subsets $\Ic \subset 2^\Gc$ satisfying the properties~\cite[Sec.~1.1]{Oxl11}:
\begin{enumerate}
	\item	$\emptyset \in \Ic$ 
	\item	If $I \in \Ic$ and $I' \subset I$, then $I' \in \Ic$
	\item	If $I_1$ and $I_2$ are in $\Ic$ and $|I_1| < |I_2|$, then there exists an element $J \in I_2 \setminus I_1$ such that $I_1 \cup J \in \Ic$.
\end{enumerate}
We refer to $\Gc$ as the \emph{ground set} of $M$ and to $\Ic$ as the \emph{independent sets} of $M$.

We say that $B \in \Ic$ is a \emph{basis} of $M = (\Gc,\Ic)$ if there is no larger $B' \in \Ic$ that contains $B$. In other words, a basis is a maximal independent set of the matroid. All bases have the same cardinality~\cite[Lem.~1.2.1, 1.2.4]{Oxl11} and a matroid is uniquely defined by the collection of its bases, which we will generally denote as $\mathscr{B}(M)$~\cite[Lem.~1.2.2, Thm.~1.2.3]{Oxl11}.

The \emph{size} of a matroid $M$ is its ground set cardinality $|\Gc|$, whereas the \emph{rank} $r(M)$ is the maximum cardinality of its bases.
\end{definition}
}

\definitionbox{
\begin{definition}[Representable matroids]   \label{def:representable_matroids}
If $\Gc$ denotes the set of column labels of a matrix $\Qs \in \mathbb{A}^{m \times n}$ over a ring or field $\mathbb{A}$, and if $\Ic$ denotes the set of subsets of $\Gc$ such that for every $I \in \Ic$, the rows of $[\Qs^\T]_I$ are linearly independent (in the vector space $\mathbb{A}^m$), then $(\Gc,\Ic)$ is a matroid, called the \emph{vector matroid} of $\Qs$ (cf.~\cite[Proposition~1.1.1]{Oxl11}) and is denoted as $M(\Qs)$.

If a matroid $M$ is isomorphic to the vector matroid of some matrix $\Qs$ over some ring $\mathbb{A}$, then we say that $M$ is \emph{representable} over $\mathbb{A}$ or $\mathbb{A}$-representable. Accordingly, $\Qs$ is a (matrix) \emph{representation} for $M$ over $\mathbb{A}$. Note that not all matroids are representable over all rings or fields, and some matroids are not representable over any field~\cite[Chapter~6]{Oxl11}.

We define $\mathscr{M}_\mathbb{A}(n)$ as the set of all representable matroids of size $n$ that have a representation over $\mathbb{A}$. Regardless of the alphabet $\mathbb{A}$, the set $\mathscr{M}_\mathbb{A}(n)$ always includes the empty matroid $([n],\{\emptyset\})$ and the full-rank matroid $([n],2^{[n]})$.

Similarly, we define $\mathscr{C}_\mathbb{A}(M)$ as the set of all matrix representations over $\mathbb{A}$ of the matroid $M$.
\end{definition}
}

\definitionbox{
\begin{definition}[Characteristic set]   \label{def:characteristic_set}
The characteristic set $\chi(M)$ of a matroid $M$ is the subset of $\mathbb{P} \cup \{0\}$ containing all values $\q$ such that $M$ is representable over some field of characteristic $\q$.
\end{definition}
}

Characteristic sets of representable matroids have the following properties:

\lemmabox{
\begin{lemma}[Rado, 1957]
\label{pro:Rado57}
If $0 \in \chi(M)$, then $\chi(M)$ contains all primes except a finite set.
\end{lemma}}

\lemmabox{
\begin{lemma}[V\'amos, 1971]
\label{pro:Vamos71}
If $\chi(M)$ is infinite, then $0 \in \chi(M)$.
\end{lemma}}
Proofs for lemmata~\ref{pro:Rado57} and \ref{pro:Vamos71} can be found, for example, in~\cite[Proposition~6.8.2]{Oxl11} and \cite[Proposition~6.8.3]{Oxl11}.
In addition, let us point out the following basic equivalence regarding representability for characteristic $0$.

\lemmabox{
\begin{lemma}   \label{lem:Q_and_Z_representability}
For a given matroid size $n$, representability over the integers and over the rationals are equivalent. That is, $\mathscr{M}_\mathbb{Z}(n) = \mathscr{M}_\mathbb{Q}(n)$.
\end{lemma}
}

\proofbox{}{
\begin{IEEEproof}
The inclusion $\mathscr{M}_\mathbb{Z}(n) \subset \mathscr{M}_\mathbb{Q}(n)$ follows immediately from $\mathbb{Z} \subset \mathbb{Q}$.
For the converse, consider a matrix $\Qs \in \mathbb{Q}^{m \times n}$ that represents a matroid $M(\Qs) \in \mathscr{M}_\mathbb{Q}(n)$. Write the entries of $\Qs$ as fractions $Q_{i,j} = n_{i,j}/D$ with $n_{i,j} \in \mathbb{Z}$ and $D$ a positive integer. Then $D\Qs$ is an integer matrix that represents $M(\Qs)$. Hence $\mathscr{M}_\mathbb{Z}(n) \supset \mathscr{M}_\mathbb{Q}(n)$.
\end{IEEEproof}
}

\begin{remark}
The equality $\mathscr{M}_\mathbb{Z}(n) = \mathscr{M}_\mathbb{Q}(n)$ does not further extend to finite fields in general. To appreciate this, consider the following matrix over $\mathbb{F}_3$:
\begin{equation}
	\Cs =
	\begin{bmatrix}
		1 & 0 & 1 & 1 \\
		0 & 1 & 1 & -1
	\end{bmatrix}.
\end{equation}
Any $2 \times 2$ submatrix of this matrix has full rank, so $\Cs$ is a representation of the so-called \emph{uniform matroid} of size $4$ and rank $2$,
\begin{equation}
	M(\Cs)
	= (\{1,2,3,4\},\{\emptyset, \{1\},\{2\},\{3\},\{4\},\{1,2\},\{1,3\},\{1,4\},\{2,3\},\{2,4\},\{3,4\}\}).
\end{equation}
However, this matroid is not representable over the binary field $\mathbb{F}_2$. In fact, one cannot create a set of four distinct $2 \times 1$ columns with binary entries such that all columns are pairwise linearly independent. Hence $\mathscr{M}_{\mathbb{F}_2}(4) \neq \mathscr{M}_{\mathbb{F}_3}(4)$, and $\mathscr{M}_{\mathbb{F}_2}(4) \neq \mathscr{M}_\mathbb{Z}(4)$.
\end{remark}

\definitionbox{
\begin{definition}[Dual matroids]
Let $M = (\Gc,\Ic)$ be a matroid and $\mathscr{B}(M)$ the collection of its bases. Then $\{\Gc \setminus B \colon B \in \mathscr{B}(M)\}$ is the set of bases of a matroid on $\Gc$, called the \emph{dual} of $M$, and denoted as $M^*$ (cf.~\cite[Thm.~2.1.1]{Oxl11}).
\end{definition}
}

Note that if $\Qs$ is a representation of $M$, then $\Qs_\perp^\T$ is a representation of $M^*$.

}

{
\section{A general formula for compute--forward achievable rates}   \label{sec:main_results}
In the following, $\mathbb{U}$ denotes a ring and $\mathbb{A} \subseteq \mathbb{U}$ denotes a discrete additive subgroup of $\mathbb{U}$, that is, a \emph{lattice} over $\mathbb{U}$. As already anticipated in Section~\ref{sec:problem_statement}, $\mathbb{U}$ represents the alphabet in which auxiliary codewords $U_k^n$ take value, such that the receiver attempts to recover linear combinations of these codewords, with coefficients in $\mathbb{A}$.

In this paper, the three main compute--forward theorems presented further below will be concerned with the following three choices of $(\mathbb{U},\mathbb{A})$, respectively:
\begin{itemize}
	\item	$(\mathbb{U},\mathbb{A}) = (\mathbb{F}_\q,\mathbb{F}_\q)$ (Theorem~\ref{thm:discrete_CF});
	\item	$(\mathbb{U},\mathbb{A}) = (\mathbb{Z},\mathbb{Z})$ (Theorem~\ref{thm:integer_CF});
	\item	$(\mathbb{U},\mathbb{A}) = (\mathbb{R},\mathbb{Z})$ (Theorem~\ref{thm:continuous_CF}).
\end{itemize}
Though we suspect that our theorems continue to hold for more general pairs $(\mathbb{U},\mathbb{A})$ of rings/lattices, we limit the scope of this paper to the three above-mentioned cases.

Let $(\uv,Y) \in \mathbb{U}^K \times \Yc$ follow a joint distribution $P_{\uv,Y} = \prod_{k=1}^K P_{U_k} P_{Y|\uv}$, where $P_{Y|\uv}=P_{Y|X_1, \dotsc, X_K}(y|x_1(u_1), \dotsc, x_K(u_K))$. In the following, for some natural numbers $1 \leq L_\Bf \leq K$,
\begin{itemize}
	\item	$\Bs$ denotes a full row-rank matrix over $\mathbb{A}$ of size $L_\Bf \times K$;
	\item	$M$ denotes a matroid of size $L_\Bf$;
	\item	$\Tc$ denotes a subset of $[K]$.
\end{itemize}

In the following, we state achievability theorems for simultaneous joint decoding, and for sequential decoding.

{
\subsection{Simultaneous joint decoding}   \label{subsec:simultaneous_joint_decoding}
With the above defined notations in mind, we first define the set
\begin{equation}
	\mathscr{Q}(\Bs,M,\Tc)
	\triangleq \Bigl\{ (R_1,\dotsc,R_K) \in \mathbb{R}_+^K \colon
	\sum_{k\in\Tc} R_k < \Hc([\uv]_\Tc) - \Hc_{\Bs}(\uv|Y) + J(\Bs,M) \Bigr\}   \label{Q_partial}
\end{equation}
where $J(\Bs,M)$ denotes a minimum-entropy term defined as
\begin{equation}   \label{def:J}
	J(\Bs,M)
	\triangleq \inf_{\Cs \in \mathscr{C}_\mathbb{A}(M)} \Hc_{\Cs\Bs}(\uv|Y)
\end{equation}
and where $\mathscr{C}_\mathbb{A}(M)$, as already defined in Section~\ref{ssec:matroids}, Definition~\ref{def:representable_matroids}, denotes the set of matrix representations over $\mathbb{A}$ of the matroid $M$.
With these definitions settled, we then define the set
\begin{equation}   \label{def:Q}
	\mathscr{Q}(\Bs)
	= \bigcap_M \; \bigcup_\Sc \; \bigcap_\Tc \mathscr{Q}(\Bs,M,\Tc)
\end{equation}
where the three nested set operations are over triples $(M,\Sc,\Tc)$ meeting the following constraints:
\begin{enumerate}
	\item $M$ iterates over all matroids of size $L_\Bf$ except the full-rank matroid, i.e., $M \in \mathscr{M}_{\mathbb{A}}(L_\Bf) \setminus ([L_\Bf],2^{[L_\Bf]})$. Henceforth, we shall denote this set of matroids (excluding the full-rank matroid) as $\mathscr{M}_{\mathbb{A}}^\circ(L_\Bf)$;
	\item $\Sc$ iterates over all index sets that correspond to bases of the dual matroid $M^*$, i.e., $\Sc \in \mathscr{B}(M^*)$;
	\item $\Tc$ iterates over all index sets that correspond to bases of the matroid of which $[\Bs]_\Sc$ (the matrix $\Bs$ with rows selected by $\Sc$) is a representation, i.e., $\Tc \in \mathscr{B}(M([\Bs]_\Sc))$.
\end{enumerate}
Finally, let us define the so-called \emph{joint simultaneous decoding} rate region.

\definitionbox{
\begin{definition}   \label{def:joint_decoding_rate_region}
The simultaneous decoding rate region for recovering the $\As$-linear combinations of auxiliary codewords is defined as
\begin{equation}   \label{def:R}
	\mathscr{R}(\As)
	\triangleq \bigcup_{\Bs} \mathscr{Q}(\Bs)
\end{equation}
where $\Bs \in \mathbb{A}^{L_\Bf\times K}$ runs over all full row-rank matrices satisfying $\Lambda_\mathbb{A}(\Bs^\T) \supseteq \Lambda_\mathbb{A}(\As^\T)$.
\end{definition}
}

The upcoming theorems will provide an operational meaning to this rate region. Meanwhile, a few remarks are in order.

\begin{remark}
Note that for some triples $(\Bs,M,\Tc)$, the region $\mathscr{Q}(\Bs,M,\Tc)$ might be an empty set because the right-hand side of the inequality in~\eqref{Q_partial} can be negative. For example, consider the two-user case ($K=2$), with $(\mathbb{U},\mathbb{A}) = (\mathbb{Z},\mathbb{Z})$ and the set $\mathscr{Q}([1, 1],(\{1\},\emptyset),\{1\})$ with an auxiliary $\uv = (U_1,U_2) \in \{0,1\}^2$ such that $U_1$ and $U_2$ are i.i.d.\ binary uniform: the right-hand side of~\eqref{Q_partial} reduces to $1 - H(U_1+U_2|Y)$, which can be negative.\footnote{A sufficient condition is for the channel's sum mutual information $I(U_1,U_2|Y)$ to be smaller than $\tfrac{1}{2}$ (bits), because it holds that $H(U_1+U_2|Y) = \tfrac{3}{2} - I(U_1+U_2;Y) > \tfrac{3}{2} - I(U_1,U_2|Y)$.}
However, $\mathscr{R}(\As)$ is never empty, because the union-taking over $\Bs$ always ensures that $\mathscr{R}(\As)$ does at least contain the region $\mathscr{Q}(\Is)$ generated by the identity $\Bs=\Is$. The latter set is never empty. In fact, $\mathscr{Q}(\Is)$ contains the corner points of the classic multiple-access rate region.
\end{remark}

\begin{remark}   \label{rmk:rate_region_invariance}
The joint decoding rate region $\mathscr{R}(\As)$ is invariant to elementary row operations, in the sense that $\mathscr{R}(\As) = \mathscr{R}(\Qs\As)$ for any full column-rank $\Qs$. This follows immediately from Definition~\ref{def:joint_decoding_rate_region}. At a high level, this is consistent with our interpretation of $\mathscr{R}(\As)$ as a compute--forward achievable rate region (as we shall make precise in the main theorem statements): one should be able to recover the $\As$-linear combinations of codewords at the receiver if and only if one can recover the $\Qs\As$-linear combinations.
\end{remark}

\begin{remark}
In light of Remark~\ref{rmk:rate_region_invariance}, it may be tempting to interpret the union-taking operation $\mathscr{R}(\As) = \bigcup_\Bs \mathscr{Q}(\Bs)$ as a means of assembling multiple rate regions $\mathscr{Q}(\Bs)$ that are \emph{individually} achievable. As a consequence, one might be led to conjecture that, much like $\mathscr{R}(\As)$, each region $\mathscr{Q}(\Bs)$ is invariant against elementary row operations in the sense that $\mathscr{Q}(\Bs) = \mathscr{Q}(\Qs\Bs)$ for a full column-rank $\Qs$. However, we have no proof or disproof of this claim, which we leave as an open question.
\end{remark}

We now state the three main achievability Theorems.

The proof is deferred to Appendix~\ref{app:proof:integer_CF}.
The proof is deferred to Appendix~\ref{app:proof:continuous_CF}.

\theorembox{\begin{theorem}[Finite-field compute--forward]   \label{thm:discrete_CF}
Let $(\mathbb{U},\mathbb{A}) = (\mathbb{F}_\q,\mathbb{F}_\q)$ for some prime field size $\q$. A rate tuple $(R_1, \dotsc, R_K)$ is achievable for decoding the $\As$-linear combinations of codewords if it is contained in $\mathscr{R}(\As)$ for some auxiliary pmf $p_{\uv}(\us) = \prod_{k=1}^K p_{U_k}(u_k)$ and modulation mappings $(x_1(u_1), \dotsc, x_K(u_K))$.
\end{theorem}}

\theorembox{\begin{theorem}[Integer compute--forward]   \label{thm:integer_CF}
Let $(\mathbb{U},\mathbb{A}) = (\mathbb{Z},\mathbb{Z})$ and assume that $\Hc(\uv)$ is finite. A tuple $(R_1, \dotsc, R_K)$ is achievable for decoding the $\As$-linear combinations of codewords if it is contained in $\mathscr{R}(\As)$ for some auxiliary pmf $p_{\uv}(\us) = \prod_{k=1}^K p_{U_k}(u_k)$ and modulation mappings $(x_1(u_1), \dotsc, x_K(u_K))$.
\end{theorem}}

\theorembox{\begin{theorem}[Continuous compute--forward]   \label{thm:continuous_CF}

Let $(\mathbb{U},\mathbb{A}) = (\mathbb{R},\mathbb{Z})$. Assume that the vector of auxiliaries $\uv \in \mathbb{R}^K$ has an absolutely continuous distribution, finite entropy $H(\floor{\uv})$ and finite differential entropy $h(\uv)$. In addition, we have either of the two (mutually exclusive) situations:
\begin{enumerate}
	\item   the mappings $\xs(\us)$ have finitely many (jump) discontinuities and images $x_k(\mathbb{R})$ of finite cardinality;
	\item   the mappings $\xs(\us)$ and the channel law $P_{Y|\xv}$ are such that the resulting channel $P_{Y|\uv}$ is given by the system equation $Y = \sum_k h_k u_k + Z$ with some real-valued coefficients $h_k$ and real-valued independent additive noise $Z$.
\end{enumerate}
Under these assumptions, a tuple $(R_1, \dotsc, R_K)$ is achievable for decoding the $\As$-linear combinations of codewords if it is contained in $\mathscr{R}(\As)$ for some auxiliary pdf $f_{\uv}(\us) = \prod_{k=1}^K f_{U_k}(u_k)$ and modulation mappings $(x_1(u_1), \dotsc, x_K(u_K))$ satisfying the above (alternative) conditions.
\end{theorem}}

\begin{remark}   \label{rmk:C_restriction}
When evaluating the term $J(\Bs,M)$ in the context of Theorem~\ref{thm:integer_CF} or~\ref{thm:continuous_CF}, we can restrict the integer matrix $\Cs$ to being right-invertible without loss of generality. This is because $\Hc_{\Cs\Bs}(\uv|Y) = \Hc_{\Ts(\Cs)\Bs}(\uv|Y)$ for any $\Cs$, and because $\Cs \in \mathscr{C}_\mathbb{Z}(M)$ if and only if $\Ts(\Cs) \in \mathscr{C}_\mathbb{Z}(M)$.
\end{remark}

}

{
\subsection{Sequential decoding}   \label{subsec:sequential_decoding}
Under sequential decoding, the receiver does not decode the linear combinations $W_{\as_1}^n, \dotsc, W_{\as_L}^n$ simultaneously, but instead it recovers them successively while using previously decoded linear combinations as side information. That is, for decoding the linear combination $W_{\as_\ell}^n$, the decoder uses $(Y^n,W_{\as_1}^n,\dotsc,W_{\as_{\ell-1}}^n)$ as side information. The decoder architecture for sequential decoding is depicted in Figure~\ref{fig:sequential_decoding}.
The generalized rate region (corresponding to Theorems~\ref{thm:discrete_CF}--\ref{thm:continuous_CF} and expressed in terms of algebraic entropies) is stated in the following.

\begin{figure}[ht!]
\centering
\begin{tikzpicture}[line width=1pt]
\node[matrix of nodes, nodes={draw, minimum width=2.5cm, minimum height=8mm}, row sep=1cm] (decoders) {
	{decoder $1$} \\
	{decoder $2$} \\[1.5cm]
	{decoder $L$} \\
};
\coordinate (l1) at ($(decoders-1-1.west)+(-1cm,0)$);
\draw[<-] (decoders-3-1.west) -| (l1) --++ (-1cm,0) node[left] {$Y^n$};
\draw[<-] (decoders-2-1.west) -| (l1);
\draw[<-] (decoders-1-1.west) -| (l1);
\coordinate (r1) at ($(decoders-1-1.east)+(1cm,0)$);
\coordinate (r2) at ($(decoders-2-1.east)+(1cm,0)$);
\coordinate (rL) at ($(decoders-3-1.east)+(1cm,0)$);
\draw[->] (decoders-1-1.east) -- (r1) |- ($(decoders-2-1.north)+(0,5mm)$) -- (decoders-2-1.north);
\draw[->] (decoders-2-1.east) -- (r2) |- ($(decoders-2-1.south)+(0,-5mm)$) --++ (0,-5mm);
\draw[->] ($(rL)+(0,1.5cm)$) |- ($(decoders-3-1.north)+(0,5mm)$) -- (decoders-3-1.north);
\draw[->] (r1) --++ (1cm,0) node[right] {$\hat{W}_{\as_1}^n$};
\draw[->] (r2) --++ (1cm,0) node[right] {$(\hat{W}_{\as_1}^n, \hat{W}_{\as_2}^n)$};
\draw[->] (decoders-3-1.east) --++ (2cm,0) node[right] {$(\hat{W}_{\as_1}^n, \dotsc, \hat{W}_{\as_L}^n)$};
\draw[fill] ($($(decoders-2-1.south)+(0,-5mm)$)!.42!(decoders-3-1.north)$) circle (.3pt);
\draw[fill] ($($(decoders-2-1.south)+(0,-5mm)$)!.50!(decoders-3-1.north)$) circle (.3pt);
\draw[fill] ($($(decoders-2-1.south)+(0,-5mm)$)!.58!(decoders-3-1.north)$) circle (.3pt);
\node[fit=(decoders), draw, dashed, label=above:decoder] {};
\end{tikzpicture}
\caption{Sequential decoder for successively recovering multiple linear combinations.}
\label{fig:sequential_decoding}
\end{figure}
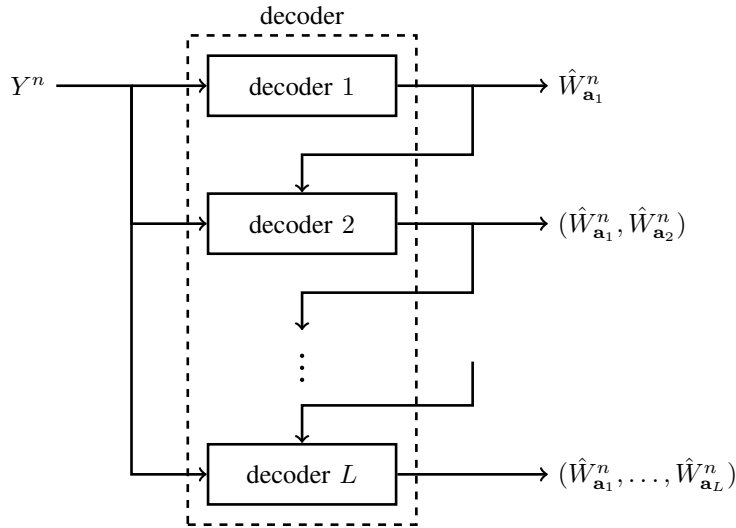

Let $\bs_j^\T = [b_{j,1} \hdots b_{j,K}]$ denote the $j$-th row of $\Bs$ and let $\Bs_{[j]} = [\bs_1, \dotsc, \bs_j]^\T = [\Bs]_{[j]}$ denote the submatrix formed by the first $j$ rows of $\Bs$.

We define $\mathscr{Q}_\mathrm{seq}(\Bs)$ as the simplex set of non-negative rate tuples satisfying
\begin{equation}   \label{sequential_decoding_rate_inequality}
	R_k
	< \Hc(U_k) + \Hc_{\Bs_{[j-1]}}(\uv|Y) - \Hc_{\Bs_{[j]}}(\uv|Y)
\end{equation}
for all $j \in [L_\Bf]$ and $k \in \{ k' \colon b_{j,k'} \neq 0 \}$. By convention, $\Hc_{\Bs_{[0]}}(\uv|Y) = 0$.

\definitionbox{
\begin{definition}   \label{def:sequential_rate_region}
The \emph{sequential decoding rate region} for recovering the $\As$-linear combinations of auxiliary codewords is defined as
\begin{equation}\label{eq:region_sequential}
	\mathscr{R}_\mathrm{seq}(\As)
	\triangleq \bigcup_{\Bs} \mathscr{Q}_\mathrm{seq}(\Bs)
\end{equation}
where $\Bs \in \mathbb{A}^{L_\Bf\times K}$ runs over all full row-rank matrices satisfying $\Lambda_\mathbb{A}(\Bs^\T) \supseteq \Lambda_\mathbb{A}(\As^\T)$.
\end{definition}
}

\begin{remark}
Note that, unlike its counterpart $\mathscr{Q}(\Bs)$ for \emph{simultaneous} decoding [cf.~\eqref{def:Q}], which is invariant against permutations of the rows of $\Bs$, the \emph{sequential} decoding rate region $\mathscr{Q}_\mathrm{seq}(\Bs)$ depends on the order of rows, which represents the decoding order. The union operation in~\eqref{eq:region_sequential}, however, restores this invariance in the order of the rows of the coefficient matrix $\As$.
\end{remark}

The following three theorems are counterparts to Theorems~\ref{thm:discrete_CF}, \ref{thm:integer_CF} and \ref{thm:continuous_CF}, but for sequential decoding instead of simultaneous joint decoding.
Theorem~\ref{thm:discrete_CF_sequential} is proven in \cite[Thm.~2]{LiFePaNaGa20}, whereas Theorems~\ref{thm:integer_CF_sequential} and \ref{thm:continuous_CF_sequential} are proven in Appendices~\ref{app:proof:integer_CF_sequential} and \ref{app:proof:continuous_CF_sequential}, respectively.

\theorembox{\begin{theorem}[Finite-field compute--forward under sequential decoding]   \label{thm:discrete_CF_sequential}
Let $(\mathbb{U},\mathbb{A}) = (\mathbb{F}_\q,\mathbb{F}_\q)$ for some prime field size $\q$. A rate tuple $(R_1, \dotsc, R_K)$ is achievable for sequentially decoding the $\As$-linear combinations of codewords if it is contained in $\mathscr{R}_\mathrm{seq}(\As)$ for some auxiliary pmf $p_{\uv}(\us) = \prod_{k=1}^K p_{U_k}(u_k)$ and modulation mappings $(x_1(u_1), \dotsc, x_K(u_K))$.
\end{theorem}}

\theorembox{\begin{theorem}[Integer compute--forward under sequential decoding]   \label{thm:integer_CF_sequential}
Let $(\mathbb{U},\mathbb{A}) = (\mathbb{Z},\mathbb{Z})$. A rate tuple $(R_1, \dotsc, R_K)$ is achievable for sequentially decoding the $\As$-linear combinations of codewords if it is contained in $\mathscr{R}_\mathrm{seq}(\As)$ for some auxiliary pmf $p_{\uv}(\us) = \prod_{k=1}^K p_{U_k}(u_k)$ and modulation mappings $(x_1(u_1), \dotsc, x_K(u_K))$.
\end{theorem}}

\theorembox{\begin{theorem}[Continuous compute--forward under sequential decoding]   \label{thm:continuous_CF_sequential}
Let $(\mathbb{U},\mathbb{A}) = (\mathbb{R},\mathbb{Z})$. A rate tuple $(R_1, \dotsc, R_K)$ is achievable for sequentially decoding the $\As$-linear combinations of codewords if it is contained in $\mathscr{R}_\mathrm{seq}(\As)$ for some auxiliary pdf $f_{\uv}(\us) = \prod_{k=1}^K f_{U_k}(u_k)$ and modulation mappings $(x_1(u_1), \dotsc, x_K(u_K))$.
\end{theorem}}

}

}

{
\section{Special cases}   \label{sec:special_cases}
The three main theorems stated in the previous Section have established achievable rate regions which are based on relatively complex nested set operations, and it is therefore difficult to glean any insight into their behavior. To provide more intuition, let us elaborate on some special cases:
\begin{itemize}
	\item	Subsection~\ref{subsec:two-user_case_simultaneous_joint_decoding}: The two-user case $K=L=2$, when the receiver decodes two linear combinations with \emph{simultaneous} decoding;
	\item	Subsection~\ref{subsec:two-user_case_sequential_decoding}: The two-user case $K=L=2$, when the receiver decodes two linear combinations with \emph{sequential} decoding;
	\item	Subsection~\ref{subsec:single-equation_decoding}: The two-user case when the receiver decodes a single linear combination ($K=2$, $L=1$);
	\item	Subsection~\ref{subsec:Gaussian_channels}: Theorem~3 particularized to Gaussian auxiliaries, which recovers and improves upon most lattice-based compute--forward results known in the literature.
\end{itemize}
Note that the first three special cases apply equally to finite-field, integer and continuous auxiliary computation (Theorems~\ref{thm:discrete_CF}, \ref{thm:integer_CF}, \ref{thm:continuous_CF}): the derivations are provided in a unified manner by using the properties of algebraic entropy.

{
\subsection{Decoding two linear combinations with two users under simultaneous decoding}   \label{subsec:two-user_case_simultaneous_joint_decoding}
For the case $K=L=2$, the matrix $\As \in \mathbb{A}^{2 \times 2}$ is square and full-rank.
This problem amounts to a two-user MAC setting in which the two users employ nested linear codes and attempt to recover two linearly independent combinations of messages (and hence both messages). As in previous publications~\cite{LiFePaNaGa18,LiFePaNaGa20}, we refer to this problem as the \emph{linear MAC} (LMAC) and denote the corresponding achievable rate region as $\mathscr{R}(\As) = \mathscr{R}_\LMAC$.\footnote{Note that $\mathscr{R}_\LMAC$ does not depend on any specific matrix $\As \in \mathbb{A}^{2 \times 2}$, as long as $\As$ is full-rank.}

The following lemma will be instrumental for simplifying the LMAC rate region expression.

\lemmabox{
\begin{lemma}   \label{lem:LMAC_simplification}
Consider full-rank square matrices $\Bs \in \mathbb{A}^{2 \times 2}$. When taking the union of sets $\mathscr{Q}(\Bs)$ over such matrices, it suffices to only retain the identity matrix $\Bs=\Is=\begin{bsmallmatrix} 1 & 0 \\ 0 & 1 \end{bsmallmatrix}$, in that
\begin{equation}   \label{union_over_B}
	\bigcup_{\Bs \in \mathbb{A}^{2 \times 2} \colon \text{$\Bs$ is full rank}} \mathscr{Q}(\Bs)
	= \mathscr{Q}(\Is).
\end{equation}
\end{lemma}
}

\proofbox{}{
\begin{IEEEproof}
The proof consists in an exhaustive evaluation of the union on the left-hand side of~\eqref{union_over_B}, and is deferred to Appendix~\ref{app:proof:two_user_two_equation}.
\end{IEEEproof}
}

By direct application of Lemma~\ref{lem:LMAC_simplification} to the rate region expression~\eqref{def:joint_decoding_rate_region}, we infer that for $K=L=2$,
\begin{equation}   \label{LMAC_characterization}
	\mathscr{R}_\LMAC
	= \mathscr{R}(\As)
	= \mathscr{Q}(\Is)
\end{equation}
(where $\As \in \mathbb{A}^{2 \times 2}$ is any full-rank matrix).
This rate region $\mathscr{R}_\LMAC$ can be expressed as [cf. Appendix~\ref{app:proof:two_user_two_equation}, Equations~\eqref{LMAC_sum_rate_bound}, \eqref{LMAC_rate_bound_5_0}--\eqref{LMAC_rate_bound_7_0}]
\begin{subequations}
\begin{IEEEeqnarray}{rCl}
	\mathscr{R}_\LMAC
	= \Bigl\{ (R_1, R_2) \in \mathbb{R}_+^2 \colon
	R_1 + R_2
	&<& I(U_1,U_2;Y)   \label{LMAC_eq1} \\
	R_1
	&<& I(U_1;Y,U_2) \\
	R_2
	&<& I(U_2;Y,U_1) \\
	\min\{ R_1 - \Hc(U_1) , R_2 - \Hc(U_2) \}
	&<& - \Hc(\uv|Y) + \inf_{c_1 \neq 0,\, c_2 \neq 0} \Hc_{[ c_1 \ c_2 ]}(\uv|Y) \Bigr\}.   \label{LMAC_eq4}
\end{IEEEeqnarray}
\end{subequations}
Notice that $\mathscr{R}_\LMAC$ is obviously a subset (sometimes proper) of the conventional multiple-access rate region
\begin{subequations}
\begin{IEEEeqnarray}{rCl}
	\mathscr{R}_\MAC
	= \Bigl\{ (R_1, R_2) \in \mathbb{R}_+^2 \colon
	R_1 + R_2
	&<& I(U_1,U_2;Y) \\ 
	R_1
	&<& I(U_1;Y,U_2) \\
	R_2
	&<& I(U_2;Y,U_1) \Bigr\}.   
\end{IEEEeqnarray}
\end{subequations}
The set $\mathscr{R}_\LMAC$, defined by the inequalities~\eqref{LMAC_eq1}--\eqref{LMAC_eq4}, is depicted in Figure~\ref{fig:LMAC} (with a dashed line). Notice that when the inequality~\eqref{LMAC_eq4} is inactive, $\mathscr{R}_\LMAC$ coincides with $\mathscr{R}_\MAC$, as illustrated in Figure~\ref{fig:LMAC_1}. Otherwise, $\mathscr{R}_\LMAC$ is diminished with respect to $\mathscr{R}_\MAC$ by a triangular shaped region located along the dominant face, as pictured in Figure~\ref{fig:LMAC_2}.

\begin{figure}[ht]
\centering
\subfloat[Case $\mathscr{R}_\LMAC = \mathscr{R}_\MAC$]{
\begin{tikzpicture}[scale=1.5, line width=1pt]
\filldraw[black!15] (0,0) |- (1.3,1.9) -- (2.2,1) |- cycle;
\draw[black!50] (0,1.9) -- (1.3,1.9) -- (2.2,1) -- (2.2,0);
\draw[dashed] (0,1.9) -- (1.3,1.9) -- (2.2,1) -- (2.2,0);
\draw[<->] (0,2.3) node[right] {$R_2$} |- (2.5,0) node[right] {$R_1$};
\end{tikzpicture}
\label{fig:LMAC_1}
}
\hspace{5mm}
\begin{tikzpicture}[scale=1.5, line width=1pt]
\draw[black!50] (0,0) --+ (5mm,0) node[black,right] {$\mathscr{R}_{\MAC}$};
\filldraw[dashed,fill=black!15] (0,-3mm) rectangle (5mm,-7mm) node[yshift=3mm,black,right] {$\mathscr{R}_\LMAC$};
\end{tikzpicture}
\hspace{5mm}
\subfloat[Case $\mathscr{R}_\LMAC \subsetneq \mathscr{R}_\MAC$]{
\begin{tikzpicture}[scale=1.5, line width=1pt]
\filldraw[black!15] (0,0) |- (1.3,2) -- (1.5,1.8) |- (2.2,1.1) -- (2.3,1) -- (2.3,0) |- cycle;
\draw[black!50] (0,2) -- (1.3,2) -- (2.3,1) -- (2.3,0);
\draw[dashed] (0,2) -- (1.3,2) -- (1.5,1.8) |- (2.2,1.1) -- (2.3,1) -- (2.3,0);
\draw[<->] (0,2.3) node[right] {$R_2$} |- (2.5,0) node[right] {$R_1$};
\end{tikzpicture}
\label{fig:LMAC_2}
}
\caption{Rate region $\mathscr{R}_\LMAC = \mathscr{R}(\As) = \mathscr{Q}(\Is)$ (dashed and shaded). The boundary of the MAC capacity rate region $\mathscr{R}_\MAC$ is shown for reference (solid gray).}
\label{fig:LMAC}
\end{figure}
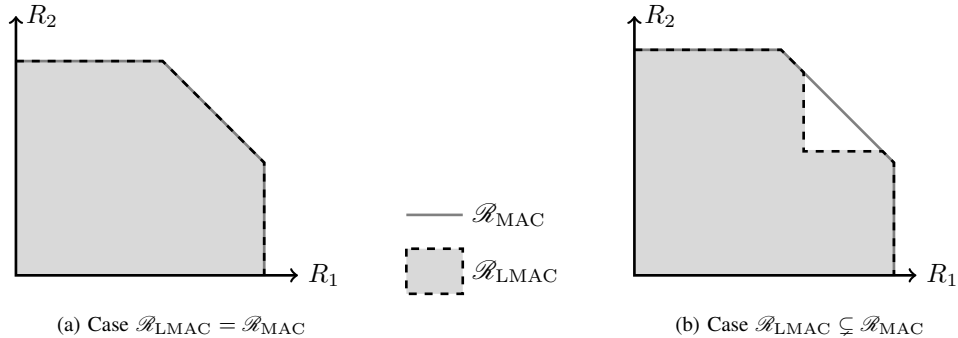

Let us derive the exact condition (in terms of an entropy inequality) under which $\mathscr{R}_\LMAC \subsetneq \mathscr{R}_\MAC$. For this purpose, let us define the set [cf.~\eqref{LMAC_eq4}]
\begin{equation}   \label{def:R_prime}
	\mathscr{R}'
	\triangleq \Bigl\{ (R_1, R_2) \in \mathbb{R}_+^2 \colon \\
	\min\{ R_1 - \Hc(U_1) , R_2 - \Hc(U_2) \}
	< - \Hc(\uv|Y) + \inf_{c_1 \neq 0,\, c_2 \neq 0} \Hc_{[ c_1 \ c_2 ]}(\uv|Y) \Bigr\}
\end{equation}
and denote its complement in $\mathbb{R}_+^2$ as $\overline{\mathscr{R}'}$.
We can now express $\mathscr{R}_\LMAC$ as
\begin{IEEEeqnarray*}{rCl}
	\mathscr{R}_\LMAC
	&=& \mathscr{R}_\MAC \cap \mathscr{R}' \\
	&=& \mathscr{R}_\MAC \setminus \overline{\mathscr{R}'} \\
	&=& \mathscr{R}_\MAC \setminus ( \mathscr{R}_\MAC \cap \overline{\mathscr{R}'} ).   \IEEEyesnumber\IEEEeqnarraynumspace\label{LMAC_as_set_difference}
\end{IEEEeqnarray*}
The condition $\mathscr{R}_\LMAC \subsetneq \mathscr{R}_\MAC$ is thus equivalent to $\mathscr{R}_\MAC \cap \overline{\mathscr{R}'}$ being non-empty, which occurs if and only if there exists a rate pair $(R_1,R_2)$ that will simultaneously satisfy the following three inequalities:
\begin{subequations}
\begin{IEEEeqnarray}{rCl}
	R_1 + R_2
	&<& I(U_1,U_2;Y)   \label{non-empty_notch_condition_1} \\
	R_1
	&\geq& \Hc(U_1) - \Hc(\uv|Y) + \inf_{c_1 \neq 0,\, c_2 \neq 0} \Hc_{[ c_1 \ c_2 ]}(\uv|Y)   \label{non-empty_notch_condition_2} \\
	R_2
	&\geq& \Hc(U_2) - \Hc(\uv|Y) + \inf_{c_1 \neq 0,\, c_2 \neq 0} \Hc_{[ c_1 \ c_2 ]}(\uv|Y).   \label{non-empty_notch_condition_3}
\end{IEEEeqnarray}
\end{subequations}
Upon subtracting~\eqref{non-empty_notch_condition_2}--\eqref{non-empty_notch_condition_3} from~\eqref{non-empty_notch_condition_1} and recalling that by the chain rule, $I(U_1,U_2;Y) = \Hc(U_1) + \Hc(U_2) - \Hc(\uv|Y)$, we infer the following lemma.

\lemmabox{
\begin{lemma}   \label{lem:notch}
The linear MAC region $\mathscr{R}_\LMAC$ is strictly contained in the MAC region $\mathscr{R}_\MAC$ if and only if there exists a coefficient vector with non-zero entries $\cs^\T = [ c_1 \ c_2 ] \in \mathbb{A}^2$ such that
\begin{equation}   \label{tip_outside_condition}
	2 \Hc_{\cs^\T}(\uv|Y)
	< \Hc(\uv|Y).
\end{equation}
Otherwise, $\mathscr{R}_\LMAC = \mathscr{R}_\MAC$.
\end{lemma}
}

}

{
\subsection{Decoding two linear combinations with two users under sequential decoding}   \label{subsec:two-user_case_sequential_decoding}
Consider the same setting $K=L=2$ as in the previous section, except that we replace \emph{simultaneous decoding} with \emph{sequential decoding}. Under these circumstances, the set $\mathscr{Q}_\mathrm{seq}(\Bs)$ [cf.~\eqref{sequential_decoding_rate_inequality}] can be expressed as follows. Let $\bs_j^\T = [b_{j,1} \ b_{j,2}]$, $j=1,2$ denote the $j$-th row of $\Bs = \begin{bsmallmatrix} b_{1,1} & b_{1,2} \\ b_{2,1} & b_{2,2} \end{bsmallmatrix}$. The set $\mathscr{Q}_\mathrm{seq}(\Bs)$ is defined as
\begin{subequations}
\begin{IEEEeqnarray}{rClCl}
	R_1
	&<& \Hc(U_1) - \Hc_{\bs_1^\T}(\uv|Y) &\qquad& \text{(if $b_{1,1} \neq 0$)}   \label{two_user_sequential_decoding_rate_inequality_1} \\
	R_2
	&<& \Hc(U_2) - \Hc_{\bs_1^\T}(\uv|Y) &\qquad& \text{(if $b_{1,2} \neq 0$)}   \label{two_user_sequential_decoding_rate_inequality_2} \\
	R_1
	&<& \Hc(U_1) + \Hc_{\bs_1^\T}(\uv|Y) - \Hc(\uv|Y) &\qquad& \text{(if $b_{2,1} \neq 0$)}   \label{two_user_sequential_decoding_rate_inequality_3} \\
	R_2
	&<& \Hc(U_2) + \Hc_{\bs_1^\T}(\uv|Y) - \Hc(\uv|Y) &\qquad& \text{(if $b_{2,2} \neq 0$).}   \label{two_user_sequential_decoding_rate_inequality_4}
\end{IEEEeqnarray}
\end{subequations}
The resulting sequential decoding rate region $\mathscr{R}_\mathrm{seq}(\As) = \bigcup_{\Bs} \mathscr{Q}_\mathrm{seq}(\Bs)$ [cf.~Definition~\ref{def:sequential_rate_region}] can be shown to simplify to a union of merely four rectangular subsets (at most), as depicted in Figure~\ref{fig:sequential_decoding_rate_region}.

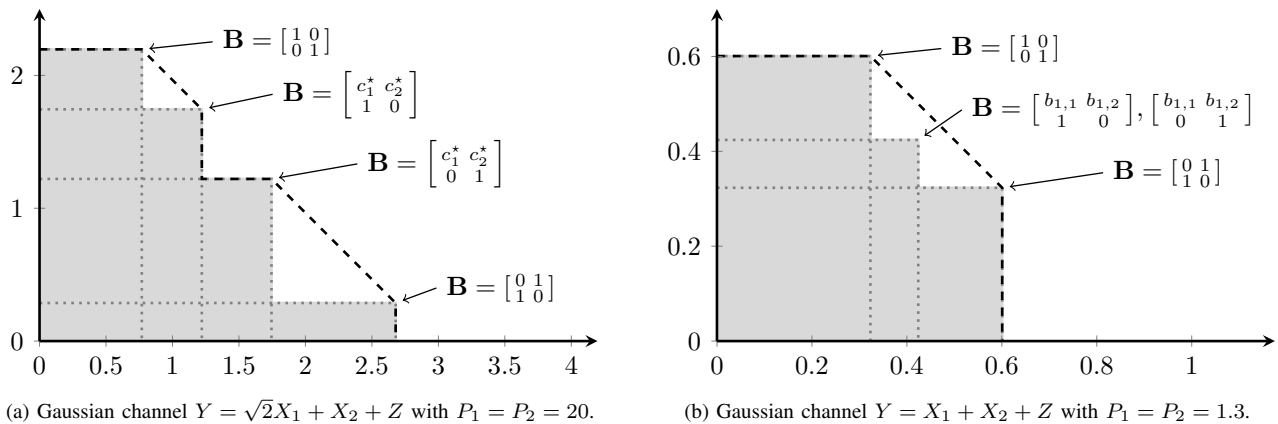
\begin{figure}[ht]
\centering
\subfloat[Gaussian channel $Y = \sqrt{2} X_1 + X_2 + Z$ with $P_1 = P_2 = 20$.]{
\begin{tikzpicture}[every path/.style={line width=1pt}]
\begin{axis}[
	axis equal = true,
	axis on top,
	axis x line = bottom,
	axis y line = left,
	axis line style = {line width=1pt},
	unit rescale keep size = true,
	xmin = 0,
	ymin = 0,
	ymax = 2.5,
	height = 6cm,
	width = 9cm,
	]
\addplot[mark=none] coordinates {(1,1)};
\coordinate (origin) at (axis cs:0,0);
\coordinate (p1) at (axis cs: 0.7692, 2.1962);
\coordinate (p2) at (axis cs: 1.2207, 1.7447);
\coordinate (p3) at (axis cs: 1.7447, 1.2207);
\coordinate (p4) at (axis cs: 2.6788, 0.2866);
\filldraw[black!15] (origin) |- (p1) |- (p2) |- (p3) |- (p4) |- cycle;
\draw[dotted,color=gray] (p1 -| origin) -| (p1 |- origin);
\draw[dotted,color=gray] (p2 -| origin) -| (p2 |- origin);
\draw[dotted,color=gray] (p3 -| origin) -| (p3 |- origin);
\draw[dotted,color=gray] (p4 -| origin) -| (p4 |- origin);
\draw[dashed, line width=1pt] (origin |- p1) -- (p1) -- (p2) |- (p3) -- (p4) -- (origin -| p4);
\node (1) at ($(p1)+2*(9mm,.5mm)$) {$\Bs = \begin{bsmallmatrix} 1 & 0 \\ 0 & 1 \end{bsmallmatrix}$};
\node (2) at ($(p2)+2*(1cm,1mm)$) {$\Bs = \begin{bsmallmatrix} c_1^\star & c_2^\star \\ 1 & 0 \end{bsmallmatrix}$};
\node (3) at ($(p3)+2*(1.1cm,1mm)$) {$\Bs = \begin{bsmallmatrix} c_1^\star & c_2^\star \\ 0 & 1 \end{bsmallmatrix}$};
\node (4) at ($(p4)+2*(7mm,1mm)$) {$\Bs = \begin{bsmallmatrix} 0 & 1 \\ 1 & 0 \end{bsmallmatrix}$};
\draw[line width=.5pt, ->, shorten >=1mm] (1.west) -- (p1);
\draw[line width=.5pt, ->, shorten >=1mm] (2.west) -- (p2);
\draw[line width=.5pt, ->, shorten >=1mm] (3.west) -- (p3);
\draw[line width=.5pt, ->, shorten >=1mm] (4.west) -- (p4);
\end{axis}
\end{tikzpicture}
\label{fig:sequential_decoding_rate_region_1}
}
\hfill
\subfloat[Gaussian channel $Y = X_1 + X_2 + Z$ with $P_1 = P_2 = 1.3$.]{
\begin{tikzpicture}[every path/.style={line width=1pt}]
\begin{axis}[
	axis equal = true,
	axis on top,
	axis x line = bottom,
	axis y line = left,
	axis line style = {line width=1pt},
	unit rescale keep size = true,
	xmin = 0,
	ymin = 0,
	ymax = 0.7,
	height = 6cm,
	width = 9cm,
	]
\addplot[mark=none] coordinates {(0.1,0.1)};
\coordinate (origin) at (axis cs:0,0);
\coordinate (p1) at (axis cs: 0.3232, 0.6008);
\coordinate (p2) at (axis cs: 0.4240, 0.4240);
\coordinate (p3) at (axis cs: 0.6008, 0.3232);
\filldraw[black!15] (origin) |- (p1) |- (p2) |- (p3) |- cycle;
\draw[dotted,color=gray] (p1 -| origin) -| (p1 |- origin);
\draw[dotted,color=gray] (p2 -| origin) -| (p2 |- origin);
\draw[dotted,color=gray] (p3 -| origin) -| (p3 |- origin);
\draw[dashed] (origin |- p1) -- (p1) -- (p3) -- (origin -| p3);
\node (1) at ($(p1)+2*(9mm,.5mm)$) {$\Bs = \begin{bsmallmatrix} 1 & 0 \\ 0 & 1 \end{bsmallmatrix}$};
\node (2) at ($(p2)+2*(1.3cm,2mm)$) {$\Bs = \begin{bsmallmatrix} b_{1,1} & b_{1,2} \\ 1 & 0 \end{bsmallmatrix}, \begin{bsmallmatrix} b_{1,1} & b_{1,2} \\ 0 & 1 \end{bsmallmatrix}$};
\node (3) at ($(p3)+2*(1.1cm,1mm)$) {$\Bs = \begin{bsmallmatrix} 0 & 1 \\ 1 & 0 \end{bsmallmatrix}$};
\draw[line width=.5pt, ->, shorten >=1mm] (1.west) -- (p1);
\draw[line width=.5pt, ->, shorten >=1mm] (2.west) -- (p2);
\draw[line width=.5pt, ->, shorten >=1mm] (3.west) -- (p3);
\end{axis}
\end{tikzpicture}
\label{fig:sequential_decoding_rate_region_2}
}
\caption{Two-user sequential-decoding achievable rate region $\mathscr{R}_\mathrm{seq}(\As)$ (shaded area) for two different Gaussian channels. The set $\mathscr{R}_\mathrm{seq}(\As)$ is expressible as the union of rectangular rate regions $\mathscr{Q}_\mathrm{seq}(\Bs)$ (dotted lines) for different two-by-two full-rank matrices $\Bs$. The simultaneous-decoding rate region $\mathscr{R}_\LMAC = \mathscr{R}(\As) = \mathscr{Q}(\Is)$ [cf.~\eqref{LMAC_characterization}] is shown for reference (dashed line).}
\label{fig:sequential_decoding_rate_region}
\end{figure}

To see why this is true, observe that the first row of any full-rank matrix $\Bs \in \mathbb{A}^{2 \times 2}$ satisfies one out of three possible cases: either $b_{1,1}$ and $b_{1,2}$ are both non-zero, or one of them is zero while the other is non-zero.

\subsubsection{Case $b_{1,1} \neq 0$ and $b_{1,2} = 0$, or $b_{1,1} = 0$ and $b_{1,2} \neq 0$}
We can assume without loss of generality that $\bs_1^\T = [ 1\ 0 ]$ or $\bs_1^\T = [ 0\ 1 ]$, since $\mathscr{Q}_\mathrm{seq}(\Bs)$ is invariant against scaling of any row of $\Bs$. Since $\mathscr{Q}_\mathrm{seq}(\Bs)$, as expressed in~\eqref{two_user_sequential_decoding_rate_inequality_1}--\eqref{two_user_sequential_decoding_rate_inequality_4}, depends on the second row of $\Bs$ only via the conditions under which~\eqref{two_user_sequential_decoding_rate_inequality_3} and \eqref{two_user_sequential_decoding_rate_inequality_4} need to hold, it is clear that setting either $b_{2,1}$ or $b_{2,2}$ to zero will remove one of two constraints, thus potentially enlarging the set $\mathscr{Q}_\mathrm{seq}(\Bs)$. Hence, without loss of generality, it suffices to consider the matrices $\Bs = \begin{bsmallmatrix} 1 & 0 \\ 0 & 1 \end{bsmallmatrix}$ and $\Bs = \begin{bsmallmatrix} 0 & 1 \\ 1 & 0 \end{bsmallmatrix}$.
For $\Bs = \begin{bsmallmatrix} 1 & 0 \\ 0 & 1 \end{bsmallmatrix}$, evaluating~\eqref{two_user_sequential_decoding_rate_inequality_1}--\eqref{two_user_sequential_decoding_rate_inequality_4} reduces to
\begin{subequations}
\begin{IEEEeqnarray}{rCl}
	R_1
	&<& \Hc(U_1) - \Hc(U_1|Y) \IEEEnonumber \\
	&=& I(U_1;Y) \\
	R_2
	&<& \Hc(U_2) + \Hc(U_1|Y) - \Hc(\uv|Y) \IEEEnonumber \\
	&=& I(U_2;Y,U_1)
\end{IEEEeqnarray}
\end{subequations}
which corresponds to the upper corner point of the conventional MAC rate region. Likewise, for the flipped identity matrix $\Bs = \begin{bsmallmatrix} 0 & 1 \\ 1 & 0 \end{bsmallmatrix}$, we obtain the lower corner point,
\begin{subequations}
\begin{IEEEeqnarray}{rCl}
	R_1
	&<& I(U_1;Y,U_2) \\
	R_2
	&<& I(U_2;Y).
\end{IEEEeqnarray}
\end{subequations}

\subsubsection{Case $b_{1,1} \neq 0$ and $b_{1,2} \neq 0$}
By a similar argument as before, $\mathscr{Q}_\mathrm{seq}(\Bs)$ is not decreased by setting either $b_{2,1}$ or $b_{2,2}$ to zero, hence we can consider $\Bs = \begin{bsmallmatrix} b_{1,1} & b_{1,2} \\ 1 & 0 \end{bsmallmatrix}$ or $\Bs = \begin{bsmallmatrix} b_{1,1} & b_{1,2} \\ 0 & 1 \end{bsmallmatrix}$ without loss of generality. We again distinguish two cases, depending on whether $\mathscr{R}_\LMAC \subsetneq \mathscr{R}_\MAC$ or $\mathscr{R}_\LMAC = \mathscr{R}_\MAC$ [cf.~Lemma~\ref{lem:notch}]:
\begin{itemize}
    \item   There exists a coefficient vector with non-zero entries $\cs^\T = [ c_1 \ c_2 ]$ such that~\eqref{tip_outside_condition} holds;
    \item   There exists no such coefficient vector.
\end{itemize}

\paragraph{Case $\mathscr{R}_\LMAC \subsetneq \mathscr{R}_\MAC$}

Consider the sets $\mathscr{Q}\bigl(\bigl[\begin{smallmatrix} c_1 & c_2 \\ 1 & 0 \end{smallmatrix}\bigr]\bigr)$ and $\mathscr{Q}\bigl(\bigl[\begin{smallmatrix} c_1 & c_2 \\ 0 & 1 \end{smallmatrix}\bigr]\bigr)$, which are delimited respectively by inequalities~\eqref{two_user_sequential_decoding_rate_inequality_1}, \eqref{two_user_sequential_decoding_rate_inequality_2}, \eqref{two_user_sequential_decoding_rate_inequality_3} for the former, and \eqref{two_user_sequential_decoding_rate_inequality_1}, \eqref{two_user_sequential_decoding_rate_inequality_2}, \eqref{two_user_sequential_decoding_rate_inequality_4} for the latter. Due to the inequality~\eqref{tip_outside_condition} holding, the right-hand sides of~\eqref{two_user_sequential_decoding_rate_inequality_1} and \eqref{two_user_sequential_decoding_rate_inequality_2} are larger than the right-hand sides of~\eqref{two_user_sequential_decoding_rate_inequality_3} and \eqref{two_user_sequential_decoding_rate_inequality_4}, respectively. Hence, the set $\mathscr{Q}\bigl(\bigl[\begin{smallmatrix} c_1 & c_2 \\ 1 & 0 \end{smallmatrix}\bigr]\bigr)$ is delimited by the inequalities
\begin{subequations}
\begin{IEEEeqnarray}{rCl}
	R_1
	&<& \Hc(U_1) + \Hc_{\cs^\T}(\uv|Y) - \Hc(\uv|Y)   \label{midway_dominant_face_point_1} \\
	R_2
	&<& \Hc(U_2) - \Hc_{\cs^\T}(\uv|Y)   \label{midway_dominant_face_point_2}
\end{IEEEeqnarray}
\end{subequations}
whereas the set $\mathscr{Q}\bigl(\bigl[\begin{smallmatrix} c_1 & c_2 \\ 0 & 1 \end{smallmatrix}\bigr]\bigr)$ is delimited by the inequalities
\begin{subequations}
\begin{IEEEeqnarray}{rCl}
	R_1
	&<& \Hc(U_1) - \Hc_{\cs^\T}(\uv|Y)   \label{midway_dominant_face_point_3} \\
	R_2
	&<& \Hc(U_2) + \Hc_{\cs^\T}(\uv|Y) - \Hc(\uv|Y).   \label{midway_dominant_face_point_4}
\end{IEEEeqnarray}
\end{subequations}
Clearly, by summing together both rate constraints~\eqref{midway_dominant_face_point_1}--\eqref{midway_dominant_face_point_2} on the one hand, and~\eqref{midway_dominant_face_point_3}--\eqref{midway_dominant_face_point_4} on the other hand, we see that these rate constraints yield two points located on the dominant face $R_1 + R_2 < I(\uv;Y)$ of the multiple-access channel, as can be observed on Figure~\ref{fig:sequential_decoding_rate_region_1}.

\paragraph{Case $\mathscr{R}_\LMAC = \mathscr{R}_\MAC$}

Consider the sets $\mathscr{Q}\bigl(\bigl[\begin{smallmatrix} b_{1,1} & b_{1,2} \\ 1 & 0 \end{smallmatrix}\bigr]\bigr)$ and $\mathscr{Q}\bigl(\bigl[\begin{smallmatrix} b_{1,1} & b_{1,2} \\ 0 & 1 \end{smallmatrix}\bigr]\bigr)$ with some $\bs_1^\T = [b_{1,1}\ b_{1,2}]$ having non-zero entries, these sets being delimited respectively by inequalities~\eqref{two_user_sequential_decoding_rate_inequality_1}, \eqref{two_user_sequential_decoding_rate_inequality_2}, \eqref{two_user_sequential_decoding_rate_inequality_3} for the former, and \eqref{two_user_sequential_decoding_rate_inequality_1}, \eqref{two_user_sequential_decoding_rate_inequality_2}, \eqref{two_user_sequential_decoding_rate_inequality_4} for the latter. Due to the inequality~\eqref{tip_outside_condition} \emph{not} holding, the right-hand sides of~\eqref{two_user_sequential_decoding_rate_inequality_1} and \eqref{two_user_sequential_decoding_rate_inequality_2} are smaller than or equal to the right-hand sides of~\eqref{two_user_sequential_decoding_rate_inequality_3} and \eqref{two_user_sequential_decoding_rate_inequality_4}, respectively. Hence, the sets $\mathscr{Q}\bigl(\bigl[\begin{smallmatrix} b_{1,1} & b_{1,2} \\ 1 & 0 \end{smallmatrix}\bigr]\bigr)$ and $\mathscr{Q}\bigl(\bigl[\begin{smallmatrix} b_{1,1} & b_{1,2} \\ 1 & 0 \end{smallmatrix}\bigr]\bigr)$ are delimited by the same two inequalities~\eqref{two_user_sequential_decoding_rate_inequality_1}--\eqref{two_user_sequential_decoding_rate_inequality_2}, namely,
\begin{subequations}
\begin{IEEEeqnarray}{rClCl}
	R_1
	&<& \Hc(U_1) - \Hc_{\bs_1^\T}(\uv|Y) &\qquad& \text{(if $b_{1,1} \neq 0$)} \\
	R_2
	&<& \Hc(U_2) - \Hc_{\bs_1^\T}(\uv|Y) &\qquad& \text{(if $b_{1,2} \neq 0$)}.
\end{IEEEeqnarray}
\end{subequations}
This leads to a rate region like the one depicted on Figure~\ref{fig:sequential_decoding_rate_region_2}.

}

{
\subsection{Decoding a single linear combination with two users}   \label{subsec:single-equation_decoding}
For the special case of $K=2$ users and $L=1$ linear combinations, the matrix $\As$ reduces to a one-by-two row vector $\as^\T = [ a_1 \ a_2 ]$. Since there is only a single linear combination to recover, the rate regions under simultaneous decoding and sequential decoding coincide, i.e., $\mathscr{R}(\as^\T) = \mathscr{R}_\mathrm{seq}(\as^\T)$. Using Lemma~\ref{lem:LMAC_simplification} and the identity~\eqref{LMAC_characterization}, one can immediately state the rate region $\mathscr{R}(\as^\T)$ as a union of two sets:
\begin{equation}
	\mathscr{R}(\as^\T)
	= \mathscr{Q}(\Is) \cup \mathscr{Q}(\as^\T)
	= \mathscr{R}_\LMAC \cup \mathscr{Q}(\as^\T).   \IEEEeqnarraynumspace\label{eq:two_user_one_equation_general_coefficients}
\end{equation}
Here, the set $\mathscr{Q}(\as^\T)$ can be expressed as [cf.~\eqref{two_user_CF_region}]
\begin{subequations}
\begin{IEEEeqnarray}{rCl}
	\mathscr{Q}(\as^\T)
	= \Bigl\{ (R_1, R_2) \in \mathbb{R}_+^2 \colon
	R_1 &<& \Hc(U_1) - \Hc_{\as^\T}(\uv|Y)   \label{CF_eq1} \\
	R_2 &<& \Hc(U_2) - \Hc_{\as^\T}(\uv|Y) \Bigr\}.   \label{CF_eq2} 
\end{IEEEeqnarray}
\end{subequations}

\lemmabox{
\begin{lemma}   \label{lem:CF_simplification}
For $(K,L)=(2,1)$, the rate region $\mathscr{R}(\as^\T)$ can be expressed as follows:
\begin{itemize}
	\item	If $\as^\T$ (with non-zero entries) satisfies
		\begin{equation}   \label{tip_outside_condition_bis}
    		2 \Hc_{\as^\T}(\uv|Y)
			< \Hc(\uv|Y)
		\end{equation}
then
		\begin{equation}   \label{eq:two_user_one_equation_simplified_1}
			\mathscr{R}(\as^\T)
			= \mathscr{R}_\MAC \cup \mathscr{Q}(\as^\T).
		\end{equation}
	\item	Otherwise,
		\begin{equation}   \label{eq:two_user_one_equation_simplified_2}
			\mathscr{R}(\as^\T)
			= \mathscr{R}_\LMAC.
		\end{equation}
\end{itemize}
\end{lemma}
}

\proofbox{}{
\begin{IEEEproof}
The proof is deferred to Appendix~\ref{app:proof:CF_simplification}.
\end{IEEEproof}
}

The geometric interpretation of Lemma~\ref{lem:CF_simplification} is shown in Figure~\ref{fig:two-user_rate_region}, where we depict the shapes of the sets $\mathscr{R}_\LMAC$ and $\mathscr{Q}(\as^\T)$ for the two different cases determined by whether~\eqref{tip_outside_condition_bis} holds or not.

\begin{figure}[ht]
\centering
\subfloat[Condition~\eqref{tip_outside_condition_bis} does not hold]{
\begin{tikzpicture}[scale=1.7, line width=1pt]
\filldraw[black!15] (0,1.9) -- (1.3,1.9) -- (2.2,1) -- (2.2,0) -| cycle;
\draw[black!50] (0,1.9) -- (1.3,1.9) -- (2.2,1) -- (2.2,0);
\draw[dashed] (0,1.9) -- (1.3,1.9) -- (2.2,1) -- (2.2,0);
\draw[dotted,gray] (0,1.3) -| (1.7,0);
\draw[<->] (0,2.3) node[right] {$R_2$} |- (2.5,0) node[right] {$R_1$};
\end{tikzpicture}
\label{fig:R_LMAC_1}
}
\hspace{5mm}
\begin{tikzpicture}[scale=1.7, line width=1pt]
\draw[black!50] (0,0) --+ (5mm,0) node[black,right] {$\mathscr{R}_\MAC$};
\draw[dashed] (0,-5mm) --+ (5mm,0) node[black,right] {$\mathscr{R}_\LMAC$};
\draw[dotted,gray] (0,-10mm) --+ (5mm,0) node[black,right] {$\mathscr{Q}(\as^\T)$};
\filldraw[black!15] (0,-13mm) rectangle (5mm,-17mm) node[yshift=3mm,black,right] {$\mathscr{R}(\as^\T)$};
\end{tikzpicture}
\hspace{5mm}
\subfloat[Condition~\eqref{tip_outside_condition_bis} holds]{
\begin{tikzpicture}[scale=1.7, line width=1pt]
\filldraw[black!15] (0,2) -- (1.3,2) -- (1.5,1.8) -| (2.2,1.1) -- (2.3,1) -- (2.3,0) -| cycle;
\draw[black!50] (0,2) -- (1.3,2) -- (2.3,1) -- (2.3,0);
\draw[dashed] (0,2) -- (1.3,2) -- (1.5,1.8) |- (2.2,1.1) -- (2.3,1) -- (2.3,0);
\draw[dotted,gray] (0,1.8) -| (2.2,0);
\draw[<->] (0,2.3) node[right] {$R_2$} |- (2.5,0) node[right] {$R_1$};
\end{tikzpicture}
\label{fig:R_LMAC_2}
}
\caption{Shapes of the sets $\mathscr{R}_\LMAC$ (dashed) and $\mathscr{Q}(\as^\T)$ (dotted), as well as the MAC capacity region $\mathscr{R}_\MAC$ (solid). These sets can be assembled by union operations to obtain $\mathscr{R}(\as^\T)$ (shaded area).}
\label{fig:two-user_rate_region}
\end{figure}
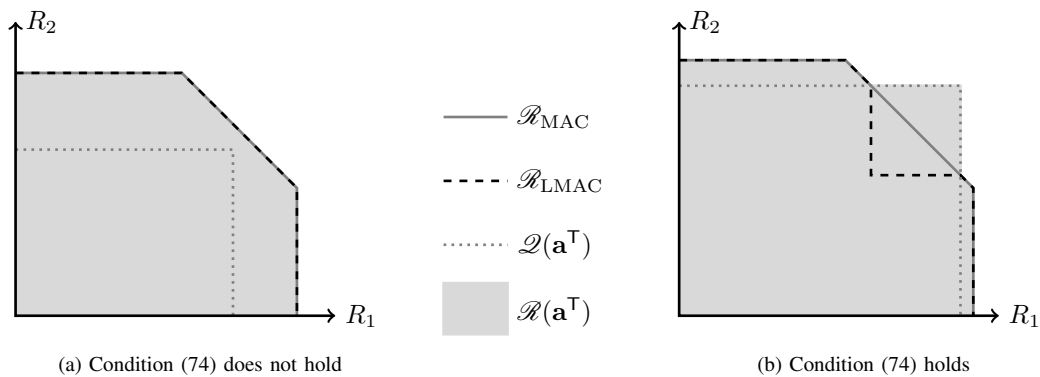

Lemma~\ref{lem:CF_simplification} states that if the coefficients $\as^\T$ and the distribution of $(\uv,Y)$ are such that the inequality~\eqref{tip_outside_condition_bis} is satisfied, the set $\mathscr{R}_\LMAC$ that appears in the (more generally valid) identity~\eqref{eq:two_user_one_equation_general_coefficients} can be swapped out for $\mathscr{R}_\MAC$. On the contrary, if the reverse of~\eqref{tip_outside_condition_bis} holds, then it is the union operation with $\mathscr{Q}(\as^\T)$, in this same identity~\eqref{eq:two_user_one_equation_general_coefficients}, that becomes redundant and can thus be removed.

}

{
\subsection{Gaussian channels}   \label{subsec:Gaussian_channels}
Arguably the most important special case of Theorem~\ref{thm:continuous_CF} is its evaluation for Gaussian distributions. Consider a multiple access channel with $M$ receiver antennas which obeys the system equation
\begin{equation}   \label{eq:GMAC}
	\yv = \Hs\xv + \zv
\end{equation}
where $\xv = \bigl[ X_1, \dotsc, X_K \bigr]^\T$ represents the vector or channel inputs, $\Hs \in \mathbb{R}^{M \times K}$ stands for the channel gain matrix, $\yv \in \mathbb{R}^M$ is the vector of channel outputs and $\zv \sim \Nc_\mathbb{R}(\mathbf{0},\Is)$ is i.i.d.~additive Gaussian noise.
We assume that the channel inputs are subject to average power constraints $\E[X_k^2] \leq P_k$, $k=1,\dotsc,K$. We define $\Ps = \diag(P_1,\dotsc,P_K)$ as the covariance matrix of $\xv$.
 
The following corollary to Theorem~\ref{thm:continuous_CF} provides a generalization of the compute--forward rate region from~\cite{NaGa11} for the Gaussian channel~\eqref{eq:GMAC}, to the effect of simultaneously computing an \emph{arbitrary} number of linearly independent combinations of Gaussian codewords, rather than only a single linear combination. This improves on the best-known rate region from prior work, which was based on nested lattice encoding and sequential decoding~\cite{NaCaNtCa16}.

\theorembox{\begin{corollary}[Gaussian compute--forward]   \label{cor:gaussian_CF}
Let $(\mathbb{U},\mathbb{A}) = (\mathbb{R},\mathbb{Z})$. We evaluate Theorem~\ref{thm:continuous_CF} for the Gaussian channel~\eqref{eq:GMAC}, auxiliary variables $U_k \sim \Nc_\mathbb{R}(0,\beta_k^2 P_k)$ with scaling parameters $\beta_k > 0$ and modulation mappings $X_k = U_k/\beta_k$ to satisfy the power constraints $\E[X_k^2] = P_k$.
Then, $\mathscr{Q}(\Bs,M,\Tc)$ in Theorem~\ref{thm:continuous_CF} specializes to the set of rate tuples $(R_1, \dotsc, R_K)$ such that
\begin{align} \label{eq:gaussian-rate}
	\sum_{k\in\Tc} R_k
	< \frac{1}{2}\sum_{k\in\Tc} \log(\beta_k^2 P_k) - \frac{1}{2}\log\frac{\det\bigl(\Bs\Ks\Bs^\T\bigr)}{\det(\Sigmas(\Bs))^2} + J(\Bs,M)
\end{align}
where $\Ks$ stands for the conditional covariance matrix
\begin{equation}
	\Ks
	= \E\bigl[ \uv\uv^\T \big| Y\bigr]
	= \diag(\betas) \left(\Ps^{-1}+\Hs^\T\Hs\right)^{-1} \diag(\betas)
\end{equation}
and where
\begin{equation}   \label{J_Gaussian}
	J(\Bs,M)
	= \inf_{\Cs \in \mathscr{C}_\mathbb{Z}(M)} \frac{1}{2}\log\frac{\det\bigl(\Cs\Bs\Ks\Bs^\T\Cs^\T\bigr)}{\det(\Sigmas(\Cs\Bs))^2}.
\end{equation}
\end{corollary}}

Note that Corollary~\ref{cor:gaussian_CF} is obtained directly from evaluating~\eqref{Q_partial}--\eqref{def:R} by means of the Gaussian entropy formula and~\eqref{algebraic_entropy_for_full_rank_Q}. As has been argued in Remark~\ref{rmk:C_restriction}, the matrix $\Cs$ in the infimum~\eqref{J_Gaussian} can be restricted to right-invertible. If, in addition, $\Bs$ is also right-invertible, then both $\Bs$ and $\Cs\Bs$ are right-invertible and thus have elementary divisors all equal to one, hence the denominators on the right-hand sides of~\eqref{eq:gaussian-rate} and \eqref{J_Gaussian} disappear, thus leading to simpler expressions. Note, however, that we have no proof nor disproof\footnote{Except for the two-user case $K=2$, for which we can show that the two right-invertible matrices $\Bs = \begin{bsmallmatrix} 1 & 0 \\ 0 & 1 \end{bsmallmatrix}$ [cf.~Subsection~\ref{subsec:two-user_case_simultaneous_joint_decoding}] and $\Bs = [b_1 \ b_2]$ (with coprime $b_1$ and $b_2$) [cf.~Subsection~\ref{subsec:single-equation_decoding}] suffice to attain the entire rate region $\mathscr{R}(\As)$.} that in the union over matrices $\Bs$ in~\eqref{def:R}, $\Bs$ can be restricted to being right-invertible without loss of optimality. This stands as an open problem.

Finally, let us particularize~\eqref{eq:gaussian-rate} to the case of two users ($K=2$) and a single linear combination ($L=1$), setting all parameters $\beta_k$ to one and equal power constraints $P_1 = P_2 = P$, and evaluating the union~\eqref{def:R} only for $\Bs = [a_1\ a_2]$. Denote the latter row vector as $\as^\T$ and the channel matrix $\Hs \in \mathbb{R}^{1 \times 2}$ as the row vector $\hs^\T$.
With these choices, we can readily recover the well-known compute--forward rectangular rate region due to Nazer and Gastpar~\cite{NaGa11}:
\begin{equation}
    \max\{R_1,R_2\}
	< \frac{1}{2} \log\left(\frac{P}{\as^\T\left(P^{-1}\Is+\hb\hb^\T\right)^{-1}\as}\right) + \log\gcd{(|a_1|,|a_2|)}.
\end{equation}
We observe how in this special case, the term $\Sigmas(\Bs)$ from~\eqref{eq:gaussian-rate}, which here reduces to a scalar, evaluates to the greatest common divisor of $|a_2|$ and $|a_2|$.
Also note that for $L=1$, the term $J(\Bs,M)$ disappears, which further contributes to simplification.

\theorembox{
\begin{corollary}[Gaussian compute--forward under sequential decoding]   \label{cor:gaussian_sequential}
Let $(\mathbb{U},\mathbb{A}) = (\mathbb{R},\mathbb{Z})$. We evaluate Theorem~\ref{thm:continuous_CF_sequential} for the Gaussian channel~\eqref{eq:GMAC}, auxiliary variables $U_k \sim \Nc_\mathbb{R}(0,\beta_k^2 P_k)$ with scaling parameters $\beta_k > 0$ and modulation mappings $X_k = U_k/\beta_k$ to satisfy the power constraints $\E[X_k^2] = P_k$.
Then, $\mathscr{Q}_\mathrm{seq}(\Bs)$ in Theorem~\ref{thm:continuous_CF_sequential} specializes to the set of rate tuples $(R_1, \dotsc, R_K)$ such that
\begin{equation}    \label{gaussian_sequential}
	R_k
	< \frac{1}{2}\log(\beta_k^2 P_k) - \frac{1}{2}\log\frac{\det\bigl(\Bs_{[j]}\Ks\Bs_{[j]}^\T\bigr)}{\det(\Sigmas(\Bs_{[j]}))^2} + \frac{1}{2}\log\frac{\det\bigl(\Bs_{[j-1]}\Ks\Bs_{[j-1]}^\T\bigr)}{\det(\Sigmas(\Bs_{[j-1]}))^2}.
\end{equation}
\end{corollary}
}

}

{
In the following, we provide some numerical evaluations for the Gaussian case in the context of Corollary~\ref{cor:gaussian_CF}.

\begin{example}   \label{example:gaussian}
Consider a $K = 3$ user Gaussian multiple-access channel with $M = 1$ receive antenna and channel gain matrix $\Hs = [1\ 1\ 1]$, symmetric transmit power $P = 3$ (for all users), and $\As = [1\ 1\ 1]$. The rate region from Corollary~\ref{cor:gaussian_CF} is shown in Figure~\ref{fig:gmac}. The sequential decoding points in Corollary~\ref{cor:gaussian_sequential} with $\Bs = [1\ 1\ 1]$ and 
\begin{equation}   
	\Bs = 
	\begin{bmatrix}
		1 & 0 & 0 \\
		0 & 1 & 1 
	\end{bmatrix},
\end{equation} 
are labelled as $A$ and $B$, respectively.
\end{example}

\begin{figure}[ht!]
\begin{center}
	\includegraphics[width=0.5\textwidth]{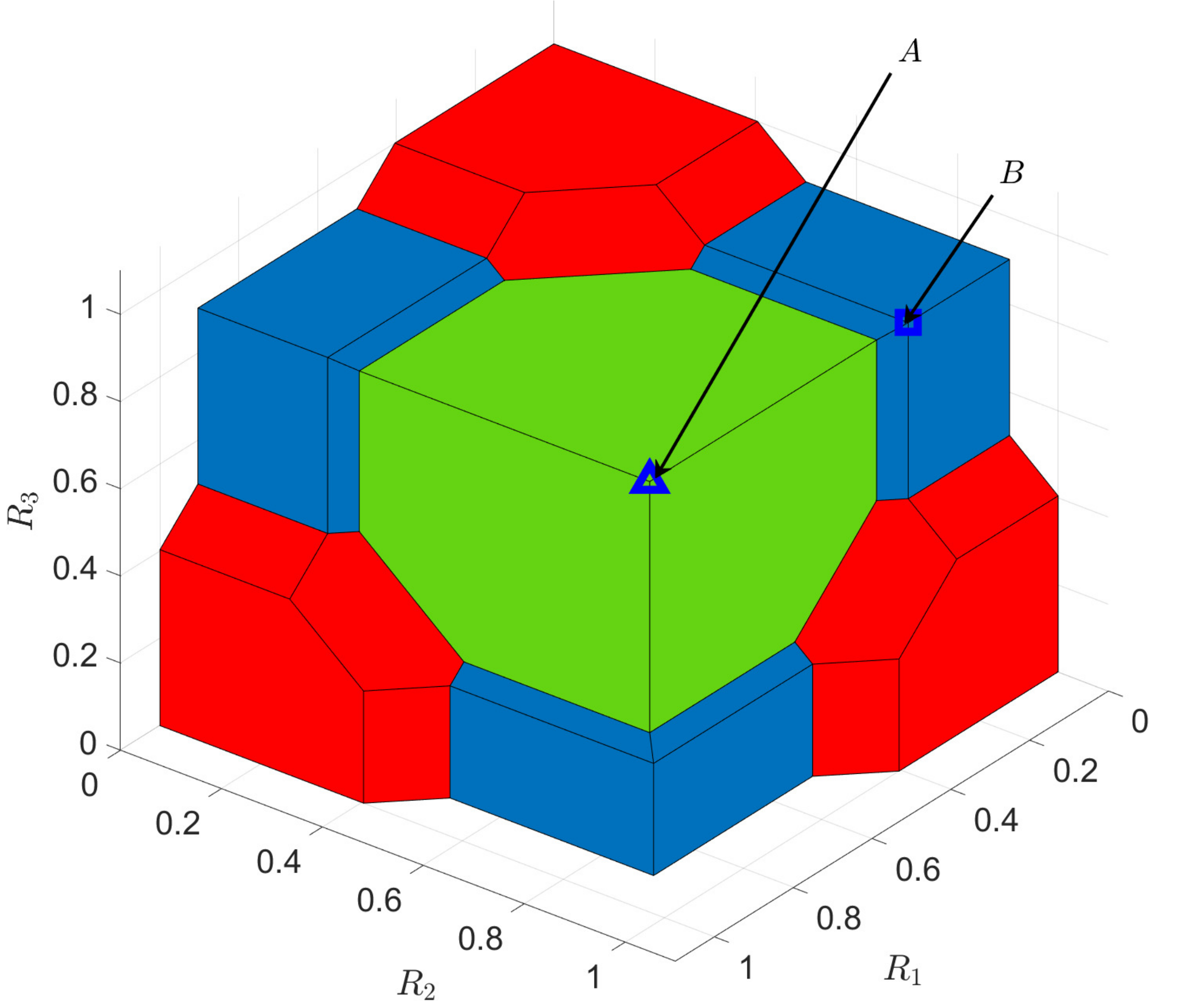}%
	\caption{Three-user simultaneous decoding rate region from Example~\ref{example:gaussian}.}
	\label{fig:gmac}
\end{center}
\end{figure}

\begin{example}   \label{example:gaussian-mimo}
Consider a $K=3$ user Gaussian multiple-access channel with $M=3$ receive antennas and channel gain matrix
\begin{equation}
	\mathbf{H} = 
	\begin{bmatrix}
		1 & 1.5 & 0.75 \\
		0.75 & 1 & 1.5 \\
		1.5 & 0.75 & 1
	\end{bmatrix},
\end{equation} 
symmetric transmit power $P=2$, and $\As = \left[ 1 \ 1 \ 1 \right]$. Then, the rate region in Corollary~\ref{cor:gaussian_CF} is evaluated in Figure~\ref{fig:gmac-MIMO}. The sequential decoding points in Corollary~\ref{cor:gaussian_sequential} with $\Bs = \left[ 1 \ 1 \ 1 \right]$ and 
\begin{equation}
	\Bs = 
	\begin{bmatrix}
		1 & 0 & 0 \\
		0 & 1 & 1 
	\end{bmatrix},
\end{equation} 
are marked as $A$ and $B$, respectively.
\end{example}

\begin{figure}[ht!]
\begin{center}
	\includegraphics[width=0.5\textwidth]{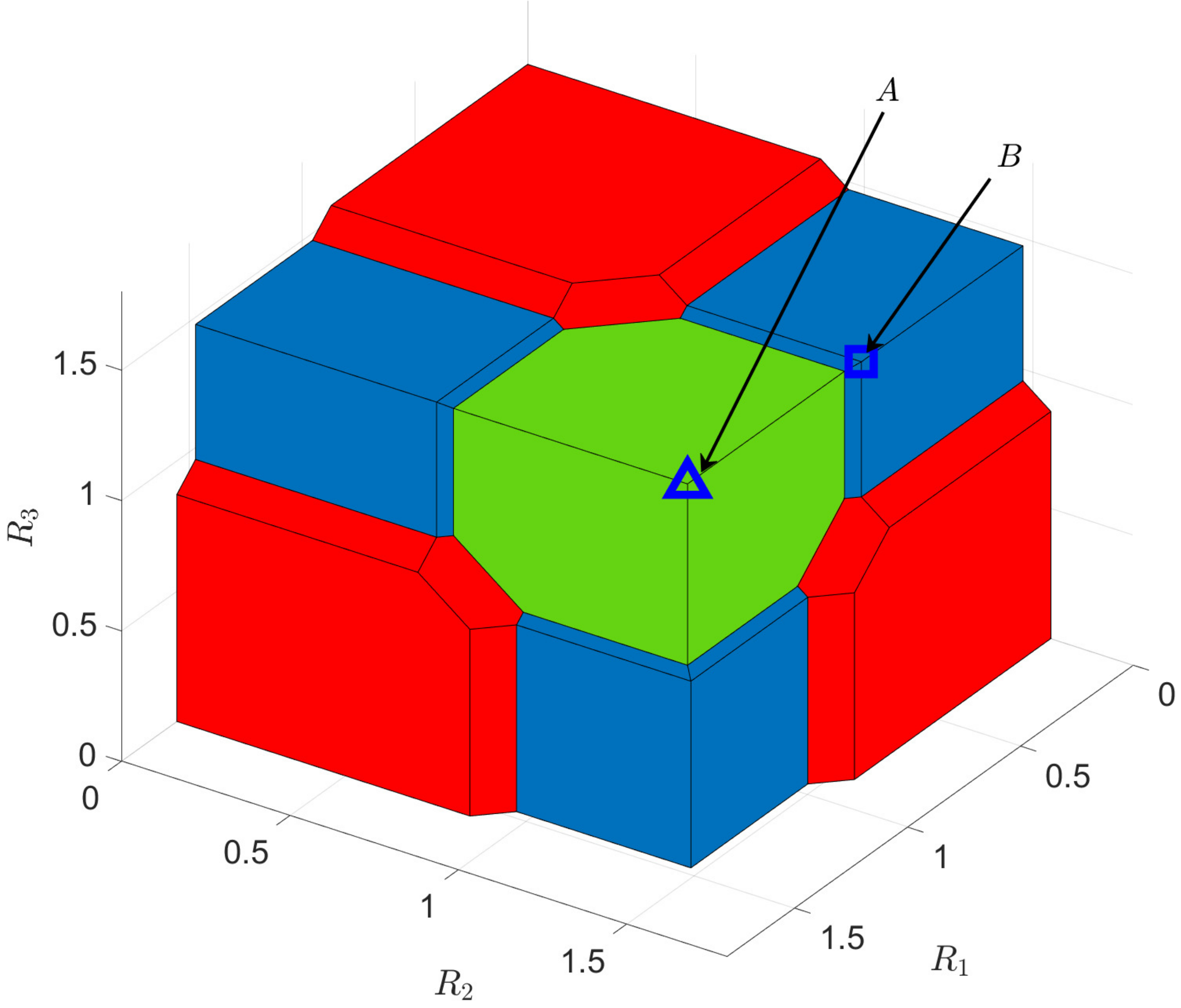}%
	\caption{Three-user simultaneous decoding rate region from Example~\ref{example:gaussian-mimo}.}
	\label{fig:gmac-MIMO}
\end{center}
\end{figure}

}

}

{
\section{Conclusion}   \label{sec:conclusion}
In this paper, we proposed a discretization approach to translate compute--forward achievability results from discrete to continuous-valued channels. This enabled us to obtain a simultaneous-decoding, compute--forward rate region for a class of continuous channels, including those of the form $Y = \sum_{k}h _k X_k + Z$ for some independent additive noise. As a special case, we obtained the simultaneous-decoding rate region for Gaussian channels, which had been an open question. Since these rate regions are expressed in terms of differences of entropies, special care was needed to overcome the fact that differential entropy is not scale-invariant. To this end, we introduced algebraic generalizations of R\'enyi's information dimension and $d$-dimensional entropy, and used these to obtain a single, unified expression for both discrete and continuous compute--forward rate regions. Our discretization approach is well-suited for generalizing other algebraic network information theory results from discrete to continuous-valued channels, and many of our supporting lemmas may be directly employed to these problems. 

For two users, it is now known that the simultaneous-decoding, compute--forward rate region for discrete memoryless channels is optimal, under the critical assumption that the transmitters employ random nested linear codes~\cite{SeLiKi20}. An open question is whether this result can be extended to $K$ users for both the discrete and the Gaussian setting. Another interesting direction for future study is to develop an analogue of an information-density decoder (see, e.g.,~\cite[Theorem 18.5]{PoWu}) and use it to directly derive compute--forward rate regions for a broader class of channels.

}

\section*{Acknowledgment}
The authors would like to thank Tobias Koch for helpful discussions on Lemma~\ref{lem:truncation}.
The work of Adriano Pastore was supported in part by the Spanish Ministry of Economy and Competitiveness through project RTI2018-099722-B-I00 (ARISTIDES).
The work of Sung Hoon Lim was supported by the National Research Foundation of Korea (NRF) under Grant NRF-2020R1F1A1074926.
The work of Chen Feng was supported by NSERC Discovery Grant RGPIN-2016-05310.
The work of Bobak Nazer was supported by National Science Foundation Grant CCF-1618800.
The work of Michael Gastpar was supported in part by the Swiss National Science Foundation under Grant 200364.

\appendices

{
\section{Proof of Lemma~\ref{lem:chain_rule}}   \label{app:proof:chain_rule}
Given that $I(\uv_1,\uv_2;Y)$ is finite by assumption, the limit
\begin{IEEEeqnarray*}{rCl}
	\lim_{\nu \to \infty} \Bigl\{ H(\floor{\nu\uv_2}|Y) - H(\floor{\nu\uv_2}|Y,\floor{\nu\uv_1}) \Bigr\}
	&=& \lim_{\nu \to \infty} I(\floor{\nu\uv_2};Y,\floor{\nu\uv_1}) \\
	&=& I(\uv_2;Y,\uv_1)   \IEEEeqnarraynumspace\IEEEyesnumber
\end{IEEEeqnarray*}
is finite too, since it is upper-bounded by $I(\uv_1,\uv_2;Y)$. The last equality is due to \cite[Lem.~7.18]{Gr11}.
Likewise, we have
\begin{IEEEeqnarray*}{rCl}
	\lim_{\nu \to \infty} \Bigl\{ H(\floor{\nu\uv_2}|Y) - H(\floor{\nu\uv_2}|Y,\uv_1) \Bigr\}
	&=& \lim_{\nu \to \infty} I(\floor{\nu\uv_2};Y,\uv_1) \\
	&=& I(\uv_2;Y,\uv_1).   \IEEEeqnarraynumspace\IEEEyesnumber
\end{IEEEeqnarray*}
Combining the above two limits by subtraction, we get
\begin{IEEEeqnarray*}{rCl}
	\lim_{\nu \to \infty} \Bigl\{ H(\floor{\nu\uv_2}|Y,\floor{\nu\uv_1}) - H(\floor{\nu\uv_2}|Y,\uv_1) \Bigr\}
	&=& 0.   \IEEEeqnarraynumspace\IEEEyesnumber\label{quantization_in_condition}
\end{IEEEeqnarray*}
If the information dimensions $d(\uv_1|Y)$ and $d(\uv_2|Y,\uv_1)$ exist, then the claimed chain rule for information dimension is an immediate consequence of the chain rule for discrete entropy and of~\eqref{quantization_in_condition}:
\begin{IEEEeqnarray*}{rCl}
	d(\uv_1,\uv_2|Y)
	&=& \lim_{\nu \to \infty} \frac{H(\floor{\nu\uv_1}|Y) + H(\floor{\nu\uv_2}|Y,\floor{\nu\uv_1})}{\log(\nu)} \\
	&=& \lim_{\nu \to \infty} \biggl\{ \frac{H(\floor{\nu\uv_1}|Y)}{\log(\nu)} + \frac{H(\floor{\nu\uv_2}|Y,\uv_1)}{\log(\nu)} \biggr\} \\
	&=& d(\uv_1|Y) + d(\uv_2|Y,\uv_1).   \IEEEeqnarraynumspace\IEEEyesnumber
\end{IEEEeqnarray*}
Similarly, for the entropy $\Hc(\uv_1,\uv_2|Y)$ we have
\begin{IEEEeqnarray*}{rCl}
	\Hc(\uv_1,\uv_2|Y)
	&=& \lim_{\nu \to \infty} H(\floor{\nu\uv_1}|Y) - d(\uv_1|Y)\log(\nu) + H(\floor{\nu\uv_2}|Y,\floor{\nu\uv_1}) - d(\uv_2|Y,\uv_1)\log(\nu) \\
	&=& \lim_{\nu \to \infty} H(\floor{\nu\uv_1}|Y) - d(\uv_1|Y)\log(\nu) + H(\floor{\nu\uv_2}|Y,\nu\uv_1) - d(\uv_2|Y,\uv_1)\log(\nu) \\
	&=& \Hc(\uv_1|Y) + \Hc(\uv_2|Y,\uv_1).   \IEEEeqnarraynumspace\IEEEyesnumber
\end{IEEEeqnarray*}
This concludes the proof.

}

{
\section{Proof of Lemma~\ref{lem:mutual_information}}   \label{app:proof:mutual_information}
First note that due to~\cite[Lem.~7.18]{Gr11}
\begin{equation}
	\lim_{\nu \to \infty} I(\floor{\nu\uv};Y)
	= I(\uv;Y)
\end{equation}
being finite, we have
\begin{IEEEeqnarray*}{rCl}
	d(\uv|Y)
	&=& \lim_{\nu \to \infty} \frac{H(\floor{\nu\uv}|Y)}{\log(\nu)} \\
	&=& \lim_{\nu \to \infty} \frac{H(\floor{\nu\uv}|Y) + I(\floor{\nu\uv};Y) - I(\uv;Y)}{\log(\nu)} \\
	&=& \lim_{\nu \to \infty} \frac{H(\floor{\nu\uv}) - I(\uv;Y)}{\log(\nu)} \\
	&=& \lim_{\nu \to \infty} \frac{H(\floor{\nu\uv})}{\log(\nu)} \\
	&=& d(\uv).   \IEEEeqnarraynumspace\IEEEyesnumber
\end{IEEEeqnarray*}
Hence,
\begin{IEEEeqnarray*}{rCl}
	\Hc(\uv) - \Hc(\uv|Y)
	&=& \lim_{\nu \to \infty} \Bigl\{ H(\floor{\nu\uv}) - d(\uv)\log(\nu) - H(\floor{\nu\uv}|Y) + d(\uv|Y)\log(\nu) \Bigr\} \\
	&=& \lim_{\nu \to \infty} I(\floor{\nu\uv};Y) \\
	&=& I(\uv;Y).   \IEEEeqnarraynumspace\IEEEyesnumber
\end{IEEEeqnarray*}
This concludes the proof.

}

{
\section{Proof of Lemma~\ref{lem:discrete_algebraic_entropy}}   \label{app:proof:discrete_algebraic_entropy}
Since $H(\Qs\lfloor \nu \uv \rfloor) \leq H(\uv)$, clearly $d_\Qs(\uv) = 0$.
By the definition of algebraic entropy, $\Hc_\Qs(\uv) = \liminf_{\nu \to \infty} H(\Qs\lfloor \nu \uv \rfloor)$.
Defining $\dv_\nu = \Qs\lfloor \nu \uv \rfloor - \lfloor \nu \Qs\uv \rfloor$ we have
\begin{IEEEeqnarray*}{rCl}
	H(\Qs\lfloor \nu \uv \rfloor)
	&=& H(\lfloor \nu \Qs\uv \rfloor + \dv_\nu) \\
	&=& H\bigl(\Qs\uv + \tfrac{1}{\nu}\dv_\nu\bigr).
\end{IEEEeqnarray*}
Denote the $(i,j)$-th entry of $\Qs$ as $Q_{i,j}$, and the $i$-th entry of $\dv_\nu$ as $[\dv_\nu]_i$. Since $\dv_\nu$ can be written as
\begin{equation}
	\dv_\nu
	= \lfloor \Qs(\lfloor \nu\uv \rfloor - \nu\uv) \rfloor
\end{equation}
we can readily derive the bound $\bigl| [\dv_\nu]_i \bigr| \leq \sum_{j=1}^n |Q_{i,j}|$. Hence, for $\nu > 2 \max_{i} \sum_{j=1}^n |Q_{i,j}|$, the support of $\tfrac{1}{\nu}\dv_\nu$ is entirely contained in the open cube $( -\tfrac{1}{2}, \tfrac{1}{2} )^m$. In these circumstances, the sum $\Qs\uv + \tfrac{1}{\nu}\dv_\nu$ can be one-to-one mapped to the pair $(\Qs\uv,\dv_\nu)$. Therefore, if $\nu > 2 \max_{i} \sum_{j=1}^n |Q_{i,j}|$, then
\begin{IEEEeqnarray*}{rCl}
	H(\Qs\lfloor \nu \uv \rfloor)
	&=& H(\Qs\uv,\dv_\nu) \\
	&\geq& H(\Qs\uv).
\end{IEEEeqnarray*}
Hence $\Hc_\Qs(\uv) \geq H(\Qs\uv)$. On the other hand, if $\nu$ is integer, we have $\lfloor \nu \uv \rfloor = \nu \uv$, since $\uv$ is integer. It follows that $H(\Qs\lfloor \nu \uv \rfloor) = H(\Qs\uv)$ for integer $\nu$, so we conclude that $\Hc_\Qs(\uv) = H(\Qs\uv)$.

}

{
\section{Proof of Lemma~\ref{lem:right_invertible_matrices}}   \label{app:proof:right_invertible_matrices}
Let us denote the reduced Smith normal decompositions of $\Qs$ and $\Qs^\sharp$ as $\Qs = \Ss(\Qs)\Sigmas(\Qs)\Ts(\Qs)$ and $\Qs^\sharp = \hat{\Ss}(\Qs)\hat{\Sigmas}(\Qs)\hat{\Ts}(\Qs)$ (assuming $\Qs^\sharp$ exists), respectively.
By definition, \textit{1)} is equivalent to there existing an integer $\Qs^\sharp$ such that $\Qs\Qs^\sharp = \Is$. The equivalence \textit{1)} $\Leftrightarrow$ \textit{2)} is obvious by transposition. The implication \textit{3)} $\Rightarrow$ \textit{1)} becomes clear when noticing that, if $\Sigmas(\Qs)$ is the identity, then $\Ts(\Qs)^\sharp\Ss(\Qs)^{-1}$ is a right inverse of $\Qs$. Conversely, given \textit{1)}, i.e., given that $\Qs^\sharp = \hat{\Ss}(\Qs)\hat{\Sigmas}(\Qs)\hat{\Ts}(\Qs)$ exists, we have that $\Sigmas(\Qs)\Ts(\Qs)\hat{\Ss}(\Qs)\hat{\Sigmas}(\Qs) = \Ss(\Qs)^{-1}\hat{\Ts}(\Qs)^{-1}$ is unimodular, hence $\det(\Sigmas(\Qs)) = \bigl|\det(\Ts(\Qs)\hat{\Ss}(\Qs))\bigr| = \det(\hat{\Sigmas}(\Qs)) = 1$, so $\Sigmas(\Qs)$ and $\hat{\Sigmas}(\Qs)$ are diagonal non-negative and unimodular, and therefore $\Sigmas(\Qs) = \hat{\Sigmas}(\Qs) = \Is$. Thus, the reverse implication \textit{1)} $\Rightarrow$ \textit{3)} also holds. Now, assuming $\Sigmas(\Qs) = \Is$, consider the (non-reduced) Smith normal decomposition $\Qs = \tilde{\Ss}(\Qs)\tilde{\Sigmas}(\Qs)\tilde{\Ts}(\Qs) = \tilde{\Ss}(\Qs)\bigl[ \Is \ \mynull \bigr] \tilde{\Ts}(\Qs)$ with unimodular $\tilde{\Ts}(\Qs) = \bigl[ \Ts(\Qs)^\T \ \Rs^\T \bigr]^\T$, which gives us $\Qs = \tilde{\Ss}(\Qs)\Ts(\Qs)$. The matrix $\Rs$ completes $\Qs$ to a unimodular matrix, because $\bigl[ \Qs^\T \ \Rs^\T \bigr] = \Bigl[ \begin{smallmatrix} \tilde{\Ss}(\Qs) & \mynull \\ \mynull & \Is \end{smallmatrix} \Bigr] \tilde{\Ts}(\Qs)$, which is a product of unimodular matrices, and therefore itself unimodular. Hence \textit{3)} $\Rightarrow$ \textit{4)}. The implication \textit{4)} $\Rightarrow$ \textit{1)} is obvious, since the first $n$ columns of $\bigl(\bigl[ \Qs^\T \ \Rs^\T \bigr]^\T\bigr)^{-1}$ are clearly a right inverse of $\Qs$. Finally, from~\eqref{elementary_divisors} it is clear that the greatest common divisor of the $n \times n$ minors, i.e., $d_n(\Qs)$, is equal to the product $\prod_i \sigma_i(\Qs) = \det(\Sigmas(\Qs))$, which equals one if and only if $\Sigmas(\Qs) = \Is$, that is, if $\Qs$ is right-invertible. This settles \textit{5)}.

}

{
\section{Proof of Lemma~\ref{lem:quantized_entropy_lower_bound}}   \label{app:proof:quantized_entropy_lower_bound}
Since $\Qc$ is countable, we can enumerate its atoms. Let $\{\Ac_i\}$ denote a sequence of all atoms of $\Sigma$ having positive Lebesgue measure $\lambda(\Ac_i) > 0$. Let $f_{\uv}$ denote the pdf of $\uv$. Then, by the concavity of $\Phi(x) = -x\log(x)$ and Jensen's inequality,
\begin{IEEEeqnarray*}{rCl}
	h(\uv)
	&=& \sum_{i \in \mathbb{N}} \int_{\Ac_i} \Phi( f_{\uv}(\us) ) \intd\us \\
	&\leq& \sum_{i \in \mathbb{N}} \lambda(\Ac_i) \Phi\left( \int_{\Ac_i} \frac{f_{\uv}(\us)}{\lambda(\Ac_i)} \intd\us \right) \\
	&=& \sum_{i \in \mathbb{N}} \lambda(\Ac_i) \Phi\left( \frac{\P\{\uv \in \Ac_i\}}{\lambda(\Ac_i)} \right) \\
	&=& H(\lceil \uv \rfloor_\Qc) + \sum_{i \in \mathbb{N}} \P\{\uv \in \Ac_i\} \log(\lambda(\Ac_i)) \\
	&\leq& H(\lceil \uv \rfloor_\Qc) + \log(\lambda_\mathrm{max}).   \IEEEeqnarraynumspace\IEEEyesnumber
\end{IEEEeqnarray*}
This concludes the proof.

}

{
\section{Proof of Lemma~\ref{lem:Renyi}}   \label{app:proof:Renyi}
Let $\nu$ and $m$ be some positive integers and let $\kv \in \mathbb{Z}^n$ and $\kv' \in \mathbb{Z}^n$ denote integer vectors. Define
\begin{equation}
	p_{\kv,\kv',m}(\nu\uv)
	\triangleq \P\bigl\{ \nu\uv \in \kv + \tfrac{\kv'}{m} + \bigl[0,\tfrac{1}{m}\bigr)^n \bigr\}.
\end{equation}
Note that
\begin{equation}
	\sum_{\kv' \in [0:m-1]^n} p_{\kv,\kv',m}(\nu\uv)
	= \P\bigl\{ \nu\uv \in \kv + [0,1)^n \bigr\}
\end{equation}
Due to the concavity of $\Phi(x) = -x\log(x)$ and Jensen's inequality,\footnote{This is also known as the log-sum inequality.}
\begin{IEEEeqnarray*}{rCl}
	S_{m\nu} + n\log(m\nu)
	&=& H(\floor{m\nu\uv}) \\
	&=& \sum_{\kv \in \mathbb{Z}^n} \sum_{\kv' \in [0:m-1]^n} \Phi(p_{\kv,\kv',m}(\nu\uv)) \\
	&\leq& m^n \sum_{\kv \in \mathbb{Z}^n} \Phi\bigl( m^{-n} \P\bigl\{ \nu\uv \in \kv + [0,1)^n \bigr\} \bigr) \\
	&\stackrel{(a)}{=}& m^n \Phi(m^{-n}) + \sum_{\kv \in \mathbb{Z}^n} \Phi\bigl( \P\bigl\{ \nu\uv \in \kv + [0,1)^n \bigr\} \bigr) \\
	&=& H(\floor{\nu\uv}) + n\log(m) \\
	&=& S_\nu + n\log(m\nu)
\end{IEEEeqnarray*}
hence $S_{m\nu} \leq S_\nu$. Here, the equality $(a)$ is due to $\Phi(xy) = x\Phi(y) + y\Phi(x)$. It immediately follows that $S_\nu \leq S_1 = H(\floor{\uv})$, which settles the right-hand inequality of~\eqref{sandwich_bounds}.
As to the lower bound in~\eqref{sandwich_bounds}, it follows from particularizing Lemma~\ref{lem:quantized_entropy_lower_bound} to the $\sigma$-algebra generated by the cubes $\nu^{-1}(\iv + [0,1)^n), \iv \in \mathbb{Z}^n$. It remains to prove the limit~\eqref{diff_entropy_as_a_limit}, for which we refer to~\cite[Thm.~1]{Renyi1959}.

}

{
\section{Proof of Lemma~\ref{lem:quantized_entropy_difference}}   \label{app:proof:quantized_entropy_difference}
Let us denote the $(i,j)$-th entry of $\Ts$ as $T_{i,j}$, and define the random vector
\begin{IEEEeqnarray*}{rCl}
	\dv
	&\triangleq& \floor{\Ts\vv} - \Ts\floor{\vv} \\
	&=& \floor{\Ts\vv - \Ts\floor{\vv} + \Ts\floor{\vv}} - \Ts\floor{\vv} \\
	&=& \floor{ \Ts \left(\vv-\floor{\vv}\right) }.   \IEEEeqnarraynumspace\IEEEyesnumber
\end{IEEEeqnarray*}
Notice that each entry of $\vv-\floor{\vv}$ takes value in $[0;1)$. Hence the $i$-th entry of $\dv$ takes value in a finite set
\begin{equation}   \label{D_i_case_distinction}
	\Dc_i
	= \mathbb{Z} \cap
	\begin{cases}
		\left( \textstyle\sum_j^- T_{i,j} ; \textstyle\sum_j^+ T_{i,j} \right) & \text{if $\textstyle\sum_j^- T_{i,j} \neq 0$ and $\textstyle\sum_j^+ T_{i,j} \neq 0$} \\
		\left[ \textstyle\sum_j^- T_{i,j} ; \textstyle\sum_j^+ T_{i,j} \right) & \text{if $\textstyle\sum_j^- T_{i,j} = 0$ and $\textstyle\sum_j^+ T_{i,j} \neq 0$} \\
		\left( \textstyle\sum_j^- T_{i,j} ; \textstyle\sum_j^+ T_{i,j} \right] & \text{if $\textstyle\sum_j^- T_{i,j} \neq 0$ and $\textstyle\sum_j^+ T_{i,j} = 0$.}
	\end{cases}
\end{equation}
Here, we have denoted
\begin{align}
	\textstyle\sum_j^- T_{i,j}
	&= \sum\limits_{j \colon T_{i,j} \leq 0} T_{i,j}
	&
	\textstyle\sum_j^+ T_{i,j}
	&= \sum\limits_{j \colon T_{i,j} \geq 0} T_{i,j}
\end{align}
for notational brevity. Note that in all three cases considered in~\eqref{D_i_case_distinction}, the cardinality of $\Dc_i$ is upper-bounded by $\sum_j |T_{i,j}|$.
Hence, we have the uniform upper bound
\begin{IEEEeqnarray*}{rCl}
	\bigl| H(\Ts \floor{\vv}) - H(\floor{\Ts\vv}) \bigr|
	&=& \bigl| I(\Ts \floor{\vv} ; \dv) - I(\floor{\Ts\vv} ; \dv) \bigr| \\
	&\leq& H(\dv) \\
	&\leq& \sum_{i=1}^m \log \left| \Dc_i \right| \\
	&\leq& \sum_{i=1}^m \log\left( \sum_{j=1}^n |T_{i,j}| \right) \\
	&\leq& m \log\left( \frac{\lVert \Ts \rVert_1}{m} \right)   \IEEEeqnarraynumspace\IEEEyesnumber
\end{IEEEeqnarray*}
where the last inequality follows from the arithmetic--geometric mean inequality.

}

{
\section{Proof of Lemma~\ref{lem:entropy_finiteness}}   \label{app:proof:entropy_finiteness}
Proving the finiteness of $H(\floor{\Qs\vv})$ is straightforward, since by Lemma~\ref{lem:quantized_entropy_difference}, we have
\begin{IEEEeqnarray*}{rCl}
	0
	\leq H(\floor{\Qs \vv})
	&\leq& H(\Qs \floor{\vv}) + m \log\left( \frac{\lVert \Qs \rVert_1}{m} \right) \\
	&\leq& H(\floor{\vv}) + m \log\left( \frac{\lVert \Qs \rVert_1}{m} \right).
\end{IEEEeqnarray*}
By Lemma~\ref{lem:Renyi}, we immediately conclude that $h(\Qs\vv) < +\infty$ due to the upper bound $h(\Qs\vv) \leq H(\floor{\Qs\vv})$.
 
To determine a lower bound on $h(\Qs\vv)$, take $\tilde{\Qs} = \Rs\Qs$ to be in reduced row echelon form, where $\Rs \in \mathbb{R}^{m \times m}$ is full-rank. Let $V_j$ denote the $j$-th entry of $\vv$ and let $\tilde{Q}_{i,\pi(i)}$ denote the pivot element\footnote{In a matrix in row echelon form, the pivot element of the $i$-th row is the first non-zero entry of the $i$-th row, read from left to right.} in the $i$-th row of $\tilde{\Qs}$. Using the chain rule for differential entropy, we can obtain a lower bound on $h(\tilde{\Qs} \vv)$ as follows:
\begin{IEEEeqnarray*}{rCl}
	h(\tilde{\Qs} \vv)
	&=& h\bigl( [\tilde{\Qs}]_{\{m\}} \vv \bigr) + h\bigl( [\tilde{\Qs}]_{\{m-1\}} \vv \big| [\tilde{\Qs}]_{\{m\}} \vv \bigr) + \dotsc + h\bigl( [\tilde{\Qs}]_{\{1\}} \vv \big| [\tilde{\Qs}]_{[2:m]} \vv \bigr) \\
	&=& \sum_{i=1}^m h\bigl( [\tilde{\Qs}]_{\{m-i+1\}} \vv \big| [\tilde{\Qs}]_{[m-i+2:m]} \vv \bigr) \\
	&\geq& \sum_{i=1}^m h\bigl( [\tilde{\Qs}]_{\{m-i+1\}} \vv \big| [\tilde{\Qs}]_{[m-i+2:m]} \vv, [\vv]_{[\pi(m-i+1)+1{:}n]} \bigr) \\
	&=& \sum_{i=1}^m h\bigl( [\tilde{\Qs}]_{\{m-i+1\}} \vv \big| [\vv]_{[\pi(m-i+1)+1{:}n]} \bigr) \\
	&=& \sum_{i=1}^m h\bigl( V_{\pi(m-i+1)} \big| [\vv]_{[\pi(m-i+1)+1{:}n]} \bigr) \\
	&=& \sum_{i=1}^m h\bigl( V_{\pi(i)} \big| [\vv]_{[\pi(i)+1{:}n]} \bigr).
\end{IEEEeqnarray*}
This lower bound is finite because
\begin{IEEEeqnarray*}{rCl}
	h\bigl( V_{\pi(i)} \big| [\vv]_{[\pi(i)+1{:}n]} \bigr)
	&=& h\bigl( V_{\pi(i)} \big| V_{\pi(i)+1}, \dotsc, V_{n} \bigr) \\
	&=& h\bigl( V_{\pi(i)}, \dotsc, V_{n} \bigr) - h\bigl(  V_{\pi(i)+1}, \dotsc, V_{n}\bigr)
\end{IEEEeqnarray*}
and these differential entropies are finite by assumption. Therefore, $h(\Qs\vv)$ has a finite lower bound
\begin{equation}   \label{diff_entropy_LB}
	h(\Qs\vv)
	\geq -\left|\log\det(\Rs)\right| + \sum_{i=1}^m h\bigl( V_{\pi(i)} \big| V_{\pi(i)+1}, \dotsc, V_{n} \bigr).
\end{equation}
This concludes the proof.

}

{
\section{Proof of Lemma~\ref{lem:Makkuva_Wu}}   \label{app:proof:Makkuva_Wu}
Lemma~\ref{lem:Makkuva_Wu} is a multivariate generalization of a result by Makkuva and Wu~\cite[Lemma~1]{MaWu18}. Our proof borrows their line of reasoning. First off, notice that
\begin{equation}   \label{MI_diff}
	H(\Ts \floor{\nu \uv}) - H(\floor{\nu \Ts \uv})
	= I(\floor{\nu \Ts \uv}; \dv_\nu) - I(\Ts \floor{\nu \uv}; \dv_\nu)
\end{equation}
where
\begin{equation}
	\dv_\nu
	= \floor{\nu \Ts \uv} - \Ts \floor{\nu \uv}.
\end{equation}
It thus suffices to prove that both mutual information terms on the right-hand side of~\eqref{MI_diff} vanish as $\nu \to \infty$. Since
\begin{IEEEeqnarray*}{rCl}
	\dv_\nu
	&=& \bigfloor{\Ts \floor{\nu \uv} + \nu \Ts \uv - \Ts \floor{\nu \uv}} - \Ts \floor{\nu \uv} \\
	&=& \bigfloor{\Ts \bigl(\nu \uv - \floor{\nu \uv}\bigr)}   \IEEEeqnarraynumspace\IEEEyesnumber\label{difference_term}
\end{IEEEeqnarray*}
the second mutual information term on the right-hand side of~\eqref{MI_diff} can be upper-bounded using the data-processing inequality as
\begin{IEEEeqnarray*}{rCl}
	I(\Ts \floor{\nu \uv}; \dv_\nu)
	&=& I\bigl(\Ts \floor{\nu \uv}; \bigfloor{\Ts \bigl(\nu\uv-\floor{\nu \uv}\bigr)}\bigr) \\
	&\leq& I(\floor{\nu \uv}; \nu\uv-\floor{\nu \uv})   \IEEEeqnarraynumspace\IEEEyesnumber\label{first_MI_vanishes}
\end{IEEEeqnarray*}
the right-hand side of which tends to zero as $\nu \to \infty$ by the following lemma.

\lemmabox{
\begin{lemma}[Asymptotic independence of quantization output and error] \label{lem:asym-error}
For any random vector $\Uv \in \mathbb{R}^K$ we have
\begin{align*}
	\lim_{\nu \to \infty} I\bigl(\floor{\nu \uv}; \nu\uv - \floor{\nu \uv}\bigr) = 0.
\end{align*}
\end{lemma}
}

\proofbox{}{
\begin{IEEEproof}
By~\cite[Lemma~7.18]{Gray1990_A}, we have
\begin{align*}
	I(\floor{\nu\uv} ; \nu\uv-\floor{\nu\uv})
	= \lim_{m\to\infty} I\bigl(\floor{\nu\uv}; \bigfloor{m(\nu\uv-\floor{\nu\uv})}\bigr).
\end{align*}
The mutual information $I\bigl(\floor{\nu\uv}; \bigfloor{m(\nu\uv-\floor{\nu\uv})}\bigr)$ can be upper-bounded as follows:
\begin{IEEEeqnarray*}{rCl}
	\IEEEeqnarraymulticol{3}{l}{
		I\bigl(\floor{\nu\uv}; \bigfloor{m(\nu\uv-\floor{\nu\uv})}\bigr)
	} \\ \quad
	&=& H(\floor{\nu\uv}) + H\bigl(\bigfloor{m\nu\uv - m\floor{\nu\uv}}\bigr) - H\bigl(\bigfloor{m\nu\uv - m\floor{\nu\uv}}, \floor{\nu\uv}\bigr) \\
	&\stackrel{(a)}{=}& H(\floor{\nu\uv}) + H\bigl(\bigfloor{m\nu\uv - m\floor{\nu\uv}}\bigr) - H\bigl(\floor{m\nu\uv}\bigr) \\
	&\stackrel{(b)}{\leq}& H(\floor{\nu\uv}) + n\log(m+1) - H\bigl(\floor{m\nu\uv}\bigr)
\end{IEEEeqnarray*}
\sloppy
where step $(a)$ follows because there is a one-to-one correspondence between $\bigl(\bigfloor{m\nu\uv - m\floor{\nu\uv}}, \floor{\nu\uv}\bigr)$ and $\floor{m\nu\uv}$ and step $(b)$ follows because each of the $n$ coordinates of $\bigfloor{m\nu\uv - m\floor{\nu\uv}}$ takes at most $m+1$ values. Taking the limit as $\nu \to \infty$, we thus get an upper bound
\begin{IEEEeqnarray*}{rCl}
	I(\floor{\nu\uv} ; \nu\uv - \floor{\nu\uv})
	&\leq& H(\floor{\nu\uv}) + \lim_{m\to\infty} \bigl\{ n\log(m+1) - H\bigl(\floor{m\nu\uv}\bigr) \bigr\} \\
	&=& H(\floor{\nu\uv}) - n\log(\nu) - h(\uv).   \IEEEeqnarraynumspace\IEEEyesnumber\label{asym_err_UB}
\end{IEEEeqnarray*}
Finally, taking the limit as $\nu \to \infty$ we obtain the desired result
\begin{IEEEeqnarray*}{rCl}
	\lim_{\nu \to \infty} I\bigl( \floor{\nu\uv} ; \nu\uv - \floor{\nu\uv} \bigr)
	&=& 0.
\end{IEEEeqnarray*}
This concludes the proof.

\end{IEEEproof}
}

Prior to studying the first mutual information term in \eqref{MI_diff}, we introduce some auxiliary definitions and state Lemmata~\ref{lem:Fano-Pinsker}, \ref{lem:marginalized_TV} and \ref{lem:TV_continuity}, which will be needed in the subsequent proof.

\begin{definition}
Let $P_X$ and $P_Y$ denote two probability measures defined on a common probability space $\Omega$. The {\em total variation distance} between $P_X$ and $P_Y$ is defined as
\begin{equation}
	d_\mathsf{TV}(P_X \Vert P_Y)
	\triangleq \sup_{\mathcal{A} \subset \Omega} |P_X(\mathcal{A})-P_Y(\mathcal{A})|.
\end{equation}
Following the terminology of~\cite{PoWu16}, we define the {\em $T$-information} between random random variables $X$ and $Y$, with joint distribution $P_{XY}$, as
\begin{equation}
	T(X;Y)
	\triangleq d_\mathsf{TV}(P_{XY} \Vert P_X \times P_Y).
\end{equation}
Note that $T$-information is upper-bounded by one.
\end{definition}

\lemmabox{
\begin{lemma}   \label{lem:Fano-Pinsker}
For any probability measure $P_{XY}$ on a probability space $\mathcal{X} \times \mathcal{Y}$, we have
\begin{subequations}
\begin{IEEEeqnarray}{rCl}
	I(X;Y) &\leq& \log(\min(|\mathcal{X}|,|\mathcal{Y}|)-1) T(X;Y) + h_\mathsf{b}(T(X;Y))   \label{T_Fano} \\
	I(X;Y) &\geq& 2 T(X;Y)^2   \label{Pinsker}
\end{IEEEeqnarray}
\end{subequations}
where $h_\mathsf{b}(x) = -x\log(x)-(1-x)\log(1-x)$ denotes the binary entropy function, and where \eqref{T_Fano} only holds if either $\mathcal{X}$ or $\mathcal{Y}$ is of finite cardinality.
\end{lemma}
}

\proofbox{}{
\begin{IEEEproof}
Proofs of~\eqref{T_Fano} and~\eqref{Pinsker} can be found in~\cite[Proposition~12]{PoWu16} (via Fano's inequality) and~\cite{Pi05}, respectively. The inequality~\eqref{Pinsker} is commonly known as the {\em Pinsker-Csisz\'ar} or {\em Pinsker} inequality.
\end{IEEEproof}
}

\lemmabox{
\begin{lemma}   \label{lem:marginalized_TV}
Let the triple $(X,Y,Z)$ of random variables satisfy $Z = f(X) = f(Y)$ where $f$ denotes some measurable function. Then for any measurable set $\mathcal{B}$ such that $\P\{Z \in \mathcal{B}\} > 0$,
\begin{equation}
	d_\mathsf{TV}\left(P_{X|Z \in \mathcal{B}} \middle\Vert P_{Y|Z \in \mathcal{B}}\right)
	\leq \frac{d_\mathsf{TV}\left(P_X \middle\Vert P_Y\right)}{\P\{Z \in \mathcal{B}\}}.
\end{equation}
\end{lemma}
}

\proofbox{}{
\begin{IEEEproof}
For any measurable $\mathcal{B}$ and $\mathcal{A}$,
\begin{IEEEeqnarray*}{rCl}
	\IEEEeqnarraymulticol{3}{l}{
		\bigl| \P\{X \in \mathcal{A}|Z \in \mathcal{B}\} - \P\{Y \in \mathcal{A}|Z \in \mathcal{B}\} \bigr|
	} \\ \qquad
	&=& \frac{\bigl| \P\{X \in \mathcal{A},f(X) \in \mathcal{B}\} - \P\{Y \in \mathcal{A},f(Y) \in \mathcal{B}\} \bigr|}{\P\{Z \in \mathcal{B}\}} \\
	&\leq& \frac{\sup_{\mathcal{A}',\mathcal{B}'} \bigl| \P\{X \in \mathcal{A}',f(X) \in \mathcal{B}'\} - \P\{Y \in \mathcal{A}',f(Y) \in \mathcal{B}'\} \bigr|}{\P\{Z \in \mathcal{B}\}} \\
	&=& \frac{d_\mathsf{TV}(P_X \Vert P_Y)}{\P\{Z \in \mathcal{B}\}}.
\end{IEEEeqnarray*}
\end{IEEEproof}
}

\lemmabox{
\begin{lemma}[Continuity of total variation distance]   \label{lem:TV_continuity}
Let $\uv \in \mathbb{R}^K$ be a continuous random vector with pdf $f_{\uv}$. Then it holds that
\begin{equation}
	\lim_{\lVert \Deltav \rVert_2 \to 0} d_\mathsf{TV}\left(f_{\uv} \middle\Vert f_{\uv + \Deltav}\right)
	= 0.
\end{equation}
\end{lemma}
}

\proofbox{}{
\begin{IEEEproof}
Let $L^1(\mathbb{R}^K)$ denote the space of $L^1$-integrable functions $\mathbb{R}^K \to \mathbb{R}$, endowed with the $L^1$-norm $\lVert f \rVert_1 = \int_{\mathbb{R}^K} |f(x)| \intd x$. Since the set of compactly supported continuous functions is dense in $L^1(\mathbb{R}^K)$, for any $\epsilon > 0$ there exists a continuous and compactly supported function $q \in L^1(\mathbb{R}^K)$ such that $\lVert f_{\uv}-q \rVert_1 < \epsilon$. Due to the Heine--Cantor Theorem \cite[Theorem~4.19]{Ru76}, continuous and compactly supported functions are uniformly continuous. Therefore, there exists $\delta(\epsilon) > 0$ such that for all vectors $\Deltav \in \mathbb{R}^K$ of norm $\lVert \Deltav \rVert_2 < \delta(\epsilon)$, we have $\lVert q(\cdot + \Deltav) - q(\cdot) \rVert_1 < \epsilon$. By the triangle inequality, $\lVert f_{\uv}(\cdot + \Deltav) - f_{\uv}(\cdot) \rVert_1 \leq 2\lVert f_{\uv}(\cdot)-q(\cdot) \rVert_1 + \lVert q(\cdot + \Deltav) - q(\cdot) \rVert_1 < 3\epsilon$. Hence the claim follows.
\end{IEEEproof}
}

With the above lemmata, we can proceed with the proof of Lemma~\ref{lem:Makkuva_Wu}. 
The random vector $\nu\uv-\floor{\nu\uv}$ is supported on the hypercube\footnote{In case that $\uv$ is not supported on $\mathbb{R}^K$ but a subset thereof, this is true provided that $\nu$ is sufficiently large} $[0,1)^K$ and is known to converge weakly to the uniform distribution over said hypercube \cite[Theorem~4.1]{JiWaWa07}, i.e., for any Lebesgue measurable set $\Vc \subset \mathbb{R}^K$, we have
\begin{equation}
	\lim_{\nu \to \infty} \P\left\{ \nu\uv - \floor{\nu\uv} \in \Vc \right\}
	= \lambda\Bigl( \Vc \cap [0,1)^K \Bigr)
\end{equation}
where $\lambda(\cdot)$ denotes the Lebesgue measure.
Let $\mathcal{S}$ denote the support of $\Ts(\nu\uv-\floor{\nu\uv})$. Let $\xiv$ denote a uniformly distributed random variable over the hypercube $[0,1)^K$, and let us define $\Dc \subset \mathbb{Z}^{L_\Qf}$ as the finite set of integer vectors such that
\begin{equation}
	\Dc = \Bigl\{ \dv \in \mathbb{Z}^{L_\Qf} \colon \P\{ \floor{\Ts\xiv} = \ds \} > 0 \Bigr\}.
\end{equation}
In particular, we have for every $\ds \in \Dc$ that
\begin{equation}
	\lim_{\nu \to \infty} \P\bigl\{ \dv_\nu = \ds \bigr\}
	= \lim_{\nu \to \infty} \P\Bigl\{ \bigfloor{\Ts(\nu \uv-\floor{\nu \uv})} = \ds \Bigr\}
	= \P\{ \floor{\Ts\xiv} = \ds \}
	> 0.
\end{equation}
Using the inequality \eqref{T_Fano} from Lemma~\ref{lem:Fano-Pinsker}, we have the upper bound
\begin{IEEEeqnarray*}{rCl}
	I\bigl(\floor{\nu \Ts \uv} ; \dv_\nu\bigr)
	&\leq& \log(|\Dc|-1) T\bigl(\floor{\nu \Ts \uv} ; \dv_\nu\bigr) + h_\mathsf{b}\bigl(T\bigl(\floor{\nu \Ts \uv} ; \dv_\nu\bigr)\bigr).
\end{IEEEeqnarray*}
It now suffices to show that
\begin{IEEEeqnarray*}{rCl}
	T\bigl(\floor{\nu \Ts \uv} ; \dv_\nu\bigr)
	&=& \sum_{\ds \in \Dc} \P\{\dv_\nu=\ds\} \; d_\mathsf{TV}\bigl(P_{\floor{\nu \Ts \uv}} \big\Vert P_{\floor{\nu \Ts \uv}|\dv_\nu=\ds}\bigr) \\
	&\leq& \sum_{\ds \in \Dc} d_\mathsf{TV}\bigl(P_{\floor{\nu \Ts \uv}} \big\Vert P_{\floor{\nu \Ts \uv}|\dv_\nu=\ds}\bigr)
\end{IEEEeqnarray*}
tends to zero as $\nu \to \infty$. By the convexity of total variation distance,
\begin{IEEEeqnarray*}{rCl}
	\IEEEeqnarraymulticol{3}{l}{
		d_\mathsf{TV}\bigl(P_{\floor{\nu \Ts \uv}} \big\Vert P_{\floor{\nu \Ts \uv}|\dv_\nu=\ds}\bigr)
	} \\ \quad
	&\leq& \sum_{\ds' \in \Dc} \P\{\dv_\nu=\ds'\} \, d_\mathsf{TV}\bigl(P_{\floor{\nu \Ts \uv}|\dv_\nu=\ds'} \big\Vert P_{\floor{\nu \Ts \uv}|\dv_\nu=\ds}\bigr) \\
	&\leq& \sum_{\ds' \in \Dc} d_\mathsf{TV}\bigl(P_{\floor{\nu \Ts \uv}|\dv_\nu=\ds'} \big\Vert P_{\floor{\nu \Ts \uv}|\dv_\nu=\ds}\bigr).
\end{IEEEeqnarray*}
Hence,
\begin{equation}   \label{limsup_T}
	T\bigl(\floor{\nu \Ts \uv} ; \dv_\nu\bigr)
	\leq \sum_{\ds \in \Dc} \sum_{\ds' \in \Dc} d_\mathsf{TV}\bigl(P_{\floor{\nu \Ts \uv}|\dv_\nu=\ds'} \big\Vert P_{\floor{\nu \Ts \uv}|\dv_\nu=\ds}\bigr).
\end{equation}
Next, we upper-bound the total variation term appearing on the right-hand side of \eqref{limsup_T} by means of the triangle inequality as follows:
\begin{IEEEeqnarray*}{rCl}
	\IEEEeqnarraymulticol{3}{l}{
		d_\mathsf{TV}\bigl(P_{\floor{\nu \Ts \uv}|\dv_\nu=\ds'} \big\Vert P_{\floor{\nu \Ts \uv}|\dv_\nu=\ds}\bigr)
	} \\ \quad
	&=& d_\mathsf{TV}\bigl(P_{\Ts \floor{\nu \uv}+\ds'|\dv_\nu=\ds'} \big\Vert P_{\Ts \floor{\nu \uv}+\ds|\dv_\nu=\ds}\bigr) \\
	&\leq& d_\mathsf{TV}\bigl(P_{\Ts \floor{\nu \uv}+\ds'|\dv_\nu=\ds'} \big\Vert P_{\Ts \floor{\nu \uv}+\ds|\dv_\nu=\ds'}\bigr) + d_\mathsf{TV}\bigl(P_{\Ts \floor{\nu \uv}+\ds|\dv_\nu=\ds'} \big\Vert P_{\Ts \floor{\nu \uv}+\ds|\dv_\nu=\ds}\bigr) \\
	&=& d_\mathsf{TV}\bigl(P_{\Ts \floor{\nu \uv}+\ds'|\dv_\nu=\ds'} \big\Vert P_{\Ts \floor{\nu \uv}+\ds|\dv_\nu=\ds'}\bigr) + d_\mathsf{TV}\bigl(P_{\Ts \floor{\nu \uv}|\dv_\nu=\ds'} \big\Vert P_{\Ts \floor{\nu \uv}|\dv_\nu=\ds}\bigr).   \IEEEeqnarraynumspace\IEEEyesnumber\label{TV_UB}
\end{IEEEeqnarray*}
The second term on the right-hand side of \eqref{TV_UB} can be upper-bounded as
\begin{IEEEeqnarray*}{rCl}
	\IEEEeqnarraymulticol{3}{l}{
		d_\mathsf{TV}\bigl(P_{\Ts \floor{\nu \uv}|\dv_\nu=\ds'} \big\Vert P_{\Ts \floor{\nu \uv}|\dv_\nu=\ds}\bigr)
	} \\ \quad
	&\leq& d_\mathsf{TV}\bigl(P_{\Ts \floor{\nu \uv}|\dv_\nu=\ds'} \big\Vert P_{\Ts \floor{\nu \uv}}\bigr) + d_\mathsf{TV}\bigl(P_{\Ts \floor{\nu \uv}|\dv_\nu=\ds} \big\Vert P_{\Ts \floor{\nu \uv}}\bigr) \\
	&\leq& \sum_{\ds'' \in \Dc} d_\mathsf{TV}\bigl(P_{\Ts \floor{\nu \uv}|\dv_\nu=\ds''} \big\Vert P_{\Ts \floor{\nu \uv}}\bigr) \\
	&\leq& \frac{1}{\min_{\tilde{\ds} \in \Dc} \P\{\dv_\nu = \tilde{\ds}\}} \sum_{\ds'' \in \Dc} \P\{ \dv_\nu = \ds'' \} \, d_\mathsf{TV}\bigl(P_{\Ts \floor{\nu \uv}|\dv_\nu=\ds''} \big\Vert P_{\Ts \floor{\nu \uv}}\bigr) \\
	&=& \frac{1}{\min_{\tilde{\ds} \in \Dc} \P\{\dv_\nu = \tilde{\ds}\}} T\bigl(\Ts \floor{\nu \uv};\dv_\nu\bigr) \\
	&\leq& \frac{1}{\min_{\tilde{\ds} \in \Dc} \P\{\dv_\nu = \tilde{\ds}\}} \sqrt{\frac{1}{2} I\bigl(\Ts \floor{\nu \uv};\dv_\nu\bigr)}
\end{IEEEeqnarray*}
where the last step is Pinsker's inequality~\eqref{lem:Fano-Pinsker}. This upper bound tends to zero as $\nu \to \infty$ because on one hand, by definition of the finite set $\Dc$, the limit $\lim_{\nu \to \infty} \P\{\dv_\nu = \tilde{\ds}\}$ is positive for all $\tilde{\ds} \in \Dc$ by definition of $\Dc$, while on the other hand, the mutual information $I\bigl(\Ts \floor{\nu \uv};\dv_\nu\bigr)$ vanishes, as already argued in~\eqref{first_MI_vanishes}.
Since, by assumption, $\Ts$ is right-invertible in the integers, let us denote a right-inverse as $\Ts^\sharp \in \mathbb{Z}^{K \times L_\Qf}$, so that we can write
\begin{equation}
	\Ts \floor{\nu \uv} + \ds' - \ds
	= \Ts\bigl( \floor{\nu \uv} + \Ts^\sharp(\ds'-\ds) \bigr).
\end{equation}
The first total variation term on the right-hand side of \eqref{TV_UB} can thus be upper-bounded as follows:
\begin{IEEEeqnarray*}{rCl}
	\IEEEeqnarraymulticol{3}{l}{
		d_\mathsf{TV}\bigl(P_{\Ts \floor{\nu \uv}+\ds'|\dv_\nu=\ds'} \big\Vert P_{\Ts \floor{\nu \uv}+\ds|\dv_\nu=\ds'}\bigr)
	} \\ \quad
	&=& d_\mathsf{TV}\bigl(P_{\Ts \floor{\nu \uv}+(\ds'-\ds)|\dv_\nu=\ds'} \big\Vert P_{\Ts \floor{\nu \uv}|\dv_\nu=\ds'}\bigr) \\
	&=& d_\mathsf{TV}\bigl(P_{\Ts(\floor{\nu \uv} + \Ts^\sharp(\ds'-\ds))|\dv_\nu=\ds'} \big\Vert P_{\Ts \floor{\nu \uv}|\dv_\nu=\ds'}\bigr) \\
	&\stackrel{(a)}{\leq}& d_\mathsf{TV}\bigl(P_{\floor{\nu \uv} + \Ts^\sharp(\ds'-\ds)|\dv_\nu=\ds'} \big\Vert P_{\floor{\nu \uv}|\dv_\nu=\ds'}\bigr) \\
	&\stackrel{(b)}{\leq}& \frac{1}{\P\{\dv_\nu=\ds'\}} d_\mathsf{TV}\bigl(P_{\floor{\nu \uv} + \Ts^\sharp(\ds'-\ds)} \big\Vert P_{\floor{\nu \uv}}\bigr).   \IEEEeqnarraynumspace\IEEEyesnumber\label{TV_UB_2}
\end{IEEEeqnarray*}
Here, the bounding step $(a)$ follows from the data-processing inequality for $T$-information, whereas step $(b)$ follows from Lemma~\ref{lem:marginalized_TV}, which is applicable because $\dv_\nu$
results from either [cf.~\eqref{difference_term}] $\uv+\tfrac{1}{\nu}\Ts^\sharp(\dv'-\dv)$ or $\uv$, by an application of the function
\begin{equation}
	f \colon \mathbb{R}^K \to \mathbb{R}^{L_\Qf}, \ \xv \mapsto \floor{\Ts(\nu\xv-\floor{\nu \xv})}.
\end{equation}
Indeed, one can readily verify that
\begin{IEEEeqnarray*}{rCl}
	f\bigl(\uv+\tfrac{1}{\nu}\Ts^\sharp(\dv'-\dv)\bigr)
	&=& \bigfloor{\Ts\bigl(\nu\uv+\Ts^\sharp(\dv'-\dv)-\bigfloor{\nu\uv+\Ts^\sharp(\dv'-\dv)}\bigr)} \\
	&=& \bigfloor{\Ts(\nu\uv-\floor{\nu\uv})} \\
	&=& f(\uv) = \dv_\nu.   \IEEEeqnarraynumspace\IEEEyesnumber
\end{IEEEeqnarray*}
Taking the limit as $\nu \to \infty$ on both sides of \eqref{TV_UB_2}, and recalling that by definition of $\Dc$, we have that $\lim_{\nu \to \infty} \P\{\dv_\nu=\ds'\} > 0$ for any $\ds' \in \Dc$, by Lemma~\ref{lem:TV_continuity} we infer that \eqref{TV_UB_2} tends to zero as $\nu \to \infty$. This finalizes the proof of Lemma~\ref{lem:Makkuva_Wu}.

}

{
\section{Proof of Theorem~\ref{thm:discrete_CF}}   \label{app:proof:discrete_CF}
We start by restating the compute--forward achievable region from \cite[Thm.~1]{LiFePaNaGa20} and then prove step by step that it is equal to the rate region $\mathscr{R}(\As)$ from Theorem~\ref{thm:discrete_CF}.

Let $(\mathbb{U},\mathbb{A}) = (\mathbb{F}_\q,\mathbb{F}_\q)$. For matrices $\Bs \in \mathbb{F}_\q^{L_\Bf \times K}$ and $\Cs \in \mathbb{F}_\q^{L_\Cf \times L_\Bf}$ and an index set $\Tc \subset [K]$, let us define
\begin{equation}
	\widetilde{\mathscr{Q}}(\Bs,\Cs,\Tc)
	\triangleq \Bigl\{ (R_1,\dotsc,R_K) \in \mathbb{R}_+^K \colon
	\sum_{k\in\Tc} R_k < H([\uv]_\Tc) - H(\Cs\Bs\uv|Y) + H(\Bs\uv|Y) \Bigr\}   \label{partial_rate_region_legacy_1}
\end{equation}
as well as
\begin{equation}   \label{partial_rate_region_legacy_2}
	\widetilde{\mathscr{Q}}(\Bs)
	\triangleq \bigcap_{\Cs} \bigcup_{\Sc} \bigcap_{\Tc} \widetilde{\mathscr{Q}}(\Bs,\Cs,\Tc)
\end{equation}
where the three nested set operations are over triples $(\Cs,\Sc,\Tc)$ meeting the following constraints:
\begin{enumerate}
	\item $\Cs \in \mathbb{F}_\q^{L_\Cf\times L_\Bf}$ is iterated over all full-rank matrices (including empty matrices)\footnote{As in~\cite{LiFePaNaGa20}, here we invoke the notion of an \emph{empty matrix,} which is a matrix with zero rows (resp.~zero columns). By convention, we will consider that an empty matrix has \emph{full row rank} (resp.~\emph{full column rank}), its rank being $0$. The product of an empty matrix with another matrix is an empty matrix, e.g., if $\As$ is a $0 \times 3$ empty matrix and $\Bs$ is a $3 \times 5$ matrix, then $\As \Bs$ is an empty matrix of size $0 \times 5$.} such that $0 \le L_\Cf < L_\Bf$,
	\item $\Sc \subseteq [L_\Bf]$ is iterated over all index sets of size $|\Sc| = L_\Bf - L_\Cf$ satisfying
\begin{equation}   \label{rank_condition_1}
	\rank\left(\begin{bmatrix} \Cs \\ [\Is]_\Sc \end{bmatrix}\right) = L_\Bf.
\end{equation}
	\item $\Tc \subseteq [K]$ is iterated over all index sets of size $|\Tc| = L_\Bf - L_\Cf$ satisfying
\begin{equation}   \label{rank_condition_2}
	\rank\left(\begin{bmatrix} [\Bs]_\Sc \\ [\Is]_{[K] \setminus \Tc} \end{bmatrix}\right) = K.
\end{equation}
\end{enumerate}
Finally, let us define the joint rate region
\begin{equation}   \label{joint_rate_region_legacy}
	\widetilde{\mathscr{R}}(\As)
	\triangleq \bigcup_{\Bs} \widetilde{\mathscr{Q}}(\Bs)
\end{equation}
where $\Bs \in \mathbb{F}_\q^{L_\Bf \times K}$ is iterated over all full-rank matrices satisfying $\Lambda_{\mathbb{F}_\q}(\Bs) \supseteq \Lambda_{\mathbb{F}_\q}(\As)$.

Notice that $\widetilde{\mathscr{R}}(\As)$ is the achievable rate--region from~\cite[Thm.~1]{LiFePaNaGa20}. To see this, bear in mind that in~\eqref{partial_rate_region_legacy_1} one can rewrite the difference of the last two entropy terms as a single one to match the expression from~\cite[Eq.~(5)]{LiFePaNaGa20}:
\begin{equation}
	H(\Bs\uv|Y) - H(\Cs\Bs\uv|Y)
	= H(\Bs\uv|Y,\Cs\Bs\uv).
\end{equation}
We will now transform the expression of $\widetilde{\mathscr{R}}(\As)$ in order to prove its equality with $\mathscr{R}(\As)$ from Theorem~\ref{thm:discrete_CF}. In particular, we will introduce the concept of matroids, as the set over which $\Cs$, $\Sc$ and $\Tc$ are iterated is more naturally represented in terms of matroids.

Note that we can express the full-rank conditions~\eqref{rank_condition_1} and \eqref{rank_condition_2}, which respectively define the sets over which $\Sc$ and $\Tc$ are iterated, in terms of matroids. Specifically,
\begin{enumerate}[i)]
	\item	condition \eqref{rank_condition_1} holds if and only if $\Sc \in \mathscr{B}(M^*(\Cs))$;
	\item	condition \eqref{rank_condition_2} holds if and only if $\Tc \in \mathscr{B}(M([\Bs]_\Sc))$.
\end{enumerate}
The set $\widetilde{\mathscr{Q}}(\Bs)$ as defined in~\eqref{partial_rate_region_legacy_2}--\eqref{rank_condition_2} can thus be expressed more concisely as
\begin{equation}   \label{partial_rate_region_legacy_2_bis}
	\widetilde{\mathscr{Q}}(\Bs)
	= \bigcap_{\Cs} \; \bigcup_{\Sc \in \mathscr{B}(M^*(\Cs))} \; \bigcap_{\Tc \in \mathscr{B}([\Bs]_\Sc)} \mathscr{Q}(\Bs,\Cs,\Tc)
\end{equation}
where for the first intersection, $\Cs \in \mathbb{F}_\q^{L_\Cf \times L_\Bf}$ is iterated over all full-rank matrices such that $0 \leq L_\Cf < L_\Bf$ (including empty matrices).
Recall that the set of all full row-rank matrices $\Qs \in \mathbb{F}_\q^{r \times L_\Bf}$ that represent a given matroid $M \in \mathscr{M}_{\mathbb{F}_\q}(L_\Bf)$ of some rank $0 \leq r \leq L_\Bf$ is denoted as $\mathscr{C}_{\mathbb{F}_\q}(M)$ [cf.~Section~\ref{ssec:matroids}, Definition~\ref{def:representable_matroids}].
In the outmost intersection of~\eqref{partial_rate_region_legacy_2}, as $\Cs$ is iterated over all matrices from $\mathbb{F}_\q^{L_\Cf \times L_\Bf}$ with $0 \leq L_\Cf < L_\Bf$, the corresponding matroids $M(\Cs)$ will be iterated over all size-$L_\Bf$ matroids representable over $\mathbb{F}_\q$ except the full-rank matroid $\{[L_\Bf],2^{[L_\Bf]}\}$, i.e., over the set
\begin{equation}
	\{ M(\Cs) \colon \Cs \in \mathbb{F}_\q^{L_\Cf \times L_\Bf}, 0 \leq L_\Cf < L_\Bf, \text{$\Cs$ has full column rank} \}
	= \mathscr{M}_{\mathbb{F}_\q}(L_\Bf) \setminus \{[L_\Bf],2^{[L_\Bf]}\}.
\end{equation}
Let us denote the latter set as $\mathscr{M}_{\mathbb{F}_\q}^\circ(L_\Bf)$.
We can now rearrange the intersection operations over matrices $\Cs$ in~\eqref{partial_rate_region_legacy_2} so as to group them according to their associated matroids, as follows:
\begin{IEEEeqnarray*}{rCl}
	\widetilde{\mathscr{Q}}(\Bs)
	&=& \bigcap_{M \in \mathscr{M}_{\mathbb{F}_\q}^\circ(L_\Bf)} \; \bigcap_{\Cs \in \mathscr{C}_{\mathbb{F}_\q}(M)} \; \bigcup_{\Sc \in \mathscr{B}(M^*)} \; \bigcap_{\Tc \in \mathscr{B}([\Bs]_\Sc)} \widetilde{\mathscr{Q}}(\Bs,\Cs,\Tc).   \IEEEeqnarraynumspace\IEEEyesnumber\label{four_fold_set_operation}
\end{IEEEeqnarray*}
Now, from the definition of $\widetilde{\mathscr{Q}}(\Bs,\Cs,\Tc)$ [cf.~\eqref{partial_rate_region_legacy_1}] it is manifest that for any two matrices $\Cs$ and $\Cs'$ that represent the same matroid $M$, that is, $\Cs, \Cs' \in \mathscr{C}_{\mathbb{F}_\q}(M)$, an inequality $H(\Cs\Bs\uv|Y) \leq H(\Cs'\Bs\uv|Y)$ will imply an inclusion relation $\widetilde{\mathscr{Q}}(\Bs,\Cs,\Tc) \subset \widetilde{\mathscr{Q}}(\Bs,\Cs',\Tc)$, from which it follows that
\begin{equation}
	\bigcup_{\Sc \in \mathscr{B}(M^*)} \; \bigcap_{\Tc \in \mathscr{B}([\Bs]_\Sc)} \widetilde{\mathscr{Q}}(\Bs,\Cs,\Tc)
	\subset \bigcup_{\Sc \in \mathscr{B}(M^*)} \; \bigcap_{\Tc \in \mathscr{B}([\Bs]_\Sc)} \widetilde{\mathscr{Q}}(\Bs,\Cs',\Tc).
\end{equation}
Therefore, if we denote the intersection of the sets $\widetilde{\mathscr{Q}}(\Bs,\Cs,\Tc)$ when $\Cs$ runs over all members of $\mathscr{C}_{\mathbb{F}_\q}(M)$ as
\begin{IEEEeqnarray*}{rCl}
	\widetilde{\mathscr{Q}}(\Bs,M,\Tc)
	&\triangleq& \bigcap_{\Cs \in \mathscr{C}_{\mathbb{F}_\q}(M)} \widetilde{\mathscr{Q}}(\Bs,\Cs,\Tc) \\ &=& \Bigl\{ (R_1,\dotsc,R_K) \in \mathbb{R}_+^K \colon \sum_{k\in\Tc} R_k < \qquad\qquad\qquad\qquad\qquad\qquad  \\
	\IEEEeqnarraymulticol{3}{r}{
		\qquad\qquad\qquad\qquad\qquad\qquad H([\uv]_\Tc) + J(\Bs,M) - H(\Bs\uv|Y) \Bigr\}
	}   \IEEEeqnarraynumspace\IEEEyesnumber\label{Q_partial_tilde}
\end{IEEEeqnarray*}
where
\begin{equation}
	J(\Bs,M)
	= \inf_{\Cs \in \mathscr{C}_{\mathbb{F}_\q}(M)} H(\Cs\Bs\uv|Y)
\end{equation}
then the order of set operations in~\eqref{four_fold_set_operation} can be exchanged in such way that the intersection over $\Cs$ is moved to the innermost position (inside the intersection over $\Tc$), leading to
\begin{IEEEeqnarray*}{rCl}
	\widetilde{\mathscr{Q}}(\Bs)
	&=& \bigcap_{M \in \mathscr{M}_{\mathbb{F}_\q}^\circ(L_\Bf)} \; \bigcup_{\Sc \in \mathscr{B}(M^*)} \; \bigcap_{\Tc \in \mathscr{B}([\Bs]_\Sc)} \widetilde{\mathscr{Q}}(\Bs,M,\Tc).   \IEEEeqnarraynumspace\IEEEyesnumber\label{partial_rate_region_in_terms_of_rank_patterns_tilde}
\end{IEEEeqnarray*}
By comparison between \eqref{Q_partial} and \eqref{Q_partial_tilde}, as well as between \eqref{def:Q} and \eqref{partial_rate_region_in_terms_of_rank_patterns_tilde}, we clearly identify that for $(\mathbb{U},\mathbb{A}) = (\mathbb{F}_\q,\mathbb{F}_\q)$ we have $\widetilde{\mathscr{Q}}(\Bs,M,\Tc) = \mathscr{Q}(\Bs,M,\Tc)$ and $\widetilde{\mathscr{Q}}(\Bs) = \mathscr{Q}(\Bs)$. This proves Theorem~\ref{thm:discrete_CF}.

}

{
\section{Proof of Theorem~\ref{thm:integer_CF}}   \label{app:proof:integer_CF}
In order to extend the finite-field result of Theorem~\ref{thm:discrete_CF} to the integers, we set the auxiliary alphabet to be $\mathbb{U} = \mathbb{Z}$ and fix an auxiliary distribution $P_{\uv} = \prod_{k=1}^K P_{U_k}$, an integer coefficient matrix $\As \in \mathbb{Z}^{L \times K}$ and a channel law $P_{Y|X_1,\dotsc,X_K}$. Given these, we will show how one can construct a sequence of rate regions that are achievable by Theorem~\ref{thm:discrete_CF} such that asymptotically, its limit represents a rate region for reliably recovering the $\As$-linear combination of auxiliary codewords, and is given by $\mathscr{R}(\As)$ as evaluated in the context of Theorem~\ref{thm:integer_CF}.

We will prove Theorem~\ref{thm:integer_CF} in two major steps: in a first part, we will assume that the integer auxiliary vector $\uv = (U_1,\dotsc,U_K) \in \mathbb{Z}^K$ is finitely supported, and prove that under this assumption, as long as we let the field size $\q$ be large enough, all finite-field operations (i.e., with modulo-$\q$ reduction) involved in the evaluation of $\mathscr{R}(\phi_\q(\As))$ become isomorphic to integer operations (i.e., \emph{without} modulo-$\q$ operations) almost surely. Thereby we prove Theorem~\ref{thm:integer_CF} for the special case of finitely supported integer auxiliaries. In a second and final part, we will relax this assumption of finite support so as to extend the validity of Theorem~\ref{thm:integer_CF} to infinitely supported, finite-entropy integer auxiliaries.

In essence, the proof consists in determining an inner bound on the set limit
\begin{equation}
	\mathscr{R}^\infty(\As)
	\triangleq \varliminf_{\q \to \infty} \mathscr{R}(\phi_\q(\As))
\end{equation}
that will turn out to be precisely $\mathscr{R}(\As)$ (from Theorem~\ref{thm:integer_CF}). A first inner-bounding step is taken by reducing the union over $\Bs$ that appears in
\begin{equation}
	\mathscr{R}(\phi_\q(\As))
	= \bigcup_{\Bs} \mathscr{Q}(\phi_\q(\Bs)).
\end{equation}
Recall that here, $\Bs$ runs over all matrices satisfying $\Lambda_{\mathbb{F}_\q}(\phi_\q(\Bs)^\T) \supseteq \Lambda_{\mathbb{F}_\q}(\phi_\q(\As)^\T)$.
This inclusion relation is implied by $\Lambda_\mathbb{Z}(\Bs^\T) \supseteq \Lambda_\mathbb{Z}(\As^\T)$. In fact, the relationship $\Lambda_\mathbb{Z}(\Bs^\T) \supseteq \Lambda_\mathbb{Z}(\As^\T)$ is equivalent to there existing a (tall) integer matrix $\Rs^\T \in \mathbb{Z}^{L_\Bf \times L}$ such that $\Bs^\T\Rs^\T = \As^\T$. If such a matrix exists, then clearly there also exists a finite-field matrix $\Rs_\q^\T \in \mathbb{F}_\q^{L_\Bf \times L}$ such that $\phi_\q(\Bs^\T)\Rs_\q^\T = \phi_\q(\As^\T)$, for it suffices to set $\Rs_\q = \phi_\q(\Rs)$. Hence, by replacing the constraint $\Lambda_{\mathbb{F}_\q}(\phi_\q(\Bs)^\T) \supseteq \Lambda_{\mathbb{F}_\q}(\phi_\q(\As)^\T)$ with the constraint $\Lambda_\mathbb{Z}(\Bs^\T) \supseteq \Lambda_\mathbb{Z}(\As^\T)$, the union over matrices $\Bs$ is not increased. It follows that
\begin{IEEEeqnarray*}{rCl}
	\mathscr{R}^\infty(\As)
	&\supseteq& \varliminf_{\q \to \infty} \bigcup_{\Bs \colon \Lambda(\Bs^\T) \supseteq \Lambda(\As^\T)} \mathscr{Q}(\phi_\q(\Bs)) \\
	&\supseteq& \bigcup_{\Bs \colon \Lambda(\Bs^\T) \supseteq \Lambda(\As^\T)} \varliminf_{\q \to \infty} \mathscr{Q}(\phi_\q(\Bs))   \IEEEeqnarraynumspace\IEEEyesnumber\label{R_infty_first_inner_bound}
\end{IEEEeqnarray*}
(where we have omitted the subscript $\mathbb{Z}$ in $\Lambda_\mathbb{Z}(\cdot)$ to alleviate notation).
The exchange of limit and union in the last step is due to~\eqref{lims_of_union_and_intersections}.
In~\eqref{R_infty_first_inner_bound}, the set $\mathscr{Q}(\phi_\q(\Bs))$ is expressible as [cf.~\eqref{def:Q}]
\begin{equation}   \label{set_operation_exchange}
	\mathscr{Q}(\phi_\q(\Bs))
	= \bigcap_{M^* \in \mathscr{M}_{\mathbb{F}_\q}^\circ(L_\Bf)} \; \bigcup_{\Sc \in \mathscr{B}(M^*)} \; \bigcap_{\Tc \in \mathscr{B}([\phi_\q(\Bs)]_\Sc)} \mathscr{Q}(\phi_\q(\Bs),M,\Tc)
\end{equation}
where $\mathscr{Q}(\phi_\q(\Bs),M,\Tc)$, in turn, can be expressed as [cf.~\eqref{Q_partial}]
\begin{multline}
	\mathscr{Q}(\phi_\q(\Bs),M,\Tc)
	= \Bigl\{ (R_1,\dotsc,R_K) \in \mathbb{R}_+^K \colon \sum_{k\in\Tc} R_k < \\
		H(\phi_\q([\uv]_\Tc)) + \min_{\phi_\q(\Cs) \in \mathscr{C}_{\mathbb{F}_\q}(M)} H(\phi_\q(\Cs\Bs\uv)|Y_\q) - H(\phi_\q(\Bs\uv)|Y_\q) \Bigr\}   \IEEEeqnarraynumspace\label{R_widetilde}
\end{multline}
since $\phi_\q(\Cs)\phi_\q(\Bs)\phi_\q(\uv)=\phi_\q(\Cs\Bs\uv)$ where $Y_\q$ denotes the channel output induced by the auxiliary $\phi_\q(\uv)$. In writing~\eqref{R_widetilde}, we have replaced the algebraic entropies that appear in~\eqref{Q_partial} with discrete entropies. This replacement is legitimate due to all random variables taking value in finite fields (cf.~Lemma~\ref{lem:discrete_algebraic_entropy}).

For now, let us assume that $\uv \in \mathbb{Z}^K$ has finite support and let $\tau \in \mathbb{N}$ be some natural number large enough to satisfy
\begin{equation}   \label{finite_support_assumption}
	\uv \in \llbracket 2\tau+1 \rrbracket^K
	\quad \text{(almost surely)}
\end{equation}
or equivalently, $\lVert \uv \rVert_\infty \leq \tau$ almost surely.

We will prove that for any given $\Bs$, when $\q$ is larger than a certain constant $\bar{\q}$ (that possibly depends on $\Bs$ and $\tau$), we have
\begin{equation}   \label{Q_region_equality}
	\mathscr{Q}(\phi_\q(\Bs),M,\Tc)
	= \mathscr{Q}(\Bs,M,\Tc)
\end{equation}
wherein on the left-hand side, the rate region $\mathscr{Q}(\cdot,M,\Tc)$ is evaluated with the first argument taking value in the finite field ($\mathbb{A}=\mathbb{F}_\q$), whereas on the right-hand side, it takes value in the integers ($\mathbb{A}=\mathbb{Z}$).
For a more compact notation, we henceforth refer to the min-entropy term appearing in~\eqref{R_widetilde} as\footnote{The only innovation in notation as compared to~\eqref{def:J} is the subscript $\q$, to emphasize the dependency of this quantity on the field size $\q$}
\begin{equation}   \label{def:Jq}
	J_\q(\Bs,M)
	\triangleq \min_{\phi_\q(\Cs) \in \mathscr{C}_{\mathbb{F}_\q}(M)} H(\phi_\q(\Cs\Bs\uv)|Y_\q).
\end{equation}
Without loss of generality, we will assume throughout that the matrix $\Cs \in \mathbb{F}_\q^{L_\Cf \times L_\Bf}$ has full column rank, i.e., that its number of rows $L_\Cf$ is always equal to the rank of the matroid $M$.

In the following, we will prove in successive steps that the multiple dependencies on the field size $\q$ in the expressions~\eqref{set_operation_exchange}, \eqref{R_widetilde} and \eqref{def:Jq} can all be removed for sufficiently large $\q$. This will eventually allow us to evaluate $\varliminf_{\q \to \infty} \mathscr{Q}(\phi_\q(\Bs))$. Specifically, we will show that~\eqref{Q_region_equality} holds for any prime $\q$ larger than the prime number $\bar{\q}$, which shall be defined as the smallest prime larger than $\max_{i=1,\dotsc,10} \bar{\q}_i$, where
\begin{subequations}
\begin{IEEEeqnarray}{rCl}
	\bar{\q}_1
	&\triangleq& (2\tau + 1) \lVert \Bs \rVert_\infty,   \label{q_1} \\
	\bar{\q}_2
	&\triangleq& 1 + 2 \prod_{i=1}^{L_\Bf} \lVert \bs_i \rVert_2,   \label{q_2} \\
	\bar{\q}_3
	&\triangleq& 9^{L_\Bf L_\Cf},  \label{q_3} \\
	\bar{\q}_4
	&\triangleq& \psi_{L_\Cf^2}^2 L_\Cf^8,   \label{q_4} \\
	\bar{\q}_5
	&\triangleq& \tau,   \label{q_5} \\
	\bar{\q}_6
	&\triangleq& \left( 4 \left\lVert \Bs \right\rVert_\infty \tau \right)^{2 L_\Bf L_\Cf},   \label{q_6} \\
    \bar{\q}_7
    &\triangleq& \min\{ \q \in \mathbb{P} \colon \forall M \in \mathscr{M}_{\mathbb{F}_\q}(L_\Bf) \colon |\chi(M)| = \infty \},   \label{q_7} \\
    \bar{\q}_8
    &\triangleq& \max\bigl\{ \max\{ \mathbb{P} \setminus \chi(M) \} \colon M \in \mathscr{M}_\infty(L_\Bf) \bigr\}   \label{q_8}, \\
    \bar{\q}_9
    &\triangleq& 2^{2^{2^{L_\Bf^5}}}   \label{q_9}, \\
    \bar{\q}_{10}
    &\triangleq& \bar{\theta}(\tau \lVert \Bs \rVert_\infty)^{4/3}.   \label{q_10}
\end{IEEEeqnarray}
\end{subequations}
Here, $\bs_i^\T$ denotes the $i$-th row of $\Bs = [\bs_1 \dotso \bs_{L_\Bf}]^\T$,  $\psi_\ell$, $\chi(\cdot)$, $\mathscr{M}_\infty(\cdot)$, and $\bar{\theta}(\cdot)$ denote specific sequences, functions and sets, all of which will be duly defined in what follows. This constant $\bar{\q}$ depends solely on $\Bs$ and its dimensions $(K,L_\Bf)$, as well as on the parameter $\tau$. Each constant $\bar{\q}_i, i=1,\dotsc,10$ will be specified and motivated along the proof steps that will be detailed in the following.
For clarity, we will structure these successive proof steps into five separate paragraphs. Namely, we will prove that for $\q > \bar{\q}$,
\begin{itemize}
	\item	the entropies $H(\phi_\q([\uv]_\Tc))$ and $H(\phi_\q(\Bs\uv)|Y_\q)$ appearing in~\eqref{R_widetilde} become equal to $H([\uv]_\Tc)$ and $H(\Bs\uv|Y)$, respectively (Step~1);
	\item	the index set $\mathscr{B}([\phi_\q(\Bs)]_\Sc)$ appearing in~\eqref{set_operation_exchange} becomes equal to $\mathscr{B}([\Bs]_\Sc)$ (Step~2);
	\item	the min-entropy term $J_\q(\Bs,M)$ appearing in~\eqref{R_widetilde}, which involves an entropy $H(\phi_\q(\Cs\Bs\uv)|Y_\q)$ of matrix-vector products over the finite field $\mathbb{F}_\q$, can be expressed in terms of the entropy $H(\Cs\Bs\uv|Y)$ which only involves matrix-vector products over the integers, and does not depend on $\q$ (Step~3);
	\item	the set $\mathscr{M}_{\mathbb{F}_\q}^\circ(L_\Bf)$ appearing in~\eqref{set_operation_exchange} in the intersection over matroids $M^*$ becomes equal to $\mathscr{M}_{\mathbb{Z}}^\circ(L_\Bf)$ (Step~4);
	\item	in the evaluation of $J_\q(\Bs,M)$ [cf.~\eqref{def:Jq}], we can have the integer matrix $\Cs$ run over the set $\mathscr{C}_\mathbb{Z}(M)$ instead of having $\phi_\q(\Cs)$ run over $\mathscr{C}_{\mathbb{F}_\q}(M)$, without loss of generality (Step~5).
\end{itemize}
We argue in conclusion that the entire finite-field rate region tends (in the sense of an inner set limit) to its integer counterpart as $\q \to \infty$.
Since all the above steps rely on the assumption of a \emph{finitely supported} integer auxiliary $\uv \in \mathbb{Z}^K$, in a final proof step we will relax this assumption by another limit-taking, so as to allow for integer auxiliaries with \emph{infinite support} (under mild assumptions).

\paragraph{Step 1}
Notice that since, by assumption, the support of $\uv$ is contained in $\llbracket 2\tau + 1 \rrbracket^K$ almost surely, for $\q > 2\tau + 1$ we have $H(\phi_\q([\uv]_\Tc))) = H([\uv]_\Tc)$.
Likewise, for $\q \geq (2\tau+1) \lVert \Bs \rVert_\infty$, there is a one-to-one correspondence between $\phi_\q(\Bs\uv)$ and $\Bs\uv$ such that all instances of $Y_\q$ can be replaced by $Y$, hence $H(\phi_\q(\Bs\uv)|Y_\q) = H(\Bs\uv|Y)$.
Since $\q > (2\tau+1) \lVert \Bs \rVert_\infty \geq 2\tau+1$ is implied by $\q > \bar{\q}$ [cf.~\eqref{q_1}],  we have that~\eqref{R_widetilde} simplifies to
\begin{multline}
	\mathscr{Q}(\phi_\q(\Bs),M,\Tc)
	= \Bigl\{ (R_1,\dotsc,R_K) \in \mathbb{R}_+^K \colon \\
	\sum_{k \in \Tc} R_k <
		H([\uv]_\Tc) + J_\q(\Bs,M) - H(\Bs\uv|Y) \Bigr\},
	\quad \textnormal{(for $\q \geq \bar{\q}$)}.   \label{R_widetilde_bis}
\end{multline}
Likewise, $H(\phi_\q(\Cs\Bs\uv)|Y_\q) = H(\phi_\q(\Cs\Bs\uv)|Y)$, so for sufficiently large $\q$, \eqref{def:Jq} may be written as
\begin{equation}
	J_\q(\Bs,M)
	\triangleq \min_{\phi_\q(\Cs) \in \mathscr{C}_{\mathbb{F}_\q}(M)} H(\phi_\q(\Cs\Bs\uv)|Y),
	\quad \textnormal{(for $\q \geq \bar{\q}$)}.   \IEEEeqnarraynumspace\label{Jq_bis}
\end{equation}

\paragraph{Step 2}
Using the following Lemma~\ref{lem:mathscr_T}, we can establish that $\mathscr{B}([\phi_\q(\Bs)]_\Sc) = \mathscr{B}([\Bs]_\Sc)$ for sufficiently large $\q$.
\lemmabox{
\begin{lemma}   \label{lem:mathscr_T}
For any full row-rank integer matrix $\Qs \in \mathbb{Z}^{m \times n}$ and an odd prime number $\q$ satisfying
$\q > 1 + 2 \sqrt{\det(\Qs\Qs^\T)}$, we have $M(\phi_\q(\Qs)) = M(\Qs)$.
\end{lemma}
}

By applying Lemma~\ref{lem:mathscr_T} to the matrix $\Qs = [\Bs]_\Sc$, we conclude that $\mathscr{B}([\phi_\q(\Bs)]_\Sc) = \mathscr{B}([\Bs]_\Sc)$ if
\begin{equation}   \label{q_2_sufficient_condition}
	\q
	\geq 1 + 2 \max_{\Sc \subset [L_\Bf]} \sqrt{\det([\Bs]_\Sc [\Bs]_\Sc^\T)}.
\end{equation}
Due to Hadamard's inequality applied on the rows of $\Bs = [\bs_1 \dotso \bs_{L_\Bf}]^\T$, all of which have Euclidean norm $\lVert \bs_i \rVert_2 \geq 1$, a sufficient condition for~\eqref{q_2_sufficient_condition} is
\begin{equation}
	\q
	\geq 1 + 2 \prod_{i=1}^{L_\Bf} \lVert \bs_i \rVert_2
\end{equation}
which is implied by $\q > \bar{\q}$ [cf.~\eqref{q_2}].

\proofbox{}{
\begin{IEEEproof}
Recall that $M(\phi_\q(\Qs))$ is characterized by its collection of bases $\mathscr{B}(\phi_\q(\Qs))$, which contains all index sets $\Bc \subset [n]$ of cardinality $m$ such that the submatrix $[\Qs^\T]_\Bc$ has full rank. Equivalently, we can test this full-rank condition by establishing whether $\det(\phi_\q([\Qs^\T]_\Bc)) \neq 0$, which is further equivalent to
\begin{equation}   \label{non-zero_det}
	\mymod_\q\bigl( \det([\Qs^\T]_\Bc) \bigr)
	\neq 0.
\end{equation}
By the Cauchy--Binet formula~\cite[Sec. 0.8.7]{HornJ1985},
\begin{IEEEeqnarray*}{rCl}
	\det(\Qs\Qs^\T)
	&=& \sum_{\Bc \subset [n]} \det([\Qs^\T]_\Bc^\T [\Qs^\T]_\Bc) \\
	&\geq& \det([\Qs^\T]_\Bc^\T [\Qs^\T]_\Bc) \\
	&=& \det([\Qs^\T]_\Bc)^2.   \IEEEeqnarraynumspace\IEEEyesnumber
\end{IEEEeqnarray*}
We infer that, if $\q$ is large enough to satisfy
\begin{equation*}
	\left(\frac{\q-1}{2}\right)^2
	\geq \det(\Qs\Qs^\T)
\end{equation*}
then $\left| \det([\Qs^\T]_\Bc) \right| \leq (\q-1)/2$ and therefore $\mymod_\q\bigl( \det([\Qs^\T]_\Bc) \bigr) = \det([\Qs^\T]_\Bc)$, which implies that~\eqref{non-zero_det} is equivalent to
\begin{equation*}
	\det([\Qs^\T]_\Bc) \neq 0
\end{equation*}
for any $\Bc \subset [n]$. Consequently, $M(\phi_\q(\Qs))$ and $M(\Qs)$ have the same collection of bases, and are therefore equal matroids.
\end{IEEEproof}
}

\paragraph{Step 3}
We now turn our attention to the min-entropy term $J_\q(\Bs,M)$ as defined in~\eqref{def:Jq}. For $\q > \bar{\q}$, the term $J_\q(\Bs,M)$ as expressed in~\eqref{Jq_bis} can be written as
\begin{IEEEeqnarray*}{rCl}
	J_\q(\Bs,M)
	&\triangleq& \min_{\Cs \colon \phi_\q(\Cs) \in \mathscr{C}_{\mathbb{F}_\q}(M)} H(\phi_\q(\Cs\Bs\uv)|Y) \\
	&=& \min_{\substack{\Cs \in \llbracket \q \rrbracket^{L_\Cf \times L_\Bf} \colon \\ \phi_\q(\Cs) \in \mathscr{C}_{\mathbb{F}_\q}(M)}} H(\mymod_\q(\Cs\Bs\uv)|Y), \quad \textnormal{(for $\q > \bar{\q}$)}.   \IEEEeqnarraynumspace\IEEEyesnumber\label{J_recall}
\end{IEEEeqnarray*}
The last equality holds because $\phi_\q(x)$ is isomorphic to $\mymod_\q(x)$ [see definitions of $\mymod_\q(\cdot)$ and $\phi_\q(\cdot)$ in \eqref{def:mod_q}, \eqref{def:natural_mapping}] and thus switching from $\phi_\q(\cdot)$ to $\mymod_\q(\cdot)$ leaves the entropy value unchanged, as well as because the entries of $\Cs$ may be restricted from $\mathbb{Z}$ to the finite set $\llbracket \q \rrbracket$ without loss of generality.

In what follows, we will show that the optimization domain in~\eqref{J_recall} can be further reduced, in that the entries of $\Cs$ may be restricted to a subset of $\llbracket \q \rrbracket$, and that as a consequence, for sufficiently large $\q$, the modulo operation $\mymod_\q(\cdot)$ in the entropy term of~\eqref{J_recall} can be removed---loosely speaking, because no ``wrap-arounds'' by the modulo operation ever occur.

The first thing to observe in~\eqref{J_recall} is that, by the definition of $\mathscr{C}_{\mathbb{F}_\q}(M)$ [cf.~Section~\ref{ssec:matroids}, Definition~\ref{def:representable_matroids}], for any full-rank $\Qs \in \mathbb{F}_\q^{L_\Cf \times L_\Cf}$ and any given matroid $M$, it holds that $\phi_\q(\Cs) \in \mathscr{C}_{\mathbb{F}_\q}(M)$ if and only if $\Qs\phi_\q(\Cs) \in \mathscr{C}_{\mathbb{F}_\q}(M)$. In addition, the entropy term, which constitutes the minimization's objective function, is invariant against invertible row operations on $\phi_\q(\Cs)$, in the sense that
\begin{IEEEeqnarray*}{rCl}
	H(\Qs\phi_\q(\Cs\Bs\uv)|Y)
	&=& H(\Qs\phi_\q(\Cs)\phi_\q(\Bs\uv)|Y) \\
	&=& H(\phi_\q(\Cs)\phi_\q(\Bs\uv)|Y) \\
	&=& H(\phi_\q(\Cs\Bs\uv)|Y).   \IEEEeqnarraynumspace
\end{IEEEeqnarray*}
This means that the set $\mathscr{C}_{\mathbb{F}_\q}(M)$, which determines the optimization domain, can be partitioned into \emph{equivalence classes}. Specifically, we define the equivalence class of some full row-rank $\phi_\q(\Cs) \in \mathbb{F}_\q^{L_\Cf \times L_\Bf}$ as the set
\begin{equation}   \label{def:equivalence_class}
	\mathscr{E}[\Cs]
	\triangleq \Bigl\{ \Qs\phi_\q(\Cs) \colon \textnormal{$\Qs \in \mathbb{F}_\q^{L_\Cf \times L_\Cf}$ is full-rank} \Bigr\}
\end{equation}
and say that $\Cs$ and $\Cs'$ are equivalent if and only if $\mathscr{E}[\Cs] = \mathscr{E}[\Cs']$.
By contrast, let us define the lattice\footnote{Since this is a lattice of \emph{matrices} rather than \emph{vectors}, we deviate from the previously used notation $\Lambda(\Cs)$ (with round parentheses) to avoid confusion.}
\begin{equation}
	\Lambda[\Cs]
	\triangleq \bigl\{ \Qs\phi_\q(\Cs) \colon \Qs \in \mathbb{F}_\q^{L_\Cf \times L_\Cf} \bigr\}.
\end{equation}
For any $\Cs \in \llbracket \q \rrbracket^{L_\Cf \times L_\Bf}$ such that $\phi_\q(\Cs)$ has full row rank, we shall choose as the \emph{representative} of its equivalence class $\mathscr{E}[\Cs]$ the $\Cs_\star \in \mathscr{E}[\Cs]$ with smallest infinity norm, i.e.,
\begin{equation}
	\Cs_\star
	\triangleq \argmin_{\Cs' \colon \mathscr{E}[\Cs'] = \mathscr{E}[\Cs]} \left\lVert \Cs' \right\rVert_\infty
\end{equation}
(ties being broken arbitrarily).\footnote{Although $\Cs_\star$ is technically a function of $\Cs$, we refrain from writing $\Cs_\star(\Cs)$ to alleviate notation.}
Then, by definition of the equivalence classes, the matrix $\Cs$ in the minimization problem~\eqref{def:Jq} can be replaced by its small-norm equivalent $\Cs_\star$ without loss of generality, yielding
\begin{equation}   \label{def:Jq_bis1}
	J_\q(\Bs,M)
	= \min_{\phi_\q(\Cs) \in \mathscr{C}_{\mathbb{F}_\q}(M)} H(\mymod_\q(\Cs_\star\Bs\uv)|Y),
	\quad \textnormal{(for $\q > \bar{\q}$)}.
\end{equation}
Next, we will derive an upper bound on the norm of $\Cs_\star$. Using the $\vectorize(\cdot)$ operation that stacks the columns of a matrix on top of each other into a single column vector, let us define the vectorized versions of the sets $\mathscr{E}[\Cs]$ and $\Lambda[\Cs]$ as
\begin{subequations}
\begin{IEEEeqnarray}{rCl}   \label{def:equivalence_class_vectorized}
	\mathscr{E}_\mathrm{vec}[\Cs]
	&\triangleq& \Bigl\{ \mathrm{vec}(\Qs\phi_\q(\Cs)) \colon \textnormal{$\Qs \in \mathbb{F}_\q^{L_\Cf \times L_\Cf}$ is full-rank} \Bigr\}   \IEEEnonumber\\
	&=& \Bigl\{ (\phi_\q(\Cs)^\T \otimes \Is_{L_\Cf}) \mathrm{vec}(\Qs) \colon \textnormal{$\Qs \in \mathbb{F}_\q^{L_\Cf \times L_\Cf}$ is full-rank} \Bigr\} \\
	\Lambda_\mathrm{vec}[\Cs]
	&\triangleq& \bigl\{ \mathrm{vec}(\Qs\phi_\q(\Cs)) \colon \Qs \in \mathbb{F}_\q^{L_\Cf \times L_\Cf} \bigr\}   \IEEEnonumber\\
	&=& \bigl\{ (\phi_\q(\Cs)^\T \otimes \Is_{L_\Cf}) \mathrm{vec}(\Qs) \colon \Qs \in \mathbb{F}_\q^{L_\Cf \times L_\Cf} \bigr\}.
\end{IEEEeqnarray}
\end{subequations}
Note that $\phi_\q^{-1}\left( \Lambda_\mathrm{vec}[\Cs] \right)$
of the lattice $\Lambda_\mathrm{vec}[\Cs]$ forms a sublattice of $\mathbb{Z}^{L_\Bf L_\Cf}$ that can be expressed as follows:
\begin{IEEEeqnarray*}{rCl}
	\phi_\q^{-1}\left( \Lambda_\mathrm{vec}[\Cs] \right)
	&=& \Bigl\{ \phi_\q^{-1} \bigl( \phi_\q(\Cs^\T \otimes \Is_{L_\Cf}) \kv \bigr) \colon \kv \in \mathbb{Z}^{L_\Cf^2} \Bigr\} \\
	&=& \left\{ \begin{bmatrix} (\Cs^\T \otimes \Is_{L_\Cf}) & \q\Is_{L_\Bf L_\Cf} \end{bmatrix} \kv \colon \kv \in \mathbb{Z}^{L_\Cf^2 + L_\Bf L_\Cf} \right\} \\
	&=& \Lambda(\hat{\Cs})   \IEEEeqnarraynumspace\IEEEyesnumber
\end{IEEEeqnarray*}
where $\hat{\Cs} \in \mathbb{Z}^{L_\Bf L_\Cf \times L_\Bf L_\Cf}$ is a full-rank integer matrix whose columns form a basis of the lattice generated by the columns of $\bigl[ (\Cs^\T \otimes \Is_{L_\Cf}) \ \ \q\Is_{L_\Bf L_\Cf} \bigr]$.
By a counting argument, the determinant of $\Lambda(\hat{\Cs})$ equals
\begin{equation}
	\det(\Lambda(\hat{\Cs}))
	= \bigl| \det(\hat{\Cs}) \bigr|
	= \frac{\left| \llbracket \q \rrbracket^{L_\Bf L_\Cf} \right|}{\left| \Lambda_\mathrm{vec}[\Cs] \right|}
	= \q^{L_\Bf L_\Cf - L_\Cf^2}.
\end{equation}
Note that since $\mathscr{E}_\mathrm{vec}[\Cs] \subset \Lambda_\mathrm{vec}[\Cs]$, we have
\begin{IEEEeqnarray}{rCl}
	\phi_\q^{-1}\left( \mathscr{E}_\mathrm{vec}[\Cs] \right)
	\subset \Lambda(\hat{\Cs}).
\end{IEEEeqnarray}
On the other hand, it becomes apparent from their definitions that the sets $\mathscr{E}_\mathrm{vec}[\Cs]$ and $\mathscr{E}[\Cs]$ have a cardinality $\left| \mathscr{E}_\mathrm{vec}[\Cs] \right| = \left| \mathscr{E}[\Cs] \right|$ equal to the number of full-rank matrices in $\mathbb{F}_\q^{L_\Cf \times L_\Cf}$, which is given by a polynomial (in $\q$) of degree $L_\Cf^2$ with integer coefficients~\cite{Lan1893}:
\begin{IEEEeqnarray*}{rCl}
	\left| \mathscr{E}[\Cs] \right|
	&=& \prod_{i=0}^{L_\Cf-1}(\q^{L_\Cf}-\q^i)
	= \sum_{i=0}^{L_\Cf(L_\Cf+1)/2} \gamma_i \q^{L_\Cf^2-i}.   \IEEEeqnarraynumspace\IEEEyesnumber
\end{IEEEeqnarray*}
We can readily check that $\gamma_0 = 1$. For other coefficients $\gamma_i$, it is easy to see that $|\gamma_i|$ is upper-bounded by the number of ways $\psi_i$ in which $i$ can be written as the sum of distinct natural numbers~\cite[p.~836, Table~24.5]{AbSt72} (see also~\cite{OEIS_A000009} and references therein). For the evaluation of $\psi_i$, a recurrence relation and generating function are known~\cite{Ew73}.
Hence, a simple lower bound on $\left| \mathscr{E}[\Cs] \right|$ is given by
\begin{IEEEeqnarray*}{rCl}
	\left| \mathscr{E}[\Cs] \right|
	&=& \q^{L_\Cf^2} + \sum_{i=1}^{L_\Cf(L_\Cf+1)/2} \gamma_i \q^{L_\Cf^2-i} \\
	&\geq& \q^{L_\Cf^2} - \sum_{i=1}^{L_\Cf(L_\Cf+1)/2} \psi_i \q^{L_\Cf^2-i} \\
	&\geq& \q^{L_\Cf^2} - \psi_{L_\Cf(L_\Cf+1)/2} \sum_{i=1}^{L_\Cf(L_\Cf+1)/2} \q^{L_\Cf^2-i} \\
	&\geq& \q^{L_\Cf^2} - 2 \psi_{L_\Cf^2} L_\Cf^4 \q^{L_\Cf^2-1}.   \IEEEeqnarraynumspace\IEEEyesnumber\label{E_UB}
\end{IEEEeqnarray*}
For the last bounding step, we have used that $\psi_{L_\Cf(L_\Cf+1)/2} \leq \psi_{L_\Cf^2}$ (because $\psi_i$ is a monotone sequence) and that $\q^{L_\Cf^2-i} \leq \q^{L_\Cf^2-1}$, as well as the fact that the sum of integers from $1$ to $L_\Cf(L_\Cf+1)/2$ is upper-bounded by $2 L_\Cf^4$.

Note that since $\left| \Lambda[\Cs] \right| = \left| \Lambda_\mathrm{vec}[\Cs] \right| = \q^{L_\Cf^2}$, which equals the minuend on the right-hand side of~\eqref{E_UB}, the cardinality of the complement
\begin{equation}
	\bar{\mathscr{E}}_\mathrm{vec}[\Cs]
	\triangleq \Lambda_\mathrm{vec}[\Cs] \setminus \mathscr{E}_\mathrm{vec}[\Cs]
\end{equation}
is upper-bounded as
\begin{equation}   \label{E_bar_UB}
	\left| \bar{\mathscr{E}}_\mathrm{vec}[\Cs] \right|
	\leq 2 \psi_{L_\Cf^2} L_\Cf^4 \q^{L_\Cf^2-1}.
\end{equation}
Using van der Corput's Convex Body Theorem [cf.~Lem.~\ref{thm:van_der_Corput}, Subsection~\ref{ssec:euclidean_lattices}] in its form~\eqref{van_der_Corput_inequality}, we can lower-bound the number of lattice points from $\Lambda(\hat{\Cs})$ enclosed in the closed convex body $\kappa\mathscr{S}_\infty^{L_\Bf L_\Cf} = [-\kappa,\kappa]^{L_\Bf L_\Cf}$ for any $\kappa > 0$ as follows:
\begin{IEEEeqnarray*}{rCl}   \label{vanderCorput}
	\bigl| \Lambda(\hat{\Cs}) \cap \kappa\mathscr{S}_\infty^{L_\Bf L_\Cf} \bigr| + 1
	&>& \frac{\vol(\kappa \mathscr{S}_\infty^{L_\Bf L_\Cf})}{2^{L_\Bf L_\Cf - 1} \det(\Lambda(\hat{\Cs}))} \\
	&=& 2 \frac{\kappa^{L_\Bf L_\Cf}}{\q^{L_\Bf L_\Cf - L_\Cf^2}}.   \IEEEeqnarraynumspace\IEEEyesnumber
\end{IEEEeqnarray*}
Hereinafter, let us set the parameter $\kappa$ to be a function of $\q$, namely,
\begin{equation}
	\kappa(\q)
	\triangleq \q^{1-\frac{1}{2 L_\Bf L_\Cf}}.
\end{equation}
On the one hand, this choice of $\kappa(\q)$ ensures that $\kappa(\q)\mathscr{S}_\infty^{L_\Bf L_\Cf} \cap \mathbb{Z}^{L_\Bf L_\Cf}$ (whose diameter grows linearly in $\q$) is eventually contained in $\llbracket \q \rrbracket^{L_\Bf L_\Cf}$ (whose diameter grows sublinearly in $\q$) for sufficiently large $\q$. Specifically, this inclusion holds as soon as $\q > 9^{L_\Bf L_\Cf}$, which is implied by $\q > \bar{\q}$ [cf.~\eqref{q_3}]. In fact, $\q > 9^{L_\Bf L_\Cf}$ is equivalent to $\q^{1-1/(2 L_\Bf L_\Cf)} < \q/3$, which implies $\q^{1-1/(2 L_\Bf L_\Cf)} < (\q-1)/2$ (since $\q/3 \leq (\q-1)/2$ due to $\q \geq 3$, because $\q$ is an odd prime). In summary,
\begin{IEEEeqnarray*}{rCl}
	\Lambda(\hat{\Cs}) \cap \kappa(\q)\mathscr{S}_\infty^{L_\Bf L_\Cf}
	&\subset& \Lambda(\hat{\Cs}) \cap \llbracket \q \rrbracket^{L_\Bf L_\Cf} \\
	&=& \phi_\q^{-1}\bigl( \Lambda_\mathrm{vec}[\Cs] \bigr), \quad \textnormal{(for $\q > \bar{\q}$)}.   \IEEEeqnarraynumspace\IEEEyesnumber
\end{IEEEeqnarray*}
On the other hand, our choice $\kappa(\q) = \q^{1-1/(2 L_\Bf L_\Cf)}$ ensures that for sufficiently large $\q$, the scaled cube $\kappa(\q)\mathscr{S}_\infty^{L_\Bf L_\Cf}$ contains more lattice points from $\Lambda(\hat{\Cs})$ than does $\phi_\q^{-1}(\bar{\mathscr{E}}_\mathrm{vec}[\Cs]) \cap \llbracket \q \rrbracket^{L_\Bf L_\Cf}$. In fact, by combining the inequalities \eqref{E_bar_UB} and \eqref{vanderCorput} for our choice of $\kappa(\q)$,
\begin{IEEEeqnarray*}{rCl}
	\bigl| \Lambda(\hat{\Cs}) \cap \kappa(\q)\mathscr{S}_\infty^{L_\Bf L_\Cf} \bigr|
	&\geq& -1 + \frac{2 \q^{L_\Bf L_\Cf-1/2}}{\q^{L_\Bf L_\Cf - L_\Cf^2}} \\
	&=& -1 + 2\q^{L_\Cf^2 - 1/2} \\
	&>& 2 \psi_{L_\Cf^2} L_\Cf^4 \q^{L_\Cf^2-1} \\
	&\geq& \left| \bar{\mathscr{E}}_\mathrm{vec}[\Cs] \right|,
	\quad \textnormal{(for $\q > \bar{\q}$)}   \IEEEeqnarraynumspace\IEEEyesnumber
\end{IEEEeqnarray*}
where the second (strict) inequality certainly holds if $\sqrt{\q} > 2 \psi_{L_\Cf^2} L_\Cf^4$, which is implied by $\q > \bar{\q}$ [cf.~\eqref{q_4}]. To see this, notice that $\sqrt{\q} > 2 \psi_{L_\Cf^2} L_\Cf^4$ is equivalent to $\q^{L_\Cf^2 - 1/2} > 2 \psi_{L_\Cf^2} L_\Cf^4 \q^{L_\Cf^2-1}$, which due to $\q^{L_\Cf^2-1/2} > 1$ (because of $\q \geq 3$ and $L_\Cf \geq 1$) in turn implies said inequality $-1 + 2\q^{L_\Cf^2 - 1/2} > 2 \psi_{L_\Cf^2} L_\Cf^4 \q^{L_\Cf^2-1}$.

It follows from this inequality that, for $\q > \bar{\q}$, the lattice points in the intersection $\Lambda(\hat{\Cs}) \cap \kappa(\q)\mathscr{S}_\infty^{L_\Bf L_\Cf}$ outnumber the points in the complement of $\mathscr{E}[\Cs]$, and therefore the cube $\kappa(\q)\mathscr{S}_\infty^{L_\Bf L_\Cf}$ must contain at least one point from $\phi_\q^{-1}(\mathscr{E}_\mathrm{vec}[\Cs]) \cap \llbracket \q \rrbracket^{L_\Bf L_\Cf}$. Hence the infinity norm of $\Cs_\star \in \phi_\q^{-1}(\mathscr{E}[\Cs]) \cap \llbracket \q \rrbracket^{L_\Cf \times L_\Bf}$ is at most $\kappa(\q)$. This in turn means that in the expression of $J_\q(\Bs,M)$ from \eqref{def:Jq_bis1}, without loss of generality we can restrain the domain of $\Cs$ to the scaled cube $\kappa(\q)\mathscr{S}_\infty^{L_\Bf L_\Cf}$, whose side length grows sublinearly in $\q$, i.e.,
\begin{equation}   \label{def:Jq_bis2}
	J_\q(\Bs,M)
	= \min_{\substack{\Cs \colon \phi_\q(\Cs) \in \mathscr{C}_{\mathbb{F}_\q}(M) \colon \\ \lVert \Cs \rVert_\infty \leq \kappa(\q)}} H(\mymod_\q(\Cs\Bs\uv)|Y).
\end{equation}
Recall that, by assumption~[cf.~\eqref{finite_support_assumption}] we have $\uv \in \tau\mathscr{S}_\infty^K \cap \mathbb{Z}^K$ almost surely, so the argument of the entropy term in~\eqref{def:Jq_bis2} has infinity norm upper-bounded (almost surely) as
\begin{IEEEeqnarray*}{rCl}
	\left\lVert \Cs\Bs\uv \right\rVert_\infty &\leq& \kappa(\q) \left\lVert \Bs \right\rVert_\infty \lVert \uv \rVert_\infty \\
	&\leq& \left\lVert \Bs \right\rVert_\infty \tau \kappa(\q), \quad \textnormal{(for $\q > \bar{\q}$).}   \IEEEeqnarraynumspace\IEEEyesnumber\label{support_bound}
\end{IEEEeqnarray*}
Since the right-hand side of~\eqref{support_bound} grows sublinearly in $\q$, we infer that it is eventually upper-bounded by $\frac{\q-1}{2}$ as $\q$ grows large. Specifically, a sufficient condition for $\left\lVert \Bs \right\rVert_\infty \tau \kappa(\q) \leq \frac{\q-1}{2}$ to hold is that
\begin{equation}
	\q
	> \left( 4 \left\lVert \Bs \right\rVert_\infty \tau \right)^{2 L_\Bf L_\Cf}
\end{equation}
which is implied by $\q > \bar{\q}$ [cf.~\eqref{q_6}].
Hence, if $\q > \bar{\q}$ the modulo operation $\mymod_\q(\cdot)$ becomes superfluous, in the sense that for $\lVert \Cs \rVert_\infty \leq \kappa(\q)$,
\begin{equation}
	\mymod_\q(\Cs\Bs\uv)
	= \Cs\Bs\uv,
	\quad \textnormal{(for $\q > \bar{\q}$, almost surely).}
\end{equation}
This identity finally allows us to write
\begin{IEEEeqnarray*}{rCl}   \label{def:Jq_bis4}
	J_\q(\Bs,M)
	&=& \min_{\substack{\Cs \in \llbracket 2\floor{\kappa(\q)} + 1 \rrbracket^{L_\Cf \times L_\Bf} \colon \\ \phi_\q(\Cs) \in \mathscr{C}_{\mathbb{F}_\q}(M)}} H(\Cs\Bs\uv|Y),
	\quad \textnormal{(for $\q > \bar{\q}$).}   \IEEEeqnarraynumspace\IEEEyesnumber
\end{IEEEeqnarray*}
This way, transitioning from~\eqref{def:Jq_bis2} to~\eqref{def:Jq_bis4}, we have removed the $\mymod_\q(\cdot)$ operation that appeared in the minimization's objective.

\paragraph{Step 4}
Consider the outer set limit
\begin{equation}
	\mathscr{M}_\infty(L_\Bf)
	\triangleq \varlimsup_{\q \to \infty} \mathscr{M}_{\mathbb{F}_\q}(L_\Bf).
\end{equation}
Recalling the definition of the outer set limit given in~\eqref{def:limsup}, we can make the outer intersection start above any arbitrary offset $\bar{\q} > 0$ without changing the value of the set limit, i.e.,
\begin{equation}
	\mathscr{M}_\infty(L_\Bf)
	= \bigcap_{\{\q_n\colon \q_n \geq \bar{\q}\}_{n \in \mathbb{N}} } \bigcup_{n \in \mathbb{N}} \mathscr{M}_{\mathbb{F}_{\q_n}}(L_\Bf)
\end{equation}
where the intersection is over all growing sequences $\{\q_n\}_{n \in \mathbb{N}}$ of prime numbers (starting at a value larger or equal to $\bar{\q}$).

Every matroid in $\mathscr{M}_\infty(L_\Bf)$ has infinite characteristic set. In fact, if some matroid $M \in \mathscr{M}_\infty(L_\Bf)$ had a finite characteristic set $|\chi(M)| < \infty$, we could pick the largest element of $\chi(M)$ and set the offset $\bar{\q}$ to be larger than that element. Since there are only finitely many matroids in $\mathscr{M}_\infty(L_\Bf)$ (in any case, no more than $2^{L_\Bf}$), this process of increasing the offset $\bar{\q}$ can be repeated a finite number of times for all other matroids $M \in \mathscr{M}_\infty(L_\Bf)$ with $|\chi(M)| < \infty$ until no more such matroids exist in $\{ \mathscr{M}_{\mathbb{F}_\q}(L_\Bf) \colon \q > \bar{\q} \}$, thus resulting in a contradiction. Hence, the prime number
\begin{equation}
	\bar{\q}_7
	\triangleq \min\{ \q \in \mathbb{P} \colon \forall M \in \mathscr{M}_{\mathbb{F}_\q}(L_\Bf) \colon |\chi(M)| = \infty \}
\end{equation}
is well defined. Since by lemma~\ref{pro:Vamos71} (cf.~Definition~\ref{def:characteristic_set}), $|\chi(M)| = \infty$ implies $0 \in \chi(M)$, we conclude that every $M \in \mathscr{M}_\infty(L_\Bf)$ is representable over $\mathbb{Q}$. In conclusion, for $\q > \bar{\q}_7$ (which is implied by $\q > \bar{\q}$ [cf.~\eqref{q_7}]), we have
\begin{equation}   \label{matroid_set_inclusions}
	\mathscr{M}_{\mathbb{F}_\q}(L_\Bf)
	\subset \mathscr{M}_\infty(L_\Bf)
	\subset \mathscr{M}_\mathbb{Q}(L_\Bf), \quad \text{(for $\q > \bar{\q}$)}
\end{equation}
Furthermore, by lemma~\ref{pro:Rado57}, every $M \in \mathscr{M}_\infty(L_\Bf)$ is representable over all prime fields $\mathbb{F}_\q$ except for a finite set of primes. Therefore, there exists a prime number
\begin{equation}
	\bar{\q}_8
	\triangleq \max\bigl\{ \max\{ \mathbb{P} \setminus \chi(M) \} \colon M \in \mathscr{M}_\infty(L_\Bf) \bigr\}
\end{equation}
such that for $\q > \max\{ \bar{\q}_7, \bar{\q}_8 \}$ (which is implied by $\q > \bar{\q}$ [cf.~\eqref{q_7}--\eqref{q_8}]) the left-hand inclusion relation in~\eqref{matroid_set_inclusions} holds with equality, $\mathscr{M}_{\mathbb{F}_\q}(L_\Bf) = \mathscr{M}_\infty(L_\Bf)$. On the other hand, we have the following lemma.

\lemmabox{
\begin{lemma}[Bell \emph{et al}, 2020]   \label{pro:Bell2020}
Let $M$ be an $n$-element matroid representable over a field of characteristic $0$, and let $\q$ be a prime satisfying
\begin{equation}
	\log_2 \log_2 \log_2(\q) > n^5.
\end{equation}
Then $M$ is representable over $\mathbb{F}_\q$.
\end{lemma}}
A proof is provided in~\cite{Bel20}.

It follows from lemma~\ref{pro:Bell2020} that for $\q$ larger than $\bar{\q}_9 \triangleq 2^{2^{2^{L_\Bf^5}}}$ (which is implied by $\q > \bar{\q}$ [cf.~\eqref{q_9}]), we have $\mathscr{M}_\mathbb{Q}(L_\Bf) \subset \mathscr{M}_{\mathbb{F}_\q}(L_\Bf)$. In conclusion, for $\q > \max\{\bar{\q}_7,\bar{\q}_8,\bar{\q}_9\}$ (implied by $\q > \bar{\q}$), we have
\begin{equation}
	\mathscr{M}_{\mathbb{F}_\q}(L_\Bf)
	= \mathscr{M}_\mathbb{Q}(L_\Bf)
	= \mathscr{M}_\mathbb{Z}(L_\Bf),
	\quad \text{(for $\q > \bar{\q}$)}
\end{equation}
(the last equality being Lemma~\ref{lem:Q_and_Z_representability}).

\paragraph{Step 5}
We will now show that in the minimization problem that defines the quantity $J_\q(\Bs,M)$ [cf.~\eqref{def:Jq_bis4}], the constraint $\phi_\q(\Cs) \in \mathscr{C}_{\mathbb{F}_\q}(M)$ can be replaced by $\Cs \in \mathscr{C}_{\mathbb{Z}}(M)$. We will leverage the fact that matrices $\Cs$ with a covolume $\det(\Cs\Cs^\T)$ exceeding a certain constant, can be removed from the minimization domain defining $J_\q(\Bs,M)$ [as expressed in~\eqref{def:Jq_bis4}] without loss of generality. Recall that $\Cs$ is right-invertible by assumption and that $\Cs_\perp$ is left-invertible by definition of the orthogonal operation `$\perp$' (see Subsection~\ref{ssec:euclidean_lattices}). Hence, we have the covolume identity~\cite[Thm.~1]{Ngu97}
\begin{equation}
	\det(\Cs\Cs^\T)
	 = \det(\Cs_\perp^\T\Cs_\perp).
\end{equation}
Let us denote $\vv = \Bs\uv$, which due to $\uv$ having bounded support $\P\{\lVert \uv \rVert_\infty \leq \tau\} = 1$, also has bounded support:
\begin{IEEEeqnarray*}{rCl}
	\lVert \vv \rVert_\infty
	&=& \lVert \Bs\uv \rVert_\infty \\
	&\leq& \lVert \Bs \rVert_\infty \lVert \uv \rVert_\infty \\
	&\leq& \lVert \Bs \rVert_\infty \tau, \quad \text{(almost surely).}   \IEEEeqnarraynumspace\IEEEyesnumber\label{V_bounded_support}
\end{IEEEeqnarray*}
In other words, this means that $\vv \in \tau \lVert \Bs \rVert_\infty \mathscr{S}_\infty^{L_\Bf}$ almost surely. Let $\hat{\tau} = \tau \Vert \Bs \rVert_\infty$ stand for the radius (with respect to the infinity norm) of the ball $\tau \lVert \Bs \rVert_\infty \mathscr{S}_\infty^{L_\Bf}$.

As a first step, we argue that the minimization~\eqref{def:Jq_bis4} can be restricted to only those matrices $\Cs$ satisfying $\lambda_{\infty,1}(\Cs_\perp) \leq 2\hat{\tau}$. In fact, the objective $H(\Cs\Bs\uv|Y) = H(\Cs\vv|Y)$ can be expressed by means of
\begin{IEEEeqnarray*}{rCl}
	H(\Cs\vv|Y=y)
	&=& \sum_{\ts \in \mathbb{Z}^{L_\Cf}} \Phi\bigl( \P\{ \Cs\vv = \ts | Y=y \} \bigr) \\
	&=& \sum_{\Lambda' \in \mathbb{Z}^{L_\Bf}/\Lambda(\Cs_\perp)} \Phi\bigl( \P\{ \vv \in \Lambda' | Y=y \} \bigr) \\
	&=& \sum_{\Lambda' \in \mathbb{Z}^{L_\Bf}/\Lambda(\Cs_\perp)} \Phi\bigl( \P\{ \vv \in \Lambda' \cap \hat{\tau} \mathscr{S}_\infty^{L_\Bf} | Y=y \} \bigr)   \IEEEeqnarraynumspace\IEEEyesnumber
\end{IEEEeqnarray*}
where $A/B$ denotes the quotient set of $A$ by $B$.
The last equality holds because the support of $\vv$ is contained in $\mathscr{S}_\infty^{L_\Bf}$. It becomes apparent that if the shortest vector of $\Lambda(\Cs_\perp)$ is larger than the diameter of the cube $\mathscr{S}_\infty^{L_\Bf}$, i.e., if $\lambda_{\infty,1}(\Cs_\perp) > 2\hat{\tau}$, then the intersection $\Lambda' \cap \hat{\tau} \mathscr{S}_\infty^{L_\Bf}$ is either an empty set or a singleton set. This entails that
\begin{IEEEeqnarray*}{rCl}
	H(\Cs\vv|Y=y)
	&=& \sum_{\Lambda' \in \mathbb{Z}^{L_\Bf}/\Lambda(\Cs_\perp)} \Phi\bigl( \P\{ \vv \in \Lambda' \cap \hat{\tau} \mathscr{S}_\infty^{L_\Bf} | Y=y \} \bigr) \\
	&=& \sum_{\ts \in \mathbb{Z}^{L_\Cf}} \Phi\bigl( \P\{ \vv = \ts | Y=y \} \bigr) \\
	&=& H(\vv|Y=y).   \IEEEeqnarraynumspace\IEEEyesnumber
\end{IEEEeqnarray*}
By the data-processing inequality $H(\Cs\vv|Y=y) \leq H(\vv|Y=y)$, we conclude that, if $\lambda_{\infty,1}(\Cs_\perp) > 2\hat{\tau}$, the objective $H(\Cs\vv|Y)$ of the minimization equals its upper bound $H(\vv|Y)$. Consequently, we can assume that $\lambda_{\infty,1}(\Cs_\perp) \leq 2\hat{\tau}$ in the minimization~\eqref{def:Jq_bis4} without loss of generality. In other words,
\begin{IEEEeqnarray*}{rCl}   \label{def:Jq_bis5}
	J_\q(\Bs,M)
	&=& \min_{\substack{ \Cs \in \llbracket 2\floor{\kappa(\q)} + 1 \rrbracket^{L_\Cf \times L_\Bf} \colon \\ \phi_\q(\Cs) \in \mathscr{C}_{\mathbb{F}_\q}(M) \\ \lambda_{\infty,1}(\Cs_\perp) \leq 2\hat{\tau} }} H(\Cs\Bs\uv|Y),
	\quad \textnormal{(for $\q > \bar{\q}$).}   \IEEEeqnarraynumspace\IEEEyesnumber
\end{IEEEeqnarray*}

Next, we show that we can further restrict the minimization domain, in that $\det(\Cs\Cs^\T)$ need not be larger than a certain constant. In the special case where $L_\Bf - L_\Cf = 1$, the proof is immediate, since the orthogonal $\Cs_\perp$ reduces to a column vector, so we directly have $\det(\Cs\Cs^\T) = \lVert \Cs_\perp \rVert_2^2 \leq L_\Bf \lVert \Cs_\perp \rVert_\infty^2 = L_\Bf \lambda_{\infty,1}(\Cs_\perp) \leq 2 L_\Bf \hat{\tau}$ from the above assumption. 
For $L_\Bf - L_\Cf > 1$, with a more elaborate reasoning we can still show that $\det(\Cs\Cs^\T)$ can be similarly upper-bounded: for the sake of argument, let us assume the \emph{opposite}, namely, that $\det(\Cs\Cs^\T)$ is strictly \emph{larger} than some positive constant $\theta$ satisfying
\begin{equation}   \label{theta_LB}
	\theta 
	\geq (L_\Bf-L_\Cf)! (2 \hat{\tau})^{L_\Bf-L_\Cf}.
\end{equation}
Since $\det(\Cs\Cs^\T) = \det(\Cs_\perp^\T\Cs_\perp)$, we immediately have $\det(\Cs_\perp^\T\Cs_\perp) > \theta$.
Minkowski's Second Theorem (Subsection~\ref{ssec:euclidean_lattices}, Thm.~\ref{thm:Minkowski_Second_Theorem}) applied to successive minima $\lambda_{\infty,\ell}(\Cs_\perp), \ell=1,\dotsc,L_\Bf-L_\Cf$ defined with respect to the convex body $\mathscr{S}_\infty^{L_\Bf}$ yields
\begin{equation}
	\prod_{\ell=1}^{L_\Bf-L_\Cf} \lambda_{\infty,\ell}(\Cs_\perp)
	\geq \frac{1}{(L_\Bf-L_\Cf)!}\det(\Cs_\perp^\T\Cs_\perp)
\end{equation}
and therefore,
\begin{IEEEeqnarray*}{rCl}
	\lambda_{\infty,L_\Bf-L_\Cf}(\Cs_\perp)
	&\geq& \prod_{\ell=1}^{L_\Bf-L_\Cf} \lambda_{\infty,\ell}(\Cs_\perp)^{1/(L_\Bf-L_\Cf)} \\
	&\geq& \frac{1}{(L_\Bf-L_\Cf)!^{1/(L_\Bf-L_\Cf)}}\det(\Cs_\perp^\T\Cs_\perp)^{1/(L_\Bf-L_\Cf)} \\
	&>& \frac{\theta^{1/(L_\Bf-L_\Cf)}}{(L_\Bf-L_\Cf)!^{1/(L_\Bf-L_\Cf)}} \\
	&\geq& 2 \hat{\tau}.   \IEEEeqnarraynumspace\IEEEyesnumber
\end{IEEEeqnarray*}
The above inequality $\lambda_{\infty,L_\Bf-L_\Cf}(\Cs_\perp) > 2 \hat{\tau}$ expresses that the last successive minimum of the lattice $\Lambda(\Cs_\perp)$ strictly exceeds the diameter (in terms of infinity norm) of the ball $\hat{\tau} \mathscr{S}_\infty^{L_\Bf}$, which contains the support of $\vv$. Consequently, as long as $\det(\Cs\Cs^\T) > \theta$ (and the inequality $\lambda_{\infty,L_\Bf-L_\Cf}(\Cs_\perp) > 2 \hat{\tau}$ thus being satisfied), it is always guaranteed that there exists a largest index $\ell_0 \in [L_\Bf-L_\Cf-1]$ such that all those successive minima of $\Lambda(\Cs_\perp)$ with index up to $\ell_0$, do not exceed $2 \hat{\tau}$ (while those with index $\ell_0+1$ or higher exceed $2\hat{\tau}$). That is,
\begin{equation}   \label{lambda_inequalities}
	\lambda_{\infty,1}(\Cs_\perp)
	\leq \dotso
	\leq \lambda_{\infty,\ell_0}(\Cs_\perp)
	\leq 2 \hat{\tau}
	< \lambda_{\infty,\ell_0+1}(\Cs_\perp)
	\leq \dotso
	\leq \lambda_{\infty,L_\Bf-L_\Cf}(\Cs_\perp).
\end{equation}
Recall from~\eqref{def:Jq_bis5} that we have assumed that $\Cs$ always satisfies $\lambda_{\infty,1}(\Cs_\perp) \leq 2\hat{\tau}$, and hence not all successive minima can simultaneously exceed $2\hat{\tau}$. Also note that the $\ell_0$ first columns of the Korkin--Zolotarev reduced basis matrix $\Cs_\perp$ form themselves a Korkin--Zolotarev reduced basis $[\Cs_\perp]_{[\ell_0]} \in \mathbb{Z}^{L_\Bf \times \ell_0}$. Let us define the orthogonal $\bar{\Cs} \in \mathbb{Z}^{(L_\Bf-\ell_0) \times L_\Bf}$ of the latter, i.e.,
\begin{equation}
	\bar{\Cs}
	\triangleq \bigl( [\Cs_\perp]_{[\ell_0]} \bigr)_\perp
\end{equation}
and note that the row space of $\bar{\Cs}$ contains the row space of $\Cs$, i.e., $\Lambda(\Cs^\T) \subset \Lambda(\bar{\Cs}^\T)$.

Let us now define a partitioned integer matrix
\begin{equation}
	\Zs(\Xs,\Xs')
	\triangleq \begin{bmatrix} \Xs & \Xs' \end{bmatrix}
\end{equation}
of same dimension as $\Cs_\perp$, and which we assume to be in Korkin--Zolotarev reduced form. Its left-hand part $\Xs \in \mathbb{Z}^{L_\Bf \times \ell_0}$ contains the first $\ell_0$ columns of $\Zs(\Xs,\Xs')$, whereas the right-hand part $\Xs' \in \mathbb{Z}^{L_\Bf \times (L_\Bf-L_\Cf-\ell_0)}$ contains the remaining columns.

Let us now fix the left-hand matrix to $\Cs_\perp$, so as to make $\Zs(\Xs,\Xs')$ and $\Cs_\perp$ coincide on the first $\ell_0$ columns. Consider $\mathscr{X}(\bar{\Cs}_\perp,\hat{\tau})$, which we shall define as the set of all matrices $\Xs' \in \mathbb{Z}^{L_\Bf \times (L_\Bf - L_\Cf - \ell_0)}$ such that the following three properties are simultaneously satisfied:
\begin{enumerate}
	\item	$\Zs(\bar{\Cs}_\perp,\Xs')$ is in Korkin--Zolotarev reduced form;
	\item	the largest index $\ell$ such that $\lambda_{\infty,\ell}\bigl(\Zs(\bar{\Cs}_\perp,\Xs')\bigr) \leq 2\hat{\tau}$, is equal to $\ell_0$;
	\item	$\Zs(\bar{\Cs}_\perp,\Xs')_\perp$ is a representation of $M(\Cs)$, or equivalently, $\Zs(\bar{\Cs}_\perp,\Xs')^\T$ is a representation of $M^*(\Cs) = M(\Cs_\perp^\T)$.
\end{enumerate}
Since plugging $\Xs' = [\Cs_\perp]_{[L_\Bf-L_\Cf] \setminus [\ell_0]}$ into $\Zs(\bar{\Cs}_\perp,\cdot)$ obviously reconstitutes the original matrix $\Cs_\perp$, i.e.,
\begin{equation}
	\Zs(\bar{\Cs}_\perp,[\Cs_\perp]_{[L_\Bf-L_\Cf] \setminus [\ell_0]})
	= \Zs([\Cs_\perp]_{[\ell_0]},[\Cs_\perp]_{[L_\Bf-L_\Cf] \setminus [\ell_0]})
	= \Cs_\perp
\end{equation}
and since this choice of $\Xs'$ will by definition satisfy the above three conditions, it is clear that the set $\mathscr{X}(\bar{\Cs}_\perp,\hat{\tau})$ always has at least one element.

Having fixed the left-hand part of $\Zs(\Xs,\Xs')$ to be $\Xs = \bar{\Cs}_\perp$, let us view its orthogonal $\Zs(\Xs,\Xs')_\perp$ as a function of the right-hand matrix $\Xs' \in \mathscr{X}(\bar{\Cs}_\perp,\hat{\tau})$ and accordingly denote it as
\begin{equation}
	\Cs(\Xs')
	\triangleq \Zs(\bar{\Cs}_\perp,\Xs')_\perp.
\end{equation}
Note that since $\mathscr{X}(\bar{\Cs}_\perp,\hat{\tau})$ is non-empty and since $\Cs(\Xs')$ is an integer matrix (so the determinant $\det\bigl( \Cs(\Xs')\Cs(\Xs')^\T \bigr)$ is a non-negative integer), the minimum
\begin{equation}
	\mu(\Cs,\hat{\tau})
	\triangleq \min_{\substack{\Xs' \in \mathscr{X}(\bar{\Cs}_\perp,\hat{\tau}) \colon \\ M(\Cs(\Xs')) = M(\Cs)}} \det\bigl( \Cs(\Xs')\Cs(\Xs')^\T \bigr)
\end{equation}
is well defined and achieved by a (not necessarily unique) minimizing matrix $\Xs'_\star(\bar{\Cs}) \in \mathscr{X}(\bar{\Cs}_\perp,\hat{\tau})$, i.e.,
\begin{equation}
	\mu(\bar{\Cs}_\perp,\hat{\tau})
	= \det\bigl( \Cs(\Xs'_\star(\bar{\Cs}_\perp))\Cs(\Xs'_\star(\bar{\Cs}_\perp))^\T \bigr).
\end{equation}
We further define the following constant:
\begin{equation}
	\theta^\star(\hat{\tau})
	\triangleq \max_{1 \leq \ell_0 \leq L_\Bf-L_\Cf} \sup_{\substack{\bar{\Cs}_\perp \in \mathbb{Z}^{L_\Bf \times \ell_0} \colon \\ \lambda_{\infty,\ell_0}(\bar{\Cs}_\perp) \leq 2\hat{\tau}}} \mu(\bar{\Cs}_\perp,\hat{\tau}).
\end{equation}
This constant is well defined, because the supremum is over a finite set (and can thus be replaced by a maximum). To see this, bear in mind that the constraint $\lambda_{\infty,\ell_0}(\bar{\Cs}_\perp) \leq 2\hat{\tau}$ implies that all successive minima of $\bar{\Cs}_\perp$ are upper-bounded by $2\hat{\tau}$. By virtue of~\eqref{KZ_slack}, the norms of the columns of $\bar{\Cs}_\perp$ are upper-bounded as
\begin{IEEEeqnarray*}{rCl}
	\left\lVert [\bar{\Cs}_\perp]_i \right\rVert_2^2
	&\leq& \frac{i+3}{4} \lambda_{2,i}(\bar{\Cs}_\perp)^2 \\
	&\leq& \frac{i+3}{4} L_\Bf \lambda_{\infty,i}(\bar{\Cs}_\perp)^2 \\
	&\leq& (i+3) L_\Bf \hat{\tau}^2
	\quad \textnormal{(for $1 \leq i \leq \ell_0$)}   \IEEEeqnarraynumspace\IEEEyesnumber
\end{IEEEeqnarray*}
whence we conclude, for example, that the Frobenius norm of $\bar{\Cs}_\perp$ is upper-bounded as
\begin{IEEEeqnarray*}{rCl}
	\left\lVert \bar{\Cs}_\perp \right\rVert_{\Frob}^2
	&\leq& \sum_{i=1}^{\ell_0} (i+3) L_\Bf \hat{\tau}^2 \\
	&\leq& \left( \frac{\ell_0+1}{2} + 3 \right) \ell_0 L_\Bf \hat{\tau}^2 \\
	&\leq& \left( \frac{L_\Bf-L_\Cf}{2} + 3 \right) \max\{1,L_\Bf-L_\Cf-1\} L_\Bf \hat{\tau}^2.   \IEEEeqnarraynumspace\IEEEyesnumber
\end{IEEEeqnarray*}
where the last step follows because $1 \leq \ell_0 \leq L_\Bf-L_\Cf-1$.

Next, we will argue that the entropy $H(\Cs(\Xs')\vv|Y)$, when viewed as a function of $\Xs' \in \mathscr{X}(\bar{\Cs}_\perp,\hat{\tau})$, is constant.
To begin with, note that this entropy can be written as
\begin{equation}   \label{entropy_written_out}
	H(\Cs(\Xs')\vv|Y)
	= \int_\Yc \sum_{\ts \in \mathbb{Z}^{L_\Cf}} \Phi\bigl( \P\{ \Cs(\Xs')\vv = \ts | Y=y \} \bigr) \intd P_Y(y)
\end{equation}
and since $\Cs(\Xs')$ is right-invertible, the (unconstrained) linear equation $\Cs(\Xs')\vv = \ts$ (in $\vv$) has a translated lattice $\mathscr{V}(\Cs(\Xs'),\ts)$ as its solution set, which can be parametrized as
\begin{IEEEeqnarray*}{rCl}
	\mathscr{V}(\Cs(\Xs'),\ts)
	&\triangleq& \bigl\{ \Cs(\Xs')^\sharp\ts + \Zs(\Xs)\ws \colon \ws \in \mathbb{Z}^{L_\Bf-L_\Cf} \bigr\} \\
	&=& \Cs(\Xs')^\sharp\ts + \Lambda(\Zs(\Xs)).   \IEEEeqnarraynumspace\IEEEyesnumber
\end{IEEEeqnarray*}
Note that $\Cs(\Xs') = (\Zs(\Xs))_\perp$ has been defined via the $(\cdot)_\perp$ operation, which is why a right-inverse $\Cs(\Xs')^\sharp$ exists.
In the above equation, the translation vector $\Cs(\Xs')^\sharp\ts$ can be replaced, without loss of generality, by a smallest-norm equivalent shift vector
\begin{equation}
	\ds(\Cs(\Xs'),\ts)
	\triangleq \Cs(\Xs')^\sharp\ts + \Zs(\Xs)\ws_0
\end{equation}
where $\ws_0$ stands for
\begin{equation}
	\ws_0
	= \argmin_{\ws \in \mathbb{Z}^{L_\Bf-L_\Cf}} \big\lVert \Cs(\Xs')^\sharp\ts + \Zs(\Xs)\ws \big\rVert_\infty.
\end{equation}
(ties being resolved arbitrarily).
Hence, $\mathscr{V}(\Cs(\Xs'),\ts) = \ds(\Cs(\Xs'),\ts) + \Lambda(\Zs(\Xs))$. Furthermore, notice that for any shift vector $\ds \in \hat{\tau} \mathscr{S}_\infty^{L_\Bf}$ and any $\Xs \in \mathscr{X}(\bar{\Cs}_\perp,\hat{\tau})$, we have that, under the intersection with the scaled ball $\hat{\tau} \mathscr{S}_\infty^{L_\Bf}$, the shifted lattice $\ds + \Lambda(\Zs(\Xs))$ is the same as the shifted sublattice $\ds + \Lambda([\Cs_\perp]_{[\ell_0-1]}) = \ds + \Lambda(\bar{\Cs}_\perp)$, namely,
\begin{IEEEeqnarray*}{rCl}
	\mathscr{V}(\Cs(\Xs'),\ts) \cap \hat{\tau} \mathscr{S}_\infty^{L_\Bf}
	&=& \bigl( \ds(\Cs(\Xs'),\ts) + \Lambda(\Zs(\Xs)) \bigr) \cap \hat{\tau} \mathscr{S}_\infty^{L_\Bf} \\
	&=& \bigl( \ds(\Cs(\Xs'),\ts) + \Lambda(\bar{\Cs}_\perp) \bigr) \cap \hat{\tau} \mathscr{S}_\infty^{L_\Bf}.
\end{IEEEeqnarray*}
Note that for any pair $(\ts,\ws)$ such that $\Cs(\Xs')^\sharp\ts + \Zs(\Xs)\ws$ belongs to the cube $\hat{\tau} \mathscr{S}_\infty^{L_\Bf}$, we have that the last $L_\Bf-L_\Cf-\ell_0+1$ entries of $\ws_0 = \begin{bmatrix} \bar{\ws}_0^\T & \mathbf{0}^\T \end{bmatrix}^\T$ are zeros, in such way that $\Zs(\Xs)\ws_0 = \bar{\Cs}_\perp\bar{\ws}_0$. On the other hand, since $\begin{bmatrix} \bar{\Cs}^\sharp & \bar{\Cs}_\perp \end{bmatrix}$ is unimodular, there exists a vector $\bar{\ts} \in \mathbb{Z}^{L_\Bf-\ell_0+1}$ such that $\Cs(\Xs')^\sharp\ts + \Zs(\Xs)\ws_0 = \bar{\Cs}^\sharp\bar{\ts} + \bar{\Cs}_\perp\bar{\ws}_0$. Hence, if we define the set
\begin{equation}
	\mathscr{D}
	\triangleq \{ \ds(\Cs(\Xs'),\ts) \colon \ts \in \mathbb{Z}^{L_\Cf} \} \cap \hat{\tau} \mathscr{S}_\infty^{L_\Bf}
\end{equation}
then this set is equal to
\begin{equation}
	\mathscr{D}
	= \{ \ds(\bar{\Cs},\ts) \colon \ts \in \mathbb{Z}^{L_\Cf} \} \cap \hat{\tau} \mathscr{S}_\infty^{L_\Bf}
\end{equation}
and thus is not a function of $\Xs$.
In the derivation~\eqref{entropy_invariance} underneath, let us adopt the convention that the probability that a random variable belongs to an empty set shall be equal to zero. With the notations defined above, the entropy term $H(\Cs(\Xs')\vv|Y)$ can now be rewritten by means of
\begin{IEEEeqnarray*}{rCl}
	H(\Cs(\Xs')\vv|Y=y)
	&=& \sum_{\ts \in \mathbb{Z}^{L_\Cf}} \Phi\bigl( \P\{ \Cs(\Xs')\vv = \ts | Y=y \} \bigr) \\
	&=& \sum_{\ts \in \mathbb{Z}^{L_\Cf}} \Phi\bigl( \P\{ \vv \in \mathscr{V}(\Cs(\Xs'),\ts) \cap \hat{\tau} \mathscr{S}_\infty^{L_\Bf} | Y=y \} \bigr) \\
	&=& \sum_{\ds \in \mathscr{D}} \Phi\bigl( \P\{ \vv \in (\ds + \Lambda(\bar{\Cs}_\perp)) \cap \hat{\tau} \mathscr{S}_\infty^{L_\Bf} | Y=y \} \bigr) \\
	&=& \sum_{\ts \in \mathbb{Z}^{L_\Cf}} \Phi\bigl( \P\{ \vv \in (\bar{\Cs}^\sharp\ts + \Lambda(\bar{\Cs}_\perp)) \cap \hat{\tau} \mathscr{S}_\infty^{L_\Bf} | Y=y \} \bigr) \\
	&=& H(\bar{\Cs}\vv|Y=y)   \IEEEeqnarraynumspace\IEEEyesnumber\label{entropy_invariance}
\end{IEEEeqnarray*}
whereby we infer that $H(\Cs(\Xs')\vv|Y)$ is constant in $\Xs \in \mathscr{X}(\bar{\Cs}_\perp,\hat{\tau})$.

In summary, we have shown that if the (right-invertible) matrix $\Cs$ is large in the precise sense that $\det(\Cs\Cs^\T) > \theta$ with some integer $\theta \geq \bar{\theta}(\hat{\tau})$ with
\begin{equation}   \label{theta_LB_bis}
	\bar{\theta}(\hat{\tau})
	\triangleq \max\bigl\{ (L_\Bf-L_\Cf)! (2 \hat{\tau})^{L_\Bf-L_\Cf} , \theta^{\star}(\hat{\tau}) \bigr\}
\end{equation}
then one can construct a matrix $\Cs^\star \triangleq \tilde{\Cs}(\Xs^\star)$ with same matroid $M(\Cs) = M(\Cs^\star)$ which is ``smaller'' than $\Cs$ in the sense that $\det(\Cs^\star(\Cs^\star)^\T) \leq \min\{ \det(\Cs\Cs^\T), \theta^\star(\hat{\tau}) \}$ and such that $H(\Cs\vv|Y) = H(\Cs^\star\vv|Y)$. As a consequence, we can restrict the minimization domain on which $\Cs$ takes value, namely,
\begin{IEEEeqnarray*}{rCl}
	J_\q(\Bs,M)
	&=& \min_{\substack{\Cs \in \llbracket 2\floor{\kappa(\q)} + 1 \rrbracket^{L_\Cf \times L_\Bf} \colon \\ \text{$\phi_\q(\Cs) \in \mathscr{C}_{\mathbb{F}_\q}(M)$ and $\det(\Cs\Cs^\T) \leq \theta$}}} H(\Cs\Bs\uv|Y),
	\quad \textnormal{(for $\q > \bar{\q}$, $\theta \geq \bar{\theta}(\hat{\tau})$).}   \IEEEeqnarraynumspace\IEEEyesnumber\label{def:Jq_bis6}
\end{IEEEeqnarray*}
Now $\det(\Cs\Cs^\T)$ can be lower-bounded as follows:
\begin{IEEEeqnarray*}{rCl}
	\det(\Cs\Cs^\T)
	&\geq& \prod_{\ell=1}^{L_\Cf} \lambda_{\infty,\ell}(\Cs) \\
	&\geq& \max_{\ell} \lambda_{\infty,\ell}(\Cs) \\
	&\geq& \frac{1}{\sqrt{L_\Bf}} \max_{\ell} \lambda_{2,\ell}(\Cs) \\
	&\geq& \max_{\ell} \frac{2}{\sqrt{L_\Bf(\ell+3)}} \bigl\lVert [\Cs]_\ell \bigr\rVert_2 \\
	&\geq& \frac{2}{\sqrt{L_\Bf(L_\Cf+3)}} \max_{\ell} \bigl\lVert [\Cs]_\ell \bigr\rVert_\infty \\
	&=& \frac{2}{\sqrt{L_\Bf(L_\Cf+3)}} \lVert \Cs \rVert_\infty.   \IEEEeqnarraynumspace\IEEEyesnumber
\end{IEEEeqnarray*}
Hence $\det(\Cs\Cs^\T) \leq \theta$ implies $\lVert \Cs \rVert_\infty \leq \Theta$ where $\Theta = \frac{\theta}{2} \sqrt{L_\Bf(L_\Cf+3)}$. We can thus replace $2\floor{\kappa(\q)} + 1 $ in the minimization domain from~\eqref{def:Jq_bis6} by $2\min\{\Theta,\floor{\kappa(\q)}\} + 1$. Hence, if $\q$ is so large as to satisfy $\kappa(\q) \geq \Theta$, then we can further modify the expression of $J_\q(\Bs,M)$ to the following:
\begin{equation}
	J_\q(\Bs,M)
	= \min_{\substack{\Cs \in \llbracket 2\Theta + 1 \rrbracket^{L_\Cf \times L_\Bf} \colon \\ \text{$\phi_\q(\Cs) \in \mathscr{C}_{\mathbb{F}_\q}(M)$ and $\det(\Cs\Cs^\T) \leq \Theta$}}} H(\Cs\Bs\uv|Y),
	\qquad \textnormal{(for $\q > \max\{\bar{\q}, \Theta^{4/3}\}$, $\Theta \geq \bar{\theta}(\hat{\tau})$).}   \label{def:Jq_bis7}
\end{equation}
Here, we have used that $\q \geq \Theta^{4/3}$ is a sufficient condition for $\kappa(\q) \geq \Theta$. In fact, since $L_\Bf L_\Cf \geq 2$, we have $\Theta^{4/3} \geq \Theta^{\frac{2 L_\Bf L_\Cf}{2 L_\Bf L_\Cf - 1}}$, hence $\q \geq \Theta^{4/3}$ implies $\q \geq \Theta^{\frac{2 L_\Bf L_\Cf}{2 L_\Bf L_\Cf - 1}}$, the latter being equivalent to $\kappa(\q) \geq \Theta$.

On the other hand, if $\q > \sqrt{\Theta}$, then any square submatrix $[\Cs]_\Lc$ of a matrix $\Cs$ from the minimization domain in~\eqref{def:Jq_bis7} has a determinant whose absolute value is strictly less than $\q$, because
\begin{equation}
	\det([\Cs]_\Lc)^2
	= \det([\Cs]_\Lc[\Cs]_\Lc^\T)
	\leq \det(\Cs\Cs^\T)
	\leq \Theta
	< \q^2.
\end{equation}
Now, recall that the (finite-field) submatrix $\phi_\q([\Cs]_\Lc)$ is rank deficient if and only if $\det([\Cs]_\Lc)$ is a multiple of $\q$. However, under the assumption $\q > \sqrt{\Theta}$, the determinant $\det([\Cs]_\Lc)$ can be a multiple of $\q$ only if $\det([\Cs]_\Lc) = 0$. Therefore, if $\q > \sqrt{\Theta}$, we have that $\phi_\q(\Cs) \in \mathscr{C}_{\mathbb{F}_\q}(M)$ if and only if $\Cs \in \mathscr{C}_{\mathbb{Z}}(M)$. However, $\q > \sqrt{\Theta}$ is already implied by $\q > \Theta^{4/3}$ because $\theta^\star(\hat{\tau}) \geq 1$.
As a consequence, we can write
\begin{equation}
	J_\q(\Bs,M)
	= \min_{\substack{\Cs \in \llbracket 2\Theta + 1 \rrbracket^{L_\Cf \times L_\Bf} \colon \\ \text{$\Cs \in \mathscr{C}_\mathbb{Z}(M)$ and $\det(\Cs\Cs^\T) \leq \Theta$}}} H(\Cs\Bs\uv|Y),
	\qquad \textnormal{(for $\q > \max\{\bar{\q}, \Theta^{4/3}\}$, $\Theta \geq \theta^\star(\hat{\tau})$).}   \label{def:Jq_bis8}
\end{equation}
Recall that, on the right-hand side of~\eqref{def:Jq_bis8}, $\Theta$ stands for an arbitrary parameter that can be set to any integer not smaller than $\bar{\theta}(\hat{\tau})$. Setting it equal to $\bar{\theta}(\hat{\tau})$, and observing that the right-hand side of~\eqref{def:Jq_bis8} is not a function of $\q$, we infer that $J_\q(\Bs,M)$ is constant in $\q$ for any prime $\q$ larger than $\max\{\bar{\q},\bar{\theta}(\hat{\tau})^{4/3}\} = \bar{\q}$ [cf.~\eqref{q_10}]. That is,
\begin{equation}
	J_\q(\Bs,M)
	= J_{\bar{\q}}(\Bs,M)
	= \lim_{\q \to \infty} J_\q(\Bs,M),
	\quad \textnormal{(for $\q \geq \bar{\q}$).}
\end{equation}
Furthermore, this limit is equal to
\begin{equation}
	J_{\bar{\q}}(\Bs,M)
	= \min_{\substack{\Cs \in \llbracket 2\Theta + 1 \rrbracket^{L_\Cf \times L_\Bf} \colon \\ \text{$\Cs \in \mathscr{C}_\mathbb{Z}(M)$ and $\det(\Cs\Cs^\T) \leq \Theta$}}} H(\Cs\Bs\uv|Y),
	\quad \textnormal{(for $\Theta \geq \theta^\star(\hat{\tau})$).}   \label{def:Jq_bis8_limit}
\end{equation}
Recall that $\Theta > \theta^\star(\hat{\tau})$ is an arbitrary parameter. Hence, since the left-hand side of~\eqref{def:Jq_bis8_limit} does not depend on $\Theta$, neither does the right-hand side. We can thus let $\Theta$ tend to infinity to obtain the simpler expression\footnote{Since the right-hand side of~\eqref{def:Jq_bis8_limit} is constant in $\theta$, we know that the infimum in~\eqref{quod_erat_demonstrandum} is in fact achieved by a minimizing matrix $\Cs$, and is therefore a minimum.}
\begin{equation}   \label{quod_erat_demonstrandum}
	J_\q(\Bs,M)
	= \inf_{\Cs \in \mathscr{C}_\mathbb{Z}(M)} H(\Cs\Bs\uv|Y),
	\quad \textnormal{(for $\q \geq \bar{\q}$).}
\end{equation}
This concludes Step~5.

By combining Steps 1 through 5, we conclude that for any prime $\q > \bar{\q}$, the equality~\eqref{set_operation_exchange} can be expressed as
\begin{IEEEeqnarray*}{rCl}
	\mathscr{Q}(\phi_\q(\Bs))
	&\overset{(\q > \bar{\q})}{=}& \bigcap_{M^* \in \mathscr{M}_\mathbb{Z}^\circ(L_\Bf)} \; \bigcup_{\Sc \in \mathscr{B}(M^*)} \; \bigcap_{\Tc \in \mathscr{B}([\Bs]_\Sc)} \mathscr{Q}(\Bs,M,\Tc) \\
	&=& \mathscr{Q}(\Bs).
	\IEEEeqnarraynumspace\IEEEyesnumber
\end{IEEEeqnarray*}
This concludes the proof of Theorem~\ref{thm:integer_CF} for finitely supported auxiliaries.

Next, as a final proof step, we will lift this finite-support restriction by a ``clipping'' argument. Let us define the clipping operator $\langle\cdot\rangle_\tau$ for real vectors $\us \in \mathbb{R}^n$ as follows:
\begin{equation}
	\langle \us \rangle_\tau
	=
	\begin{cases}
		\us & \textnormal{if $\lVert \us \rVert \leq \tau$} \\
		\mynull & \textnormal{if $\lVert \us \rVert > \tau$.}
	\end{cases}
\end{equation}
In the following, let us incorporate the dependency on the auxiliary distribution into the rate region notations, by writing $\mathscr{Q}(\Bs;\uv)$ and $\mathscr{Q}(\Bs,M,\Tc;\uv)$ in lieu of $\mathscr{Q}(\Bs)$ and $\mathscr{Q}(\Bs,M,\Tc)$, respectively. By Theorem~\ref{thm:integer_CF}, which thus far we have proven to hold for finitely supported auxiliaries, we can assert that $\mathscr{Q}(\Bs;\langle \uv \rangle_\tau)$ is achievable. Therefore, the inner set limit $\varliminf_{\tau \to \infty} \mathscr{Q}(\Bs;\langle \uv \rangle_\tau)$ is achievable. Upon writing out $\mathscr{Q}(\Bs;\langle \uv \rangle_\tau)$ according to its definition~\eqref{def:Q} as
\begin{equation}
	\mathscr{Q}(\Bs;\langle \uv \rangle_\tau)
	= \bigcap_{M \in \mathscr{M}^\circ_{\mathbb{Z}}(L_\Bf)} \bigcup_{\Sc \in \mathscr{B}(M^*)} \bigcap_{\Tc \in \mathscr{B}([\Bs]_\Sc)} \mathscr{Q}(\Bs,M,\Tc;\langle \uv \rangle_\tau)
\end{equation}
and observing that this expression involves a finite number of set operations that are independent of $\tau$, we can argue that the inner set limit can be exchanged with the other set operations, so as to yield
\begin{equation}   \label{inner_set_limit_of_rate_region_with_clipping}
	\varliminf_{\tau \to \infty} \mathscr{Q}(\Bs;\langle \uv \rangle_\tau)
	= \bigcap_{M \in \mathscr{M}^\circ_{\mathbb{Z}}(L_\Bf)} \bigcup_{\Sc \in \mathscr{B}(M^*)} \bigcap_{\Tc \in \mathscr{B}([\Bs]_\Sc)} \varliminf_{\tau \to \infty} \mathscr{Q}(\Bs,M,\Tc;\langle \uv \rangle_\tau).
\end{equation}
The set on the right-hand side of the last equation can in turn be expressed as
\begin{multline}
	\mathscr{Q}(\Bs,M,\Tc;\langle \uv \rangle_\tau)
	= \Bigl\{ (R_1,\dotsc,R_K) \in \mathbb{R}_+^K \colon \sum_{k\in\Tc} R_k < \\
		H([\langle \uv \rangle_\tau]_\Tc) + \inf_{\Cs \in \mathscr{C}_\mathbb{Z}(M)} H(\Cs\Bs\langle \uv \rangle_\tau|Y_\tau) - H(\Bs\langle \uv \rangle_\tau|Y_\tau) \Bigr\}
\end{multline}
where $Y_\tau$ denotes the channel output induced by an auxiliary $\langle \uv \rangle_\tau$. Specifically, conditioned on any given $\uv = \us$ and given a modulation mapping $\xs(\us) = (x_1(u_1),\ldots, x_K(u_K))$, the variable $Y_\tau$ is distributed as $Y_\tau \sim P_{Y|\xv}(\cdot|\xs(\langle\us\rangle_\tau))$.

\lemmabox{
\begin{lemma}[Clipping lemma]   \label{lem:clipping}
Given some real matrix $\Qs \in \mathbb{R}^{m \times n}$, an integer random vector $\uv \in \mathbb{Z}^n$ with finite entropy $H(\uv)$ and any modulation mapping $\xs(\us)$, we have that, as $\tau \to \infty$, the entropy $H(\Qs \langle \uv \rangle_\tau | Y_\tau)$ converges to $H(\Qs\uv | Y)$ uniformly in $\Qs$, in the sense that
\begin{equation}
	\bigl| H(\Qs \langle \uv \rangle_\tau | Y_\tau) - H(\Qs\uv | Y) \bigr|
	\leq \delta(\tau)
\end{equation}
for all $\Qs \in \mathbb{R}^{m \times n}$, where the function $\delta(\tau)$ only depends on the distribution of $(\uv,Y)$ (but not on $\Qs$) and satisfies $\lim_{\tau \to \infty} \delta(\tau) = 0$.
\end{lemma}
}

\proofbox{}{
\begin{IEEEproof}
Let us define the binary random variable
\begin{equation}
	E_\tau
	= \mathds{1}\{ \lVert \uv \rVert_\infty \leq \tau \}
	= \begin{cases}
		1 & \text{if $\lVert \uv \rVert_\infty \leq \tau$} \\
		0 & \text{if $\lVert \uv \rVert_\infty > \tau$}.
	\end{cases}
\end{equation}
The conditional mutual informations $I(\Qs \langle \uv \rangle_\tau; E_\tau | Y_\tau)$ and $I(\Qs \uv; E_\tau | Y)$ can be expanded in two alternative ways, respectively. Namely,
\begin{subequations}
\begin{IEEEeqnarray}{rCl}
	H(\Qs \langle \uv \rangle_\tau | Y_\tau) - H(\Qs \langle \uv \rangle_\tau | E_\tau, Y_\tau )
	&=& H(E_\tau | Y_\tau) - H(E_\tau | \Qs \langle \uv \rangle_\tau, Y_\tau)   \label{entropy_expansion_1} \\
	H(\Qs \uv | Y) - H(\Qs \uv | E_\tau, Y)
	&=& H(E_\tau | Y) - H(E_\tau | \Qs \uv, Y).   \label{entropy_expansion_2}
\end{IEEEeqnarray}
\end{subequations}
Since the right-hand sides of~\eqref{entropy_expansion_1} and \eqref{entropy_expansion_2} are each non-negative and upper-bounded by $H(E_\tau)$, upon taking the difference between~\eqref{entropy_expansion_1} and \eqref{entropy_expansion_2}, we obtain the bound
\begin{equation}   \label{entropy_difference_UB_1}
	\bigl| H(\Qs \langle \uv \rangle_\tau | Y_\tau) - H(\Qs \uv | Y) \bigr|
	\leq H(E_\tau) + \bigl| H(\Qs \langle \uv \rangle_\tau | E_\tau, Y_\tau) - H(\Qs \uv | E_\tau, Y) \bigr|.
\end{equation}
The conditional entropies $H(\Qs \langle \uv \rangle_\tau | E_\tau, Y_\tau)$ and $H(\Qs \uv | E_\tau, Y)$ can be expressed as
\begin{subequations}
\begin{IEEEeqnarray*}{rCl}
	H(\Qs \langle \uv \rangle_\tau | E_\tau, Y_\tau)
	&=& H(\Qs \langle \uv \rangle_\tau | E_\tau = 0, Y_\tau) \P\{ E_\tau = 0 \} + {} \qquad\qquad\qquad\qquad \\
	\IEEEeqnarraymulticol{3}{r}{
		{} + H(\Qs \langle \uv \rangle_\tau | E_\tau = 1, Y_\tau) \P\{ E_\tau = 1 \}
	}   \IEEEeqnarraynumspace\IEEEyesnumber\label{conditional_entropy_expansion_1}\\
	H(\Qs \uv | E_\tau, Y)
	&=& H(\Qs \uv | E_\tau = 0, Y) \P\{ E_\tau = 0 \} + {} \\
	\IEEEeqnarraymulticol{3}{r}{
		{} + H(\Qs \uv | E_\tau = 1, Y) \P\{ E_\tau = 1 \}.
	}   \IEEEeqnarraynumspace\IEEEyesnumber\label{conditional_entropy_expansion_2}
\end{IEEEeqnarray*}
\end{subequations}
Since $E_\tau = 1$ implies $\langle \uv \rangle_\tau = \uv$, it follows that, conditioned on $E_\tau = 1$, the pair $(\langle \uv \rangle_\tau, Y_\tau)$ has the same distribution as $(\uv,Y)$, hence $H(\Qs \langle \uv \rangle_\tau | E_\tau = 1, Y_\tau) = H(\Qs \uv | E_\tau = 1, Y)$. On the other hand, conditioned on $E_\tau = 0$, the clipped variable $\langle \uv \rangle_\tau$ is deterministically equal to the zero vector, hence $H(\Qs \langle \uv \rangle_\tau | E_\tau = 0, Y_\tau) = 0$. Subtracting~\eqref{conditional_entropy_expansion_1} from~\eqref{conditional_entropy_expansion_2}, we thus get
\begin{IEEEeqnarray*}{rCl}
	H(\Qs \uv | E_\tau, Y) - H(\Qs \langle \uv \rangle_\tau | E_\tau, Y_\tau)
	&=& H(\Qs \uv | E_\tau = 0, Y) \P\{ E_\tau = 0 \}.
\end{IEEEeqnarray*}
The non-negativity of the right-hand side of the last equality reveals that the difference of entropies on the left-hand side is non-negative. Hence the absolute value of said difference can be bounded as
\begin{IEEEeqnarray*}{rCl}
	\bigl| H(\Qs \uv | E_\tau, Y) - H(\Qs \langle \uv \rangle_\tau | E_\tau, Y_\tau) \bigr|
	&=& H(\Qs \uv | E_\tau = 0, Y) \P\{ E_\tau = 0 \} \\
	&\leq& H(\uv | E_\tau = 0) \P\{ E_\tau = 0 \} \\
	&=& H(\uv | E_\tau) - H(\uv | E_\tau = 1) \P\{ E_\tau = 1 \} \\
	&\leq& H(\uv) - H(\uv | E_\tau = 1) \P\{ E_\tau = 1 \}.   \IEEEeqnarraynumspace\IEEEyesnumber\label{entropy_difference_UB_2}
\end{IEEEeqnarray*}
Next, note that
\begin{IEEEeqnarray*}{rCl}
	H(\uv | E_\tau = 1)
	&=& - \sum_{\substack{\us \in \mathbb{Z}^K \colon \\ \lVert \us \rVert_\infty \leq \tau}} \frac{\P\{ \uv = \us \}}{\P\{ E_\tau = 1 \}} \log \frac{\P\{ \uv = \us \}}{\P\{ E_\tau = 1 \}} \\
	&=& \log\P\{ E_\tau = 1 \} - \frac{1}{\P\{ E_\tau = 1 \}} \sum_{\substack{\us \in \mathbb{Z}^K \colon \\ \lVert \us \rVert_\infty \leq \tau}} \P\{ \uv = \us \} \log \P\{ \uv = \us \}
\end{IEEEeqnarray*}
from which, since $\lim_{\tau \to \infty} \P\{ \lVert \uv \rVert_\infty \leq \tau \} = \lim_{\tau \to \infty} \P\{ E_\tau = 1 \} = 1$, it becomes manifest that $\lim_{\tau \to \infty} H(\uv | E_\tau = 1) \P\{ E_\tau = 1\} = H(\uv)$. Hence, taking limits as $\tau \to \infty$ we get
\begin{IEEEeqnarray*}{rCl}
	\lim_{\tau \to \infty} \bigl\{ H(\Qs \langle \uv \rangle_\tau | E_\tau, Y_\tau) - H(\Qs \uv | E_\tau, Y) \bigr\}
	&=& 0
\end{IEEEeqnarray*}
which concludes the proof.
In addition, it is easy to see that this convergence is uniform by combining \eqref{entropy_difference_UB_1} and \eqref{entropy_difference_UB_2}, which yields an upper bound $\bigl| H(\Qs \langle \uv \rangle_\tau | E_\tau, Y_\tau) - H(\Qs \uv | E_\tau, Y) \bigr| \leq \delta(\tau)$ which only depends on the distribution of $\uv$, but not on $\Qs$.

\end{IEEEproof}
}

By Lemma~\ref{lem:clipping}, we have the limits
\begin{subequations}
\begin{IEEEeqnarray}{rCl}
	\lim_{\tau \to \infty} H([\langle \uv \rangle_\tau]_\Tc)
	&=& H([\uv]_\Tc)   \label{clipping_limit_1} \\
	\lim_{\tau \to \infty} H(\Cs\Bs \langle \uv \rangle_\tau|Y_\tau)
	&=& H(\Cs\Bs\uv|Y)   \label{clipping_limit_2} \\
	\lim_{\tau \to \infty} H(\Bs \langle \uv \rangle_\tau|Y_\tau)
	&=& H(\Bs\uv|Y)   \label{clipping_limit_3}
\end{IEEEeqnarray}
\end{subequations}
all of which are finite because $H(\uv)$, which upper bounds all three of them, is finite by assumption. Here, establishing the last two equalities~\eqref{clipping_limit_2}--\eqref{clipping_limit_3} additionally requires the Dominated Convergence Theorem~\cite[Theorem~1.6.9]{AsDo00}, which is applicable due to $H(\uv|Y=y)$ being an upper bound both on $H(\Cs\Bs\langle \uv \rangle_\tau|Y=y)$ and on $H(\Bs\langle \uv \rangle_\tau|Y=y)$, and this upper bound is integrable with respect to the distribution of $Y$, since $H(\uv|Y) \leq H(\uv) < \infty$ by assumption.
As regards the infimum term, by Lemma~\ref{lem:clipping} we know that $H(\Cs\Bs\langle \uv \rangle_\tau|Y)$ converges to $H(\Cs\Bs\uv|Y)$ uniformly in $\Cs$, as $\tau \to \infty$. It follows that the order of limit-taking and infimization can be exchanged, yielding
\begin{equation}   \label{clipping_limit_2_bis}
	\lim_{\tau \to \infty} \inf_{\Cs \in \mathscr{C}_\mathbb{Z}(M)} H(\Cs\Bs\langle \uv \rangle_\tau|Y_\tau)
	= \inf_{\Cs \in \mathscr{C}_\mathbb{Z}(M)} H(\Cs\Bs\uv|Y).
\end{equation}
Furthermore, by Lemma~\ref{lem:discrete_algebraic_entropy} we can revert all entropies back to the algebraic entropy notation:
\begin{subequations}   \label{back_to_algebraic_entropy_notation}
\begin{IEEEeqnarray}{rCl}
	H([\uv]_\Tc)
	&=& \Hc([\uv]_\Tc) \\
	H(\Cs\Bs\uv|Y)
	&=& \Hc_{\Cs\Bs}(\uv|Y) \\
	H(\Bs\uv|Y)
	&=& \Hc_{\Bs}(\uv|Y).
\end{IEEEeqnarray}
\end{subequations}
Combining~\eqref{clipping_limit_1}, \eqref{clipping_limit_3} and \eqref{clipping_limit_2_bis}, and using algebraic entropy notation~\eqref{back_to_algebraic_entropy_notation}, we conclude that
\begin{equation}
	\varliminf_{\tau \to \infty} \mathscr{Q}(\Bs,M,\Tc;\langle \uv \rangle_\tau)
	= \mathscr{Q}(\Bs,M,\Tc;\uv)
\end{equation}
and therefore, substituting this back into~\eqref{inner_set_limit_of_rate_region_with_clipping}, we finally end up with
\begin{equation}   \label{quod_erat_demonstrandum_2}
	\varliminf_{\tau \to \infty} \mathscr{Q}(\Bs;\langle \uv \rangle_\tau)
	= \mathscr{Q}(\Bs;\uv).
\end{equation}
This finalizes the proof of Theorem~\ref{thm:integer_CF}.\footnote{Note that actually, inner and outer set limits in $\tau \to \infty$ coincide, so we could as well strengthen the convergence claim~\eqref{quod_erat_demonstrandum_2} to $\lim_{\tau \to \infty} \mathscr{Q}(\Bs;\langle \uv \rangle_\tau) = \mathscr{Q}(\Bs;\uv)$.}

}

{
\section{Proof of Theorem~\ref{thm:continuous_CF}}   \label{app:proof:continuous_CF}
We fix the coefficient matrix $\As \in \mathbb{Z}^{L \times K}$, the channel law $P_{Y|X_1,\dotsc,X_K} = P_{Y|\Xv}$ (satisfying the technical conditions in the statement of Theorem~\ref{thm:continuous_CF}), modulation mappings $\xs(\us) := (x_1(u_1), \dotsc, x_K(u_K))$ and the distribution of an auxiliary vector $\uv = \bigl[ U_1, \dotsc, U_K \bigr]^\T \in \mathbb{R}^K$ with mutually independent entries and an absolutely continuous distribution (with respect to the Lebesgue measure).

In what follows, let $\nu > 0$ stand for a quantization resolution parameter. We will now append a subscript `$\nu$' to the notations of rate regions and quantities defined in \eqref{Q_partial}, \eqref{def:J}, \eqref{def:Q} and Definition~\ref{def:joint_decoding_rate_region}, so as to signify that these are generated by an integer auxiliary $\floor{\nu\uv} \in \mathbb{Z}^K$ and modulation mapping $\us \mapsto \xs(\tfrac{1}{\nu}\us)$. Specifically, we will write these as $\mathscr{R}_\nu(\As)$, $\mathscr{Q}_\nu(\Bs)$, $\mathscr{Q}_\nu(\Bs,M,\Tc)$ and $J_\nu(\Bs,M)$, all of which being evaluated as per Theorem~\ref{thm:integer_CF}.

Each member of the sequence of rate regions $\left\{ \mathscr{R}_\nu(\As) \right\}_{\nu \in \mathbb{N}}$ is achievable as per Theorem~\ref{thm:integer_CF}. The proof of Theorem~\ref{thm:continuous_CF} consists in showing that this sequence has an inner limit, as $\nu$ tends to infinity, that contains the rate region $\mathscr{R}(\As)$ (evaluated as per Theorem~\ref{thm:continuous_CF}) with the continuous auxiliary $\uv \in \mathbb{R}^K$ and modulation mapping $\xs(\uv)$. That is, we will show that
\begin{equation}   \label{quod_est_demonstrandum_1}
	\mathscr{R}(\As)
	\subseteq \varliminf_{\nu \to \infty} \mathscr{R}_\nu(\As).
\end{equation}

The channel output variable induced by the continuous auxiliary $\uv$ and modulation mapping $\uv \mapsto \xs(\uv)$ shall be denoted as $Y$, in accordance with the system model from Section~\ref{sec:problem_statement}. By contrast, the channel output induced by the {\em quantized} auxiliary $\floor{\nu\uv}$ and modulation mapping $\us \mapsto \xs(\tfrac{1}{\nu}\us)$ shall be denoted as $Y_\nu$. This means that the joint distributions of $(\uv,Y)$ and $(\uv,Y_\nu)$ are specified as follows: conditioned on $\uv=\us$, we have that $Y$ and $Y_\nu$ are distributed respectively as
\begin{align}
	Y &\sim P_{Y|\Xv}(\cdot|\xs(\us))
	&
	Y_\nu &\sim P_{Y|\Xv}\left(\cdot \middle| \xs(\tfrac{1}{\nu}\floor{\nu\us})\right)
\end{align}
Thus far, we have merely fixed the conditional distributions $P_{Y|\uv=\us}$ and $P_{Y_\nu|\uv=\us}$ and thus, together with the given distribution of $\uv \sim P_{\uv}$, we have determined the joint distributions of the pairs $(\uv,Y)$ and $(\uv,Y_\nu)$, respectively. But we have \emph{not} specified a joint distribution on the triple $(\uv,Y,Y_\nu)$. Since all informational quantities (entropies and mutual information) involved in the present proof only depend on either $(\uv,Y)$ or $(\uv,Y_\nu)$, we remain free to choose an arbitrary coupling between $Y$ and $Y_\nu$ as long as the marginals are given by the distributions of $(\uv,Y)$ and $(\uv,Y_\nu)$ specified above. This degree of freedom will be exploited in the proof of Lemma~\ref{lem:exchange_condition_on_Y} stated further below.

The proof follows a similar two-stage outline as that of Theorem~\ref{thm:integer_CF}: we initially prove the theorem under the assumption of a compactly supported auxiliary, to then relax this assumption in a second and final step. To start off, let us state three auxiliary lemmata that will be needed at a later stage.

Since $\mathscr{R}_\nu(\As) = \bigcup_\Bs \mathscr{Q}_\nu(\Bs)$ where the union runs over $\Lambda_\mathbb{Z}(\Bs) \supseteq \Lambda_\mathbb{Z}(\As)$, and due to~\eqref{lims_of_union_and_intersections}, in order to prove \eqref{quod_est_demonstrandum_1} it will suffice to prove
\begin{equation}   \label{quod_est_demonstrandum_2}
	\mathscr{Q}(\Bs)
	\subseteq \varliminf_{\nu \to \infty} \mathscr{Q}_\nu(\Bs).
\end{equation}
Recall that $\mathscr{Q}_\nu(\Bs)$ is given by [cf.~\eqref{def:Q}]
\begin{IEEEeqnarray}{rCl}   \label{Q_nu}
	\mathscr{Q}_\nu(\Bs)
	&=& \bigcap_{M \in \mathscr{M}_\mathbb{Z}^\circ(L_\Bf)} \; \bigcup_{\Sc \in \mathscr{B}(M^*)} \; \bigcap_{\Tc \in \mathscr{B}([\Bs]_\Sc)} \mathscr{Q}_\nu(\Bs,M,\Tc).
\end{IEEEeqnarray}
Since, on the right-hand side of~\eqref{Q_nu}, the triple $(M,\Sc,\Tc)$ is iterated over a finite set that does not depend on $\nu$, it follows that the order of set limit (as $\nu \to \infty$) and the set operations (unions and intersections) can be exchanged, leading to
\begin{IEEEeqnarray}{rCl}   \label{Q_nu_liminf}
	\varliminf_{\nu \to \infty} \mathscr{Q}_\nu(\Bs)
	&=& \bigcap_{M \in \mathscr{M}_\mathbb{Z}^\circ(L_\Bf)} \; \bigcup_{\Sc \in \mathscr{B}(M^*)} \; \bigcap_{\Tc \in \mathscr{B}([\Bs]_\Sc)} \varliminf_{\nu \to \infty} \mathscr{Q}_\nu(\Bs,M,\Tc)
\end{IEEEeqnarray}
where $\mathscr{Q}_\nu(\Bs,M,\Tc)$ is given by [cf.~\eqref{Q_partial}]
\begin{equation}
	\mathscr{Q}_\nu(\Bs,M,\Tc)
	= \Bigl\{ (R_1,\dotsc,R_K) \in \mathbb{R}_+^K \colon
	\sum_{k\in\Tc} R_k < H([\floor{\nu\uv}]_\Tc) - H(\Bs\floor{\nu\uv}|Y_\nu) + J_\nu(\Bs,M) \Bigr\}
\end{equation}
and where [cf.~\eqref{def:J}]
\begin{equation}   \label{J_quantized}
	J_\nu(\Bs,M)
	= \inf_{\Cs \in \mathscr{C}_\mathbb{Z}(M)} H(\Cs\Bs\floor{\nu\uv}|Y_\nu).
\end{equation}
It will thus suffice to prove
\begin{IEEEeqnarray*}{rCl}
	\IEEEeqnarraymulticol{3}{l}{
		\varliminf_{\nu \to \infty} \mathscr{Q}_\nu(\Bs,M,\Tc)
	} \\ \qquad\qquad
	&\supseteq& \mathscr{Q}(\Bs,M,\Tc) \\
	&=& \Bigl\{ (R_1,\dotsc,R_K) \in \mathbb{R}_+^K \colon
	\sum_{k\in\Tc} R_k < \Hc([\uv]_\Tc) - \Hc_{\Bs}(\uv|Y) + J(\Bs,M) \Bigr\}.  \IEEEeqnarraynumspace\IEEEyesnumber
\end{IEEEeqnarray*}
For this purpose we will prove the following three limits:
\begin{subequations}
\begin{IEEEeqnarray}{rCl}
	\lim_{\nu \to \infty} \bigl\{ H([\floor{\nu\uv}]_\Tc) - (L_\Bf-L_\Cf) \log(\nu) \bigr\}
	&=& h([\uv]_\Tc)   \label{quod_est_demonstrandum_limit_1} \\
	\lim_{\nu \to \infty} \bigl\{ H(\Bs\floor{\nu\uv}|Y_\nu) - L_\Bf \log(\nu) \bigr\}
	&=& h(\Bs\uv|Y)   \label{quod_est_demonstrandum_limit_2} \\
	\liminf_{\nu \to \infty} \bigl\{ J_\nu(\Bs,M) - L_\Cf \log(\nu) \bigr\}
	&\geq& J(\Bs,M).   \label{quod_est_demonstrandum_limit_3}
\end{IEEEeqnarray}
\end{subequations}
In summary, it suffices to prove~\eqref{quod_est_demonstrandum_limit_1}--\eqref{quod_est_demonstrandum_limit_3} in order to prove Theorem~\ref{thm:continuous_CF}.

The equality~\eqref{quod_est_demonstrandum_limit_1} follows directly from Lemma~\ref{lem:continuous_algebraic_entropy} applied to the right-invertible coefficient matrix $\Qs = [\Is]_\Tc \in \{0,1\}^{(L_\Bf-L_\Cf) \times K}$. As for~\eqref{quod_est_demonstrandum_limit_2}, the proof is more involved and will require several steps. We will first make use of the following lemma.

\lemmabox{
\begin{lemma}   \label{lem:exchange_condition_on_Y}
For any auxiliary variables $\uv \in \mathbb{R}^K$, modulation mappings $\xs(\uv)$ and a channel law $P_{Y|X_1,\dotsc,X_K}$ that all conform with the assumptions of Theorem~\ref{thm:continuous_CF}, it holds that
\begin{IEEEeqnarray}{rCl}
	\lim_{\nu \to \infty} \Bigl\{ H\bigl( \Qs \floor{\nu\uv} \big| Y_\nu \bigr) - H\bigl( \Qs \floor{\nu\uv} \big| Y \bigr) \Bigr\}
	= 0
\end{IEEEeqnarray}
for any constant real-valued matrix $\Qs$. Furthermore, this convergence is uniform in $\Qs$, in that for any $\epsilon > 0$, there exists an $\nu_0(\epsilon) > 0$ such that for any $\nu > \nu_0(\epsilon)$ and any $\Qs$, we have
\begin{equation}
	\bigl| H\bigl( \Qs \floor{\nu\uv} \big| Y_\nu \bigr) - H\bigl( \Qs \floor{\nu\uv} \big| Y \bigr) \bigr|
	< \epsilon.
\end{equation}
\end{lemma}
}

\proofbox{}{
\begin{IEEEproof}
See Appendix~\ref{app:proof:exchange_condition_on_Y}.
\end{IEEEproof}
}

Lemma~\ref{lem:exchange_condition_on_Y} allows us to swap out $Y_\nu$ for $Y$ in the conditional entropy term on the left-hand side of~\eqref{quod_est_demonstrandum_limit_2}, yielding
\begin{IEEEeqnarray*}{rCl}
	\lim_{\nu \to \infty} \bigl\{ H(\Bs\floor{\nu\uv}|Y_\nu) - L_\Bf \log(\nu) \bigr\}
	&=& \lim_{\nu \to \infty} \bigl\{ H(\Bs\floor{\nu\uv}|Y) - L_\Bf \log(\nu) \bigr\} \\
	&=& \int_\Yc \lim_{\nu \to \infty} \Bigl\{ H\bigl( \Bs \floor{\nu\uv} \big| Y=y \bigr) - L_\Bf \log(\nu) \Bigr\} \intd P_Y(y) \\
	&=& \int_\Yc \mathcal{H}_{\Bs}(\uv|Y=y) \intd P_Y(y) \\
	&=& \mathcal{H}_{\Bs}(\uv|Y).   \IEEEyesnumber\label{H_B_limit}
\end{IEEEeqnarray*}
The exchange of limit and integration carried out in the second equality of~\eqref{H_B_limit} follows from the Dominated Convergence Theorem~\cite[Theorem~1.6.9]{AsDo00}, which is applicable because by Lemmata~\ref{lem:Renyi} and \ref{lem:quantized_entropy_difference}, the integrand can be bounded from above by
\begin{IEEEeqnarray*}{rCl}
	H\bigl( \Bs \floor{\nu\uv} \big| Y=y \bigr) - L_\Bf \log(\nu)
	&\leq& H\bigl( \floor{\nu\Bs\uv} \big| Y=y \bigr) + L_\Bf \log\left( \frac{\lVert \Bs \rVert_1}{\nu L_\Bf} \right) \\
	&\leq& H\bigl( \floor{\Bs\uv} \big| Y=y \bigr) + L_\Bf \log\left( \frac{\lVert \Bs \rVert_1}{L_\Bf} \right)   \IEEEyesnumber\label{DCT_UB}
\end{IEEEeqnarray*}
and from below by
\begin{IEEEeqnarray*}{rCl}
	H\bigl( \Bs \floor{\nu\uv} \big| Y=y \bigr) - L_\Bf \log(\nu)
	&\geq& H\bigl( \floor{\nu\Bs\uv} \big| Y=y \bigr) - L_\Bf \log\left( \frac{\lVert \Bs \rVert_1}{\nu L_\Bf} \right) \\
	&\geq& h(\Bs\uv|Y=y) - L_\Bf \log\left( \frac{\lVert \Bs \rVert_1}{L_\Bf} \right).   \IEEEyesnumber\label{DCT_LB}
\end{IEEEeqnarray*}
Both the upper bound~\eqref{DCT_UB} and lower bound~\eqref{DCT_LB} are integrable in $y$ because $H(\floor{\Bs\uv}|Y)$ and $h(\Bs\uv|Y)$ are finite. The former is finite because $0 \leq H(\floor{\Bs\uv}|Y) \leq H(\floor{\Bs\uv})$ where $H(\floor{\Bs\uv})$ is finite by Lemma~\ref{lem:entropy_finiteness}, whereas the latter equals $h(\Bs\uv|Y) = h(\Bs\uv) - I(\Bs\uv;Y)$, where $h(\Bs\uv)$ is finite by Lemma~\ref{lem:entropy_finiteness} and $I(\Bs\uv;Y)$ is finite by the theorem's assumption of a finite sum mutual information $I(\uv;Y)$.

Now, only the inequality~\eqref{quod_est_demonstrandum_limit_3} remains to be proven. Due to the uniform convergence stated in Lemma~\ref{lem:exchange_condition_on_Y}, we can swap out $Y_\nu$ for $Y$ in the limit on the left-hand side of~\eqref{quod_est_demonstrandum_limit_3}, i.e., if we define
\begin{equation}   \label{J_tilde_quantized}
	\tilde{J}_\nu(\Bs,M)
	\triangleq \inf_{\Cs \in \mathscr{C}_\mathbb{Z}(M)} H(\Cs\Bs\floor{\nu\uv}|Y)
\end{equation}
then $\lim_{\nu \to \infty} \bigl\{ J_\nu(\Bs,M) - \tilde{J}_\nu(\Bs,M) \bigr\} = 0$, hence
\begin{equation}   \label{liminf_J}
	\liminf_{\nu \to \infty} \bigl\{ J_\nu(\Bs,M) - L_\Cf \log(\nu) \bigr\}
	 = \liminf_{\nu \to \infty} \bigl\{ \tilde{J}_\nu(\Bs,M) - L_\Cf \log(\nu) \bigr\}.
\end{equation}
Recall that the matrix $\Cs\Bs$ can be represented by Smith normal decomposition as [cf.~Definition~\ref{def:SNF}]
\begin{equation}   \label{SNF_of_CB}
	\Cs\Bs = \Ss(\Cs\Bs) \Sigmas(\Cs\Bs) \Ts(\Cs\Bs)
\end{equation}
where $\Ss(\Cs\Bs) \in \mathbb{Z}^{L_\Cf \times L_\Cf}$ is unimodular, $\Sigmas(\Cs\Bs) \in \mathbb{Z}^{L_\Cf \times L_\Cf}$ is diagonal with positive diagonal entries $\sigma_1(\Cs\Bs),\dotsc,\sigma_{L_\Cf}(\Cs\Bs)$, and $\Ts(\Cs\Bs) \in \mathbb{Z}^{L_\Cf \times K}$ is right-invertible.

Let $\Omegas^\T \in \mathbb{Z}^{K \times L_\Cf}$ denote the matrix obtained from the Korkin--Zolotarev reduction of $\Ts(\Cs\Bs)^\T$. In other words, the columns of $\Omegas^\T$ are in Korkin--Zolotarev reduced form, and generate the same lattice as the columns of $\Ts(\Cs\Bs)^\T$. Consider partition cells
\begin{equation}
	\Pc_\is
	= \bigl\{ \us \in \mathbb{R}^K \colon \Omegas\floor{\us} = \is \bigr\}
\end{equation}
indexed by $\is \in \mathbb{Z}^{L_\Cf}$, which can be equivalently parametrized as
\begin{IEEEeqnarray}{rCl}
	\Pc_\is
	&=& \bigcup_{\js \in \mathbb{Z}^{K-L_\Cf}} \bigl\{ \us \in \mathbb{R}^K \colon \floor{\us} = \Omegas^\sharp\is + \Omegas_\perp\js \bigr\} \\
	&=& \Omegas^\sharp\is + \Lambda(\Omegas_\perp) + [0,1)^K.
\end{IEEEeqnarray}
We have that the sets $\Pc_\is$ are mutually disjoint and $\{ \Pc_\is \colon \is \in \mathbb{Z}^{L_\Cf} \}$ constitutes a partition of $\mathbb{R}^K$, in that $\mathbb{R}^K = \bigcup_{\is \in \mathbb{Z}^{L_\Cf}} \Pc_\is$. Consider the entropy $H(\Cs\Bs\floor{\nu\uv}|Y)$, which can be written out as
\begin{IEEEeqnarray*}{rCl}
	H(\Cs\Bs\floor{\nu\uv}|Y)
	&=& H(\Omegas\floor{\nu\uv}|Y) \\
	&=& \int_\Yc \sum_{\is \in \mathbb{Z}^{L_\Cf}} \Phi\bigl( \P\{ \Omegas\floor{\nu\uv} = \is | Y=y \} \bigr) \intd P_Y(y) \\
	&=& \int_\Yc \sum_{\is \in \mathbb{Z}^{L_\Cf}} \Phi\bigl( P_{\uv|Y=y}(\nu^{-1}\Pc_\is) \bigr) \intd P_Y(y)
\end{IEEEeqnarray*}
where $\Phi(x) = -x\log(x)$. Let $\Qc_\nu$ denote the $\sigma$-algebra generated by the partition $\{ \nu^{-1}\Pc_\is \colon \is \in \mathbb{Z}^{L_\Cf} \}$. Then the operation $\us \mapsto \Omegas\floor{\nu\us}$ (i.e., scaling by $\nu$, followed by flooring and left-multiplication with $\Omegas$) can be interpreted as a quantization operation with respect to the atoms of $\Sigma_\nu$. Accordingly, we can write
\begin{equation}   \label{entropy_as_quantization}
	H(\Cs\Bs\floor{\nu\uv}|Y)
	= H(\Omegas\floor{\nu\uv}|Y)
	= H(\lceil \uv \rfloor_{\Qc_\nu}|Y).
\end{equation}
Let us assume from now on that $\uv$ has compact support, with a support set contained in the cube $\tau \mathscr{S}_\infty^K$
and let $\tilde{\Qc}_\nu$ denote the $\sigma$-algebra generated by the quantization cells $\{ \nu^{-1}\Pc_\is \cap \tau \mathscr{S}_\infty^K \colon \is \in \mathbb{Z}^{L_\Cf} \}$. Then
\begin{equation}   \label{entropy_as_quantization_bis}
	H(\Cs\Bs\floor{\nu\uv}|Y)
	= H(\lceil \uv \rfloor_{\tilde{\Qc}_\nu}|Y).
\end{equation}
Next, we will upper-bound the volume of the $\is$-th quantization cell $\nu^{-1}\Pc_\is \cap \tau \mathscr{S}_\infty^K$.
For this purpose, we first outer-bound said quantization cell as follows:
\begin{IEEEeqnarray*}{rCl}
    \nu^{-1}\Pc_\is \cap \tau\mathscr{S}_\infty^K
    &=& \nu^{-1} \left( \Omegas^\sharp\is + \Lambda(\Omegas_\perp) + [0,1)^K \right) \cap \tau\mathscr{S}_\infty^K \\
    &\subseteq& \bigl( \nu^{-1} ( \Omegas^\sharp\is + \Lambda(\Omegas_\perp) ) \cap (\tau+\nu^{-1})\mathscr{S}_\infty^K \bigr) + [0,\nu^{-1})^K.   \IEEEeqnarraynumspace\IEEEyesnumber\label{quantization_cell_outer_bound}
\end{IEEEeqnarray*}
To see why this holds, bear in mind that a vector $\us$ belongs to the quantization cell $\nu^{-1}\Pc_\is \cap \tau\mathscr{S}_\infty^K$ if and only if there exists a pair of vectors $(\js,\ts) \in \mathbb{Z}^{K-L_\Cf} \times [0,1)^K$ such that $\us = \nu^{-1}(\Omegas^\sharp\is + \Omegas_\perp\js + \ts)$ and $\lVert \us \rVert_\infty \leq \tau$. The latter implies that $\nu^{-1} \lVert \Omegas^\sharp\is + \Omegas_\perp\js \rVert_\infty \leq \tau + \nu^{-1}$. Hence $\us = \us_1 + \us_2$ is the sum of a vector $\us_1 = \nu^{-1}(\Omegas^\sharp\is + \Omegas_\perp\js)$ satisfying $\us_1 \in \nu^{-1}(\Omegas^\sharp\is + \Lambda(\Omegas_\perp))$ as well as $\lVert \us_1 \rVert_\infty \leq \tau + \nu^{-1}$, and a vector $\us_2 = \nu^{-1}\ts$ satisfying $\us_2 \in [0,\nu^{-1})^K$. This means that $\us$ belongs to the set written out on the right-hand side of~\eqref{quantization_cell_outer_bound}. Hence with the help of the inclusion relation~\eqref{quantization_cell_outer_bound}, we can now upper-bound the volume of $\nu^{-1}\Pc_\is \cap \tau\mathscr{S}_\infty^K$ as follows:
\begin{IEEEeqnarray*}{rCl}
    \vol\bigl(\nu^{-1}\Pc_\is \cap \tau\mathscr{S}_\infty^K\bigr)
    &=& \nu^{-K}\vol\bigl(\Pc_\is \cap \nu\tau\mathscr{S}_\infty^K\bigr) \\
    &\leq& \nu^{-K} \vol{\Bigl( \left( ( \Omegas^\sharp\is + \Lambda(\Omegas_\perp) ) \cap (\nu\tau+1)\mathscr{S}_\infty^K \right) + [0,1)^K \Bigr)} \\
    &=& \nu^{-K} \Bigl| ( \Omegas^\sharp\is + \Lambda(\Omegas_\perp) ) \cap (\nu\tau+1)\mathscr{S}_\infty^K \Bigr| \\
    &=& \nu^{-K} \Bigl| \Lambda(\Omegas_\perp) \cap \bigl( (\nu\tau+1)\mathscr{S}_\infty^K - \Omegas^\sharp\is \bigr) \Bigr| \\
    &\leq& \nu^{-K} \Bigl| \Lambda(\Omegas_\perp) \cap \bigl( 2\nu\tau \mathscr{S}_\infty^K - \Omegas^\sharp\is \bigr) \Bigr|, 
	\IEEEeqnarraynumspace\IEEEyesnumber\label{quantization_cell_upper_bound}
\end{IEEEeqnarray*}
where the last inequality holds for $\nu\tau \geq 1$.
To further upper-bound the volume of quantization cells $\nu^{-1}\Pc_\is \cap \tau\mathscr{S}_\infty^K$, we will now elaborate an upper bound on the right-hand side of~\eqref{quantization_cell_upper_bound} via the following lemma.

\lemmabox{\begin{lemma}[Point-counting bound]   \label{lem:point-counting_bound}
Let $\Fs = [ \fs_1 , \dotsc , \fs_L ] \in \mathbb{R}^{K \times L}$ denote a tall matrix with linearly independent columns, and let $\Gs = [ \gs_1 , \dotsc , \gs_L ] \in \mathbb{R}^{K \times L}$ denote a Gram--Schmidt orthogonalization of $\Fs$, with orthogonal columns $\gs_\ell, \ell=1,\dotsc,L$ constructed from successive orthogonal projections as follows:
\begin{subequations}
\begin{IEEEeqnarray}{rCl}
	\gs_1
	&=& \fs_1   \label{Gram-Schmidt_1} \\
	\gs_\ell
	&=& \fs_\ell - \sum_{\ell'=1}^{\ell-1} \frac{\fs_\ell^\T\gs_{\ell'}}{\lVert \gs_{\ell'} \rVert_2^2} \gs_{\ell'},
	\quad \text{$\ell = 2, \dotsc, L$.}   \label{Gram-Schmidt_2}
\end{IEEEeqnarray}
\end{subequations}
The number of lattice points from $\Lambda(\Fs)$ that lie within a $K$-dimensional infinity-norm ball of radius $\rho$ and shifted by an arbitrary vector $\ts \in \mathbb{R}^K$ is upper-bounded as
\begin{IEEEeqnarray}{rCl}
    \left| \Lambda(\Fs) \cap (\rho\mathscr{S}_\infty^K + \ts) \right|
    &\leq& \prod_{\ell=1}^L \left\lceil \frac{2\sqrt{K}\rho}{\lVert \gs_\ell \rVert_2} \right\rceil.
\end{IEEEeqnarray}
\end{lemma}}

\proofbox{}{
\begin{IEEEproof}
Let us define the $L$-dimensional cube
\begin{equation}
	\tilde{\mathscr{S}}_\infty^L
	= \left\{ \sum_{\ell=1}^L \alpha_\ell \frac{\gs_\ell}{\lVert \gs_\ell \rVert_2} \colon (\alpha_1, \dotsc, \alpha_L)^\T \in [-1,1]^L \right\}
	\subset \mathbb{R}^K
\end{equation}
which results from rotating the unit cube $\mathscr{S}_\infty^L = [-1,1]^L$ (embedded in $\mathbb{R}^K$ by zero-padding) about the origin to have its vertices aligned with the orthogonal basis $\Gs$. Alternatively, to represent $\tilde{\mathscr{S}}_\infty^L$ we may use the sumset notation
\begin{equation}
	\tilde{\mathscr{S}}_\infty^L
	= \sum_{\ell=1}^L [-1,1] \frac{\gs_\ell}{\lVert \gs_\ell \rVert_2}.
\end{equation}
We have the inclusion relations
\begin{IEEEeqnarray*}{rCl}
	\Lambda(\Fs) \cap (\rho\mathscr{S}_\infty^K + \ts)
	&\stackrel{(a)}{\subset}& \Span(\Gs) \cap (\rho\mathscr{S}_\infty^K + \ts) \\
	&\stackrel{(b)}{\subset}& \Span(\Gs) \cap (\rho\sqrt{K}\mathscr{S}_2^K + \ts) \\
	&\stackrel{(c)}{\subset}& \Span(\Gs) \cap (\rho\sqrt{K}\mathscr{S}_2^K + \Gs(\Gs^\T\Gs)^{-1}\Gs^\T\ts) \\
	&\stackrel{(d)}{\subset}& \Span(\Gs) \cap (\rho\sqrt{K}\tilde{\mathscr{S}}_\infty^L + \Gs(\Gs^\T\Gs)^{-1}\Gs^\T\ts).   \IEEEeqnarraynumspace\IEEEyesnumber\label{truncated_lattice_bounding}
\end{IEEEeqnarray*}
Here, Step $(a)$ is simply due to $\Lambda(\Fs) \subset \Span(\Fs) = \Span(\Gs)$; Step $(b)$ follows from $\mathscr{S}_\infty^K \subset \sqrt{K}\mathscr{S}_2^K$; Step $(c)$ holds because, if we decompose $\ts = \ts_\perp + \ts_{\|}$ into a vector $\ts_\perp = (\Is - \Gs(\Gs^\T\Gs)^{-1}\Gs^\T)\ts$ that is orthogonal to $\Span(\Gs)$ and a vector $\ts_{\|} = \Gs(\Gs^\T\Gs)^{-1}\Gs^\T\ts$ that lies in $\Span(\Gs)$, then the intersection of the $L$-dimensional plane $\Span(\Gs)$ and the $K$-dimensional sphere $\mathscr{S}_2^K$ shifted by $\ts$, is maximal when the orthogonal component $\ts_\perp$ is zero; Step $(d)$ follows because $\tilde{\mathscr{S}}_\infty^L \subset \Span(\Gs)$ includes all elements of $\Gs$ with Euclidean norm smaller or equal to unity. Intersecting both sides of~\eqref{truncated_lattice_bounding} with $\Lambda(\Fs)$ we end up with
\begin{equation}   \label{truncated_lattice_outer_bounding}
	\Lambda(\Fs) \cap (\rho\mathscr{S}_\infty^K + \ts)
	\subset \Lambda(\Fs) \cap (\rho\sqrt{K}\tilde{\mathscr{S}}_\infty^L + \ts_{\|})
\end{equation}
where $\ts_{\|}$ is an abbreviation for the projection of $\ts$ onto the column space of $\Gs$, as defined above. Hence, from~\eqref{truncated_lattice_outer_bounding} we infer that
\begin{equation}   \label{truncated_lattice_cardinality_upper_bound}
	\bigl| \Lambda(\Fs) \cap (\rho\mathscr{S}_\infty^K + \ts) \bigr|
	\leq \bigl| \Lambda(\Fs) \cap (\rho\sqrt{K}\tilde{\mathscr{S}}_\infty^L + \ts_{\|}) \bigr|.
\end{equation}
Since the lattice $\Lambda(\Fs)$ can be expressed as an iterated union
\begin{equation}
	\Lambda(\Fs)
	= \bigcup_{\alpha_L \in \mathbb{Z}} \dotso \bigcup_{\alpha_1 \in \mathbb{Z}} \left\{ \sum_{\ell=1}^L \alpha_\ell \fs_\ell \right\}
\end{equation}
we can evaluate the cardinality $\bigl| \Lambda(\Fs) \cap (\rho\sqrt{K}\tilde{\mathscr{S}}_\infty^L + \ts_{\|}) \bigr|$ by an iterated sum as
\begin{IEEEeqnarray}{rCl}
	\bigl| \Lambda(\Fs) \cap (\rho\sqrt{K}\tilde{\mathscr{S}}_\infty^L + \ts_{\|}) \bigr|
	&=& \sum_{\alpha_L \in \mathbb{Z}} \dotso \sum_{\alpha_1 \in \mathbb{Z}} \mathds{1}{\left\{ \sum_{\ell=1}^L \alpha_\ell \fs_\ell \in (\rho\sqrt{K}\tilde{\mathscr{S}}_\infty^L + \ts_{\|}) \right\}}   \label{cardinality_bound_as_iterated_sum}
\end{IEEEeqnarray}
where the indicator function $\mathds{1}\{\cdot\}$ equals one if the statement in braces is true, and zero otherwise. The iterated sum~\eqref{cardinality_bound_as_iterated_sum} can be decomposed into partial summations
\begin{IEEEeqnarray}{rCl}
	S_1(\alpha_2,\dotsc,\alpha_L)
	&=& \sum_{\alpha_1 \in \mathbb{Z}} \mathds{1}{\left\{ \sum_{\ell=1}^L \alpha_\ell \fs_\ell \in \bigl( \rho\sqrt{K}\tilde{\mathscr{S}}_\infty^L + \ts_{\|} \bigr) \right\}} \\
	S_{\ell+1}(\alpha_{\ell+2},\dotsc,\alpha_L)
	&=& \sum_{\alpha_{\ell+1} \in \mathbb{Z}} S_\ell(\alpha_{\ell+1},\dotsc,\alpha_L),
	\quad \text{for $\ell = 1,\dotsc,L-1$}
\end{IEEEeqnarray}
such that the last sum $S_L = \sum_{\alpha_L \in \mathbb{Z}} S_{L-1}(\alpha_L)$ equals the cardinality $\bigl| \Lambda(\Fs) \cap (\rho\sqrt{K}\tilde{\mathscr{S}}_\infty^L + \ts_{\|}) \bigr|$, which we seek to upper-bound. We will now prove by induction that the following upper bound holds:
\begin{equation}   \label{upper_bound_for_induction}
	S_\ell(\alpha_{\ell+1},\dotsc,\alpha_L)
	\leq \prod_{k=1}^\ell \left\lceil \frac{2\rho\sqrt{K}}{\lVert \gs_k \rVert_2} \right\rceil \times \mathds{1}{\left\{ \sum_{\ell'=\ell+1}^L \alpha_{\ell'} \sum_{\ell''=\ell+1}^L \frac{\fs_{\ell'}^\T\gs_{\ell''}}{\lVert \gs_{\ell''} \rVert_2^2} \gs_{\ell''} \in \bigl( \rho\sqrt{K}\tilde{\mathscr{S}}_\infty^L + \ts_{\|} \bigr) \right\}}.
\end{equation}
For the first partial sum, we can prove the bound~\eqref{upper_bound_for_induction} as follows:
\begin{IEEEeqnarray*}{rCl}
	S_1(\alpha_2,\dotsc,\alpha_L)
	&\stackrel{(a)}{=}& \sum_{\alpha_1 \in \mathbb{Z}} \mathds{1}{\left\{ \left( \alpha_1 + \sum_{\ell=2}^L \alpha_\ell \frac{\fs_\ell^\T\gs_1}{\lVert \gs_1 \rVert_2^2} \right) \gs_1 + \sum_{\ell=2}^L \alpha_\ell \sum_{\ell'=2}^L \frac{\fs_\ell^\T\gs_{\ell'}}{\lVert \gs_{\ell'} \rVert_2^2} \gs_{\ell'} \in \bigl( \rho\sqrt{K}\tilde{\mathscr{S}}_\infty^L + \ts_{\|} \bigr) \right\}} \\
	&\stackrel{(b)}{=}& \sum_{\alpha_1 \in \mathbb{Z}} \mathds{1}{\left\{ \left( \alpha_1 + \sum_{\ell=2}^L \alpha_\ell \frac{\fs_\ell^\T\gs_1}{\lVert \gs_1 \rVert_2^2} \right) \gs_1 - \frac{\gs_1^\T\ts}{\lVert \gs_1 \rVert_2^2 }\gs_1 \in [-1,1] \frac{\rho\sqrt{K}\gs_1}{\lVert \gs_1 \rVert_2} \right\}} \times \\
	&& \qquad {} \times \mathds{1}{\left\{ \sum_{\ell=2}^L \alpha_\ell \sum_{\ell'=2}^L \frac{\fs_\ell^\T\gs_{\ell'}}{\lVert \gs_{\ell'} \rVert_2^2} \gs_{\ell'} - \sum_{\ell'=2}^L \frac{\gs_{\ell'}^\T\ts}{\lVert \gs_{\ell'} \rVert_2^2 }\gs_{\ell'} \in \sum_{\ell'=2}^L [-1,1] \frac{\rho\sqrt{K}\gs_{\ell'}}{\lVert \gs_{\ell'} \rVert_2} \right\}} \\
	&\stackrel{(c)}{\leq}& \left\lceil \frac{2\rho\sqrt{K}}{\lVert \gs_1 \rVert_2} \right\rceil \mathds{1}{\left\{ \sum_{\ell=2}^L \alpha_\ell \sum_{\ell'=2}^L \frac{\fs_\ell^\T\gs_{\ell'}}{\lVert \gs_{\ell'} \rVert_2^2} \gs_{\ell'} \in \bigl( \rho\sqrt{K}\tilde{\mathscr{S}}_\infty^L + \ts_{\|} \bigr) \right\}}.   \IEEEeqnarraynumspace\IEEEyesnumber
\end{IEEEeqnarray*}
Here, Step $(a)$ consists in expanding $\sum_{\ell=1}^L \alpha_\ell \fs_\ell$ along the basis $\Gs$; in Step $(b)$ we have exploited the orthogonality of $\Gs$ and the geometry of $\tilde{\Bs}_\infty^L$ to factorize the indicator function; in Step $(c)$ we have upper-bounded the first indicator function by the ceiling term.
By comparison with~\eqref{upper_bound_for_induction}, we confirm that the upper bound holds indeed for $\ell = 1$.
If the bound~\eqref{upper_bound_for_induction} holds for some $\ell \geq 1$, we prove that it must also hold for $\ell+1$, as follows:
\begin{IEEEeqnarray*}{rCl}
	\IEEEeqnarraymulticol{3}{l}{
		S_{\ell+1}(\alpha_{\ell+2},\dotsc,\alpha_L)
	} \\ \
	&=& \sum_{\alpha_{\ell+1} \in \mathbb{Z}} S_\ell(\alpha_{\ell+1},\dotsc,\alpha_L) \\
	&\leq& \prod_{k=1}^\ell \left\lceil \frac{2\rho\sqrt{K}}{\lVert \gs_k \rVert_2} \right\rceil \sum_{\alpha_{\ell+1} \in \mathbb{Z}} \mathds{1}{\left\{ \sum_{\ell'=\ell+1}^L \alpha_{\ell'} \sum_{\ell''=\ell+1}^L \frac{\fs_{\ell'}^\T\gs_{\ell''}}{\lVert \gs_{\ell''} \rVert_2^2} \gs_{\ell''} \in \bigl( \rho\sqrt{K}\tilde{\mathscr{S}}_\infty^L + \ts_{\|} \bigr) \right\}} \\
	&\stackrel{(a)}{=}& \prod_{k=1}^\ell \left\lceil \frac{2\rho\sqrt{K}}{\lVert \gs_k \rVert_2} \right\rceil \sum_{\alpha_{\ell+1} \in \mathbb{Z}} \mathds{1}\Biggl\{ \alpha_{\ell+1} \left( \gs_{\ell+1} + \sum_{\ell''=\ell+2}^L \frac{\fs_{\ell+1}^\T\gs_{\ell''}}{\lVert \gs_{\ell''} \rVert_2^2} \gs_{\ell''} \right) + \sum_{\ell'=\ell+2}^L \alpha_{\ell'} \sum_{\ell''=\ell+1}^L \frac{\fs_{\ell'}^\T\gs_{\ell''}}{\lVert \gs_{\ell''} \rVert_2^2} \gs_{\ell''} \\
	\IEEEeqnarraymulticol{3}{r}{
		\in \bigl( \rho\sqrt{K}\tilde{\mathscr{S}}_\infty^L + \ts_{\|} \bigr) \Biggr\}
	} \\
	&\stackrel{(b)}{=}& \prod_{k=1}^\ell \left\lceil \frac{2\rho\sqrt{K}}{\lVert \gs_k \rVert_2} \right\rceil \sum_{\alpha_{\ell+1} \in \mathbb{Z}} \mathds{1}\Biggl\{ \alpha_{\ell+1} \gs_{\ell+1} + \sum_{\ell'=\ell+2}^L \alpha_{\ell'} \frac{\fs_{\ell'}^\T\gs_{\ell+1}}{\lVert \gs_{\ell+1} \rVert_2^2} \gs_{\ell+1} - \frac{\gs_{\ell+1}^\T\ts}{\lVert \gs_{\ell+1} \rVert_2^2} \gs_{\ell+1} \\
	\IEEEeqnarraymulticol{3}{r}{
		\in [-1,1] \frac{\rho\sqrt{K}\gs_{\ell+1}}{\lVert \gs_{\ell+1} \rVert_2} \Biggr\} \times
	} \\
	&& \qquad {} \times \mathds{1}\Biggl\{ \alpha_{\ell+1} \sum_{\ell''=\ell+2}^L \frac{\fs_{\ell+1}^\T\gs_{\ell''}}{\lVert \gs_{\ell''} \rVert_2^2} \gs_{\ell''} + \sum_{\ell'=\ell+2}^L \alpha_{\ell'} \sum_{\ell''=\ell+2}^L \frac{\fs_{\ell'}^\T\gs_{\ell''}}{\lVert \gs_{\ell''} \rVert_2^2} \gs_{\ell''} - \sum_{\ell''=\ell+2}^L \frac{\gs_{\ell''}^\T\ts}{\lVert \gs_{\ell''} \rVert_2^2} \gs_{\ell''} \\
	\IEEEeqnarraymulticol{3}{r}{
		{} \in \sum_{\ell''=\ell+2}^L [-1,1] \frac{\rho\sqrt{K}\gs_{\ell''}}{\lVert \gs_{\ell''} \rVert_2} \Biggr\}
	} \\
	&\stackrel{(c)}{\leq}& \prod_{k=1}^{\ell+1} \left\lceil \frac{2\rho\sqrt{K}}{\lVert \gs_k \rVert_2} \right\rceil \mathds{1}\Biggl\{ \sum_{\ell'=\ell+1}^L \alpha_{\ell'} \sum_{\ell''=\ell+2}^L \frac{\fs_{\ell'}^\T\gs_{\ell''}}{\lVert \gs_{\ell''} \rVert_2^2} \gs_{\ell''} - \sum_{\ell''=\ell+2}^L \frac{\gs_{\ell''}^\T\ts}{\lVert \gs_{\ell''} \rVert_2^2} \gs_{\ell''} \\
	\IEEEeqnarraymulticol{3}{r}{
		\in \sum_{\ell''=\ell+2}^L [-1,1] \frac{\rho\sqrt{K}\gs_{\ell''}}{\lVert \gs_{\ell''} \rVert_2} \Biggr\}
	} \\
	&=& \prod_{k=1}^{\ell+1} \left\lceil \frac{2\rho\sqrt{K}}{\lVert \gs_k \rVert_2} \right\rceil \mathds{1}\Biggl\{ \sum_{\ell'=\ell+1}^L \alpha_{\ell'} \sum_{\ell''=\ell+2}^L \frac{\fs_{\ell'}^\T\gs_{\ell''}}{\lVert \gs_{\ell''} \rVert_2^2} \gs_{\ell''} \in \bigl( \rho\sqrt{K}\tilde{\mathscr{S}}_\infty^L + \ts_{\|} \bigr) \Biggr\}.   \IEEEeqnarraynumspace\IEEEyesnumber
\end{IEEEeqnarray*}
The steps are similar as before: in Step $(a)$ we expand along the basis $\Gs$; in Step $(b)$ we factorize the indicator function; in Step $(c)$ we upper-bound the first indicator function. This concludes the argument by induction. Therefore, the inequality~\eqref{upper_bound_for_induction} holds for $\ell = L-1$, hence
\begin{equation}
	S_L
	= \bigl| \Lambda(\Fs) \cap (\rho\sqrt{K}\tilde{\mathscr{S}}_\infty^L + \ts_{\|}) \bigr|
	\leq \prod_{k=1}^L \left\lceil \frac{2\rho\sqrt{K}}{\lVert \gs_k \rVert_2} \right\rceil
\end{equation}
which in combination with~\eqref{truncated_lattice_cardinality_upper_bound} concludes the proof of Lemma~\ref{lem:point-counting_bound}.
\end{IEEEproof}
}

Let $\Gs = [ \gs_1, \dotsc, \gs_{K-L_\Cf} ] \in \mathbb{R}^{K \times (K-L_\Cf)}$ denote the Gram--Schmidt orthogonalization of
\begin{equation}
	\Omegas_\perp = \bigl[ \omegas_1^\perp, \dotsc, \omegas_{K-L_\Cf}^\perp \bigr] \in \mathbb{Z}^{K \times (K-L_\Cf)}.
\end{equation}
That is, $\gs_\ell$ is constructed from the vectors $(\omegas_1^\perp, \dotsc, \omegas_\ell^\perp)$ by successive projections as described in~\eqref{Gram-Schmidt_1}--\eqref{Gram-Schmidt_2}.
Applying Lemma~\ref{lem:point-counting_bound} on the lattice $\Lambda(\Omegas_\perp)$ intersected with a ball of radius $\rho$ and arbitrary shift $-\ts$, we get
\begin{IEEEeqnarray*}{rCl}
    \left| \Lambda(\Omegas_\perp) \cap (\rho\mathscr{S}_\infty^K - \ts) \right|   
    &\leq& \prod_{\ell=1}^{K-L_\Cf} \left\lceil \frac{2\sqrt{K} \rho}{\lVert \gs_\ell \rVert_2} \right\rceil \\
    &=& \frac{(2\sqrt{K} \rho)^{K-L_\Cf}}{\sqrt{\det(\Omegas\Omegas^\T)}}
	\qquad \text{(for $2\sqrt{K} \rho \geq \max_{\ell \in [K-L_\Cf]} \lVert \gs_\ell \rVert_2$)}   \IEEEeqnarraynumspace\IEEEyesnumber\label{weakened_point-counting_bound}
\end{IEEEeqnarray*}
because $\prod_{\ell=1}^{K-L_\Cf} \lVert \gs_\ell \rVert_2^2 = \det(\Gs^\T\Gs) = \det(\Omegas_\perp^\T\Omegas_\perp) = \det(\Omegas\Omegas^\T)$.
Note that
\begin{IEEEeqnarray*}{rCl}
	\max_{\ell \in [K-L_\Cf]} \lVert \gs_\ell \rVert_2
	&\stackrel{(a)}{\leq}& \max_{\ell \in [K-L_\Cf]} \lVert \omegas_\ell^\perp \rVert_2 \\
	&\stackrel{(b)}{\leq}&  \max_{\ell \in [K-L_\Cf]} \sqrt{\frac{\ell+3}{4}} \lambda_{2,\ell}(\Omegas_\perp) \\
	&\leq& \sqrt{\frac{K-L_\Cf+3}{4}} \lambda_{2,K-L_\Cf}(\Omegas_\perp) \\
	&\stackrel{(c)}{\leq}& \sqrt{\frac{K-L_\Cf+3}{4}} \prod_{\ell=1}^{K-L_\Cf} \lambda_{2,\ell}(\Omegas_\perp) \\
	&\stackrel{(d)}{\leq}& \sqrt{\frac{K-L_\Cf+3}{4}} 2^{K-L_\Cf} \det(\Omegas_\perp^\T\Omegas_\perp) \\
	&\stackrel{(e)}{\leq}& \sqrt{K} 2^K \det(\Omegas\Omegas^\T).
\end{IEEEeqnarray*}
Here, Step $(a)$ holds because $\gs_\ell$ results from a (contractive) projection of $\omegas_\ell^\perp$, hence $\lVert \gs \rVert_2 \leq \lVert \omegas_\ell^\perp \rVert_2$ for all $\ell$; Step $(b)$ involves the fact that $\Omegas_\perp$ is in reduced Korkin--Zolotarev form and the inequality~\eqref{KZ_slack}; Step $(c)$ makes use of the fact that the successive minima of $\Omegas_\perp$ are lower-bounded by unity because $\Omegas_\perp$ is an integer lattice; Step $(d)$ involves Lemma~\ref{thm:Minkowski_Second_Theorem}; Step $(e)$ uses $\det(\Omegas_\perp^\T\Omegas_\perp) = \det(\Omegas\Omegas^\T)$ and $L_\Cf \geq 0$ as well as $3 \leq 3K$.
From this inequality we can deduce a sufficient condition for the inequality~\eqref{weakened_point-counting_bound} to hold, leading to a weaker version of~\eqref{weakened_point-counting_bound}, namely
\begin{IEEEeqnarray*}{rCl}
    \left| \Lambda(\Omegas_\perp) \cap (\rho\mathscr{S}_\infty^K - \ts) \right|   
    &\leq& \frac{(2\sqrt{K} \rho)^{K-L_\Cf}}{\sqrt{\det(\Omegas\Omegas^\T)}}
	\qquad \text{(for $\rho \geq 2^{K-1} \det(\Omegas\Omegas^\T)$).}   \IEEEeqnarraynumspace\IEEEyesnumber\label{weakened_point-counting_bound_2}
\end{IEEEeqnarray*}
Evaluating~\eqref{weakened_point-counting_bound_2} for $\rho = 2\nu\tau$ and noticing that the condition $2\nu\tau \geq 2^{K-1} \det(\Omegas\Omegas^\T)$ in~\eqref{weakened_point-counting_bound_2} already implies the condition $\nu\tau \geq 1$ from~\eqref{quantization_cell_upper_bound} (due to $K \geq 2$ and $\det(\Omegas\Omegas^\T) \geq 1$ since $\Omegas$ is an integer lattice), we can combine~\eqref{quantization_cell_upper_bound} and~\eqref{weakened_point-counting_bound_2} to obtain
\begin{IEEEeqnarray*}{rCl}
    \vol\bigl(\nu^{-1}\Pc_\is \cap \tau\mathscr{S}_\infty^K\bigr)
    &\leq& \frac{(4\sqrt{K}\tau)^{K-L_\Cf}}{\sqrt{\det(\Omegas\Omegas^\T)}} \nu^{-L_\Cf}
	\qquad \text{(for $\nu\tau \geq 2^{K-2} \det(\Omegas\Omegas^\T)$).}   \IEEEeqnarraynumspace\IEEEyesnumber\label{weakened_point-counting_bound_3}
\end{IEEEeqnarray*}
Based on~\eqref{entropy_as_quantization_bis}, Lemma~\ref{lem:quantized_entropy_lower_bound} and the bound~\eqref{weakened_point-counting_bound_3}, we can state the following lower bound on the entropy $H(\Cs\Bs\floor{\nu\uv}|Y)$:
\begin{IEEEeqnarray*}{rCl}
	\IEEEeqnarraymulticol{3}{l}{
		H(\Cs\Bs\floor{\nu\uv}|Y)
	} \\ \quad
	&=& H(\lceil \uv \rfloor_{\tilde{\Qc}_\nu}|Y) \\
	&\geq& h(\uv|Y) - \max_{\is \in \mathbb{Z}^{L_\Cf}} \log\bigl(\vol{\bigl(\nu^{-1}\Pc_{\is} \cap \tau\mathscr{S}_\infty^K\bigr)}\bigr) \\
	&\geq& h(\uv|Y) - \log\left( \frac{(4\sqrt{K}\tau)^{K-L_\Cf}}{\sqrt{\det(\Omegas\Omegas^\T)}} \right) + L_\Cf\log(\nu)
	\qquad \text{(for $\nu\tau \geq 2^{K-2} \det(\Omegas\Omegas^\T)$).}   \IEEEeqnarraynumspace\IEEEyesnumber\label{entropy_LB}
\end{IEEEeqnarray*}
On the other hand, we can derive an upper bound on $H(\Cs\Bs\floor{\nu\uv}|Y)$ by means of Lemmata~\ref{lem:Renyi} [Inequality~\eqref{sandwich_bounds}] and \ref{lem:quantized_entropy_difference}, as follows:
\begin{IEEEeqnarray*}{rCl}
	H(\Cs\Bs\floor{\nu\uv}|Y) - L_\Cf\log(\nu)
	&\leq& H(\floor{\Cs\Bs\nu\uv}) - L_\Cf\log(\nu) + L_\Cf \log{\left( \frac{\lVert \Cs\Bs \rVert_1}{L_\Cf} \right)}\\
	&\leq& H(\floor{\Cs\Bs\uv}) + L_\Cf \log{\left( \frac{\lVert \Cs\Bs \rVert_1}{L_\Cf} \right)}.   \IEEEeqnarraynumspace\IEEEyesnumber\label{entropy_UB}
\end{IEEEeqnarray*}
Let us define the quantity
\begin{equation}
	H_0(\Bs,M)
	= \min_{\Cs \in \mathscr{C}(M)} \left\{ H(\floor{\Cs\Bs\uv}) + L_\Cf \log{\left( \frac{\lVert \Cs\Bs \rVert_1}{L_\Cf} \right)} \right\}
\end{equation}
which is an upper bound on $\tilde{J}_\nu(\Bs,M) - L_\Cf\log(\nu)$.
In the optimization problem that defines the quantity $\tilde{J}_\nu(\Bs,M)$ [cf.~\eqref{J_tilde_quantized}], a matrix $\Cs \in \mathscr{C}(M)$ is \emph{strictly} suboptimal, in the sense that $H(\Cs\Bs\floor{\nu\uv}|Y) > \tilde{J}_\nu(\Bs,M)$, if the following holds:
\begin{equation}   \label{sufficient_condition_for_suboptimality}
	H(\Cs\Bs\floor{\nu\uv}|Y) - L_\Cf\log(\nu)
	> H_0(\Bs,M).
\end{equation}
We will now derive a sufficient condition on $\Cs$ for being strictly suboptimal, so as to eventually reduce the domain of optimization in~\eqref{J_tilde_quantized} to a finite set. For this purpose, consider that~\eqref{sufficient_condition_for_suboptimality} certainly holds if [cf.~\eqref{entropy_LB}] the following two conditions on $\nu$ and $\Omegas$ are satisfied:
\begin{subequations}
\begin{IEEEeqnarray}{rCl}   \label{sufficient_condition_for_suboptimality_2}
	h(\uv|Y) - \log\left( \frac{(4\sqrt{K}\tau)^{K-L_\Cf}}{\sqrt{\det(\Omegas\Omegas^\T)}} \right)
	&>& H_0(\Bs,M)   \label{sufficient_condition_for_suboptimality_2_1} \\
	\nu\tau
	&\geq& 2^{K-2} \det(\Omegas\Omegas^\T).   \label{sufficient_condition_for_suboptimality_2_2}
\end{IEEEeqnarray}
\end{subequations}
(Recall that $\Omegas$ is a function of $\Bs$ and $\Cs$).
As these two conditions~\eqref{sufficient_condition_for_suboptimality_2_1}--\eqref{sufficient_condition_for_suboptimality_2_2} involve the determinant $\det(\Omegas\Omegas^\T)$, we now wish to translate these into sufficient conditions that involve $\det(\Cs\Cs^\T)$.
Consider the following inequality, which relates $\det(\Omegas\Omegas^\T)$ and $\det(\Cs\Cs^\T)$:
\begin{IEEEeqnarray*}{rCl}
	\lambda_\mathrm{min}(\Bs\Bs^\T)^{L_\Cf} \det(\Cs\Cs^\T)
	&\stackrel{(a)}{\leq}& \det(\Cs\Bs\Bs^\T\Cs^\T) \\
	&\stackrel{(b)}{=}& \det(\Sigmas(\Cs\Bs))^2 \det(\Ts(\Cs\Bs)\Ts(\Cs\Bs)^\T) \\
	&=& \det(\Sigmas(\Cs\Bs))^2 \det(\Omegas\Omegas^\T) \\
	&\stackrel{(c)}{\leq}& \det(\Sigmas(\Bs))^2 \det(\Omegas\Omegas^\T).   \IEEEeqnarraynumspace\IEEEyesnumber\label{link_C_with_Lambda}
\end{IEEEeqnarray*}
The inequality $(a)$ is obtained by upper-bounding the positive semidefinite matrix $\Bs\Bs^\T$ by the scaled identity matrix $\lambda_\mathrm{max}(\Bs\Bs^\T) \Is$, where $\lambda_\mathrm{max}(\Bs\Bs^\T) = \max_{\xs \in \mathbb{R}^{L_\Bf}} \xs^\T\Bs\Bs^\T \xs / \lVert \xs \rVert_2^2$ denotes the maximum eigenvalue of $\Bs\Bs^\T$; the equality $(b)$ results from applying the Smith normal decomposition~\eqref{SNF_of_CB} and recalling that $\Ss(\Cs\Bs)$ is unimodular; Step $(c)$, in turn, is an application of the following auxiliary result.

\lemmabox{
\begin{lemma}   \label{lem:invariant_factors}
For any integer matrix $\Bs \in \mathbb{Z}^{\ell \times k}$ with $\ell \leq k$ and full column rank, and for any right-invertible matrix $\Cs \in \mathbb{Z}^{r \times n}$ ($r \leq n$) we have $\det(\Sigmas(\Cs\Bs)) \leq \det(\Sigmas(\Bs))$.
\end{lemma}
}

\proofbox{}{
\begin{IEEEproof}
Consider the Smith normal decomposition $\Cs = \tilde{\Ss}(\Cs) \tilde{\Sigmas}(\Cs) \tilde{\Ts}(\Cs)$ and the reduced Smith normal decomposition $\Bs = \Ss(\Bs) \Sigmas(\Bs) \Ts(\Bs)$. Since $\Cs$ is right-invertible by assumption, we have $\tilde{\Sigmas}(\Cs) = [\Is_{r \times r} \ \mynull_{r \times (n-r)}]$.
Since $\tilde{\Ts}(\Cs)$ and $\tilde{\Ss}(\Cs)$ are unimodular, on the one hand, $\det(\Sigmas(\Bs)) = \det(\tilde{\Ts}(\Cs)\Bs)$ and on the other hand, $\det(\Sigmas(\Cs\Bs)) = \det(\tilde{\Sigmas}(\Cs)\tilde{\Ts}(\Cs)\Bs) = \det([\tilde{\Ts}(\Cs)\Bs]_{[r]})$.
The product $\tilde{\Ts}(\Cs)\Bs \in \mathbb{Z}^{\ell \times k}$ has full column rank, much like $\Bs$.
Therefore, in the statement of Lemma~\ref{lem:invariant_factors} we can assume without loss of generality that $\Cs$ is of the form $[\Is_{r \times r} \ \mynull_{r \times (n-r)}]$ and thus restate the lemma as the inequality
\begin{equation}
	\det(\Sigmas([\Bs]_{[r]})) \leq \det(\Sigmas(\Bs)).
\end{equation}
To prove the latter, in turn, it will suffice to prove that, if $\Bs' \in \mathbb{Z}^{(\ell-1) \times k}$ denotes the matrix $\Bs$ with one row removed, we have
\begin{equation}
	\det(\Sigmas(\Bs')) \leq \det(\Sigmas(\Bs)).
\end{equation}
Without loss of generality, let $\Bs' = [\Bs]_{[2{:}\ell]}$, that is, $\Bs'$ is obtained from removing the first row of $\Bs$. Let us denote the $\ell \times \ell$ submatrices of $\Bs$ as $\Bs_i \in \mathbb{Z}^{\ell \times \ell}$, of which there are $\binom{k}{\ell}$. As a direct consequence of the definition of elementary divisors~\eqref{elementary_divisors}, the determinant $\det(\Sigmas(\Bs))$ is equal to the (positive) greatest common divisor of all $\ell \times \ell$ minors of $\Bs$, i.e.,
\begin{equation}
	\det(\Sigmas(\Bs))
	= \gcd\left( \det(\Bs_1), \det(\Bs_2), \dotsc \right).
\end{equation}
Let us now develop each of these determinants along the elements of the first row of $\Bs_i$, that is,
\begin{equation}
	\det(\Bs_i)
	= \sum_{j=1}^{\ell} (-1)^{j+1} [\Bs_i]_{1,j} \det([\Bs_i]_{\setminus 1, \setminus j})
\end{equation}
where $[\Bs_i]_{\setminus 1, \setminus j}$ denotes the matrix $\Bs_i$ with the first line and $j$-th column removed. Note that these submatrices $[\Bs_i]_{\setminus 1, \setminus j}$ are in fact $(\ell-1) \times (\ell-1)$ submatrices of $\Bs'$. Clearly, by iterating over all $(i,j) \in [\binom{\ell}{k}] \times [\ell-1]$, the submatrices $[\Bs_i]_{\setminus 1, \setminus j}$ will reach all submatrices of $\Bs'$ (each submatrix multiple times). Consequently, the greatest common divisor of the minors of $\Bs'$, which is 
\begin{multline}
	\det(\Sigmas(\Bs'))
	= \gcd\Bigl( \det([\Bs_1]_{\setminus 1, \setminus 1}), \dotsc, \det([\Bs_1]_{\setminus 1, \setminus (\ell-1)}), \det([\Bs_2]_{\setminus 1, \setminus 1}), \dotsc \\
	\dotsc, \det([\Bs_2]_{\setminus 1, \setminus (\ell-1)}), \dotsc \Bigr)
\end{multline}
divides every $\det(\Bs_i)$, and is therefore a divisor of $\det(\Sigmas(\Bs))$. Since $\Bs$ has full row rank, we have $\det(\Sigmas(\Bs)) \neq 0$ and therefore the claim $\det(\Sigmas(\Bs')) \leq \det(\Sigmas(\Bs))$ follows, thereby concluding the proof of Lemma~\ref{lem:invariant_factors}.
\end{IEEEproof}
}

To complement~\eqref{link_C_with_Lambda}, one can obtain inequality in reverse direction as follows:
\begin{IEEEeqnarray*}{rCl}
	\lambda_\mathrm{max}(\Bs\Bs^\T)^{L_\Cf} \det(\Cs\Cs^\T)
	&\geq& \det(\Cs\Bs\Bs^\T\Cs^\T) \\
	&=& \det(\Sigmas(\Cs\Bs))^2 \det(\Omegas\Omegas^\T) \\
	&\geq& \det(\Omegas\Omegas^\T)   \IEEEeqnarraynumspace\IEEEyesnumber\label{link_C_with_Lambda_reverse}
\end{IEEEeqnarray*}
where the last inequality follows from $\Sigmas(\Cs\Bs)$ being an integer non-singular matrix.
Combining~\eqref{sufficient_condition_for_suboptimality_2_1}--\eqref{sufficient_condition_for_suboptimality_2_2} with \eqref{link_C_with_Lambda} and \eqref{link_C_with_Lambda_reverse}, and upon rearranging the terms in~\eqref{sufficient_condition_for_suboptimality_2_1}, we obtain a new pair of sufficient conditions (on $\Cs$ and $\nu$) that together imply~\eqref{sufficient_condition_for_suboptimality_2_1}--\eqref{sufficient_condition_for_suboptimality_2_2}, namely
\begin{subequations}
\begin{IEEEeqnarray}{rCl}   \label{sufficient_condition_for_suboptimality_3}
	\frac{\lambda_\mathrm{min}(\Bs\Bs^\T)^{L_\Cf/2} \sqrt{\det(\Cs\Cs^\T)}}{\det(\Sigmas(\Bs))}
	&>& (4\sqrt{K}\tau)^{K-L_\Cf} e^{ H_0(\Bs,M) - h(\uv|Y) }   \label{sufficient_condition_for_suboptimality_3_1} \\
	\nu\tau
	&\geq& 2^{K-2} \lambda_\mathrm{max}(\Bs\Bs^\T)^{L_\Cf} \det(\Cs\Cs^\T).   \label{sufficient_condition_for_suboptimality_3_2}
\end{IEEEeqnarray}
\end{subequations}
In summary, we have shown that if $\Cs$ is such that $\det(\Cs\Cs^\T)$ is so large as to satisfy~\eqref{sufficient_condition_for_suboptimality_3_1}, then for all $\nu$ large enough to satisfy~\eqref{sufficient_condition_for_suboptimality_3_2}, the matrix $\Cs$ is not a solution to the minimization problem. In other words, matrices $\Cs$ satisfying~\eqref{sufficient_condition_for_suboptimality_3_1} can be excluded from the minimization domain, to such effect that the minimization domain can be reduced to a finite set and thus the order of limit-taking and minimization can be exchanged. Specifically, if we denote the constant
\begin{equation}   \label{def:C_0}
	C_0(\Bs,M)
	\triangleq \frac{(4\sqrt{K}\tau)^{K-L_\Cf} \det(\Sigmas(\Bs))}{\lambda_\mathrm{min}(\Bs\Bs^\T)^{L_\Cf/2}} e^{ H_0(\Bs,M) - h(\uv|Y) }
\end{equation}
then we have [cf.~\eqref{J_tilde_quantized}--\eqref{liminf_J}]
\begin{IEEEeqnarray*}{rCl}   \label{liminf_J_bis}
	\liminf_{\nu \to \infty} \bigl\{ J_\nu(\Bs,M) - L_\Cf \log(\nu) \bigr\}
	&=& \liminf_{\nu \to \infty} \bigl\{ \tilde{J}_\nu(\Bs,M) - L_\Cf \log(\nu) \bigr\} \\
	&=& \liminf_{\nu \to \infty} \Bigl\{ \inf_{\Cs \in \mathscr{C}_\mathbb{Z}(M)} H(\Cs\Bs\floor{\nu\uv}|Y) - L_\Cf \log(\nu) \Bigr\} \\
	&=& \min_{\Cs \in \mathscr{C}_\mathbb{Z}(M) \colon \det(\Cs\Cs^\T) \leq \Theta} \liminf_{\nu \to \infty} \bigl\{ H(\Cs\Bs\floor{\nu\uv}|Y) - L_\Cf \log(\nu) \bigr\} \\
	&=& \min_{\Cs \in \mathscr{C}_\mathbb{Z}(M) \colon \det(\Cs\Cs^\T) \leq \Theta} \Hc_{\Cs\Bs}(\uv|Y)   \IEEEeqnarraynumspace\IEEEyesnumber
\end{IEEEeqnarray*}
for any $\Theta \geq C_0(\Bs,M)$. Taking the limit as $\Theta \to \infty$, we finally end up with
\begin{IEEEeqnarray}{rCl}   \label{liminf_and_infimum_exchanged}
	\liminf_{\nu \to \infty} \bigl\{ J_\nu(\Bs,M) - L_\Cf \log(\nu) \bigr\}
	&=& \inf_{\Cs \in \mathscr{C}_\mathbb{Z}(M)} \Hc_{\Cs\Bs}(\uv|Y).
\end{IEEEeqnarray}

This completes the proof of~\eqref{quod_est_demonstrandum_limit_3} and thus of Theorem~\ref{thm:continuous_CF} for absolutely continuous auxiliaries of bounded support. To remove this limitation, we will now use a ``truncation'' argument, similar to the ``clipping'' argument used in the proof of Theorem~\ref{thm:integer_CF}. Based on an absolutely continuous auxiliary $\uv \in \mathbb{R}^K$ with distribution $P_{\uv}$, let us define the truncated auxiliary $\uv_\tau \in \mathbb{R}^K$ (for any $\tau$ sufficiently large to satisfy $P_{\uv}(\tau\mathscr{S}_\infty^K) > 0$) as

\begin{equation}   \label{truncated_auxiliary}
	P_{\uv_\tau}(V) = \frac{P_{\uv}(V \cap \tau\mathscr{S}_\infty^K)}{P_{\uv}(\tau\mathscr{S}_\infty^K)}
	\quad \text{(for all measurable $V$).}
\end{equation}
In other words, if $f_{\uv}$ and $f_{\uv_\tau}$ denote the probability density functions of $\uv$ and $\uv_\tau$, respectively, then $f_{\uv_\tau}$ is given by
\begin{equation}
	f_{\uv_\tau}(\us) = \frac{f_{\uv}(\us)}{\int_{\tau\mathscr{S}_\infty^K} f_{\uv}(\us) \intd\us} \mathds{1}\bigl\{ \us \in \tau\mathscr{S}_\infty^K \bigr\}.
\end{equation}
As already done in the proof of Theorem~\ref{thm:integer_CF}, let us incorporate the dependency on the auxiliary distribution into the rate region notations, by writing $\mathscr{Q}(\Bs;\uv)$ and $\mathscr{Q}(\Bs,M,\Tc;\uv)$ in lieu of $\mathscr{Q}(\Bs)$ and $\mathscr{Q}(\Bs,M,\Tc)$, respectively. By Theorem~\ref{thm:continuous_CF}, which thus far we have proven to hold for auxiliaries with bounded support, we can assert that $\mathscr{Q}(\Bs;\uv_\tau)$ is achievable. Therefore, the inner set limit $\varliminf_{\tau \to \infty} \mathscr{Q}(\Bs;\uv_\tau)$ is achievable. Upon writing out $\mathscr{Q}(\Bs;\uv_\tau)$ according to its definition~\eqref{def:Q} as
\begin{equation}
	\mathscr{Q}(\Bs;\uv_\tau)
	= \bigcap_{M \in \mathscr{M}^\circ_{\mathbb{Z}}(L_\Bf)} \bigcup_{\Sc \in \mathscr{B}(M^*)} \bigcap_{\Tc \in \mathscr{B}([\Bs]_\Sc)} \mathscr{Q}(\Bs,M,\Tc;\uv_\tau)
\end{equation}
and observing that this expression involves a finite number of set operations that are independent of $\tau$, we can argue that the inner set limit can be exchanged with the other set operations, so as to yield
\begin{equation}   \label{inner_set_limit_of_rate_region_with_truncation}
	\varliminf_{\tau \to \infty} \mathscr{Q}(\Bs;\uv_\tau)
	= \bigcap_{M \in \mathscr{M}^\circ_{\mathbb{Z}}(L_\Bf)} \bigcup_{\Sc \in \mathscr{B}(M^*)} \bigcap_{\Tc \in \mathscr{B}([\Bs]_\Sc)} \varliminf_{\tau \to \infty} \mathscr{Q}(\Bs,M,\Tc;\uv_\tau).
\end{equation}
The set on the right-hand side of the last equation can in turn be expressed as [cf.~\eqref{algebraic_entropy_for_continuous}]
\begin{multline}
	\mathscr{Q}(\Bs,M,\Tc;\uv_\tau)
	= \biggl\{ (R_1,\dotsc,R_K) \in \mathbb{R}_+^K \colon \sum_{k\in\Tc} R_k < \\
		h([\uv_\tau]_\Tc) + \inf_{\Cs \in \mathscr{C}_\mathbb{Z}(M)} h(\Ts(\Cs\Bs)\uv_\tau|Y_\tau) - h(\Ts(\Bs)\uv_\tau|Y_\tau) \biggr\}
\end{multline}
where $Y_\tau$ denotes the channel output induced by the auxiliary $\uv_\tau$. Specifically, for any value $\us$ belonging to the support of $\uv_\tau$, conditioned on $\uv_\tau = \us$ and given a modulation mapping $\xs(\us)$, the variable $Y_\tau$ is distributed as $Y_\tau \sim P_{Y|\xv}(\cdot|\xs(\us))$.

\lemmabox{
\begin{lemma}[Truncation lemma]   \label{lem:truncation}
Let $(\uv,Y) \in \mathbb{R}^K \times \Yc$ and $(\uv_\tau,Y_\tau) \in \mathbb{R}^K \times \Yc$ denote pairs of random variables as defined above, where $\uv \sim P_{\uv}$ is absolutely continuous with finite differential entropy $h(\uv)$ and finite entropy $H(\floor{\uv})$, and where $\uv_\tau \sim P_{\uv_\tau}$ denotes its truncated counterpart as defined in~\eqref{truncated_auxiliary}.
Given some real matrix $\Qs \in \mathbb{R}^{M \times K}$ with full row rank, we have that, as $\tau \to \infty$, the entropy $h(\Qs\uv_\tau|Y_\tau)$ converges to $h(\Qs\uv|Y)$ uniformly in $\Qs$, in the sense that there exists an upper bound on the absolute value of the difference of differential entropies
\begin{equation}
	\bigl| h(\Qs\uv_\tau|Y_\tau) - h(\Qs\uv|Y) \bigr|
	\leq \delta(\tau)
\end{equation}
that does not depend on $\Qs$, and satisfies $\lim_{\tau \to \infty} \delta(\tau) = 0$.
\end{lemma}
}

\proofbox{}{
\begin{IEEEproof}
First note that it will suffice to prove simple convergence, i.e.,
\begin{equation}   \label{truncation_quod_est}
	\lim_{\tau \to \infty} \bigl| h(\Qs\uv_\tau|Y_\tau) - h(\Qs\uv|Y) \bigr|
	= 0.
\end{equation}
In fact, the uniform convergence claimed in Lemma~\ref{lem:truncation} immediately follows from a scale invariance of the entropy difference: for any full row rank $\Qs \in \mathbb{R}^{M \times K}$, consider the factorization $\Qs = \Rs \tilde{\Qs}$ with a full-rank square diagonal matrix $\Rs \in \mathbb{R}^{M \times M}$ and a matrix $\tilde{\Qs} \in \mathbb{C}^{M \times K}$ with rows normalized such that each row has an infinity norm equal to unity. In particular, such a factorization entails $\lVert \tilde{\Qs} \rVert_\infty = 1$. We can now express the difference of differential entropies as
\begin{IEEEeqnarray*}{rCl}
	h(\Qs\uv_\tau|Y_\tau) - h(\Qs\uv|Y)
	&=& h(\tilde{\Qs}\uv_\tau|Y_\tau) + \log\left|\det(\Rs)\right| - h(\tilde{\Qs}\uv|Y) - \log\left|\det(\Rs)\right| \\
	&=& h(\tilde{\Qs}\uv_\tau|Y_\tau) - h(\tilde{\Qs}\uv|Y).   \IEEEeqnarraynumspace\IEEEyesnumber
\end{IEEEeqnarray*}
Hence, for any upper bound $\delta(\tau;\Qs)$ that might depend on $\Qs$, such that
\begin{equation}
	\bigl| h(\Qs\uv_\tau|Y_\tau) - h(\Qs\uv|Y) \bigr|
	\leq \delta(\tau;\Qs)
\end{equation}
and $\lim_{\tau \to \infty} \delta(\tau;\Qs) = 0$ (for every $\Qs$), one can define the uniform upper bound $\delta(\tau) = \max_{\tilde{\Qs}} \delta(\tau;\tilde{\Qs})$, where the maximum is taken over the compact set of full row-rank matrices $\tilde{\Qs} \in \mathbb{R}^{M \times K}$ with rows having unit infinity norm. In the following, we shall thus assume that the rows of $\Qs$ have unit infinity norm, without loss of generality. In particular, this assumption implies $\lVert \Qs \rVert_\infty = 1$.

As a first proof step, we will show convergence of mutual informations, namely,
\begin{equation}
	\lim_{\tau \to \infty} I(\Qs\uv_\tau;Y_\tau)
	= I(\Qs\uv;Y).
\end{equation}
To this end, let us define the binary random variable
\begin{equation}   \label{def:E_tau}
	E_\tau
	= \mathds{1}\{ \lVert \uv \rVert_\infty \leq \tau \}
	= \begin{cases}
		1 & \text{if $\lVert \uv \rVert_\infty \leq \tau$} \\
		0 & \text{if $\lVert \uv \rVert_\infty > \tau$}
	\end{cases}
\end{equation}
as well as the clipping operation
\begin{equation}   \label{def:clipping}
	\langle \xs \rangle_\tau
	= \mathds{1}\{ \lVert \xs \rVert_\infty \leq \tau \} \xs
	= \begin{cases}
		\xs & \text{if $\lVert \xs \rVert_\infty \leq \tau$} \\
		\mynull   & \text{if $\lVert \xs \rVert_\infty > \tau$.}
	\end{cases}
\end{equation}
In order to relate the mutual information term $I(\Qs\uv_\tau;Y_\tau)$ (based on the truncated variable $\uv_\tau$) with the mutual information $I(\Qs\langle \uv \rangle_\tau;Y)$ (based on the clipped variable $\langle \uv \rangle_\tau$), consider the following chain of equalities:
\begin{IEEEeqnarray*}{rCl}
	\IEEEeqnarraymulticol{3}{l}{
		I(\Qs\langle \uv \rangle_\tau; Y)
	} \\ \quad
	&=& I(\Qs\langle \uv \rangle_\tau, E_\tau; Y) - I(E_\tau; Y | \Qs\langle \uv \rangle_\tau) \\
	&\stackrel{(a)}{=}& I(\Qs\langle \uv \rangle_\tau, E_\tau; Y) \\
	&=& I(E_\tau; Y) + I(\Qs\langle \uv \rangle_\tau; Y | E_\tau) \\
	&=& I(E_\tau; Y) + I(\Qs\langle \uv \rangle_\tau; Y | E_\tau = 0) \P\{ E_\tau = 0 \} + I(\Qs\langle \uv \rangle_\tau; Y | E_\tau = 1) \P\{ E_\tau = 1 \} \\
	&\stackrel{(b)}{=}& I(E_\tau; Y) + I(\Qs\uv_\tau; Y_\tau) \P\{ E_\tau = 1 \}.   \IEEEeqnarraynumspace\IEEEyesnumber\label{clipped_in_terms_of_truncated}
\end{IEEEeqnarray*}
The first equality $(a)$ holds because the mutual information $I(E_\tau; Y | \Qs\langle \uv \rangle_\tau)$ is zero. In fact, its upper bound $H(E_\tau | \Qs\langle \uv \rangle_\tau)$ vanishes since $\Qs\langle \uv \rangle_\tau$ determines $E_\tau$ almost surely. This is because $\uv$ is absolutely continuous and $\Qs$ has full row rank, such that, conditional on $\Qs \langle \uv \rangle_\tau = \mynull$ (resp.~$\Qs \langle \uv \rangle_\tau \neq 0$) we have $E_\tau = 1$ (resp.~$E_\tau = 0$) with probability one. The last equality $(b)$ holds because conditioned on $E_\tau = 0$, the variable $\Qs \langle \uv \rangle_\tau$ equals to zero, hence $I(\Qs\langle \uv \rangle_\tau; Y | E_\tau = 0) = 0$, and on the other hand because, conditioned on $E_\tau = 1$, the variable $(\Qs\langle \uv \rangle_\tau; Y)$ has the same distribution as $(\Qs\uv_\tau; Y_\tau)$.

Building upon on the equality~\eqref{clipped_in_terms_of_truncated}, we can now upper-bound the mutual information $I(\Qs\langle \uv \rangle_\tau; Y)$ in terms of $I(\Qs\uv; Y)$, as follows:
\begin{IEEEeqnarray*}{rCl}
	I(\Qs\langle \uv \rangle_\tau; Y)
	&=& I(E_\tau; Y) + I(\Qs\uv_\tau; Y_\tau) \P\{ E_\tau = 1 \} \\
	&=& I(E_\tau; Y) + I(\Qs\uv; Y | E_\tau = 1) \P\{ E_\tau = 1 \} \\
	&\leq& I(E_\tau; Y) + I(\Qs\uv; Y | E_\tau) \\
	&=& I(\Qs\uv; Y) + I(E_\tau; Y | \Qs\uv).   \IEEEeqnarraynumspace\IEEEyesnumber\label{clipped_in_terms_of_plain}
\end{IEEEeqnarray*}
Bearing in mind that $I(E_\tau; Y | \Qs\uv)$ vanishes as $\tau \to \infty$ (due to $\lim_{\tau \to \infty} H(E_\tau) = 0$), and taking the superior limit on both sides of~\eqref{clipped_in_terms_of_plain}, we get
\begin{IEEEeqnarray*}{rCl}
	\limsup_{\tau \to \infty} I(\Qs\uv_\tau; Y_\tau)
	&=& \limsup_{\tau \to \infty} I(\Qs\langle \uv \rangle_\tau; Y) \\
	&\leq& I(\Qs\uv ; Y).   \IEEEeqnarraynumspace\IEEEyesnumber\label{MI_limsup}
\end{IEEEeqnarray*}

Here, the first equality is obtained by using the equality~\eqref{clipped_in_terms_of_truncated} and recalling that $\lim_{\tau \to \infty} \P\{ E_\tau = 1\} = 1$, which entails that $I(E_\tau ; Y)$, which is upper-bounded by the (binary) entropy $H(E_\tau)$, tends to zero as $\tau \to \infty$.

To derive a matching inferior limit, we show that the pair $(\Qs\uv_\tau,Y_\tau)$ converges in distribution to $(\Qs\uv,Y)$. By the data processing inequality and the chain rule of relative entropy, we can upper-bound the Kullback--Leibler divergence between $(\Qs\uv_\tau,Y_\tau)$ and $(\Qs\uv,Y)$ as follows:
\begin{IEEEeqnarray*}{rCl}
	D\bigl( P_{\Qs\uv_\tau,Y_\tau} \big\Vert P_{\Qs\uv,Y} \bigr)
	&\leq& D\bigl( P_{\uv_\tau} P_{Y|\uv} \big\Vert P_{\uv} P_{Y|\uv} \bigr) \\
	&=& D\bigl( P_{\uv_\tau} \big\Vert P_{\uv} \bigr) \\
	&=& \int_{\tau\mathscr{S}_\infty^K} \frac{f_{\uv}(\us)}{\P\bigl\{ \uv \in \tau\mathscr{S}_\infty^K \bigr\}} \log\frac{1}{\P\bigl\{ \uv \in \tau\mathscr{S}_\infty^K \bigr\}} \intd\us \\
	&=& -\log\P\bigl\{ \uv \in \tau\mathscr{S}_\infty^K \bigr\}.   \IEEEeqnarraynumspace\IEEEyesnumber
\end{IEEEeqnarray*}
Taking the limit on both sides as $\tau \to \infty$, we conclude that $D\bigl( P_{\Qs\uv_\tau,Y_\tau} \big\Vert P_{\Qs\uv,Y} \bigr)$ tends to zero, hence $(\Qs\uv_\tau,Y_\tau)$ converges to $(\Qs\uv,Y)$ in Kullback--Leibler divergence, which implies convergence in distribution. Therefore, by the lower semi-continuity of mutual information~\cite[Thm.~1]{Po75}, \cite[Thm.~19]{ErHa14},
\begin{equation}   \label{MI_liminf}
	\liminf_{\tau \to \infty} I(\Qs\uv_\tau ; Y_\tau)
	\geq I(\Qs\uv; Y).
\end{equation}
Combining the superior limit~\eqref{MI_limsup} with the inferior limit~\eqref{MI_liminf}, we obtain the limit
\begin{equation}   
	\lim_{\tau \to \infty} I(\Qs\uv_\tau ; Y_\tau)
	= I(\Qs\uv; Y).
\end{equation}

With this convergence of mutual information established, and since
\begin{equation}
	h(\Qs\uv_\tau|Y_\tau) - h(\Qs\uv|Y)
	= h(\Qs\uv_\tau) - h(\Qs\uv) - I(\Qs\uv_\tau ; Y_\tau) + I(\Qs\uv; Y)
\end{equation}
it follows that a proof of~\eqref{truncation_quod_est} (and thus of Lemma~\ref{lem:truncation}) reduces to proving
\begin{equation}   \label{truncation_quod_est_bis}
	\lim_{\tau \to \infty} \bigl\{ h(\Qs\uv_\tau) - h(\Qs\uv) \bigr\}
	= 0.
\end{equation}
For this purpose, we first show that the function
\begin{equation}
	\tilde{f}_{\tau}(\ws)
	\triangleq f_{\Qs\uv | \lVert \uv \rVert_\infty \leq \tau}(\ws) \P\{ E_\tau = 0 \}
\end{equation}
is monotone (non-decreasing) in $\tau$ and for every $\ws \in \mathbb{R}^M$, we have the limit
\begin{equation}
	\lim_{\tau \to \infty} \tilde{f}_{\tau}(\ws)
	= f_{\Qs\uv}(\ws).
\end{equation}
\begin{lemma}
Consider an absolutely continuous random variable $\uv \in \mathbb{R}^K$ with probability distribution $\mu_{\uv}$. For a set $U \subset \mathbb{R}^K$ with $\mu_{\uv}(U) > 0$, denote as $\uv_U \sim \mu_{U}$ a random variable distributed as $\uv$ restricted on $\uv \in U$, i.e., $\mu_{U}(S) = \mu_{\uv}(S \cap U) / \mu_{\uv}(U)$ for any measurable $S$. Let $g \colon \mathbb{R}^K \to \mathbb{R}^M$ be some deterministic function such that $\wv = g(\uv)$ is absolutely continuous. Then $\wv|\uv \in U$ is absolutely continuous with a density denoted as $f_{\wv|U}$, and for any measurable sets $U \subset U'$ we have
\begin{equation}
	\mu_{\uv}(U) f_{\wv|U}(w)
	\leq \mu_{\uv}(U') f_{\wv|U'}(w).
\end{equation}
\end{lemma}
\begin{IEEEproof}
Consider the variable pair $(\uv,\wv)$, which has a joint distribution $\mu_{\uv,\wv}$. Clearly, for $U \subset U'$ and any $W \subset \mathbb{R}^M$, it holds that
\begin{equation}
	\mu_{\uv,\wv}(U \times W)
	\leq \mu_{\uv,\wv}(U' \times W)
\end{equation}
or equivalently,
\begin{equation}
	\mu_{\uv}(U) \mu_{\wv|\uv}(W|U)
	\leq \mu_{\uv}(U') \mu_{\wv|\uv}(W|U')
\end{equation}
Clearly $\mu_{U} \ll \mu_{\uv} \ll \lambda$ and since $\wv = g(\uv)$ is absolutely continuous by assumption, we also have $\mu_{\wv|\uv}(\cdot|U) \ll \lambda$. The distributions $\mu_{\wv|\uv}(\cdot|U)$ and $\mu_{\wv|\uv}(\cdot|U')$ thus have densities $f_{\wv|U}$ and $f_{\wv|U'}$, respectively, so can write the latter inequality as
\begin{equation}
	\mu_{\uv}(U) \int_W f_{\wv|U}(w) \intd w
	\leq \mu_{\uv}(U') \int_W f_{\wv|U'}(w) \intd w.
\end{equation}
Since this equality holds for every $W \in \Wc$, it follows that
\begin{equation}
	\mu_{\uv}(U) f_{\wv|U}(w)
	\leq \mu_{\uv}(U') f_{\wv|U'}(w)
\end{equation}
which proves the claim.
\end{IEEEproof}

Let us express $h(\Qs\uv_\tau)$ is terms of the function $\tilde{f}_{\tau}(\ws)$:
\begin{IEEEeqnarray*}{rCl}
	h(\Qs\uv_\tau)
	&=& h(\Qs\uv | E_\tau = 0) \\
	&=& -\int_{\mathbb{R}^M} f_{\Qs\uv | E_\tau = 0}(\ws) \log f_{\Qs\uv | E_\tau = 0}(\ws) \intd\ws \\
	&=& -\int_{\mathbb{R}^M} \frac{\tilde{f}_\tau(\ws)}{\P\{E_\tau = 0\}} \log \frac{\tilde{f}_\tau(\ws)}{\P\{E_\tau = 0\}} \intd\ws \\
	&=& \log\P\{E_\tau = 0\} - \frac{1}{\P\{E_\tau = 0\}} \int_{\mathbb{R}^M} \tilde{f}_\tau(\ws) \log \tilde{f}_\tau(\ws) \intd\ws.   \IEEEeqnarraynumspace\IEEEyesnumber\label{differential_entropy_as_f_tilde}
\end{IEEEeqnarray*}
Furthermore, let us define the set
\begin{equation}
	\Rc
	= \Bigl\{ \ws \in \mathbb{R}^M \colon f_{\Qs\uv}(\ws) \geq \tfrac{1}{e} \Bigr\}
\end{equation}
where $e$ denotes the Euler constant. Clearly, the set $\Rc$ is compact because $f_{\Qs\uv}(\ws)$ is a probability density function and as such, it is non-negative and integrates to unity.
The integral in~\eqref{differential_entropy_as_f_tilde} can be split as follows:
\begin{equation}   \label{integral_split}
	\int_{\mathbb{R}^M} \tilde{f}_\tau(\ws) \log \tilde{f}_\tau(\ws) \intd\ws
	= \int_{\Rc} \tilde{f}_\tau(\ws) \log \tilde{f}_\tau(\ws) \intd\ws + \int_{\mathbb{R}^M \setminus \Rc} \tilde{f}_\tau(\ws) \log \tilde{f}_\tau(\ws) \intd\ws.   
\end{equation}
When taking the limit as $\tau \to \infty$ on both sides of~\eqref{integral_split}, one can pull the limit operation inside the first right-hand side integral due to $\Rc$ being compact, and also inside the second integral due to the Monotone Convergence Theorem~\cite[Thm.~1.6.2]{AsDo00}, which is applicable because $\tilde{f}_\tau(\ws)$ is monotone in $\tau$ (for every $\ws$) and $x \mapsto x\log(x)$ is monotonely decreasing on $[0,1/e]$. This yields
\begin{IEEEeqnarray*}{rCl}
	\lim_{\tau \to \infty} \int_{\mathbb{R}^M} \tilde{f}_\tau(\ws) \log \tilde{f}_\tau(\ws) \intd\ws
	&=& \int_{\Rc} f_{\Qs\uv}(\ws) \log f_{\Qs\uv}(\ws) \intd\ws + \int_{\mathbb{R}^M \setminus \Rc} f_{\Qs\uv}(\ws) \log f_{\Qs\uv}(\ws) \intd\ws \\
	&=& h(\Qs\uv).
\end{IEEEeqnarray*}
In combination with~\eqref{differential_entropy_as_f_tilde} and the fact that $\lim_{\tau \to \infty} \P\{E_\tau = 0\} = 1$, we conclude that
\begin{equation}
	\lim_{\tau \to \infty} h(\Qs\uv_\tau)
	= h(\Qs\uv).
\end{equation}

\end{IEEEproof}
}

By Lemma~\ref{lem:truncation}, we have the limits
\begin{subequations}
\begin{IEEEeqnarray}{rCl}
	\lim_{\tau \to \infty} h([\uv_\tau]_\Tc)
	&=& h([\uv]_\Tc)   \label{truncation_limit_1} \\
	\lim_{\tau \to \infty} h(\Ts(\Cs\Bs) \uv_\tau|Y_\tau)
	&=& h(\Ts(\Cs\Bs)\uv|Y)   \label{truncation_limit_2} \\
	\lim_{\tau \to \infty} h(\Ts(\Bs) \uv_\tau|Y_\tau)
	&=& h(\Ts(\Bs)\uv|Y).   \label{truncation_limit_3}
\end{IEEEeqnarray}
\end{subequations}
Regarding~\eqref{truncation_limit_2}, since Lemma~\ref{lem:truncation} further asserts that the convergence is uniform, we can exchange the order of limit-taking and infimization can be exchanged, yielding
\begin{equation}   \label{truncation_limit_2_bis}
	\lim_{\tau \to \infty} \inf_{\Cs \in \mathscr{C}_\mathbb{Z}(M)} h(\Ts(\Cs\Bs)\uv_\tau|Y_\tau)
	= \inf_{\Cs \in \mathscr{C}_\mathbb{Z}(M)} h(\Ts(\Cs\Bs)\uv|Y).
\end{equation}
Furthermore, by Lemma~\ref{lem:continuous_algebraic_entropy} we can revert all entropies back to the algebraic entropy notation:
\begin{subequations}   \label{back_to_algebraic_entropy_notation_bis}
\begin{IEEEeqnarray}{rCl}
	h([\uv]_\Tc)
	&=& \Hc([\uv]_\Tc) \\
	h(\Ts(\Cs\Bs)\uv|Y)
	&=& \Hc_{\Cs\Bs}(\uv|Y) \\
	h(\Ts(\Bs)\uv|Y)
	&=& \Hc_{\Bs}(\uv|Y).
\end{IEEEeqnarray}
\end{subequations}
Combining~\eqref{truncation_limit_1}, \eqref{truncation_limit_3} and \eqref{truncation_limit_2_bis}, and using algebraic entropy notation~\eqref{back_to_algebraic_entropy_notation_bis}, we conclude that
\begin{equation}
	\varliminf_{\tau \to \infty} \mathscr{Q}(\Bs,M,\Tc;\uv_\tau)
	= \mathscr{Q}(\Bs,M,\Tc;\uv)
\end{equation}
and therefore, substituting back into~\eqref{inner_set_limit_of_rate_region_with_truncation}, we finally end up with
\begin{equation}   \label{quod_erat_demonstrandum_truncation}
	\varliminf_{\tau \to \infty} \mathscr{Q}(\Bs;\uv_\tau)
	= \mathscr{Q}(\Bs;\uv).
\end{equation}
This finalizes the proof of Theorem~\ref{thm:continuous_CF}.

}

{
\section{Proof of Lemma~\ref{lem:asym-error}}
\label{app:proof:asym-error}
By~\cite[Lemma~7.18]{Gray1990_A}, we have
\begin{align*}
	I(\floor{\nu\uv} ; \nu\uv-\floor{\nu\uv})
	= \lim_{m\to\infty} I\bigl(\floor{\nu\uv}; \bigfloor{m(\nu\uv-\floor{\nu\uv})}\bigr).
\end{align*}
The mutual information $I\bigl(\floor{\nu\uv}; \bigfloor{m(\nu\uv-\floor{\nu\uv})}\bigr)$ can be upper-bounded as follows:
\begin{IEEEeqnarray*}{rCl}
	\IEEEeqnarraymulticol{3}{l}{
		I\bigl(\floor{\nu\uv}; \bigfloor{m(\nu\uv-\floor{\nu\uv})}\bigr)
	} \\ \quad
	&=& H(\floor{\nu\uv}) + H\bigl(\bigfloor{m\nu\uv - m\floor{\nu\uv}}\bigr) - H\bigl(\bigfloor{m\nu\uv - m\floor{\nu\uv}}, \floor{\nu\uv}\bigr) \\
	&\stackrel{(a)}{=}& H(\floor{\nu\uv}) + H\bigl(\bigfloor{m\nu\uv - m\floor{\nu\uv}}\bigr) - H\bigl(\floor{m\nu\uv}\bigr) \\
	&\stackrel{(b)}{\leq}& H(\floor{\nu\uv}) + n\log(m+1) - H\bigl(\floor{m\nu\uv}\bigr)
\end{IEEEeqnarray*}
\sloppy
where step $(a)$ follows because there is a one-to-one correspondence between $\bigl(\bigfloor{m\nu\uv - m\floor{\nu\uv}}, \floor{\nu\uv}\bigr)$ and $\floor{m\nu\uv}$ and step $(b)$ follows because each of the $n$ coordinates of $\bigfloor{m\nu\uv - m\floor{\nu\uv}}$ takes at most $m+1$ values. Taking the limit as $\nu \to \infty$, we thus get an upper bound
\begin{IEEEeqnarray*}{rCl}
	I(\floor{\nu\uv} ; \nu\uv - \floor{\nu\uv})
	&\leq& H(\floor{\nu\uv}) + \lim_{m\to\infty} \bigl\{ n\log(m+1) - H\bigl(\floor{m\nu\uv}\bigr) \bigr\} \\
	&=& H(\floor{\nu\uv}) - n\log(\nu) - h(\uv).   \IEEEeqnarraynumspace\IEEEyesnumber\label{asym_err_UB}
\end{IEEEeqnarray*}
Finally, taking the limit as $\nu \to \infty$ we obtain the desired result
\begin{IEEEeqnarray*}{rCl}
	\lim_{\nu \to \infty} I\bigl( \floor{\nu\uv} ; \nu\uv - \floor{\nu\uv} \bigr)
	&=& 0.
\end{IEEEeqnarray*}
This concludes the proof.

}

{
\section{Proof of Lemma~\ref{lem:exchange_condition_on_Y}}
\label{app:proof:exchange_condition_on_Y}
The statement of Theorem~\ref{thm:continuous_CF} distinguishes two mutually exclusive assumptions:
\begin{itemize}
	\item   Case~1: Staircase mappings $x_k(u_k)$ with finite images;
	\item   Case~2: Linear mappings $x_k(u_k) = \beta_k u_k$ and a linear additive noise system equation.
\end{itemize}
The proof of this lemma requires distinct treatment for each of these assumptions, and will thus be structured in two parts.

\subsection{Case~1: Staircase mappings with finite images}

Let the $|\Xc_k|$ locations of jump discontinuities of the staircase function $x_k(u_k)$ be denoted as $\upsilon_1^{(k)} < \upsilon_2^{(k)} < \dotsc$ and let
\begin{equation}
	\Upsilon_\nu^{(k)}
	= \left[ \upsilon_1^{(k)} - \tfrac{1}{\nu} ; \upsilon_1^{(k)} + \tfrac{1}{\nu} \right] \cup \left[ \upsilon_2^{(k)} - \tfrac{1}{\nu} ; \upsilon_2^{(k)} + \tfrac{1}{\nu} \right] \cup \dotso
\end{equation}
denote the closed $1/\nu$-radius neighborhood of the set of discontinuity locations. Furthermore, let $\Upsilon_\nu = \Upsilon_\nu^{(1)} \times \dotso \times \Upsilon_\nu^{(K)}$ be the Cartesian product of these neighborhoods. Let us define the binary random variable
\begin{equation}
	E_\nu
	=
	\begin{cases}
		0   &   \text{if $\uv \notin \Upsilon_\nu$} \\
		1   &   \text{if $\uv \in \Upsilon_\nu$}.
	\end{cases}
\end{equation}
Note that $\lim_{\nu \to \infty} \P\{E_\nu = 0\} = 1$ as a consequence of $\uv$ being absolutely continuous and $\Upsilon_\nu$ having vanishing Lebesgue measure as $\nu \to \infty$. Moreover, note that the event $E_\nu = 0$ implies that $x(\floor{\uv}_\nu) = x(\uv)$. As a consequence, we can choose to couple the channel output variables $Y$ and $Y_\nu$ in a way that, conditioned on $E_\nu = 0$, we have $Y = Y_\nu$. Since mutual informations $I(E_\nu ; Y_\nu)$, $I(E_\nu ; Y_\nu | \Qs\floor{\uv}_\nu)$, $I(E_\nu ; Y)$ and $I(E_\nu ; Y | \Qs\floor{\uv}_\nu)$ are all upper-bounded by $H(E_\nu)$,
we infer that
\begin{subequations}   \label{binary_entropy_squeeze}
\begin{IEEEeqnarray}{rCl}
	\bigl| I(E_\nu ; Y_\nu) - I(E_\nu ; Y_\nu | \Qs\floor{\uv}_\nu) \bigr| &\leq& H(E_\nu), \\
	\bigl| I(E_\nu ; Y) - I(E_\nu ; Y | \Qs\floor{\uv}_\nu) \bigr| &\leq& H(E_\nu)
\end{IEEEeqnarray}
\end{subequations}
whose left-hand sides tend to zero as $\nu \to \infty$ because of $\lim_{\nu \to \infty} H(E_\nu) = 0$. By the chain rule for mutual information, we have
\begin{subequations}   \label{MI_chain_rule}
\begin{IEEEeqnarray}{rCl}
	I(E_\nu ; Y_\nu) - I(E_\nu ; Y_\nu | \Qs\floor{\uv}_\nu)
	&=& I(\Qs\floor{\uv}_\nu ; Y_\nu) - I(\Qs\floor{\uv}_\nu ; Y_\nu | E_\nu), \\
	I(E_\nu ; Y) - I(E_\nu ; Y | \Qs\floor{\uv}_\nu)
	&=& I(\Qs\floor{\uv}_\nu ; Y) - I(\Qs\floor{\uv}_\nu ; Y | E_\nu).
\end{IEEEeqnarray}
\end{subequations}
Combining~\eqref{binary_entropy_squeeze} and~\eqref{MI_chain_rule}, we infer that
\begin{subequations}   \label{MI_vanishing_limits_1}
\begin{IEEEeqnarray}{rCl}
	\bigl| I(\Qs\floor{\uv}_\nu ; Y_\nu) - I(\Qs\floor{\uv}_\nu ; Y_\nu | E_\nu) \bigr|
	&\leq& H(E_\nu), \\
	\bigl| I(\Qs\floor{\uv}_\nu ; Y) - I(\Qs\floor{\uv}_\nu ; Y | E_\nu) \bigr|
	&\leq& H(E_\nu).
\end{IEEEeqnarray}
\end{subequations}
Let us now focus on the two conditional mutual information terms appearing in~\eqref{MI_vanishing_limits_1}, which can be written out respectively as
\begin{subequations}   \label{conditional_MIs_written_out}
\begin{IEEEeqnarray}{rCl}
	I(\Qs\floor{\uv}_\nu ; Y_\nu | E_\nu)
	&=& 
	I(\Qs\floor{\uv}_\nu ; Y_\nu | E_\nu = 0) \P\{E_\nu = 0\}   \IEEEnonumber\\
	&& {} + I(\Qs\floor{\uv}_\nu ; Y_\nu | E_\nu = 1) \P\{E_\nu = 1\},  \IEEEeqnarraynumspace\IEEEyesnumber\label{conditional_MIs_written_out_1} \\
	I(\Qs\floor{\uv}_\nu ; Y | E_\nu)
	&=& 
	I(\Qs\floor{\uv}_\nu ; Y | E_\nu = 0) \P\{E_\nu = 0\}   \IEEEnonumber\\
	&& {} + I(\Qs\floor{\uv}_\nu ; Y | E_\nu = 1) \P\{E_\nu = 1\}.   \IEEEeqnarraynumspace\IEEEyesnumber\label{conditional_MIs_written_out_2}
\end{IEEEeqnarray}
\end{subequations}
Due to the coupling between $Y$ and $Y_\nu$ that we have chosen, which imposes that $Y=Y_\nu$ whenever $E_\nu = 0$, it holds that $I(\Qs\floor{\uv}_\nu ; Y_\nu | E_\nu = 0) = I(\Qs\floor{\uv}_\nu ; Y | E_\nu = 0)$. Subtracting~\eqref{conditional_MIs_written_out_1} from~\eqref{conditional_MIs_written_out_2} and taking the absolute value, we obtain the equality
\begin{multline}
	\Bigl| I(\Qs\floor{\uv}_\nu ; Y | E_\nu) - I(\Qs\floor{\uv}_\nu ; Y_\nu | E_\nu) \Bigr| \\
	= \Bigl| I(\Qs\floor{\uv}_\nu ; Y | E_\nu = 1) - I(\Qs\floor{\uv}_\nu ; Y_\nu | E_\nu = 1) \Bigr| \cdot \P\{E_\nu = 1\}.   \label{MI_squeeze_final}
\end{multline}
As to the mutual information terms appearing on the right-hand side of~\eqref{MI_squeeze_final}, they can be bounded as follows:
\begin{subequations}
\begin{IEEEeqnarray*}{rCl}
	I(\Qs\floor{\uv}_\nu ; Y | E_\nu = 1)
	&=& I(\Qs\floor{\uv}_\nu ; Y | \uv \in \Upsilon_\nu) \\
	&\leq& I(\uv ; Y | \uv \in \Upsilon_\nu) \\
	&\leq& \sup_{P_U} I(\uv;Y) \\
	&=& \sup_{P_X} I(X;Y) \\
	&\leq& \log|\Xc|   \IEEEyesnumber\\
	I(\Qs\floor{\uv}_\nu ; Y_\nu | E_\nu = 1)
	&=& I(\Qs\floor{\uv}_\nu ; Y_\nu | \uv \in \Upsilon_\nu) \\
	&\leq& I(\floor{\uv}_\nu ; Y_\nu | \uv \in \Upsilon_\nu) \\
	&\leq& \sup_{P_U} I(\uv;Y) \\
	&=& \sup_{P_X} I(X;Y) \\
	&\leq& \log|\Xc|   \IEEEyesnumber
\end{IEEEeqnarray*}
\end{subequations}
where $|\Xc| = \prod_{k=1}^K |\Xc_k|$. It follows that the left-hand side of~\eqref{MI_squeeze_final} is upper-bounded by $\log|\Xc| \cdot \P\{E_\nu = 1\}$, i.e.,
\begin{equation}   \label{MI_vanishing_limits_2}
	\bigl| I(\Qs\floor{\uv}_\nu ; Y | E_\nu) - I(\Qs\floor{\uv}_\nu ; Y_\nu | E_\nu) \bigr|
	\leq \log|\Xc| \cdot \P\{E_\nu = 1\}.
\end{equation}
Combining~\eqref{MI_vanishing_limits_1} and \eqref{MI_vanishing_limits_2}, we get
\begin{IEEEeqnarray*}{rCl}
	\bigl| H(\Qs\floor{\uv}_\nu | Y) - H(\Qs\floor{\uv}_\nu | Y_\nu) \bigr|
	&=& \bigl| I(\Qs\floor{\uv}_\nu ; Y) - I(\Qs\floor{\uv}_\nu ; Y_\nu) \bigr| \\
	&\leq& \log|\Xc| \cdot \P\{E_\nu = 1\} + 2 H(E_\nu).
\end{IEEEeqnarray*}
Since this upper bound does not depend on $\Qs$ and vanishes as $\nu \to \infty$, this concludes the proof for Case~1.

\subsection{Case~2: Linear mappings and linear additive noise channel}

Now assume that we fix a linear mapping $x_k(u_k) = \beta_k u_k$. Let us define, conditioned on values $(U_1,\dotsc,U_K) = (u_1,\dotsc,u_K)$ (in short $\uv=u$) of the auxiliary variables, the outputs $Y$ and $Y_\nu$ as being
\begin{subequations}
\begin{IEEEeqnarray}{rCl}
	Y
	&=& \sum_{k=1}^K h_k \beta_k u_k + Z \\
	Y_\nu
	&=& \sum_{k=1}^K h_k \beta_k \floor{u_k}_\nu + Z
\end{IEEEeqnarray}
\end{subequations}
respectively, with real-valued weights $\beta_k$, where $Z$ is an independent (not necessarily Gaussian) additive noise. We choose the coupling of $Y$ and $Y_\nu$ to be such that they are generated by the same realization of $Z$. Specifically, this means that, conditioned on $\uv = u$, their difference is given by
\begin{equation}   \label{coupling}
	Y - Y_\nu
	= \sum_{k=1}^K h_k \beta_k (u_k - \floor{u_k}_\nu)
\end{equation}
and is thus a linear function of the quantization error vector
\begin{equation}
	\dv_\nu
	= \uv - \floor{\uv}_\nu.
\end{equation}
We first bound the entropy difference from above as follows:
\begin{IEEEeqnarray*}{rCl}
	H\bigl( \Qs \floor{\uv}_\nu \big| Y_\nu \bigr) - H\bigl( \Qs \floor{\uv}_\nu \big| Y \bigr)
	&=& I\bigl( \Qs \floor{\uv}_\nu ; Y \bigr) - I\bigl( \Qs \floor{\uv}_\nu ; Y_\nu \bigr) \\
	&\leq& I\bigl( \Qs \floor{\uv}_\nu ; Y , Y_\nu, \dv_\nu \bigr) - I\bigl( \Qs \floor{\uv}_\nu ; Y_\nu \bigr) \\
	&=& I\bigl( \Qs \floor{\uv}_\nu ; \dv_\nu \big| Y_\nu \bigr) \\
	&\leq& I\bigl( \floor{\uv}_\nu ; \dv_\nu \big| Y_\nu \bigr) \\
	&\leq& I( \floor{\uv}_\nu ; \dv_\nu ).   \IEEEeqnarraynumspace\IEEEyesnumber\label{H_diff_UB}
\end{IEEEeqnarray*}
The second equality is due to $Y$ being a deterministic function of $(Y_\nu,\dv_\nu)$; the second inequality is the data processing inequality; the third and last inequality holds because $\dv_\nu \rightarrow \uv \rightarrow Y_\nu$ forms a Markov chain.

Next, we bound the same entropy difference from below:
\begin{IEEEeqnarray*}{rCl}
	H\bigl( \Qs \floor{\uv}_\nu \big| Y_\nu \bigr) - H\bigl( \Qs \floor{\uv}_\nu \big| Y \bigr)
	&=& I\bigl( \Qs \floor{\uv}_\nu ; Y \bigr) - I\bigl( \Qs \floor{\uv}_\nu ; Y_\nu \bigr) \\
	&\geq& I\bigl( \Qs \floor{\uv}_\nu ; Y \bigr) - I\bigl( \Qs \floor{\uv}_\nu ; \dv_\nu , Y \bigr) \\
	&=& -I\bigl( \Qs \floor{\uv}_\nu ; \dv_\nu \bigl| Y \bigr) \\
	&\geq& -I\bigl( \floor{\uv}_\nu ; \dv_\nu \bigl| Y \bigr).   \IEEEeqnarraynumspace\IEEEyesnumber\label{H_diff_LB}
\end{IEEEeqnarray*}
Here, the first inequality is due to $Y_\nu$ being a deterministic function of $(\dv_\nu,Y)$; the second inequality is the data processing inequality.

The upper bound~\eqref{H_diff_UB} tends to zero as $\nu \to \infty$ as a consequence of Lemma~\ref{lem:asym-error}. As for the lower bound~\eqref{H_diff_LB}, we have
\begin{equation}
	\lim_{\nu \to \infty} I\bigl( \floor{\uv}_\nu ; \dv_\nu \bigl| Y \bigr)
	= \int_\Yc \lim_{\nu \to \infty} I\bigl( \floor{\uv}_\nu ; \dv_\nu \bigl| Y = y \bigr) P_Y(\mathrm{d}y)
	= 0
\end{equation}
by the Dominated Convergence Theorem~\cite[(1.6.9),~p.~50]{AsDo00}. The latter can be applied because $I\bigl( \floor{\uv}_\nu ; \dv_\nu \bigl| Y = y \bigr)$ has an integrable upper bound. To derive such a bound, reproducing the steps in the proof of Lemma~\ref{lem:asym-error} up to~\eqref{asym_err_UB} first leads to
\begin{IEEEeqnarray*}{rCl}
	I\bigl( \floor{\uv}_\nu ; \dv_\nu \big| Y=y \bigr)
	&\leq& H\bigl( \floor{\uv}_\nu \big| Y=y \bigr) - K \log(\nu) - h(\uv|Y=y).   \IEEEeqnarraynumspace\IEEEyesnumber\label{MI_UB}
\end{IEEEeqnarray*}
Then, the discrete entropy on the right-hand side of~\eqref{MI_UB} can be upper-bounded as follows: let $k$ and $k'$ denote $K$-dimensional integer vectors and define
\begin{equation}
	p_{k,k'}^\nu
	= \P\bigl\{ k + \tfrac{k'}{\nu} \leq \uv < k + \tfrac{k'+1}{\nu} \big| Y=y \bigr\}
\end{equation}
where inequalities on vectors are to be interpreted entrywise. Note that
\begin{equation}
	\sum_{k' \in [0:\nu-1]^K} p_{k,k'}^\nu
	= \P\bigl\{ k \leq \uv < k + 1 \big| Y=y \bigr\}.
\end{equation}
By the log-sum inequality,
\begin{IEEEeqnarray*}{rCl}
	H\bigl( \floor{\uv}_\nu \big| Y=y \bigr)
	&=& \sum_{k \in \mathbb{Z}^K} \sum_{k' \in [0:\nu-1]^K} p_{k,k'}^\nu \log \frac{1}{p_{k,k'}^\nu} \\
	&\leq& \sum_{k \in \mathbb{Z}^K} \left( \textstyle\sum_{k' \in [0:\nu-1]^K} p_{k,k'}^\nu \right) \log \left( \frac{\nu^K}{\textstyle\sum_{k' \in [0:\nu-1]^K} p_{k,k'}^\nu} \right) \\
	&=& K\log(\nu) + H\bigl( \floor{\uv} \big| Y=y \bigr).   \IEEEeqnarraynumspace\IEEEyesnumber
\end{IEEEeqnarray*}
Combining the latter with~\eqref{MI_UB}, we obtain
\begin{equation}
	I\bigl( \floor{\uv}_\nu ; \dv_\nu \big| Y=y \bigr)
	\leq H\bigl( \floor{\uv} \big| Y=y \bigr) - h(\uv|Y=y).
\end{equation}
This bound is integrable due to $I(\uv;Y)$, $H(\floor{\uv})$ and $h(\uv)$ being finite.

In summary, the upper and lower bound~\eqref{H_diff_UB} and \eqref{H_diff_LB} on the entropy difference $H\bigl( \Qs \floor{\uv}_\nu \big| Y_\nu \bigr) - H\bigl( \Qs \floor{\uv}_\nu \big| Y \bigr)$ both tend to zero. Since these bounds do not depend on $\Qs$, this convergence is uniform. This concludes the proof for Case~2.

}

{
\section{Proof of Theorem~\ref{thm:integer_CF_sequential}}   \label{app:proof:integer_CF_sequential}
This proof follows a similar logic as the proof of Theorem~\ref{thm:integer_CF} (although the absence of the minimum-entropy term $J(\Bs,M)$ drastically simplifies the proof).

We set $\mathbb{U} = \mathbb{Z}$ and fix the pmf $p_{\uv}(\us) = \prod_{k=1}^K p_{U_k}(u_k)$, the modulation mappings $\xs(\us)$ as well as the matrix $\Bs_{[j]} \in \mathbb{Z}^{j \times K}$. Let $Y_\q$ denote the channel output induced by the (finite-field) auxiliary $\phi_\q(\uv)$, in the context of Theorem~\ref{thm:discrete_CF_sequential}. We prove Theorem~\ref{thm:integer_CF_sequential} in two steps: first, assuming that $\uv$ is finitely supported, and then, we relax this finiteness assumption.

If the support of $\uv$ is contained in the finite cube $\llbracket 2\tau + 1 \rrbracket^K$ for some $\tau > 0$, then for $\q > 2\tau + 1$ we have $H(\phi_\q([\uv]_\Tc))) = H([\uv]_\Tc)$. Furthermore, for $\q \geq (2\tau+1) \lVert \Bs_{[j]} \rVert_\infty$, there is a one-to-one correspondence between $\phi_\q(\Bs_{[j]}\uv)$ and $\Bs_{[j]}\uv$ such that all instances of $Y_\q$ can be replaced by $Y$, hence $H(\phi_\q(\Bs_{[j]}\uv)|Y_\q) = H(\Bs_{[j]}\uv|Y) = \Hc_{\Bs_{[j]}}(\uv|Y)$.
Consequently, the right-hand side of~\eqref{sequential_decoding_rate_inequality}, as evaluated in the context of Theorem~\ref{thm:discrete_CF_sequential}, becomes equal to
\begin{multline}
	H(\phi_\q(U_k)) + H(\phi_\q(\Bs_{[j-1]}\uv)|Y_\q) - H(\phi_\q(\Bs_{[j]}\uv)|Y_\q) \\
	= H(U_k) + H(\Bs_{[j-1]}\uv|Y) - H(\Bs_{[j]}\uv|Y)
	\quad \text{(for $\q \geq (2\tau+1) \lVert \Bs \rVert_\infty$).}
\end{multline}
This proves Theorem~\ref{thm:integer_CF_sequential} for finitely supported auxiliaries. To lift this restriction, now consider an auxiliary $\uv \in \mathbb{Z}^K$ with arbitrary support but finite entropy $H(\uv)$. Using the clipping operation $\langle \cdot \rangle_\tau$ as defined in~\eqref{def:clipping} and denoting as $Y_\tau$ the channel output induced by the clipped auxiliary $\langle \uv \rangle_\tau$, we have the limits
\begin{subequations}
\begin{IEEEeqnarray}{rCl}
	\lim_{\tau \to \infty} H(\langle U_k \rangle_\tau)
	&=& H(U_k)
	= \Hc(U_k) \\
	\lim_{\tau \to \infty} H(\Bs_{[j-1]}\langle \uv \rangle_\tau|Y_\tau)
	&=& H(\Bs_{[j-1]}\uv|Y)
	= \Hc_{\Bs_{[j-1]}}(\uv|Y) \\
	\lim_{\tau \to \infty} H(\Bs_{[j]}\langle \uv \rangle_\tau|Y_\tau)
	&=& H(\Bs_{[j]}\uv|Y)
	= \Hc_{\Bs_{[j]}}(\uv|Y).
\end{IEEEeqnarray}
\end{subequations}
Here, the last two limits follow from Lemma~\ref{lem:clipping}. This concludes the proof of Theorem~\ref{thm:integer_CF_sequential}.

}

{
\section{Proof of Theorem~\ref{thm:continuous_CF_sequential}}   \label{app:proof:continuous_CF_sequential}
We set $(\mathbb{U},\mathbb{A}) = (\mathbb{R},\mathbb{Z})$ and fix the pdf $f_{\uv}(\us) = \prod_{k=1}^K f_{U_k}(u_k)$, the modulation mappings $\xs(\us)$ as well as the matrix $\Bs_{[j]} \in \mathbb{Z}^{j \times K}$. Let $Y_\nu$ denote the channel output induced by the integer-valued auxiliary $\lfloor \nu\uv \rfloor$, in the context of Theorem~\ref{thm:integer_CF_sequential}.

By Lemma~\ref{lem:exchange_condition_on_Y}, we have
\begin{subequations}
\begin{IEEEeqnarray}{rCl}
	\lim_{\nu \to \infty} \Bigl\{ H(\Bs_{[j-1]}\lfloor \nu\uv \rfloor|Y_\nu) - H(\Bs_{[j-1]}\lfloor \nu\uv \rfloor|Y) \Bigr\}
	&=& 0 \\
	\lim_{\nu \to \infty} \Bigl\{ H(\Bs_{[j]}\lfloor \nu\uv \rfloor|Y_\nu) - H(\Bs_{[j]}\lfloor \nu\uv \rfloor|Y) \Bigr\}
	&=& 0.
\end{IEEEeqnarray}
\end{subequations}
Hence, $Y_\nu$ can be swapped out for $Y$ in these limits, so we obtain
\begin{subequations}
\begin{IEEEeqnarray}{rCl}
	\lim_{\nu \to \infty} H(\lfloor \nu U_k \rfloor)
	&=& h(U_k)
	= \Hc(U_k) \\
	\lim_{\nu \to \infty} H(\Bs_{[j-1]}\lfloor \nu\uv \rfloor | Y_\tau)
	&=& h(\Ts(\Bs_{[j-1]})\uv|Y)
	= \Hc_{\Bs_{[j-1]}}(\uv|Y) \\
	\lim_{\nu \to \infty} H(\Bs_{[j]}\lfloor \nu\uv \rfloor | Y_\tau)
	&=& h(\Ts(\Bs_{[j]})\uv|Y)
	= \Hc_{\Bs_{[j]}}(\uv|Y).
\end{IEEEeqnarray}
\end{subequations}
This concludes the proof of Theorem~\ref{thm:continuous_CF_sequential}.

}

{
\section{Evaluation of $\mathscr{R}(\As)$ for $(K,L)=(2,2)$}
\label{app:proof:two_user_two_equation}
We evaluate $\mathscr{R}(\As) = \bigcup_{\Bs} \mathscr{Q}(\Bs)$ for the special case $K = L = 2$.
Let us denote $\Bs = \begin{bsmallmatrix} b_{2,1} & b_{1,2} \\ b_{2,1} & b_{2,2} \end{bsmallmatrix} = \begin{bsmallmatrix} \bs_1^\T \\ \bs_2^\T \end{bsmallmatrix}$. We distinguish two cases based on the rank of the matroid $M$.

\paragraph{The rank-$0$ matroid $M = (\{1,2\},\emptyset)$} In this case, the index sets $\Sc$ and $\Tc$ are equal to $\Sc = \Tc = \{1,2\}$, hence we obtain the sum-rate bound
	\begin{IEEEeqnarray*}{rCl}
		R_1 + R_2
		&<& \Hc(\uv) - \Hc_\Bs(\uv|Y) \\
		&=& \Hc(\uv) - \Hc(\uv|Y) \\
		&=& I(\uv;Y) \\
		&=& I(\xv;Y).   \IEEEeqnarraynumspace\IEEEyesnumber\label{LMAC_sum_rate_bound}
	\end{IEEEeqnarray*}

\paragraph{The rank-$1$ matroids}

Since $\Bs$ has full rank, either the diagonal entries $(b_{1,1},b_{2,2})$ or the off-diagonal entries $(b_{1,2},b_{2,1})$ are both non-zero, for otherwise $\Bs$ would have an all-zero row or column. Since we can freely permute the rows of $\Bs$ without affecting the value of $\mathscr{Q}(\Bs)$, we shall assume without loss of generality that the diagonal entries $(b_{1,1},b_{2,2})$ are non-zero. In the following, we make case distinctions based on the off-diagonals:

\begin{itemize}
	\item	The off-diagonals are $b_{1,2} = b_{2,1} = 0$:

	\begin{itemize}
		\item   Case $M = (\{1,2\},\{1\})$: the index sets are equal to $\Sc = \Tc = \{2\}$, hence
				\begin{IEEEeqnarray*}{rCl}
					R_2
					&<& \Hc(U_2) - \Hc(\uv|Y) + \Hc(U_1|Y) \\
					&=& I(U_2;Y,U_1).   \IEEEeqnarraynumspace\IEEEyesnumber\label{LMAC_rate_bound_1_0}
				\end{IEEEeqnarray*}
		\item   Case $M = (\{1,2\},\{2\})$: the index sets are equal to $\Sc = \Tc = \{1\}$, hence
				\begin{IEEEeqnarray*}{rCl}
					R_1
					&<& \Hc(U_1) - \Hc(\uv|Y) + \Hc(U_2|Y) \\
					&=& I(U_1;Y,U_2).   \IEEEeqnarraynumspace\IEEEyesnumber\label{LMAC_rate_bound_2_0}
				\end{IEEEeqnarray*}
		\item   Case $M = (\{1,2\},\{1\},\{2\})$: the index set $\Sc$ is iterated over $\Sc \in \{\{1\},\{2\}\}$:
			\begin{itemize}
				\item	For $\Sc = \{1\}$, the index set $\Tc$ is equal to $\Tc = \{1\}$, hence
					\begin{IEEEeqnarray}{rCl}
						R_1
						&<& \Hc(U_1) - \Hc(\uv|Y) + \inf_{c_1 \neq 0,\, c_2 \neq 0} \Hc_{[ c_1 b_{1,1} \ c_2 b_{2,2} ]}(\uv|Y).   \IEEEeqnarraynumspace\label{LMAC_rate_bound_3_0}
					\end{IEEEeqnarray}
				\item	For $\Sc = \{2\}$, the index set $\Tc$ is equal to $\Tc = \{2\}$, hence
					\begin{IEEEeqnarray}{rCl}
						R_2
						&<& \Hc(U_2) - \Hc(\uv|Y) + \inf_{c_1 \neq 0,\, c_2 \neq 0} \Hc_{[ c_1 b_{1,1} \ c_2 b_{2,2} ]}(\uv|Y).   \IEEEeqnarraynumspace\label{LMAC_rate_bound_4_0}
					\end{IEEEeqnarray}
			\end{itemize}
			
			Since algebraic entropy is invariant against scaling of the coefficient matrix, in the infimum term in~\eqref{LMAC_rate_bound_3_0}--\eqref{LMAC_rate_bound_4_0}, the coefficients $(b_{1,1},b_{2,2})$ can be set to $(1,1)$ without loss of generality. In other words, we can set $\Bs$ to the identity matrix.
			The inequalities~\eqref{LMAC_rate_bound_3_0}--\eqref{LMAC_rate_bound_4_0} are to be combined by a logical `or' (due to union-taking over $\Sc$), and the result is to be combined with~\eqref{LMAC_rate_bound_1_0} and \eqref{LMAC_rate_bound_2_0} by a logical `and' operation (due to the intersection over matroids), hence we obtain, all in all,
			\begin{subequations}
			\begin{IEEEeqnarray}{rCl}
				R_1
				&<& I(U_1;Y,U_2)   \IEEEeqnarraynumspace\label{LMAC_rate_bound_5_0} \\
				R_2
				&<& I(U_2;Y,U_1)   \IEEEeqnarraynumspace\label{LMAC_rate_bound_6_0} \\
				\min\{ R_1 - \Hc(U_1) , R_2 - \Hc(U_2) \}
				&<& - \Hc(\uv|Y) + \inf_{c_1 \neq 0,\, c_2 \neq 0} \Hc_{[ c_1 \ c_2 ]}(\uv|Y).   \IEEEeqnarraynumspace\label{LMAC_rate_bound_7_0}
 			\end{IEEEeqnarray}
			\end{subequations}
	\end{itemize}

	\item	The off-diagonals are $b_{1,2} \neq 0$ and $b_{2,1} = 0$:

	\begin{itemize}
		\item   Case $M = (\{1,2\},\{1\})$: the index sets are equal to $\Sc = \Tc = \{2\}$, hence we obtain the rate bound
				\begin{IEEEeqnarray*}{rCl}
					R_2 &<& \Hc(U_2) - \Hc(\uv|Y) + \Hc_{\bs_1^\T}(\uv|Y).   \IEEEeqnarraynumspace\IEEEyesnumber\label{LMAC_rate_bound_1}
				\end{IEEEeqnarray*}
		\item   Case $M = (\{1,2\},\{2\})$: The index set $\Sc$ is equal to $\Sc = \{1\}$, and $\Tc$ is iterated over $\Tc \in \{\{1\},\{2\}\}$, hence we obtain the pair of rate bounds
				\begin{subequations}
				\begin{IEEEeqnarray*}{rCl}
					R_1
					&<& \Hc(U_1) - \Hc(\uv|Y) + \Hc(U_2|Y)
					= I(U_1;Y,U_2)    \IEEEeqnarraynumspace\IEEEyesnumber\label{LMAC_rate_bound_2} \\
					R_2
					&<& \Hc(U_2) - \Hc(\uv|Y) + \Hc(U_2|Y).   \IEEEeqnarraynumspace\IEEEyesnumber\label{LMAC_rate_bound_3}
				\end{IEEEeqnarray*}
				\end{subequations}
				where the equality in~\eqref{LMAC_rate_bound_2} follows from Lemmata~\ref{lem:chain_rule} and \ref{lem:mutual_information}.
		\item   Case $M = (\{1,2\},\{1\},\{2\})$: the index set $\Sc$ is iterated over $\Sc \in \{\{1\},\{2\}\}$. We make further case distinctions:
			\begin{itemize}
				\item	For $\Sc = \{1\}$, the index set $\Tc$ is iterated over $\Tc \in \{\{1\},\{2\}\}$, hence we obtain the pair of inequalities
					\begin{subequations}
					\begin{IEEEeqnarray}{rCl}
						R_1
						&<& \Hc(U_1) - \Hc(\uv|Y) + \inf_{c_1 \neq 0,\, c_2 \neq 0} \Hc_{[c_1 b_{1,1} \ c_1 b_{1,2} + c_2 b_{2,2}]}(\uv|Y)   \IEEEeqnarraynumspace\label{LMAC_rate_bound_4} \\
						R_2
						&<& \Hc(U_2) - \Hc(\uv|Y) + \inf_{c_1 \neq 0,\, c_2 \neq 0} \Hc_{[c_1 b_{1,1} \ c_1 b_{1,2} + c_2 b_{2,2}]}(\uv|Y).   \IEEEeqnarraynumspace\label{LMAC_rate_bound_5}
					\end{IEEEeqnarray}
					\end{subequations}
				\item	For $\Sc = \{2\}$, the index set $\Tc$ is equal to $\Tc = \{2\}$, hence we obtain once more the inequality~\eqref{LMAC_rate_bound_5}. The pair of inequalities~\eqref{LMAC_rate_bound_4}--\eqref{LMAC_rate_bound_5} is thus combined by a logical `or' with~\eqref{LMAC_rate_bound_5}, hence what remains from this combination is only~\eqref{LMAC_rate_bound_5}.
			\end{itemize}
	\end{itemize}
The above three case distinctions on the matroid $M$ are to be combined by a logical `and' (due to intersections over matroids), thus leading to a set of four rate inequalities~\eqref{LMAC_rate_bound_1}, \eqref{LMAC_rate_bound_2}--\eqref{LMAC_rate_bound_3} and \eqref{LMAC_rate_bound_5} (since~\eqref{LMAC_rate_bound_4} was removed by union-taking over $\Sc$) that need to be simultaneously satisfied. Upon combining these four inequalities, we can reduce them to just two:
\begin{subequations}
\begin{IEEEeqnarray*}{rCl}
	R_1
	&<& I(U_1;Y,U_2)   \IEEEeqnarraynumspace\IEEEyesnumber\label{LMAC_rate_bound_4_simplified} \\
	R_2
	&<& \Hc(U_2) - \Hc(\uv|Y) + \inf_{(c_1,c_2) \neq (0,0)} \Hc_{[c_1 \ c_2]}(\uv|Y).   \IEEEeqnarraynumspace\IEEEyesnumber\label{LMAC_rate_bound_5_simplified}
\end{IEEEeqnarray*}
\end{subequations}
To see why~\eqref{LMAC_rate_bound_1}, \eqref{LMAC_rate_bound_3} and \eqref{LMAC_rate_bound_5} can be combined to~\eqref{LMAC_rate_bound_5_simplified}, observe that the infimum term in~\eqref{LMAC_rate_bound_5} can be simply expressed as $\inf_{(c_1,c_2) \colon c_1 \neq 0} \Hc_{[c_1 \ c_2]}(\uv|Y)$.

	\item	The off-diagonals are $b_{1,2} = 0$ and $b_{2,1} \neq 0$:

The derivations in this case are analogous to the previous case of $b_{1,2} \neq 0$ and $b_{2,1} = 0$, resulting in rate inequalities like~\eqref{LMAC_rate_bound_4_simplified}--\eqref{LMAC_rate_bound_5_simplified}, but with user indices swapped. That is,
\begin{subequations}
\begin{IEEEeqnarray*}{rCl}
	R_1
	&<& \Hc(U_1) - \Hc(\uv|Y) + \inf_{(c_1,c_2) \neq (0,0)} \Hc_{[c_1 \ c_2]}(\uv|Y)   \IEEEeqnarraynumspace\IEEEyesnumber\label{LMAC_rate_bound_4_simplified_swapped} \\
	R_2
	&<& I(U_2;Y,U_1).   \IEEEeqnarraynumspace\IEEEyesnumber\label{LMAC_rate_bound_5_simplified_swapped}
\end{IEEEeqnarray*}
\end{subequations}

	\item	The off-diagonals are $b_{1,2} \neq 0$ and $b_{2,1} \neq 0$:

	\begin{itemize}
		\item   Case $M = (\{1,2\},\{1\})$: the index set $\Sc$ is equal to $\Sc = \{2\}$, and $\Tc$ is iterated over $\Tc \in \{\{1\},\{2\}\}$, hence we obtain the pair of rate bounds
			\begin{subequations}
			\begin{IEEEeqnarray*}{rCl}
				R_1
				&<& \Hc(U_1) - \Hc(\uv|Y) + \Hc_{\bs_1^\T}(\uv|Y)   \IEEEeqnarraynumspace\IEEEyesnumber\label{LMAC_rate_bound_2_bis} \\
				R_2
				&<& \Hc(U_2) - \Hc(\uv|Y) + \Hc_{\bs_1^\T}(\uv|Y).   \IEEEeqnarraynumspace\IEEEyesnumber\label{LMAC_rate_bound_3_bis}
			\end{IEEEeqnarray*}
			\end{subequations}
		\item   Case $M = (\{1,2\},\{2\})$: the index set $\Sc$ is equal to $\Sc = \{1\}$, and $\Tc$ is iterated over $\Tc \in \{\{1\},\{2\}\}$, hence we obtain the pair of rate bounds
			\begin{subequations}
			\begin{IEEEeqnarray*}{rCl}
				R_1
				&<& \Hc(U_1) - \Hc(\uv|Y) + \Hc_{\bs_2^\T}(\uv|Y)   \IEEEeqnarraynumspace\IEEEyesnumber\label{LMAC_rate_bound_4_bis} \\
				R_2
				&<& \Hc(U_2) - \Hc(\uv|Y) + \Hc_{\bs_2^\T}(\uv|Y).   \IEEEeqnarraynumspace\IEEEyesnumber\label{LMAC_rate_bound_5_bis}
			\end{IEEEeqnarray*}
			\end{subequations}
		\item   Case $M = (\{1,2\},\{1\},\{2\})$: the index set $\Sc$ is iterated over $\Sc \in \{\{1\},\{2\}\}$:
		\begin{itemize}
			\item	For $\Sc = \{1\}$, the index set $\Tc$ is iterated over $\Tc \in \{\{1\},\{2\}\}$, hence we obtain the pair of inequalities
				\begin{subequations}
				\begin{IEEEeqnarray*}{rCl}
					R_1
					&<& \Hc(U_1) - \Hc(\uv|Y) + \inf_{c_1 \neq 0, c_2 \neq 0} \Hc_{\Cs\Bs}(\uv|Y)   \IEEEeqnarraynumspace\IEEEyesnumber\label{LMAC_rate_bound_6_bis} \\
					R_2
					&<& \Hc(U_2) - \Hc(\uv|Y) + \inf_{c_1 \neq 0, c_2 \neq 0} \Hc_{\Cs\Bs}(\uv|Y).   \IEEEeqnarraynumspace\IEEEyesnumber\label{LMAC_rate_bound_7_bis}
				\end{IEEEeqnarray*}
				\end{subequations}
			\item	For $\Sc = \{2\}$, the index set $\Tc$ is also iterated over $\Tc \in \{\{1\},\{2\}\}$ and we obtain the same inequalities as~\eqref{LMAC_rate_bound_6_bis}--\eqref{LMAC_rate_bound_7_bis}.
		\end{itemize}
	\end{itemize}
	The above three case distinctions on the matroid $M$ are to be combined by a logical `and' (due to the intersection over matroids), leading to
	\begin{subequations}
	\begin{IEEEeqnarray}{rCl}
		R_1
		&<& \Hc(U_1) - \Hc(\uv|Y) + \inf_{(c_1,c_2) \neq (0,0)} \Hc_{[c_1 \ c_2]}(\uv|Y)   \IEEEeqnarraynumspace\label{LMAC_rate_bound_8_bis} \\
		R_2
		&<& \Hc(U_2) - \Hc(\uv|Y) + \inf_{(c_1,c_2) \neq (0,0)} \Hc_{[c_1 \ c_2]}(\uv|Y).   \IEEEeqnarraynumspace\label{LMAC_rate_bound_9_bis}
	\end{IEEEeqnarray}
	\end{subequations}
\end{itemize}
Due to the union-taking over matrices $\Bs$, the above four case distinctions on the off-diagonals of $\Bs$ have to be combined by a logical `or'. Namely, the four groups of inequalities~\eqref{LMAC_rate_bound_5_0}--\eqref{LMAC_rate_bound_7_0}, \eqref{LMAC_rate_bound_4_simplified}--\eqref{LMAC_rate_bound_5_simplified}, \eqref{LMAC_rate_bound_5_simplified_swapped}--\eqref{LMAC_rate_bound_5_simplified_swapped} and \eqref{LMAC_rate_bound_8_bis}--\eqref{LMAC_rate_bound_9_bis} have to be combined by logical `or'. As we shall see, the group of inequalities~\eqref{LMAC_rate_bound_5_0}--\eqref{LMAC_rate_bound_7_0} (corresponding to a diagonal matrix $\Bs$) are implied by either of the latter three, that is, by either~\eqref{LMAC_rate_bound_4_simplified}--\eqref{LMAC_rate_bound_5_simplified}, \eqref{LMAC_rate_bound_5_simplified_swapped}--\eqref{LMAC_rate_bound_5_simplified_swapped} or \eqref{LMAC_rate_bound_8_bis}--\eqref{LMAC_rate_bound_9_bis}. To confirm this, we first argue that the two side bounds~\eqref{LMAC_rate_bound_5_0} and \eqref{LMAC_rate_bound_6_0} are implied by any of the aforementioned inequality pairs. This is because, for any $(c_1,c_2)$, we have
\begin{subequations}
\begin{IEEEeqnarray}{rCl}
	\Hc(U_1) - \Hc(\uv|Y) + \inf_{(c_1,c_2) \neq (0,0)} \Hc_{[c_1 \ c_2]}(\uv|Y)
	&\leq& I(U_1;Y,U_2)   \label{LMAC_rate_bound_comparison_1} \\
	\Hc(U_2) - \Hc(\uv|Y) + \inf_{(c_1,c_2) \neq (0,0)} \Hc_{[c_1 \ c_2]}(\uv|Y)
	&\leq& I(U_2;Y,U_1).   \label{LMAC_rate_bound_comparison_2}
\end{IEEEeqnarray}
\end{subequations}
These two bounds are obtained when upper-bounding the infimum terms on the left-hand sides of~\eqref{LMAC_rate_bound_comparison_1} and \eqref{LMAC_rate_bound_comparison_2} by setting $c_1=0$ and $c_2 \neq 0$ in the former, and $c_1 \neq 0$ and $c_2=0$ in the latter, respectively. As regards the third and last inequality~\eqref{LMAC_rate_bound_7_0}, clearly it is implied by either of the inequalities~\eqref{LMAC_rate_bound_5_simplified}, \eqref{LMAC_rate_bound_5_simplified_swapped}, \eqref{LMAC_rate_bound_8_bis} or \eqref{LMAC_rate_bound_9_bis}, because $\inf_{(c_1,c_2) \neq (0,0)} \Hc_{[c_1 \ c_2]}(\uv|Y) \leq \inf_{c_1 \neq 0,\, c_2 \neq 0} \Hc_{[c_1 \ c_2]}(\uv|Y)$. We conclude that in the union over $\Bs$, it suffices to consider diagonal matrices $\Bs$, which in turn can be reduced to the identity $\Bs=\Is$. Combining the rate inequalities~\eqref{LMAC_rate_bound_5_0}--\eqref{LMAC_rate_bound_7_0} with the sum rate bound~\eqref{LMAC_sum_rate_bound}, we finally obtain the rate region
\begin{subequations}
\begin{IEEEeqnarray}{rCl}
	R_1 + R_2
	&<& I(U_1,U_2;Y)   \IEEEeqnarraynumspace\label{LMAC_sum_rate_bound_final} \\
	R_1
	&<& I(U_1;Y,U_2)   \IEEEeqnarraynumspace\label{LMAC_rate_bound_5_final} \\
	R_2
	&<& I(U_2;Y,U_1)   \IEEEeqnarraynumspace\label{LMAC_rate_bound_6_final} \\
	\min\{ R_1 - \Hc(U_1) , R_2 - \Hc(U_2) \}
	&<& - \Hc(\uv|Y) + \inf_{c_1 \neq 0,\, c_2 \neq 0} \Hc_{[ c_1 \ c_2 ]}(\uv|Y).   \IEEEeqnarraynumspace\label{LMAC_rate_bound_7_final}
\end{IEEEeqnarray}
\end{subequations}

}

{
\section{Evaluation of $\mathscr{R}(\As)$ for $(K,L)=(2,1)$}
\label{app:proof:two_user_one_equation}
We evaluate $\mathscr{R}(\As) = \bigcup_{\Bs} \mathscr{Q}(\Bs)$ for the special case $K=2$, $L=1$. The coefficient matrix $\As = \as^\T = [a_1 \ a_2] \in \mathbb{A}^{1 \times 2}$ with $a_1 \neq 0$ and $a_2 \neq 0$ is a one-by-two row vector with non-zero entries. Since in the union operation in~\eqref{def:R}, the matrix $\Bs$ is iterated over all matrices satisfying $\Lambda_\mathbb{A}(\Bs) \supset \Lambda_\mathbb{A}(\As)$, we infer that $\Bs$ must run over all full-rank matrices $\Bs \in \mathbb{A}^{2 \times 2}$ as well as over all those $\Bs \in \mathbb{A}^{1 \times 2}$ that are scalar multiples of $\as^\T$ (for which case it suffices to consider $\Bs=\as^\T$).
We make a case distinction based on the rank of $\Bs$.

\subsubsection{The matrix $\Bs$ has rank $1$}

In this case, $\Bs$ is a non-zero multiple of the coefficient vector $\as^\T$. Due to invariance of $\mathscr{Q}(\Bs)$ against scalar multiplication of $\Bs$, if will suffice to set $\Bs = \as^\T$. By virtue of the constraints on $(M, \Sc, \Tc)$ laid out under Equation~\eqref{Q_partial}, we have that
\begin{itemize}
	\item	the matroid $M$ is the empty matroid $(\{1,2\},\emptyset)$;
	\item	the index set $\Sc$ is the singleton set $\{1\}$;
	\item	the index set $\Tc$ is iterated over $\{\{1\},\{2\}\}$.
\end{itemize}
The resulting rate region is thus
\begin{IEEEeqnarray*}{rCl}
	\mathscr{Q}(\as)
	\triangleq \Bigl\{ (R_1, R_2) \in \mathbb{R}_+^2 \colon
		R_1 &<& \Hc(U_1) - \Hc_{\as^\T}(\uv|Y) \\
		R_2 &<& \Hc(U_2) - \Hc_{\as^\T}(\uv|Y) \Bigr\}.   \IEEEeqnarraynumspace\IEEEyesnumber\label{two_user_CF_region}
\end{IEEEeqnarray*}

\subsubsection{The matrix $\Bs$ has rank $2$}

For this case, the union $\bigcup_{\Bs} \mathscr{Q}(\Bs)$ was already computed in Appendix~\ref{app:proof:two_user_two_equation} and shown to be equal to $\mathscr{Q}(\Is)$, which is described by inequalities~\eqref{LMAC_sum_rate_bound_final}--\eqref{LMAC_rate_bound_7_final}.

}

{
\section{Proof of Lemma~\ref{lem:CF_simplification}}
\label{app:proof:CF_simplification}
As a first step, we show that if the inequality~\eqref{tip_outside_condition_bis} holds, then $\as^\T$ is also the \emph{only} vector (with two non-zero entries) satisfying said inequality. As a consequence, $\as^\T$ is then necessarily the \emph{unique} minimizer that achieves the infimum in~\eqref{non-empty_notch_condition_2}--\eqref{non-empty_notch_condition_3}.

To prove this, let $\hat{\as}^\T = [ \hat{a}_1 \ \hat{a}_2 ] \in \mathbb{A}^2 $ denote some other coefficient vector with non-zero entries, that is linearly independent of $\as^\T = [ a_1 \ a_2 ]$. Then we have
\begin{IEEEeqnarray*}{rCl}
    d(\uv|Y)
    &=& \lim_{\nu \to \infty} \frac{H(\lfloor\nu\uv\rfloor|Y)}{\log(\nu)} \\
    &=& \lim_{\nu \to \infty} \biggl\{ \frac{H(a_1 \lfloor\nu U_1\rfloor + a_2 \lfloor\nu U_2\rfloor|Y)}{\log(\nu)} + \frac{H(\hat{a}_1 \lfloor\nu U_1\rfloor + \hat{a}_2 \lfloor\nu U_2\rfloor | Y, a_1 \lfloor\nu U_1\rfloor + a_2 \lfloor\nu U_2\rfloor)}{\log(\nu)} \biggr\} \\
    &=& d_{\as^\T}(\uv|Y) + d_{\hat{\as}^\T}(\uv|Y) - \lim_{\nu \to \infty} \frac{I(\hat{a}_1 \lfloor\nu U_1\rfloor + \hat{a}_2 \lfloor\nu U_2\rfloor ; a_1 \lfloor\nu U_1\rfloor + a_2 \lfloor\nu U_2\rfloor | Y)}{\log(\nu)} \\
    &=& d_{\as^\T}(\uv|Y) + d_{\hat{\as}^\T}(\uv|Y).   \IEEEeqnarraynumspace\IEEEyesnumber
\end{IEEEeqnarray*}
To prove the last equality, we argue that the mutual information $I(\hat{a}_1 \lfloor\nu U_1\rfloor + \hat{a}_2 \lfloor\nu U_2\rfloor ; a_1 \lfloor\nu U_1\rfloor + a_2 \lfloor\nu U_2\rfloor | Y)$ has a finite upper bound that is independent of $\nu$. In fact, by Lemma~\ref{lem:quantized_entropy_difference} and the data-processing inequality,
\begin{IEEEeqnarray*}{rCl}
	\IEEEeqnarraymulticol{3}{l}{
		I(\hat{a}_1 \lfloor\nu U_1\rfloor + \hat{a}_2 \lfloor\nu U_2\rfloor ; a_1 \lfloor\nu U_1\rfloor + a_2 \lfloor\nu U_2\rfloor | Y)
	} \\ \qquad
	&\leq& 2\log(\lVert \hat{\as} \rVert_1) + I(\lfloor \hat{a}_1 \nu U_1 + \hat{a}_2 \nu U_2 \rfloor ; \lfloor a_1 \nu U_1 + a_2 \nu U_2 \rfloor | Y) \\
	&\leq& 2\log(\lVert \hat{\as} \rVert_1) + I(\hat{a}_1 U_1 + \hat{a}_2 U_2 ; a_1 U_1 + a_2 U_2 | Y).  \IEEEeqnarraynumspace\IEEEyesnumber\label{rotated_coordinates_MI}
\end{IEEEeqnarray*}
It now suffices to prove the finiteness of the mutual information term on the right-hand side of~\eqref{rotated_coordinates_MI}.
We make a case distinction depending on the setting:
\paragraph{Theorems~\ref{thm:discrete_CF} and \ref{thm:integer_CF}}
In these cases, $(U_1,U_2)$ is discrete and the mutual information term can be upper-bounded by discrete entropies as
\begin{IEEEeqnarray*}{rCl}
	I(\hat{a}_1 U_1 + \hat{a}_2 U_2 ; a_1 U_1 + a_2 U_2 | Y)
	&\leq& H(\hat{a}_1 U_1 + \hat{a}_2 U_2 | Y) \\
	&\leq& H(U_1) + H(U_2)
\end{IEEEeqnarray*}
which is finite by assumption. 
\paragraph{Theorem~\ref{thm:continuous_CF}}
In this case, $(U_1,U_2) \in \mathbb{R}^2$ is absolutely continuous and $I(U_1,U_2;Y)$ is assumed finite. The mutual information term can be expressed as
\begin{IEEEeqnarray*}{rCl}
	\IEEEeqnarraymulticol{3}{l}{
		I(\hat{a}_1 U_1 + \hat{a}_2 U_2 ; a_1 U_1 + a_2 U_2 | Y)
	} \qquad \\
	&=& h(\hat{a}_1 U_1 + \hat{a}_2 U_2 | Y) + h(a_1 U_1 + a_2 U_2 | Y) - h(\hat{a}_1 U_1 + \hat{a}_2 U_2 , a_1 U_1 + a_2 U_2 | Y) \\
	&=& h(\hat{a}_1 U_1 + \hat{a}_2 U_2 | Y) + h(a_1 U_1 + a_2 U_2 | Y) - h(U_1,U_2|Y) - \log\bigl| \det\bigl( \bigl[ \hat{\as} \ \as \bigr] \bigr) \bigr|.   \IEEEeqnarraynumspace\IEEEyesnumber
\end{IEEEeqnarray*}
Given that $H(\lfloor \uv \rfloor |Y)$, $h(U_1|Y)$, $h(U_2|Y)$ and $h(U_1,U_2|Y)$ are finite, by Lemma~\ref{lem:entropy_finiteness}, the first two differential entropy terms $h(\hat{a}_1 U_1 + \hat{a}_2 U_2 | Y)$ and $h(a_1 U_1 + a_2 U_2 | Y)$ are finite too.
This closes the previous argument. We conclude that in all cases,
\begin{equation*}
    d(\uv|Y)
    = d_{\as}(\uv|Y) + d_{\hat{\as}}(\uv|Y).
\end{equation*}
Hence, by the data-processing inequality,
\begin{IEEEeqnarray*}{rCl}
    \Hc(\uv|Y)
    &=& \liminf_{\nu \to \infty} \Bigl\{ H(\lfloor\nu\uv\rfloor|Y) - d(\uv|Y) \log(\nu) \Bigr\} \\
    &=& \liminf_{\nu \to \infty} \Bigl\{ H(a_1 \lfloor\nu U_1\rfloor + a_2 \lfloor\nu U_2\rfloor|Y) + H(\hat{a}_1 \lfloor\nu U_1\rfloor + \hat{a}_2 \lfloor\nu U_2\rfloor|Y, a_1 \lfloor\nu U_1\rfloor + a_2 \lfloor\nu U_2\rfloor) \\
    \IEEEeqnarraymulticol{3}{r}{
        {} - d_{\as^\T}(\uv|Y)\log(\nu) - d_{\hat{\as}^\T}(\uv|Y)\log(\nu) \Bigr\}
    } \\
    &\leq& \liminf_{\nu \to \infty} \Bigl\{ H(a_1 \lfloor\nu U_1\rfloor + a_2 \lfloor\nu U_2\rfloor|Y) + H(\hat{a}_1 \lfloor\nu U_1\rfloor + \hat{a}_2 \lfloor\nu U_2\rfloor|Y) \\
    \IEEEeqnarraymulticol{3}{r}{
        {} - d_{\as^\T}(\uv|Y)\log(\nu) - d_{\hat{\as}^\T}(\uv|Y)\log(\nu) \Bigr\}
    } \\
    &=& \Hc_{\as^\T}(\uv|Y) + \Hc_{\hat{\as}^\T}(\uv|Y).   \IEEEeqnarraynumspace\IEEEyesnumber\label{information_preservation}
\end{IEEEeqnarray*}
As a consequence,
\begin{subequations}
\begin{IEEEeqnarray}{rCl}
    \Hc(U_1) - \Hc_{\hat{\as}^\T}(\uv|Y)
    &\leq& \Hc(U_1) - \Hc(\uv|Y) + \Hc_{\as^\T}(\uv|Y)   \label{below_sum_rate_1} \\
    \Hc(U_2) - \Hc_{\hat{\as}^\T}(\uv|Y)
    &\leq& \Hc(U_2) - \Hc(\uv|Y) + \Hc_{\as^\T}(\uv|Y).   \label{below_sum_rate_2}
\end{IEEEeqnarray}
\end{subequations}
The sum of the right-hand sides of this pair of inequalities equals
\begin{equation*}
    I(\uv;Y) + 2\Hc_{\as^\T}(\uv|Y) - \Hc(\uv|Y)
\end{equation*}
which due to~\eqref{tip_outside_condition} is upper-bounded by the sum rate $I(\uv;Y)$.
Hence $\mathscr{Q}(\hat{\as}^\T)$, which is delimited by the left-hand side quantities of~\eqref{below_sum_rate_1}--\eqref{below_sum_rate_2}, is contained in $\mathscr{R}_\MAC$.
Therefore, $\as^\T$ is the only coefficient vector such that $\mathscr{Q}(\as^\T)$ is \emph{not} contained in $\mathscr{R}_\MAC$, while for any other (linearly independent) vector $\hat{\as}^\T$, we have $\mathscr{Q}(\hat{\as}^\T) \subseteq \mathscr{R}_\MAC$. Consequently, if the inequality~\eqref{tip_outside_condition_bis} holds, then $\as^\T$ is the \emph{unique} maximizer of the set $\mathscr{Q}(\as^\T)$, i.e., the unique minimizer of $\cs^\T \mapsto \Hc_{\cs^\T}(\uv|Y)$ over vectors $\cs^\T$ with non-zero entries [cf.~\eqref{CF_eq1}--\eqref{CF_eq2}]. That is,
\begin{equation}   \label{unique_minimizer_of_entropy}
	\as^\T
	= \argmin_{c_1 \neq 0,\, c_2 \neq 0} \Hc_{[ c_1 \ c_2 ]}(\uv|Y).
\end{equation}
Since, as argued in~\eqref{LMAC_as_set_difference}, \eqref{non-empty_notch_condition_1}--\eqref{non-empty_notch_condition_3} and Lemma~\ref{lem:notch}, the condition~\eqref{tip_outside_condition} entails a non-empty set $\mathscr{R}_\MAC \cap \overline{\mathscr{R}'}$, i.e., the set described by the inequalities~\eqref{non-empty_notch_condition_1}--\eqref{non-empty_notch_condition_3}. By substracting~\eqref{non-empty_notch_condition_2} from~\eqref{non-empty_notch_condition_3}, and \eqref{non-empty_notch_condition_3} from~\eqref{non-empty_notch_condition_1}, it becomes clear that any element from $\mathscr{R}_\MAC \cap \overline{\mathscr{R}'}$ will satisfy
\begin{subequations}
\begin{IEEEeqnarray}{rCl}
	R_1
	&<& \Hc(U_1) - \Hc_{\as^\T}(\uv|Y) \\
	R_2
	&<& \Hc(U_2) - \Hc_{\as^\T}(\uv|Y)
\end{IEEEeqnarray}
\end{subequations}
and thus, we have $\mathscr{R}_\MAC \cap \overline{\mathscr{R}'} \subseteq \mathscr{Q}(\as^\T)$. As a consequence [cf.~\eqref{eq:two_user_one_equation_general_coefficients}, \eqref{LMAC_as_set_difference}],
\begin{IEEEeqnarray*}{rCl}
	\mathscr{R}(\as^\T)
	&=& \mathscr{R}_\LMAC \cup \mathscr{Q}(\as^\T) \\
	&=& \mathscr{R}_\MAC \setminus ( \mathscr{R}_\MAC \cap \overline{\mathscr{R}'} ) \cup \mathscr{Q}(\as^\T) \\
	&=& \mathscr{R}_\MAC \cup \mathscr{Q}(\as^\T).    \IEEEyesnumber\IEEEeqnarraynumspace
\end{IEEEeqnarray*}
This proves~\eqref{eq:two_user_one_equation_simplified_1} and thus the first part of Lemma~\ref{lem:CF_simplification}.

By contrast, if $\as^\T$ satisfies the reverse of the inequality~\eqref{tip_outside_condition_bis}, then we distinguish two cases:
\paragraph{Case $\mathscr{R}_\LMAC \subsetneq \mathscr{R}_\MAC$}
By Lemma~\ref{lem:notch}, there exists a vector $\cs^\T$  such that $2 \Hc_{\cs^\T}(\uv|Y) < \Hc(\uv|Y)$. As argued before [cf.~\eqref{unique_minimizer_of_entropy}], this coefficient vector is the unique minimizer of $\tilde{\cs}^\T \mapsto \Hc_{\tilde{\cs}^\T}(\uv|Y)$, hence
\begin{equation}   \label{unique_entropy_minimum}
	\Hc_{\cs^\T}(\uv|Y)
	= \min_{\tilde{c}_1 \neq 0,\, \tilde{c}_2 \neq 0} \Hc_{[ \tilde{c}_1 \ \tilde{c}_2 ]}(\uv|Y).
\end{equation}
 Yet since we have assumed, on the other hand, that $2 \Hc_{\as^\T}(\uv|Y) \geq \Hc(\uv|Y)$, it follows that $\as$ and $\cs$ are distinct, linearly independent vectors. Re-using the bounding steps performed in~\eqref{information_preservation} for a pair of linearly independent coefficient vectors, we get
\begin{IEEEeqnarray*}{rCl}
    \Hc(\uv|Y)
    &\leq& \Hc_{\as^\T}(\uv|Y) + \Hc_{\cs^\T}(\uv|Y).
\end{IEEEeqnarray*}
Similarly as in~\eqref{below_sum_rate_1}--\eqref{below_sum_rate_2},
the latter allows us to write down the following two inequalities:
\begin{subequations}
\begin{IEEEeqnarray}{rCl}
    \Hc(U_1) - \Hc_{\as^\T}(\uv|Y)
    &\leq& \Hc(U_1) - \Hc(\uv|Y) + \Hc_{\cs^\T}(\uv|Y)   \label{below_sum_rate_1_bis} \\
    \Hc(U_2) - \Hc_{\as^\T}(\uv|Y)
    &\leq& \Hc(U_2) - \Hc(\uv|Y) + \Hc_{\cs^\T}(\uv|Y).   \label{below_sum_rate_2_bis}
\end{IEEEeqnarray}
\end{subequations}
in which $\Hc_{\cs^\T}(\uv|Y)$ may be written as the minimum~\eqref{unique_entropy_minimum}.
Hence the set $\mathscr{Q}(\as^\T)$, which is delimited by the quantities on the left-hand sides of~\eqref{below_sum_rate_1_bis}--\eqref{below_sum_rate_2_bis}, is contained in $\mathscr{R}'$ [cf.~\eqref{def:R_prime}]. Given that $2 \Hc_{\as^\T}(\uv|Y) \geq \Hc(\uv|Y)$, we also have that $\mathscr{Q}(\as^\T) \subset \mathscr{R}_\MAC$. Taken together, we have $\mathscr{Q}(\as^\T) \subseteq \mathscr{R}_\MAC \cap \mathscr{R}' = \mathscr{R}_\LMAC$. Therefore,
\begin{IEEEeqnarray*}{rCl}
	\mathscr{R}(\as^\T)
	&=& \mathscr{R}_\LMAC \cup \mathscr{Q}(\as^\T) \\
	&=& \mathscr{R}_\LMAC.   \IEEEeqnarraynumspace\IEEEyesnumber
\end{IEEEeqnarray*}
\paragraph{Case $\mathscr{R}_\LMAC = \mathscr{R}_\MAC$}
By Lemma~\ref{lem:notch}, there exists no vector $\cs^\T$ such that $2 \Hc_{\cs^\T}(\uv|Y) < \Hc(\uv|Y)$. Therefore, all sets $\mathscr{Q}(\cs^\T)$ (for all $\cs^\T$ with non-zero entries) are subsets of $\mathscr{R}_\LMAC = \mathscr{R}_\MAC$. As in the previous case, it follows that $\mathscr{R}(\as^\T) = \mathscr{R}_\LMAC$.

This proves~\eqref{eq:two_user_one_equation_simplified_2} and thus concludes the proof of Lemma~\ref{lem:CF_simplification}.

}

\bibliographystyle{IEEEtran}
\bibliography{IEEEfull,references,nit}

\end{document}